\documentclass{elsart1p}

\usepackage{amsmath}
\usepackage[usenames]{color}
\usepackage{amssymb}
\usepackage{epsfig}
\usepackage{sidecap}
\usepackage{ulem}

\newcommand{\PGcomm}[1]{{\color{black} #1}}
\newcommand{\PGmod}[1]{{\color{black} #1}}

\newcommand{\EScomm}[1]{{\color{black} #1}}
\newcommand{\esmod}[1]{{\color{black} #1}}

\newcommand{\MDcomm}[1]{{\color{black} #1}}

\renewcommand{\vec}[1]{{\mbox{\boldmath $#1$}}}
\begin{document}

\begin{frontmatter}
\title{Dynamics of clusters and molecules\\in contact with an environment}
\author{P.~M.~Dinh\corauthref{cor}$^a$}
\author{, P.-G.~Reinhard$^b$, and E.~Suraud$^a$}

\corauth[cor]{Corresponding author\\{\it Email-address}~:
  dinh@irsamc.ups-tlse.fr} 
\address{$^a$Laboratoire de Physique Th\'eorique, Universit\'e Paul
  Sabatier, CNRS\\
  118 route de Narbonne F-31062 Toulouse C\'edex, France}
\address{$^b$Institut f{\"u}r Theoretische Physik, Universit{\"a}t
  Erlangen,\\
  Staudtstrasse 7 D-91058 Erlangen, Germany}

\begin{abstract}
We present recent theoretical investigations on the dynamics of metal
clusters in contact with \PGmod{an environment}, deposited or embedded. This
concerns soft deposition as well as  irradiation of the deposited/embedded
clusters by intense laser pulses.
The description of these complex and demanding compounds employs a
hierarchical model in \PGmod{an extension} of a
Quantum-Mechanical/Molecular-Mechanical (QM/MM) approach where the
cluster electrons are described by Time-Dependent Density-Functional
Theory (TDDFT) and the constituents of the more inert \esmod{environment} by
classical equations of motion. Key ingredients are the polarization
potentials where, in particular, our QM/MM implementation takes care
to include the full dynamical polarizability of the substrate. This is
crucial for an appropriate modeling of dynamical scenarios.
We discuss the observables accessible in that model, from
quantum-mechanical cluster electrons, from classical cluster ions and
from the degrees of freedom of the environment (positions, dipole
polarizabilities).

We discuss examples of applications for two typical test cases, Na
clusters deposited on MgO(001) surface and Na clusters in/on Ar
substrate. Both \PGmod{environments} are insulators with sizeable
polarizability. They differ in their geometrical and mechanical
properties.
We first survey the low-energy properties of these compounds,
structure and optical response. We work out the impact of surface
corrugation and of polarizability. We analyze the difference
between deposited and embedded clusters.

The second part discusses the dynamics of soft deposition processes,
for Na clusters impinging on Ar(001) or MgO(001) surfaces. We analyze
charge and size effects, and details of energy transfer to the
environment. We show how the deposition process can create "hot spots"
in the surface where sizeable amounts of energy are stored in internal
excitations of the substrate atoms.

At last, we consider laser irradiation of embedded/deposited Na
clusters.  These systems serve as generic test cases for chromophore
effects.  We discuss a broad range of scenarios,
from "gentle" to "strong" irradiation processes.  Key effect is the
ionization through the laser pulse.  We analyze the effect of the
substrate on angular distributions of emitted electrons and the effect
of ionization on substrate and interface interaction. The case of
strong excitations shows a dramatic change of cluster dynamics due to
the environment, in particular hindered (or delayed) Coulomb
explosion.
\end{abstract}

\begin{keyword}
Time-Dependent Density Functional Theory 
\sep 
Hierarchical method 
\sep 
QM/MM 
\sep 
deposited/embedded metal clusters 
\sep
dynamical polarization potentials
\sep
Na metal cluster
\sep
Ar(001) surface
\sep
MgO(001) surface
\sep
Ar matrix
\sep
cluster structure
\sep
surface corrugation
\sep
optical absorption 
\sep 
soft deposition process
\sep
photo-electron angular distributions 
\sep 
laser induced dynamics 
\sep
Coulomb explosion

\PACS
 34.10.+x\sep 34.35.+a\sep 34.50.-s\sep 34.50.Gb\sep 36.40.-c\sep 61.46.Bc
\end{keyword}
\end{frontmatter}

\tableofcontents
\clearpage


\section{Introduction}
\label{sec:intro}


This review deals with the dynamics of metal clusters in contact with
inert environments, either deposited on a surface or embedded inside
\esmod{a medium}. The study of clusters is a rather recent branch of
physics, developing with the steadily improving preparation methods
and laser analysis.
The case of free clusters has been extensively studied in the past and
there exists a broad literature on that topic, for books and reviews
see
\cite{Kre93,Bra93,Hee93,Hab94a,Sug98,Eka99,Bjo99,Jel99,Rei03a,Alo06aB}.
The progress of the field is also well documented in the impressive
series of ISSPIC proceedings
\cite{Iss1,Iss2,Iss3,Iss4,Iss5,Iss6,Iss7,Iss8,Iss9,Iss10,Iss11,Iss12,Iss13}.
Metal clusters play a special role in that field because of their
remarkable electronic shell structure \cite{Bra93,Hee93,Bjo99} and
pronounced Mie plasmon resonance which provides a well defined and
strong coupling to light \cite{Kre93,Eka99,Rei03a}.
\MDcomm{Bi- or trimetallic cluster design also constitute a subject of
great interest in material science, for a recent review,
see~\cite{Fer08}.} 

The case of clusters in contact with an \esmod{environment} is more involved and
covers an extremely large range of physical and chemical
situations. The field is still very much under development, see
e.g. the collections \cite{Hab94b,Mei00aB,Mei06aB,Mei07aB}.
One motivation to deal with these compounds is that many experiments
can better be performed for non-isolated systems. Substrates serve to
prepare well defined conditions of temperature and orientation, they
help to hold neutral clusters, and they allow to gather higher cluster
densities. Even the analysis of small molecules can take advantage of
handling in \PGmod{an environment} as experiments in He droplets show
\cite{Mil01aB,Sti01,Sti06,Kue07}.
A further important aspect is that contact with an environment is a
realistic scenario in applications. For example, there are promising
attempts to employ clusters in the dedicated shaping of nano-scaled
devices \cite{Mil99,Wen99a,Bin01,Fel02,Oua05a}, small Au clusters on
surfaces are found to be efficient catalysts \cite{San99}, metal
clusters are considered as nano-junctions in electrical circuitry
\cite{Jan00}, and the coupling to light is exploited in producing an
enhanced photo-current by depositing Au$_N$ on a semi-conductor
surface \cite{Sch05a}.
Metal clusters in inert substrate are also a simple model system for
chromophores where the field amplification effect has large impact on
the environment, see e.g. the study of localized melting for the
generic combination of Au clusters embedded in ice \cite{Ric06a}.
Such combinations can be used as a test system to understand the first
stages of radiation damage starting with defect formation in solids
\cite{Niv00,Bar02b,Fel96}.
Furthermore, there are promising applications in medicine where the
frequency selective optical coupling of organically coated metal
clusters attached to biological tissue may by used as tool for
diagnosis \cite{Bru98,May01,Dub02,Sim07a} and for stronger laser
fields for localized heating in therapy \cite{Khl06a}.

Last but not least, compounds of two different materials are a
research topic in its own right. 
It is interesting \esmod{and it may even be crucial} to watch
modifications within each species if the two come into contact.
\esmod{This is typically the case of biological molecules whose properties and behaviors are strongly
linked to the (often water) environment.}
Moreover, the mutual influence of the two species can create new
effects which were not possible for the isolated species.
\esmod{This aspect is {important} as it indirectly 
points out the key role of interactions between the species and its
environment and\PGmod{, correspondingly,} the importance of 
\PGmod{a proper description of}
such couplings. \PGmod{Thereby, it is essential for a proper description}
to account for the possible response of
both, species \PGmod{and} environment.}
But the combination of different materials and the typically large sizes
of the environments pose a very demanding problem for a theoretical
description. One needs to find a good compromise between
simplification and yet proper inclusion of the environment dynamical
response and relevant coupling mechanisms, see subsection
\ref{sec:intro-theo}.

Even with a simplified account for the environment, the effort remains
huge. As a first step, we shall focus on rather simple cases of
especially optically active metal clusters in contact with insulators,
taking as examples the Ar material and the MgO(001) surface.  Both
environments are insulators with a large band gap.  The metal cluster
serves as a chromophore which opens the road to a bunch of interesting
dynamical scenarios. Deposition processes are also offering surprising
scenarios which we shall equally discuss. We \PGmod{will thus discuss both
applications,  clusters embedded in a "matrix" and clusters
deposited on a "surface"}.

\subsection{Physical context}
\label{sec:phys-cont}

\begin{figure}[htbp]
\begin{center}
%
\epsfig{file=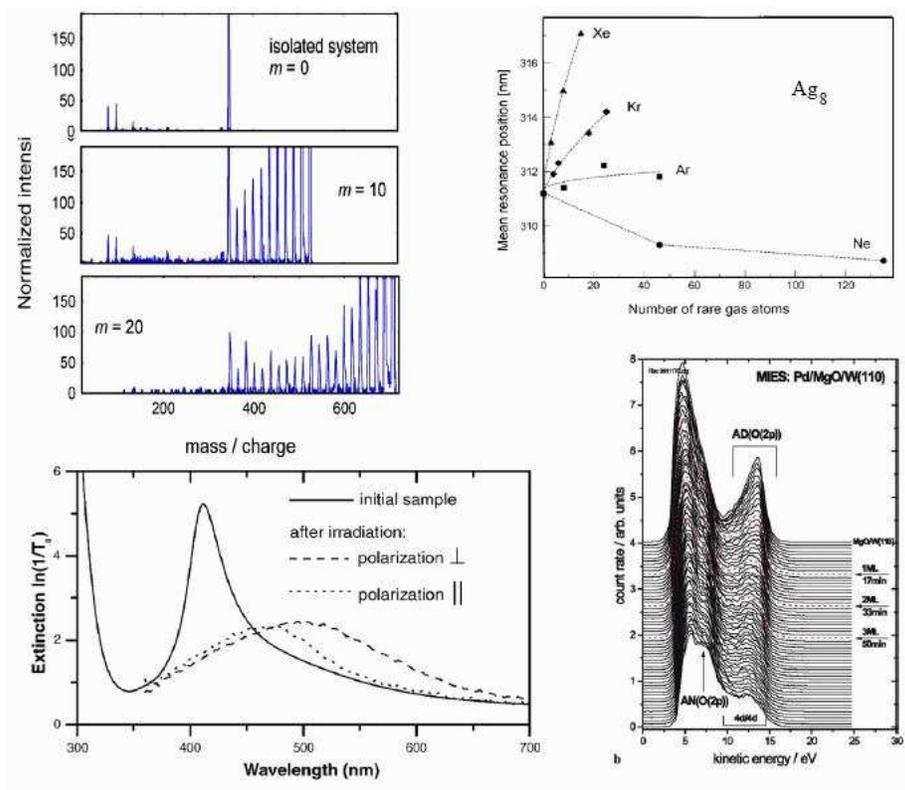,width=0.9\linewidth}
\caption{Typical examples of effect of environment in contact with
metal clusters.  \esmod{The upper left panel from \cite{Liu06}
is a an example of violent scenario of fragmentation of a biological
molecule coated by a finite number $m$ of water molecules; the
lower left panel from \cite{Sei00} also concerns a violent
excitation this time by a laser and for a silver cluster embedded in
a bulk glass matrix; the upper right panel from \cite{Die01} again
concerns Ag clusters but in the (linear) regime of optical response,
and coated by a finite number of rare gas atoms; the lower right panel
from \cite{Kri06} finally illustrates the impact of deposition on
substrate electronic properties.}   See text for details.
\label{fig:examples}}
\end{center}
\end{figure}

Fig.~\ref{fig:examples} provides a few illustrating
examples  of studies on mixed cluster-environment
systems.
Let us first focus on the upper right panel which displays the optical
response ("color") of a Ag$_8$ cluster embedded in a finite size rare
gas matrix \cite{Die01} . The evolution of the peak position with "matrix" size is
plotted. It provides an example of how the response of a given species
(here the Ag$_8$ cluster) is affected by its surroundings. The effect
is admittedly subtle (see the ordinate scale) but mind that the
"matrix" is composed of rare gas atoms, supposed to be extremely
inert. One can actually spot differences and qualitatively different
trends when considering \PGmod{different rare gases}, \EScomm{the different trends being 
closely related to the different rare gas polarizabilities}.
The lower right panel focuses the analysis, not on the
cluster, but on the surface itself. The case is deposit of Pd on an
insulating MgO surface and the energies of electronic levels of MgO
are recorded as a function of Pd coverage  \cite{Kri06}. The interesting point here
is that the MgO levels are significantly affected by the deposition
process, in spite of the fact that MgO is a well bound insulator. This
points out the fact that subtle interaction effects enter the picture
as soon as two materials are put in contact.
Both the optical response of the embedded Ag$_8$ cluster and the
photoelectron spectra of MgO concern low energy phenomena, close to
the ground state of the system. Experiments on (possibly violent)
dynamical scenarios have also been performed. We illustrate them in the
left column of Fig.~\ref{fig:examples}. The left bottom panel shows
again an optical response of an embedded cluster (Ag cluster inside
a bulk glass matrix) but this time, in relation to a violent laser
irradiation \cite{Sei00}. The optical responses prior and after irradiation with a
strong laser pulse are plotted. The spectra before and after
irradiation show significant differences, indicating that the
irradiation provoked a sizable shape variation of the embedded cluster
(see also the discussion at the beginning of section \ref{sec:light}).
Finally we consider in the upper \PGmod{left} panel an even more violent
scenario but in a somewhat different context. The system under study
is an adenosine monophosphate nucleotide molecule coated by a finite
number of water molecules \cite{Liu06}. Collisions with neutral atoms provoke the
fragmentation of the complex. The fragmentation spectrum, plotted as 
usual as a function of mass over charge ratio,
exhibits a sizable dependence on the number $m$ of coating
molecules.
The example thus demonstrates the intricate relation between system and "matrix"
in this example of biomolecular systems, even in the course of violent
disintegration.
This experiments takes care to control and vary systematically the
number of embedding water molecules.  The study of such model systems
eventually allows to decipher elementary
mechanisms responsible for DNA damages by irradiation.

\subsection{Dynamics of clusters in contact with an environment}

\subsubsection{Sizes and energies}
\label{sec:sizes}

\begin{figure}[htbp]
\centering\resizebox{1.0\columnwidth}{!}{\includegraphics{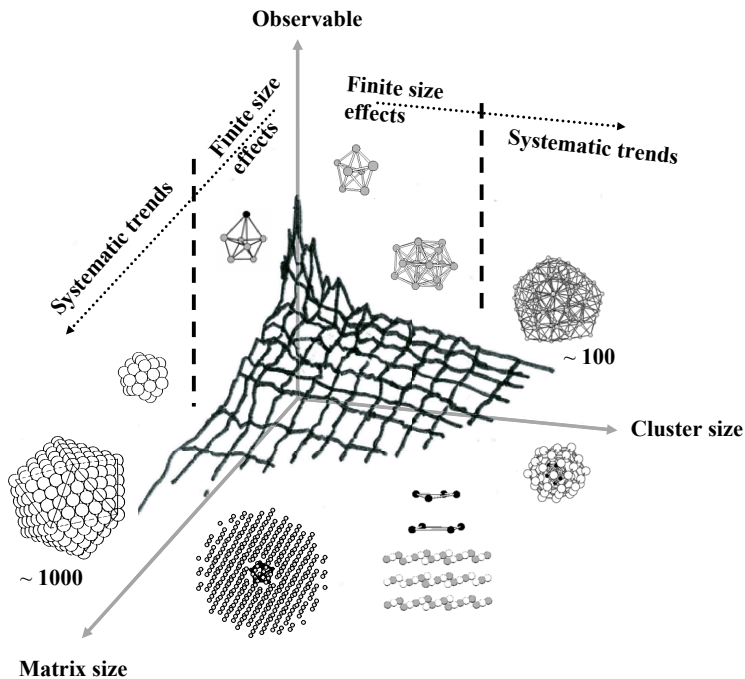}}
\caption{
Schematic panorama of clusters in contact with an environment, as
discussed in this paper. The two "horizontal" axes
correspond to cluster and environment, while the vertical axis
represents a typical observable such as the ionization
potential (IP). The observable  is
plotted as function of cluster ($n$) and environment ($N$) size and
exhibits large fluctuations for small values of $n$ and/or $N$. 
For larger values of $n$ and/or $N$,
there develops a more monotonous trend towards the asymptotic
values. A few typical structures
are, furthermore, indicated: a choice of pure Na or
Ar clusters along the two horizontal axes and of mixed systems  in
between (Na cluster
embedded in Ar matrix or deposited on MgO).
\label{fig:scales}}
\end{figure}
\esmod{As was seen in section \ref{sec:phys-cont},}
clusters in contact with an environment comprise a world of 
different situations and systems. It is thus important to clarify 
the situation by some classification and by considering limiting cases. 
The cluster-\PGmod{environment} compounds cover very different systems, from
small mixed molecules up to nanometer scale clusters deposited on an
infinite surface. 
It is well known that small systems usually exhibit specific size
effects which tend to level off for larger size. For clusters in an
environment, size effects appear twice, in terms of the cluster size
and in terms of the environment size. That is illustrated schematically
in Fig.~\ref{fig:scales} \esmod{in the case of simple metal clusters 
(\PGmod{Na}) 
in contact with insulator environments (MgO surface and Ar surface and/or matrix), which 
represent typical systems discussed in this paper.}
The transition between size-specific and generic behaviors
depends to a large extent on the considered observable.
{Size specific effects appear as fluctuations on the observed values 
of a given observable while the trend at larger sizes is more monotonous and 
exhibits a slow convergence towards the bulk value.}
This question has been addressed since long in the case of free
clusters, see e.g. \cite{McH89,Hee93,Bra93}. The impact of
environment size on cluster properties was also considered in a few
experiments as for example \cite{Die02}. 

The tour from small to large systems concerns also the way such
composite objects can be described theoretically.  Very small systems,
in practice mixed clusters, can be treated \PGmod{by} sophisticated quantum
chemistry methods, while bulk materials call for techniques from solid
state or surface science sacrificing some details.  Treating the mixed
system of a cluster in contact with an environment thus corresponds to
an intermediate situation in which one would like to combine
advantages of these two extremes: detailed treatment of the cluster
with a less detailed description of the environment. This
calls for hierarchical methods, see sections \ref{sec:intro-theo} and
\ref{sec:model}.

Fig.~\ref{fig:scales} does also show typical cluster-environment
configurations. Three free Na clusters are shown along the axis
``cluster size'', Na$_7^+$, Na$_{21}^+$ and Na$_{92}$.
The first insert along the axis ``Matrix size'' shows the very small
compound NaAr$_6$. The further figures along that axis represent a
medium size ($N=55$) and large Ar  cluster $N=561$. The plane
between the two axes shows the two typical test cases which we will
consider in the following, down left a sketch of Na$_8$Ar$_{434}$ as
an example for an embedded cluster and right of that Na$_8$@MgO(001) for
a deposited cluster, while a small mixed cluster Na$_8$Ar$_{42}$ is also 
indicated  close to the cluster axis.

\esmod{The mixing of two different systems, metal cluster with insulator
material, induces also a larger span in energy and length scales.  The
metal has strongly delocalized electrons with large mean free path and
small energy differences. On the other hand, the electrons in the
insulating materials remain tightly bound to atoms and involve large
electronic energies. This holds for rare gases as well as for the
ionic crystals in our sample (NaCl and MgO).  Thus  the
description of such mixed systems has to accommodate 
larger range of energies (with corresponding time scales) and
lengths, which complicates matters as compared to free clusters.}

\subsubsection{Time scales}
\label{sec:timescales}

\begin{figure}[htbp]
\centering\resizebox{1.0\columnwidth}{!}{\includegraphics{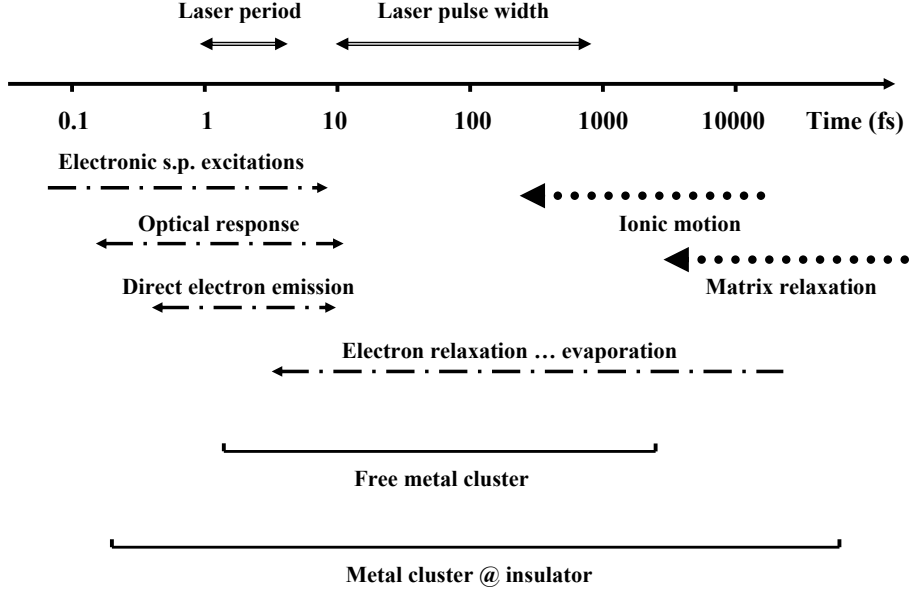}}
\caption{
Typical time scales for the dynamics of a cluster in an environment,
taking metal clusters, i.e. Na$_n$, as a prototype for the cluster and
Ar$_N$ as a prototype of environment. The uppermost block shows times
associated to fs lasers, the left middle block times related to
electronic motion including both cycle times and lifetimes due to
relaxation processes. The right middle block exhibits the corresponding
ionic (cluster) and atomic (environment) times.  All in all, as
schematically indicated at the bottom of the figure, while free
cluster dynamics covers about 5 orders of magnitude time scales,
dynamics of clusters in contact with an environment covers up to
typically 7 orders of magnitude, which makes their handling even more
demanding from the theoretical point of view in particular.
\label{fig:timescales}}
\end{figure}

As we are primarily interested in dynamical scenarios, we briefly
recall key time scales for the systems which we are considering in
this paper. These are sketched in Fig.~\ref{fig:timescales}, including
both ``intrinsic'' times of the system itself and ``external'' time
scales associated to the excitation process.
To have a specific example, we consider the system times associated to
Na clusters and for the excitation process wth an optical laser.  We
ignore the extremely short times associated with Na core electrons,
which will play no role in the following.

The pulse duration of optical lasers may be varied over a wide range
and extends in principle from fs to ps or even ns. We focus here on fs
laser with pulse widths of order a few tens to a few hundreds of fs.
The fastest cluster time scales concern the motion of the valence
electrons. As already mentioned, metal clusters act as excellent
chromophores. Coupling to light is predominantly mediated through the
Mie plasmon, whose period lies in the fs range. It corresponds to the
collective oscillations of the electron cloud with respect to the
ionic background, triggered by an external electric field. Other
single-particle excitations and direct electron escape,
i.e. single-particle excitation into the continuum, lie in the same
range, but with wider span from sub-fs to several fs.

The most widely varying times are related to electron relaxation due
to damping from electron-electron collisions and thermal electron
evaporation. Both strongly depend on the internal excitation of the
cluster which may be characterized by an average excitation energy
$E^*$. That might be expressed alternatively by an electronic
temperature $T$ which allows easier comparison between systems of
different size. The Fermi gas model then provides a simple connection
between temperature and internal excitation energy. For \mbox{$k_{\rm
B}T\ll\epsilon_{_\mathrm{F}}$}, $T$ can be estimated as \mbox{$k_{\rm
B}T=2(\epsilon_{_\mathrm{F}}E^*/N)^{1/2}/\pi$}, where
\mbox{$\epsilon_{_\mathrm{F}}=\hbar^2(9\pi^2/4)^{2/3}/(2m_er_s^2)$} is
the Fermi energy. In the case of Na at bulk density, this amounts to
$k_BT=(1.28\,\mathrm{eV}\,E^*)^{1/2}$.
Electronic thermalization is mostly mediated by electron-electron
collisions. Fermi liquid theory leads to a $T^{-2}$ law for the
corresponding collision time \cite{Pin66,Kad62}. At low temperatures,
most collisions are Pauli blocked and relaxation times become
comfortably long. Inclusion of electron-electron collisions in
dynamical scenarios is at present mostly done in semi-classical
treatments at the level of the Vlasov-\"Uhling-Uhlenbeck approach
\cite{Dom98b,Dom00c,Gig01a,Fen04}. A tractable quantum-mechanical
description of the collisions has yet to be developed.  Thus we do not
include electron-electron collisions in the studies presented here. It
ought to be kept in mind that they should be included for violent
processes in the future stages of the theory. 
The time scale for electron evaporation depends even more dramatically
on temperature (or excitation energy). It is given by the Weisskopf estimate
which predicts a trend dominated by the exponential factor
$\exp{({E_{\rm IP}}/(k_BT))}$, where $E_{\rm IP}$ denotes the value of
the ionization potential \cite{Wei37,Cal00}.
In general, electron evaporation represents a very 
efficient cooling mechanism for highly excited clusters
\cite{Sch01}.

Ionic motion runs at much slower time scales and spans a wide range of
times. Vibration frequencies typically lie in the meV regime. That
means they have cycle times of order 100\,fs to 1\,ps which can be
measured by Raman scattering (see, e.g., \cite{Por01a}). Strong
perturbations (laser, projectile) can lead to large amplitude ionic
motion and cluster explosion due to Coulomb pressure generated by
ionization and thermal excitation. The coupling between electrons and
ions proceeds at an electronic time scale, i.e. within a few fs. But
the effects on ions develops at much slower scale, typically around
100\,fs, because of the much heavier ion mass. 
Note that the time scale for explosion depends on the violence
of the process and becomes shorter with higher initial ionization.
Besides ionization effects, the further energy transfer from electrons
to ions takes much longer, up to ns range \cite{Feh06b}. Ionic
relaxation processes are even slower, e.g. thermal emission of a
monomer can easily last $\mu$s.

Relevant time scales at the side of the environment lie in the same
span.  Ionic/atomic times typically scale with the square root of
ion/atom masses and come again into the range of hundreds of fs to ps
for typical environments studied here. Moreover, it should be noted
that the slowest ionic/atomic vibration and relaxation times increase
with system size (more precisely $\propto N^{1/3}$). That size effect
has to be kept in mind when interpreting results from finite samples.
On the other hand, the electronic degrees-of-freedom of the
environment show shorter time scales than those in the cluster.  We
consider inert materials where the electrons remain rather tightly
bound to their parent atoms which produces the much shorter time
scales and which allows the simplified QM/MM treatment.
The coupling of the environment electronic degrees-of-freedom with
cluster electrons proceeds at the same short time scale.  The typical
values lie in the sub-fs range.  We thus need to resolve the dynamics
at this rather short time scales.

This quick survey shows that relevant dynamical scenarios comprise a
large span of time scales which is a great \PGmod{challenge} for the
theoretical description. Already in free clusters, ionic motion may
require a simulation time up to several ps while electronic motion has
to be resolved down to a fraction of fs. The effect \PGmod{becomes} even more
dramatic for embedded/deposited clusters. One aims to cover the
possibly very slow relaxation of the large environment while
accounting for its especially fast electronic response.  This means
that one is going to extend the typical range of time scales to be
accounted for by 2 orders of magnitude compared to the case of free
metal clusters, as illustrated in Fig.~\ref{fig:timescales}.

\subsection{Description of cluster and environment}
\label{sec:intro-theo}

There is a broad range of theoretical approaches to deal with statics
and dynamics of free clusters, from macroscopic models over shell
models using an educated guess for the cluster mean field up to fully
fledged ab initio methods, for an overview see chapter 3 of
\cite{Rei03a} and section \ref{sec:freeclust} here. The higher
complexity of mixed systems (cluster/molecule + environment) calls for
a re-examination of the modeling.

The first step is the description of static properties.  Such static
studies have been undertaken since long and have allowed to understand
many properties of such "dressed" clusters or
molecules \cite{Mei00}. All approaches which are used for free
clusters can be used for the whole combined system. One even performs
fully quantum-mechanical calculations of adsorbate and substrate
\cite{Hak96b,Jac01,Pav05a,Bue05a,Gle07a}. That, however, imposes a
heavy restriction on the "substrate" size. 
That limitation can be overcome by the sophisticated Quantum
Mechanical/Molecular Mechanical (QM/MM) approaches of quantum
chemistry. These exploit a hierarchy of importance from the active
zone of interest down to the farther outskirts of the system and
couple a quantum description of the active piece to a classical
description of the environment. That method had been developed first
for dealing with the very complex systems of bio-chemistry
\cite{War76a} and is often used in that context, see e.g
\cite{Gao02aR,Mor05,Cas07a,Alt07aR}.  But it \PGmod{is} also extremely useful in
surface chemistry \cite{Mat99,Roe04aR,Sok04a}, for a detailed
description see \cite{Gre99a}. It is the
method of choice, in particular if we have that clear distinction
between a metal cluster and inert, insulating substrate.

The next step, namely to consider dynamical situations, requires a
much larger effort. One way would be to consider small samples as
representatives of the environment. However, truly dynamical
calculations accounting for all electronic degrees of freedom in such
"small" systems are not yet available. They may show up within a few
years from now.  Even if such calculations will be available in the
near future, it is likely that they will be limited to rather small
numbers of particles. And there will remain a gap between such small
systems and very large (bulk) ones.  We shall thus not elaborate
further on these approaches and look for a description of dynamical
processes with appropriate simplifications.  The simple-most and widely
used approach is a purely classical molecular dynamical simulation
using effective force fields \cite{Hab93,Tim97,Pal99,Koh01,Xir02}.
However, there are many situations where the quantum mechanical
aspects of the cluster electrons become important.  The QM/MM \esmod{methods} offer
here \esmod{again} a powerful tool of description.  Still, these methods freeze
electronic degrees of freedom of the environment inside
phenomenological interaction potentials and thus cannot account for a
proper dynamical response of the environment.  This, however, becomes a
key issue as soon as one considers truly dynamical situations, as
strong irradiation processes in biological systems as well as for
clusters in/on a substrate.
One thus needs to go \esmod{even} one step further in order to account for electronic
response of the environment.  We have developed over the last years
such a QM/MM model augmented by a simplified treatment of electronic
degrees of freedom of the environment for Ar environments
\cite{Ger05,Feh05,Feh06a} and for MgO(001) surfaces \cite{Bae07a}.
This dynamical QM/MM model constitutes the basis of results which will
be presented in the following.

Even with the enormous savings when using QM/MM, a full dynamical,
microscopic treatment of clusters in/on a substrate is not feasible
because of the much too large number of degrees of freedom for the
environment.  There are two complementing directions in which the
problem of system size can be attacked.
One solution is to simulate the \PGmod{environment in terms of a} finite size
system. In other words, the environment is modeled by a \PGmod{finite
cluster} of the environment material.  Of course, convergence of results
from such a finite system to bulk values has to be carefully tested.
But the ``finite environment'' as such is an interesting system as it
allows to vary the size of the ``environment'' and so to analyze
theoretically the impact of embedding on the cluster and on the
environment. There \PGmod{are}, in fact, experiments performed following that
strategy in the case of solvated biomolecules, see the example in
Fig.~\ref{fig:examples}.  The study of such model systems is thus
becoming a key issue and \PGmod{they} are better accessible to a
dynamical QM/MM description.
The alternative solution for a simulation of bulk material is to
consider a finite piece of the system and to copy it to an infinite
number of similar pieces with periodic boundary conditions.
That method is well known from simulations of true bulk in
solids, liquids, and plasma \cite{Cat82,All87,Zwi99}. Surfaces can be
modeled that way with periodic copies in lateral
direction. Lattice translational symmetry is broken in vertical
direction. Thus one uses here the ``finite sample'' approach in
considering only a finite number of layers. 
A \esmod{more} detailed description of the modeling will be given in section
\ref{sec:model}.

\subsection{Outline}

The paper is orgnized as follows. Section \ref{sec:model} provides a
detailed presentation of the model used in the following, discussing
also its relations to other approaches and pointing out the importance
of properly including dynamical effects. 
Section \ref{sec:struct} focuses on structural and low energy aspects
for atoms and for clusters in contact with an environment.  It allows
to validate our generalized QM/MM approach with respect to experimental
results and other theoretical approaches, as structure properties can
be accessed at various levels of sophistication. That section contains
also a discussion of optical properties which constitute the doorway
to dynamical scenarios.
Section \ref{sec:depos} is devoted to the study of deposited clusters
and deposition scenarios. We discuss both the response of the cluster
itself and the response at the side of the environment, analyzing in
detail the excitation of internal degrees of freedom of the environment.
Section \ref{sec:light} discusses the dynamics following irradiation by
intense laser pulses, mostly for embedded clusters. We are addressing
highly non-linear situations involving sizeable ionization of the
cluster. The response of the system is analyzed in terms of all its
constituents, the electrons and ions of the cluster, the atoms of the
matrix (including their internal excitations).
Finally, section \ref{sec:concl} summarizes our major conclusions and
outlines future perspectives suggested by these studies.


\section{Model}
\label{sec:model}

\subsection{Brief review of models for clusters and environments}


The discussions in the introduction have demonstrated the difficulties
and challenges implied in a proper dynamical description of
embedded/deposited clusters subject to external perturbations. We now
want to summarize briefly the theoretical tools which are commonly
used in such problems. We start with reviewing methods for free
clusters (Sec.~\ref{sec:freeclust}), continue with discussing the
specifically new aspects coming up with embedded/deposited clusters
(Sec.~\ref{sec:embclust}), provide a graphical overview over the
various approaches (Sec.~\ref{sec:overview}) and close with
summarizing briefly the method used in the further pursuit of the
paper (Sec.~\ref{sec:ownmodel}).

\subsubsection{Free clusters}
\label{sec:freeclust}

Free clusters consist of ions and electrons. Each species calls for
its own approximation. Ions are usually treated as classical
particles.  Simpler approaches replace the ionic background by a
smooth jellium distribution, particularly for metal clusters, see,
e.g., \cite{Bec84,Eka84,Kre93}.
The ion-electron coupling is simply the Coulomb interaction in
all-electron models. But these are too bulky in applications to
clusters. One usually treats only a few valence electrons per ion and
the coupling is then described by pseudopotentials for which well
developed modeling exists in molecular and solid state physics, see
e.g. \cite{Sza85,Bac82,Goe96}.
The most demanding part is the description of the electron cloud and
thus the greatest variety of approaches is found here.
The modeling of structure and dynamics of {free} clusters covers all
methods used in quantum chemistry and/or atomic physics, from detailed
Configuration Interaction (CI) calculations \cite{Bon89} to robust
dielectric models \cite{Kre93}, for overviews see
\cite{Rei03a,Fen08a}.

Simulations of truly dynamical processes are
much more demanding. In molecular physics, there exist dynamic
extensions for the most detailed methods, e.g. exact solutions of the
time-dependent Schr\"odinger equation \cite{Par03a} or time-dependent
CI \cite{Kra05a,Sch07a}.  These methods are still confined to small
systems.  The more complex cluster systems employ mainly 
Time Dependent Density Functional Theory (TDDFT) at the
level of the Time-Dependent Local-Density Approximation (TDLDA) for
the cluster electrons augmented by Molecular Dynamics (MD)
for the motion of ionic cores \cite{Cal00}. Semi-classical approaches to TDLDA, as
Vlasov-LDA \cite{Dom97b,Dom00c,Fen04}, become valid for higher
excitation energies. For reviews see
\cite{Cal00,Rei03a,Rei06aR,Fen08a}.  We will discuss in that review
only results from TDLDA-MD and provide a short description of the
scheme later on in that section.

\subsubsection{Clusters in contact with environments}
\label{sec:embclust}


Combined cluster plus environment systems are much more complex than
free clusters. Their description is thus more demanding.
Particularly the interface requires a very careful treatment, almost
at the same level of detail as the cluster itself.  The conceptually
simplest procedure is to use the same (detailed) approach for cluster
and environment.
\PGcomm{Typical examples are calculations} of very small Na clusters on NaCl
\cite{Hak96b} or on Cu surface \cite{Pav05a}.  In these calculations, the
substrate is represented by rather small pieces of material kept close to the
known surface and bulk configuration. An alternative strategy is to consider
freely varied composite systems as a model for clusters in/on environment, as
done e.g. for Ag clusters with Ag-oxide substrate in \cite{Bue05a} or MgO
substrate in \cite{Gle07a}.
However, such a complete treatment requires small \PGcomm{samples}
for the substrate and is restricted mostly to structural studies, at
best dynamical calculations at Born-Oppenheimer level.  At the other
extreme, very simple theoretical approaches have been developed
relying on a macroscopic description of the substrate as jellium or by
dielectric theory, see e.g. \cite{Kre93,Rub93}.  Such a macroscopic
approach is valuable for first explorations of trends and orders of
magnitude.  But it becomes clearly insufficient when one aims at
describing dynamical scenarios.  An intermediate solution consists in
\PGcomm{using} highly detailed microscopic
approaches \cite{Hak96b} as a benchmark for deriving effective
interface potentials which then allow systematic studies
\cite{Koh97a,Koh98a} in rather large systems.  However, the use of
such interface potentials also imposes severe limitations on the
combination of cluster and environment.


If one wants to describe the excitation and dynamical response of
realistic systems with large numbers of degrees of freedom, one needs
to find a proper compromise between approximate and yet sufficiently
detailed description of the environment.  A complete treatment, as
mentioned above, is a successful first step, but misses the
long-range effects in the material. The \PGcomm{interface-potential}
approach is also a step into the right direction, but misses the
response of the environment. Both examples call for an approach which
takes care of the polarizability of the environment while allowing a
lower level description in other aspects of the material.
In particular, inert (i.e. insulating) environments provide a natural
hierarchy of reactivity.  The idea is thus to start from a fully
microscopic description of the electronically active cluster and to
treat the environment at a simpler level, namely as classical particles
with an internal dipole polarization, also handled as classical
degrees of freedom. Such a hierarchical approach sorting the systems
in shells of different importance and dominated by polarization
interactions was initiated in \cite{Dic58}, \PGcomm{further developed
for organic systems in \cite{War76a},} and extended to dynamical
situations in \cite{Cat82}. The method has evolved over the years and
is now often called Quantum-Mechanical/Molecular-Mechanical model
(QM/MM), for a general overview of \PGcomm{the method} see
\cite{Gre99a}. \PGcomm{QM/MM is often applied to deal with the huge
systems of bio-chemistry \cite{Gao02aR,Mor05,Cas07a,Alt07aR}. It
serves also as a powerful tool in surface chemistry see, e.g.,
\cite{Mat99,Roe04aR,,Sok04a}.}
For ionic crystals, one even adds a further outer layer with inert
ions, i.e. without the internal dipole polarizability, to account
for the long range Ewald summation in the material  \cite{Mat99}.
The calibration of the effective potentials for QM/MM models is
cumbersome. But once established, the models provide an extremely
efficient and reliable tool for mixed systems.  Well calibrated models
allow a dramatic reduction of the quantum mechanically active part of
the environment.
Earlier applications in cluster physics already took the
polarizability of the environment into account, but used a tight-binding
treatment for the cluster \cite{Gro98}. A model which maintains the
capability of dealing with clusters in possibly strong external
perturbations needs at least the cluster electrons (and ions)
explicitly, and a hierarchical treatment of the environment properly
including polarization effects.
\PGcomm{And ``properly'' means, for the dynamical applications which we
have in mind, that the \PGmod{dynamical} polarizability of the environment
atoms is taken into account, as we will outline in section
\ref{sec:ownmodel}. In that respect, the present approach goes beyond
most previous QM/MM implementations which were oriented on static
properties, at most some Born-Oppenheimer or linear-response
dynamics.

As already mentioned in Sec.~\ref{sec:intro-theo}, even QM/MM
cannot deal with the macroscopic number of atoms in true bulk. One
employs finite representatives of bulk material, either as
a finite simulation box of (periodically copied) bulk material or
as a freely varied finite ``cluster'' of bulk atoms.
}

\subsubsection{Overview of methods}
\label{sec:overview}

\begin{figure}[htbp]
\centering\resizebox{1.0\columnwidth}{!}
{\includegraphics[angle=-0]{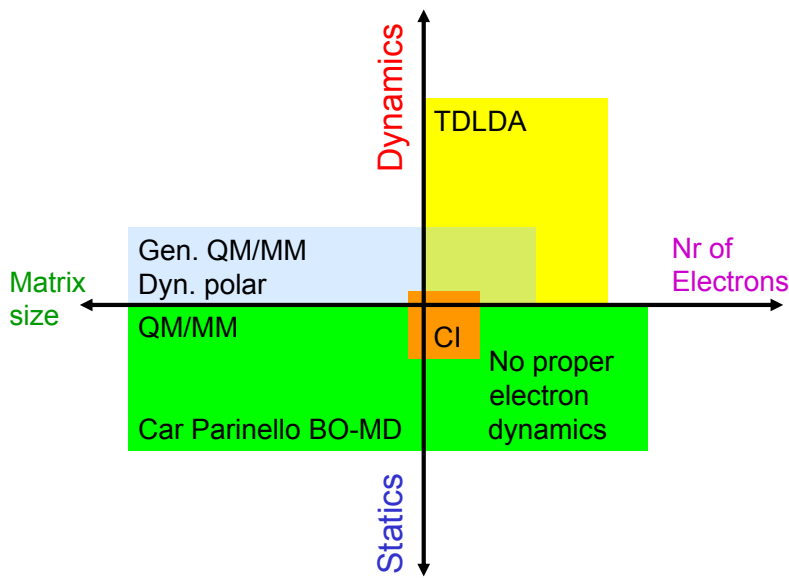}}
\caption{Diagram of approaches in a plane of system size (horizontal
axis)  and dynamical range (vertical axis).
\label{fig:theories}}
\end{figure}
Fig.~\ref{fig:theories} provides a sketch of the typical ranges of
application for the various theories used in the
field.
The horizontal axis represents the number of particles in the system,
to the right for the cluster, to the left for the environment. The
vertical axis represents an electronic "dynamical scale". Below zero,
electrons cannot be excited and are treated either as frozen (inside
potentials) or at best in Born-Oppenheimer manner. Above zero, the
axis qualitatively scales with the degree of electronic
excitation. The various \PGcomm{boxes in} Fig.~\ref{fig:theories}
represent the theories as indicated. At the side of cluster dynamics
(no environment), the TDLDA-MD approach \cite{Cal00} represents a
robust tool allowing to deal with several tens or even hundreds of
electrons and reaching up to sizable (non-linear) electronic
excitations and ionization. In turn, the more fundamental Time
Dependent Configuration Interaction methods (CI) of quantum chemistry
only allow to treat very small numbers of electrons. At the side of
the environment, small samples of substrate can still be treated with
the fully quantum mechanical methods.  Standard QM/MM (or generally
speaking Born-Oppenheimer methods) allow to consider relatively large
numbers of atoms, but cannot account for any electronic excitation,
whence they are placed towards the negative ordinate on the figure
(static or slow regime). The \PGcomm{dynamical QM/MM modeling we
propose in this paper accounts for a dynamical polarizability of the
environment and couples that to TDLDA-MD for the cluster. This allows
to explore the relevant dynamical features also at the side of the
environment}. Because of the simplified treatment in terms of
polarizabilities, one cannot consider too high excitations of
environment's electrons as compared to what could be accessible from
TDLDA-MD in a free cluster \PGcomm{(see section
\ref{sec:limits})}. But it already covers an interesting range of
possible dynamical scenarios as we shall see. And, to the best of our
knowledge, it is the first time that this domain is explored.

\subsubsection{The modeling used in that paper}
\label{sec:ownmodel}

The systems which we will review here are of the type metal cluster
combined with inert environment. We aim at resolving detailed structure
and at reaching a simultaneous dynamical description of electrons (from cluster
and to some extent atoms), ions (from cluster) and atoms.  This
excludes macroscopic (or partially macroscopic) approaches.  The
employed environments, rare gases \PGmod{(Rg)} and MgO, exhibit long-range
polarization effects. This inhibits a fully detailed description of
the environment with its limitation to small samples.  The cluster
dynamics should furthermore be tractable in a large range of excitations. These
requirements altogether point towards the best suited method: a
\PGcomm{dynamical} QM/MM approach.
We describe the cluster electrons in full quantum-mechanical detail
with Time Dependent Density Functional Theory (TD-DFT) at the level of
Local Density Approximation (LDA) coupled to Molecular Dynamics for
cluster ions (TDLDA-MD, \cite{Cal00}). The atoms/ions of the
environment are treated classically, attributing two \PGmod{($\times 3$)}
degrees-of-freedom to each: position and dipole momentum.  Due to
active dipoles, polarizability is treated fully dynamically and can
thus account for proper response to external fields, particularly
fields created by the cluster. 
As the environment is
assumed to be much more inert than the cluster, this modeling still
covers a wide range of dynamical situations as we shall illustrate
below.
The correct implementation of dynamical polarizability requires some
investment into careful calibration. Once established, the model is
rather simple to handle, and yet, it yields 
\MDcomm{in a broad range of dynamical situations a remarkable
improvement over standard approaches in which all constituents are
treated on an equal footing}. To
recall former discussions, one could say that such a modeling with an
environment of dynamically polarizable atoms represents so to say a
dynamical extension of more conventional QM/MM models.

\subsection{The constituents and degrees of freedom}

The summation over the various  elements of the environment looks very involved
when written in detail. To simplify notation, we introduce a compact
super-index $I$ which is composed as
\begin{subequations}
\begin{equation}
  I
  \equiv
  \left\{i^{(s\tau)},\tau\in\{c,v\},s\in\mbox{species}\right\}
  \quad.
\end{equation}
The coordinates and features are then associated with
\begin{eqnarray}
  s_I
  &\equiv&
  \mbox{atom species of } I
  \quad,
\\
  \tau_I
  &\equiv&
  \mbox{core $=c$ or valence cloud $=v$}
  \quad,
\\
  \vec{R}_{I}
  &\equiv&
  \vec{R}_{i^{(s_I\tau_I)}}  
  \quad.
\end{eqnarray}
\end{subequations}
The notational savings become obvious, e.g., from the summation over
all atom constituents~:
$$
  \sum_I
  \equiv
  \sum_{s\in\mbox{species}}\;\sum_{\tau\in\{c,v\}}\;
  \sum_{i^{(s\tau)}=1...N_{(s\tau})}
$$

\esmod{We present the model \PGmod{here for the case of MgO}. }
But the hierarchical modeling for Na clusters interacting with the
MgO(001) surface is developed in analogy to earlier studies of Na
clusters interacting with an Ar environment, for a detailed protocol,
see \cite{Feh05}.  Table~\ref{tab:freedom} summarizes the various
degrees of freedom.
\begin{table}[htbp]%
\begin{center}
\begin{tabular}[c]{l|l|l}\hline
\multicolumn{1}{c|}{\em d.o.f.}
 & \multicolumn{1}{|c|}{\em counter}
 & \multicolumn{1}{|c}{\em description}
\\
\hline
$\varphi_{n}(\vec{r})$ 
 \;&\; $n=1, \ldots ,N_{\mathrm{el}}$ 
 \;&\; s.p. wave function for cluster valence electrons
\\
$\vec{R}_{i^{\mathrm{(Na)}}}$
 \;&\; $i^{\mathrm{(Na)}}=1, \ldots, N_{\mathrm{Ion}}$
 \;&\; positions of the Na$^{+}$ ions
\\
\hline
$\vec{R}_{i^{(sc)}}$
 \;&\; $s\in\{\mbox{substr. species}\}\,,\,i^{(sc)}=1, \ldots, N_{(sc)}$
 \;&\; \esmod{position of the $N_{(sc)}$ cores of}
\\
 \;&\;
 \;&\; 
 \esmod{environment atoms(ions) }
\\
$\vec{R}_{i^{(sv)}}$
 \;&\; $s\in\{\mbox{substr. species}\}\,,\,i^{(sv)}=1, \ldots, N_{(sv)}$
 \;&\; \esmod{position of the $N_{(sv)}$ valence clouds of }
\\
 \;&\; 
 \;&\; \esmod{environment atoms(ions) }
\\
\hline
\end{tabular}
\\[8pt]
\end{center}
\caption{\label{tab:freedom}%
The degrees of freedom (d.o.f.) of the model. Upper block: Na
cluster. Lower block: dynamically active cell of the environment. 
}
\end{table}
The Na cluster is treated in standard fashion~\cite{Cal00,Rei03a}~:
Valence electrons are described as single-particle (s.p.)
wave functions $\varphi_{n}(\vec{r})$ and the complementing Na$^{+}$
ions are handled as charged classical point particles characterized by
their positions $\mathbf{R}_{i^{\mathrm{(Na)}}}$, see upper block of Table
\ref{tab:freedom}. 
In the following, we will present our model for the MgO substrate.
\PGcomm{(The 
case of a rare gas material is simpler, since it consists in only
one type of element, and will be introduced in the appendix,
section~\ref{sec:RG}.)}  
The MgO substrate is composed of two species whose properties
are formed in the context of bulk structure~: Mg$^{2+}$ cations and
O$^{2-}$ anions. The cations are electrically inert and can be treated
as charged point particles; they are labeled by $i^{(k)}$. The anions
are easily polarizable and are handled in terms of two constituents~: a
valence electron distribution (labeled by $i^{(v)}$) and the
complementing core (labeled by $i^{(c)}$). Each of these three types
of constituents is described in terms of positions
$\vec{R}_{i^{\mathrm{(type)}}}$. The difference $\mathbf{R}^{(c)}
-\mathbf{R}^{(v)}$ defines the electrical dipole moment of the
O$^{2-}$ anion, {which by construction is thus fully dynamical}. 
These (classical) degrees of freedom are summarized in
the lower block of Table~\ref{tab:freedom}.
 
\begin{figure}[htbp]
\centerline{\includegraphics[width=10cm]{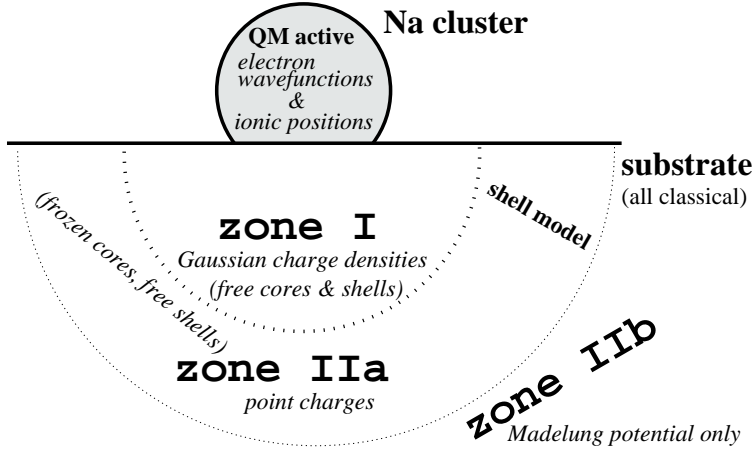}}
\caption{\label{fig:MgO_model}
Sketch of the shell model for hierarchical sorting of
quantum-mechanically treated metal cluster, classical dynamics
in the active region of the substrate and outer regions of
frozen substrate elements.
}
\end{figure}

The Mg and O ions reside in an active cell of the MgO(001) surface
region underneath the Na cluster \EScomm{in the case of a deposited species
(see also Fig.~\ref{fig:MgO_model}).
In the case of an embedded cluster, the idea remains the same with an
active cell \PGmod{around} the cluster. In the deposited case,} the
cell consists of three layers, 
each containing square arrangements of Mg$_{242}$O$_{242}$.
To avoid the Coulomb singularity and to simulate the finite
extension of these constituents, we associate a smooth charge distribution
$\rho(\vec{r})\propto\exp{(-{\vec{r}^{2}}/{\sigma^{2}})}$ with each of
these ionic centers. This yields a soft Coulomb potential [see
Eq.~(\ref{eq:Vsoft}) below] to be used for all active particles.
\MDcomm{For a rare gas environment, $\sigma_{c}$ and $\sigma_{v}$,
taken identical, are chosen so that the folded Coulomb interaction
energy of a rare gas atom corresponds (for small displacements) to the
polarization energy. They are thus related to the rare gas atom
polarizability and their values are given in
Table~\ref{tab:Rg-params} of the appendix. In the case of MgO, the
three width parameters are all taken equal to $0.6\sqrt{2}$~$a0$, so that
they comply with the grid mesh set by the Na pseudopotential [given
in Eq.(\ref{eq:locPsP}) below].}  
The active cell is surrounded by an infinitely extended outer region
of spectators, whose effect on the active part is given by a
time-independent shell-model potential, see below.

The MgO(001) surface is modeled in different stages of detail as 
sketched in Fig.~\ref{fig:MgO_model}. 
The size of the dynamically active part (zone I) depends
on the actual application. For instance, calculating the optical
response will not significantly affect the surface geometry far away
from the cluster, so that a diameter of 20 $a_0$ for zone I is
sufficient. On the other hand, deposition studies with high impact
energies require an increased diameter, since the surface will absorb
some of the impinging momentum as will be seen below.
The long-range Coulomb interaction in ionic crystals requires
inclusion of remote sites. To this end, we add an outer region of
``spectators'' (zone II).  That outer
region is built analogous to the active cell, consisting out of 
\PGmod{O$^{2-}$}
cores and O valence clouds as well as Mg$^{2+}$ cations. However,
\MDcomm{cores} are kept ``frozen'' at the positions of the free MgO
surface, \MDcomm{while O valence clouds remain fully dynamical}. The
region extends, in principle, over an infinite manifold 
of periodic copies in the two lateral directions and the vertical
layers below the surface. The \PGmod{atoms/ions in the} 
first layer of the outer region (zone
IIa) which stays in direct contact with the active part is 
frozen in position, but still allows dynamical dipole response and
exerts short-range repulsion to the active atoms.
\PGmod{The atoms/ions in the 
second layer of the outer region (zone
IIb) have all constituents frozen, including the dipoles.  
They
deliver only their Coulomb potential (adding up to the  Madelung potential).
}

\subsection{The total energy}
\label{sec:energ}

Starting point is the total energy
\begin{equation}
  E
  =
  E_{\mathrm{Na}}+E_{\mathrm{substr}}+E_{\mathrm{coupl}}
\label{eq:etot}
\end{equation}
where $E_{\mathrm{Na}}$ describes an isolated Na cluster,
$E_{\mathrm{substr}}$ the substrate, and
{$E_{\mathrm{coupl}}$} the coupling between the two
subsystems.

\subsubsection{The cluster energy}

For $E_{\mathrm{Na}}$, we take the standard TDLDA-MD functional as in
previous studies of free clusters \cite{Cal00,Rei03a,Rei06aR}. It is
composed as 
\begin{subequations}
\label{eq:Etotal}
\begin{eqnarray}
  E_\mathrm{Na}
  &=&
  E_\mathrm{kin,el}(\{\varphi_n\})
  +
  E_\mathrm{kin,ion}(\{\dot{\vec{R}}_{i^{\mathrm{(Na)}}}\})
  +
  E_\mathrm{C}(\rho)
  +
  E_\mathrm{xc}^\mathrm{LDA}(\rho^\uparrow,\rho^\downarrow)
\nonumber\\
  &&
  +
  E_\mathrm{SIC}(\{|\varphi_n|^2\})
  +
  E_\mathrm{pot,ion}(\{{\vec{R}}_{i^{\mathrm{(Na)}}}\})
  +
  E_\mathrm{PsP}(\{\varphi_n\},\{{\bf R}_I\})
  +
  E_{\rm ext}
  \quad,
\\
  E_\mathrm{kin,el}
  &=&
  \int \textrm d{\bf r}\,\sum_n
  \varphi_n^\dagger \frac{\hat{\vec{p}}^2}{2m_{\rm el}}\varphi_n^{\mbox{}}
  \quad,\quad
  E_\mathrm{kin,ion}
  =
  \frac{M_\mathrm{Na}}{2}
  \sum_{i^{\mathrm{(Na)}}=1, \ldots, N_{\mathrm{Ion}}}
   \big|\dot{\vec{R}}_{i^{\mathrm{(Na)}}}\big|^2
  \quad,
\\
  E_\mathrm{pot,ion}
  &=&
  \frac{1}{2}
  \sum_{i^{\mathrm{(Na)}}\neq j^{\mathrm{(Na)}}}
  \frac{e^2}{|\vec{R}_{i^{\mathrm{(Na)}}}-\vec{R}_{j^{\mathrm{(Na)}}}|}
  \quad,
\\
  E_\mathrm{SIC}
  &=&
  -\sum_n\left[
    E_\mathrm{C}(|\varphi_n^{\mbox{}}|^2)
    +
    E_\mathrm{xc}^\mathrm{LDA}(|\varphi_n^{\mbox{}}|^2,0)
  \right]
\label{eq:EnSIC}
\\
  E_\mathrm{PsP}
  &=&
  \sum_{i^{\mathrm{(Na)}}=1, \ldots, N_{\mathrm{Ion}}} \int \textrm
  d{\bf r}\,\sum_n 
  \varphi_n^\dagger\hat{V}^\mathrm{(PsP)}_{\vec{R}_{i^{\mathrm{(Na)}}}}
  \varphi_n^{\mbox{}}
  \quad,
\label{eq:EnPsP}
\\
  E_{\rm ext}
  &=&
  q_\mathrm{el}\int \textrm d{\bf r}\,\rho({\bf r})U_\mathrm{ext}({\bf
  r},t) +
  \sum_{i^{\mathrm{(Na)}}=1, \ldots, N_{\mathrm{Ion}}}
   q_\mathrm{Na}U_\mathrm{ext}(\vec{R}_{i^{\mathrm{(Na)}}},t)
  \quad,
\label{eq:Eext}
\\
  \rho(\vec{r})
  &=&
  \sum_n\varphi_n^\dagger({\bf r})\varphi_n^{\mbox{}}({\bf r})
  \quad,
\end{eqnarray}
\end{subequations}
where $E_\mathrm{C}$ is the direct part of the electronic Coulomb energy
which is naturally a functional of the local electron density
$\rho({\bf r})$. 
The coupling to the ions is described by pseudopotentials (PsP) as
indicated in Eq.~(\ref{eq:EnPsP}). These may be involved operators in
case of non-local PsP (see e.g. \cite{Bac82,Goe96}). Alkalines allow
to deal with local PsP for which we employ the soft Gaussian form
of \cite{Kue99}~:
\begin{eqnarray}
  V^\mathrm{PsP}_{\vec{R}}(\vec{r})
  &=&
  Q_+ V_{\rm soft} \left(|\vec r-\vec R|,\sigma_+^{(\rm Na)} \right)
  +
  Q_- V_{\rm soft} \left(|\vec r-\vec R|,\sigma_-^{(\rm Na)} \right)
  \quad,
\label{eq:locPsP}
\\
  &&
  Q_+ = 2.292
  \quad,\quad\quad
  Q_- = -3.292
  \quad,
\nonumber
\\
  &&
  \sigma_+^{(\rm Na)} = 0.6810\sqrt 2
  \quad,\quad\quad
  \sigma_-^{(\rm Na)} = 1.163\sqrt 2
  \quad,
\nonumber
\\
  &&
  V_{\mathrm{soft}}(r,\sigma ) 
  =
  e^{2}\frac{{\mathrm{erf}}(r/\sigma )}{r}
  \quad.
\nonumber
\end{eqnarray}
The term $E_\mathrm{ext}$ stands for a possible external, time-dependent
electromagnetic perturbation (e.g. a laser field or a bypassing ion) for which
we consider here simply the external Coulomb potential $U_\mathrm{ext}$.  That
is not a unique choice due the freedom of a gauge transformation. A laser
field can also be described through coupling to the vector potential as
$U_\mathrm{ext}\propto \hat{\mathbf{p}}\!\cdot\!\mathbf{A}_\mathrm{las}$
which is called the coupling in velocity gauge \cite{Fai87}. The different
gauges yield the same results but can impose different conditions on the
numerical treatment. The present choice (\ref{eq:Eext}) is simplest
to implement.
The term $E_\mathrm{xc}^\mathrm{(LDA)}$ stands for the energy-density
functional for electronic exchange and correlations.  We use here standard LDA
forms, inserting the instantaneous electron density $\rho(\vec{r},t)$
which is often called adiabatic LDA \cite{Gro96}. Actually, we will employ
the widely used exchange-correlation functional of \cite{Per92}.
Self-Interaction Correction (SIC), as introduced in \cite{Per81}, moves the
Ionization Potential (IP) and the other single-electron energies into correct
relation with the continuum threshold and is thus necessary whenever one deals
with electron emission.  The full variational treatment of the SIC energy
(\ref{eq:EnSIC}) leads to rather inconvenient mean-field equations such that
various approximations have been developed in the past. For alkaline metal
clusters, one can employ the very simple average-density SIC (ADSIC)
\cite{Fer34,Leg02} with
\begin{eqnarray}
  E_\mathrm{ADSIC}
  &=&
  - N_{\mathrm{el},\uparrow}
    \left[
      E_\mathrm{C}
      \left( \frac{\rho_\uparrow}{N_{\mathrm{el},\uparrow}} \right)
      +
      E_\mathrm{xc}^\mathrm{LDA}
      \left(\frac{\rho_\sigma}{N_{\mathrm{el},\uparrow}},0\right)
    \right]
\nonumber\\
  &&
  - N_{\mathrm{el},\downarrow}
    \left[
      E_\mathrm{C}
      \left( \frac{\rho_\downarrow}{N_{\mathrm{el},\downarrow}}\right)
      +
      E_\mathrm{xc}^\mathrm{LDA}
      \left(0,\frac{\rho_\sigma}{N_{\mathrm{el},\downarrow}}\right)
    \right]
\label{eq:ADSIC}
\end{eqnarray}
where $N_{\mathrm{el},\sigma}$ is the number of electrons with spin
orientation $\sigma$.

A word is in order about the level of DFT treatment used here. The
LDA-ADSIC is chosen for reasons of simplicity, efficiency, and
robustness. There are known (and unknown) drawbacks of such approaches
as, e.g., the lack of reproducing the derivative discontinuity
\cite{Kue04a,Mun05b} which can raise problems in reproducing the
polarizability in soft systems \cite{Gis98} and in describing details
in the dynamics of molecular fragmentation. Many attempts try to
overcome these problems to develop a DFT which employs exact exchange
\cite{Lei05a,Roh06a,Goe06b,Arm08a}, for a recent review see
\cite{Kue08}.  These extensions are, however, much more involved and
presently not released for straightforward applications to large scale
dynamics. An alternative track is a tuning of SIC for dynamical
equations \cite{Mes08b,Mes08a} \PGcomm{which allows to recycle the
well tested energy functionals for \PGmod{LDA}}.  But also this is still
under development. We keep on the safe side of well understood and
well tested methods and use the most traditional approach, namely LDA
with ADSIC.

\subsubsection{The substrate and coupling energy}

The energy of the substrate subsystem is given by
\begin{eqnarray}
  E_\mathrm{substr}^{\mbox{}} 
  &=&
  \sum_I
  \frac{M_{s_I\tau_I}}{2}\big|\dot{\vec{R}}_{I}\big|^2
  +
  \sum_{I} 
   q_{s_I\tau_I}\Phi_{\mathrm{out}}^{\prime }(\vec{R}_{I})
\nonumber
\\  
  &&
  +\frac{1}{2}\sum_{I\neq J}
   V_{s_I\tau_I,s_J\tau_J}^{\mbox{}}
   \left(|\vec{R}_I\!-\!\vec{R}_J|\right)  
\nonumber
\\
  &&
  +\sum_s\sum_{i^{(sc)}}\left[ 
    \frac{\kappa_{s}}{2}
    \left(|\vec{R}_{i^{(sc)}}\!-\!\vec{R}_{i^{(sv)}}| \right)^2
    -
    V_{sc,sv}^{\mbox{}}
    \left(|\vec{R}_{i^{(sc)}}\!-\!\vec{R}_{i^{(sv)}}|\right)\right] 
\label{eq:Esubstr} 
\end{eqnarray}
where $q_{s\tau}$ is the charge associated with subtype $\tau$ of an
atom of species $s$, and $M_{s\tau}$ the corresponding mass.
The second line refers to interaction potentials between different
atoms and consists of a soft Coulomb part and
a (mainly repulsive) short-range contribution~:
\begin{subequations}
\begin{eqnarray}
  V_{\alpha \beta }^{\mbox{}}(r) 
  &=&
  q_{\alpha }q_{\beta }V_{\mathrm{soft}}(r,\sigma _{\alpha \beta })
  +
  f_{\alpha \beta }^\mathrm{(short)}(r)
  \quad,
\label{eq:Vab} 
\\
  V_{\mathrm{soft}}(r,\sigma ) 
  &=&
  e^{2}\frac{{\mathrm{erf}}(r/\sigma )}{r}
  \quad,
\label{eq:Vsoft} 
\\
  \sigma_{\alpha\beta} 
  &=&
  \sqrt{\sigma _{\alpha}^{2}+\sigma _{\beta}^{2}}
  \quad,
\\
  \sigma _\mathrm{Mg}
  &=&
  \sigma _\mathrm{O}
  =
  0.6\sqrt{2}\, a_0
  \quad.
\end{eqnarray}
\label{eq:matrix_pot}
\end{subequations}
\noindent where $\alpha,\beta\equiv(s\tau)$ serves as combined index
simpler notations.  The explicit forms of $f_{\alpha \beta
}^\mathrm{(short)}$ is given for the specific case of MgO in
Table~\ref{tab:MgOparams} and for rare gas in
Table~\ref{tab:Rg-params}, \PGcomm{see appendix.}
The third line stands for the interaction between core and valence 
cloud of a same atom~: The interaction potential (\ref{eq:Vab})
appears again as the second term, while the first term is the potential of a
simple spring with species-dependent spring constant $\kappa_s$. This,
in turn, allows a separate tuning for the correct (dynamical)
dielectric response of the substrate material.
%

\PGcomm{Ionic crystals, here MgO, consist of ions with alternating
charge.  This case requires to take care of the long-range Coulomb
potential $\Phi_{\mathrm{out}}^\prime$ of the substrate spectator
ions} which extend to infinity in all lateral directions and
vertically downwards from the surface.  The ions far from the active
region stay practically at their crystalline equilibrium position, but
still produce a long range Coulomb field. It is given as
\begin{equation}
  \Phi_{\mathrm{out}}^{\prime}(\vec{r})
  =
  \sum_{I'\in\{\mathrm{outer}\}}
  \frac{q_{s_{I'}\tau_{I'}}e^{2}}{\left|\vec{r}-\vec{R}_{I'}\right|}
\end{equation}
where the summation runs only over $I$ which are in the outer region
of frozen crystal atoms/ions.  That Coulomb field influences the
cluster and substrate inside the active cell, particularly for ionic
crystals.  The smoothing can be ignored for these outer atoms/ions
because the effect of local smoothing is of very short range. This, in
turn, allows one to use analytical techniques similar to Ewald
summation \cite{Par75a,Par76a}. The shell potential
$\Phi_{\mathrm{out}}^{\prime}$, being time-independent, is computed in
great detail at the start of the calculations and tabulated for
frequent later use. The boundary region of the active cell will also
feel some effect of the short-range core repulsion of
neighboring atoms in the outer region \cite{Nas01a}. We take that into
account by having one boundary layer of explicit atoms/ions but frozen
positions.

\PGcomm{In all cases,} the coupling from substrate to cluster is 
dominated by the (soft) Coulomb field. At short distances, some core
repulsion is added in the form of appropriate short-range
potentials. The coupling energy becomes
\begin{eqnarray}
  E_{\mathrm{coupl}}  
  &=&
  \int \textrm d \mathbf r\ \rho(\vec{r})\left[
    \Phi_{\mathrm{out}}^{\prime}(\vec{r})
    +
    \sum_I
    V_{s_I\tau_I,\mathrm{el}}(|\vec{r}-\vec{R}_I|)\right] 
\nonumber
\\
  &&
  +
  \sum_{{{j}^{(\mathrm{Na)}}}}\Big[
    \Phi_{\mathrm{out}}^{\prime}(\vec{R}_{j^{(\mathrm{Na )}}})
    +
    \sum_{I}
    V_{s_I\tau_I,\mathrm{Na}}(|\vec{R}_I-\vec{R}_{j^{(\mathrm{Na )}}}|)
  \Big]
\nonumber
\\
  &&
  +
  E_\mathrm{VdW}
  \quad,
\label{eq:coupl}
\end{eqnarray}
where the interaction potentials $V_{\alpha \beta}$ are also given in the general form
(\ref{eq:Vab}).
The Van der Waals (VdW) energy between cluster electrons and substrate
atom \PGcomm{is negligible for ionic crystals (MgO), but crucial for
rare gases. It} reads in detail
\begin{eqnarray}
  E_\mathrm{VdW}
  &=&  
  \frac{e^2}{2} \sum_s\alpha_{s}\sum_{i^{(sc)}}
  \left[\frac{1}{N_\mathrm{el}}
    \left(\int{\textrm d \mathbf r\ \rho(\vec{r})\ 
      \mathbf{f}_s(|\vec{r}\!-\!\vec{R}^{\mbox{}}_{i^{(sc)}}|) 
         }
    \right)^2 \right.
\\
  &&\qquad\qquad\qquad\qquad
   - \left.
   \int{\textrm d \mathbf r\ \rho(\vec{r}) \
     \mathbf{f}^2_s(|\vec{r}\!-\!\vec{R}^{\mbox{}}_{i^{(sc)}}|)
       }
  \right]
  \;,
\label{eq:EvdW}
\\
  &&
  \mathbf{f}_s(\vec{r})
  =
  \nabla\frac{\mbox{erf}\left(\vec{r}/\sigma_{sv}^{\mbox{}}\right)}
             {|\vec{r}|}
  \quad,
\nonumber
\end{eqnarray}
where $\alpha_{s}$ is the polarizability of species $s$ and
$\sigma_{sv}$ the smoothing width associated with the valence cloud of
species $s$. This contribution is not always taken into account
\PGcomm{(case of MgO)}, or effectively replaced in other terms
\PGcomm{(simplified VdW for rare gases)}. It is then switched off by
letting for the polarizability $\alpha_s\longrightarrow 0$. It
should be noted that the VdW contribution provides a
sizable fraction of binding \PGcomm{of Na on rare gases.  It should
thus be} taken into account, even if its derivation is well beyond
DFT.

\subsubsection{Final calibration}
\label{sec:calib}

\PGmod{ 
The calibration of the whole model has to take care of three pieces,
the cluster as such, the environment as such, and the coupling between
both. 
The modeling for the cluster is taken over from work on pure
clusters, where calibration concerns mainly the electron-ion
pseudopotential (\ref{eq:locPsP}).
The model parameters for the pure environment are taken over from
previous studies.
\MDcomm{
The expression of the Rg-Rg core interaction is a standard
Lennard-Jones-type potential from~\cite{Ash76}, while Buckingam-type
ones are used in MgO~\cite{Lew85,Nas01}. 
The valence cloud parameters of Rg and O in Rg and MgO respectively
are adjusted to the dynamical polarization at low 
frequencies of the Rg atoms~\cite{Cha91,Cha92} and
MgO~\cite{Lew85,Ben98a,Nas01}.
}
The parameters for the coupling between environment and Na cluster are
calibrated from scratch \MDcomm{in the case of MgO}. The tuning for Na@MgO(001)
was performed using Born-Oppenheimer surfaces from \cite{Win06} (the
corresponding curves are presented in Sec.~\ref{sec:Na_MgO_PES},
Fig.~\ref{fig:PES_Na_MgO}).
\MDcomm{For a Rg environment}, the Na$^+$-Rg potential is calibrated by a fit to
scattering data \cite{Ahm95,Vie03}.  The combined Na$^+$-Rg and
electron-Rg potentials are finally tuned to ground-state and
excitation properties of NaRg dimers taken from experimental as well
as theoretical work, for NaNe \cite{Lap80,Hli85}, for NaAr from
\cite{Sax77,Las81,Sch00,Sch03}, and for NaKr \cite{Bru91}. 
A detailed description together with all model parameters is given
in appendix.
}

\begin{figure}[htbp]
\centering\resizebox{0.8\columnwidth}{!}
{\includegraphics[angle=-0]{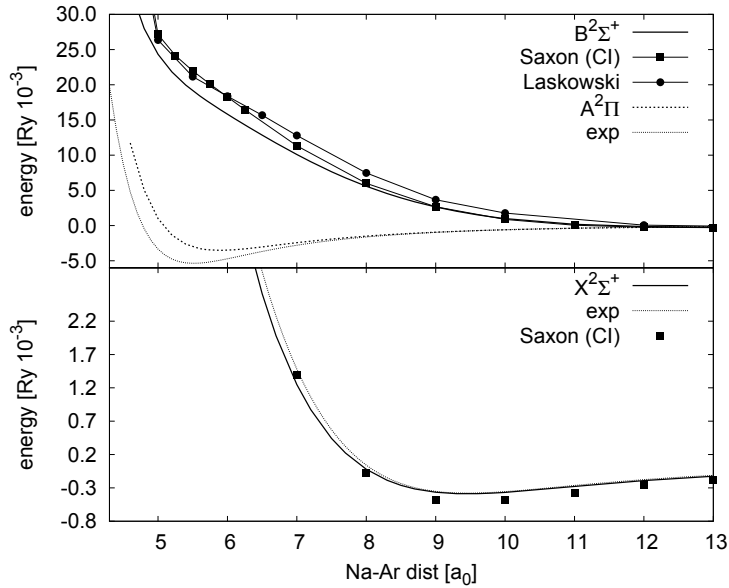}}
\caption{
Born-Oppenheimer energy curves for the NaAr molecule.
The lower panel shows the energies for the ground-state
configuration (X$^2\Sigma^+$) and the upper panel for two
excited configurations as indicated. The results from the
present modeling (thick solid lines for X$^2\Sigma^+$ and B$^2\Sigma^+$,
dashed for A$^2\Pi$) are compared with
experimental data from \cite{Sch00} and quantum-chemical CI calculations
\cite{Sax77,Las81}. The energies are scaled relative to  the
3$s$- (lower panel) or 3$p$-state (upper panel) of the Na atom.
Taken from \cite{Feh06c}.
\label{fig:na_ar_pot}}
\end{figure}

Fig.~\ref{fig:na_ar_pot} demonstrates the calibration of the model for the
NaAr system in terms of the Born-Oppenheimer energies for ground state and
excited configuration. Note that the excitation splits into an A$^2\Pi$ state
and a B$^2\Sigma$ state. Experimental data for the more critical B$^2\Sigma$
transition are not available for small distances. But the proper description
of core repulsion in that channel is crucial for the modeling of molecular
dynamics. Therefore we included also detailed quantum-chemical calculations in
the calibration. The agreement with experiment and with other calculations is
very satisfying. Similar quality has been achieved for the other rare gases
\cite{Feh06c,Feh07a}.

The VdW contribution, nevertheless,
requires a sizable numerical effort. We have thus also explored the
possibility of re-tuning the other model parameters to mock up its
effect inside model parameters. 
\PGcomm{As argued above,}
the VdW energy (\ref{eq:EvdW}) between Na and
environment atoms is a crucial ingredient for rare gases due to the very
faint general binding.
But precisely that VdW energy turns out to be very expensive in large scale
dynamical calculations due to its long range nature. On the other
hand, the faint details of the VdW interaction become unimportant in
truly dynamical situations carrying large amounts of excitation
energy. For that purpose, we have developed an alternative calibration
where the net effect of the VdW interaction is accounted for in a
modified parameterization of the electron-Rg potential. The
corresponding VdW energy is then set to zero. The results are listed
in Table~\ref{tab:effVdW} in the appendix for the case of Ar only. 
\EScomm{This \PGcomm{implicit} strategy turns out to be quite
efficient and has been adopted for obvious practical reasons in most
reported results. The implicit VdW \PGcomm{energy}
finally provides numbers very close to the ones attained with an
explicit VdW energy: \PGcomm{The small differences
remain} within the range of expected validity of the modeling,
\PGcomm{a result} which justifies such a strategy. }

\subsection{The coupled dynamical equations of motion}
\label{eq:dyn-EoM}

The total energy (\ref{eq:etot}) as defined above determines uniquely
the dynamical equations by variation. The stationary equations are
obtained by minimization of the energy, leading to the variational
conditions
\begin{subequations}
\begin{equation}
  \delta_{\varphi_n^\dag}
  \left[
   E-\sum_n\varepsilon_n(\varphi_n|\varphi_n)
  \right]
  =
  0
  \quad,\quad
  \delta_{\vec{R}_I}E
  =  
  0
  \quad.
\end{equation}
The dynamical equations can be derived through the principle of
stationary action
\begin{equation}
  \delta_{\varphi_n^\dag}S
  =
  0
  \quad,\quad
  \delta_{\vec{R}_I}S
  =  
  0
  \quad,\quad
  S
  =
  \int \textrm dt \left\{
   E
   -
   \sum_n(\varphi_n|\mathrm{i}\hbar\partial_t|\varphi_n)
   -
   \sum_I\vec{P}_I\dot{\vec{R}}_I
  \right\}
  \quad,
\end{equation}
\end{subequations}
where the classical conjugate momentum is $\vec{P}_I=\dot{\vec{R}}_I/M_I$.

The emerging stationary Kohn-Sham (KS) equations read
\begin{eqnarray}
  \hat{h}_{\mathrm{KS}}\varphi_n
  &=&
  \varepsilon_n\varphi_n
  \quad\mbox{for}\quad
  n\in\{1, \ldots, N_{\rm el}\}
  \quad,
\label{eq:statKS}
\\
  \hat{h}_{\mathrm{KS}}
  &=&
  \frac{\hat{p}^2}{2m}
  +
  U_{\mathrm{KS},\sigma_n}(\vec{r})
  \quad,\quad
  U_{\mathrm{KS},\sigma_n}(\vec{r})
  =
  \frac{1}{\varphi_n(\vec{r})}
  \frac{\delta E}{\delta\rho^{(\sigma_n)}_{\mbox{}}(\vec{r})}
  \quad,\quad
  \sigma_n\in\{\uparrow,\downarrow\}
  \quad.
\label{eq:U-KS}
\end{eqnarray}
That simple spin-density variation leading to a local mean field
potential $U_{\mathrm{KS},\sigma_n}(\vec{r})$ is justified for
the local pseudopotentials (\ref{eq:locPsP}) and the ADSIC correction
(\ref{eq:ADSIC}), our standard treatment in the subsequent
applications. An extension to non-local mean fields is
straightforward. 
The classical complement of the stationary Kohn-Sham equations is the
set of conditions $\nabla_{\vec{R}_I}E=0$. A stationary configuration
is achieved if both sets of equations are satisfied.

The time-dependent KS equations analogously read
\begin{subequations}
\begin{equation}
  \hat{h}_{\mathrm{KS}}\varphi_n
  =
  \mathrm{i}\hbar\partial_t\varphi_n
\label{eq:TDKS}
\end{equation}
where the KS Hamiltonian $\hat{h}_{\mathrm{KS}}$ is composed 
as in Eq.~(\ref{eq:U-KS}) above. 
The complementing classical, molecular mechanical (MM), equations 
are
\begin{equation}
  \dot{\vec{P}}_I
  =
  -\nabla_{\!\vec{R}_I}^{\mbox{}}E
  \quad,\quad
  \dot{\vec{R}}_I
  =
  \nabla_{\!\vec{P}_I}^{\mbox{}}E
  =
  \frac{\vec{P}}{M_I}
  \quad.
\label{eq:classEoM}
\end{equation}
\end{subequations}
The detailed expressions for the forces $\nabla_{\!\vec{R}_I}E$ are
straightforward to obtain but tedious to write explicitly. Even in practical
applications, it is safer, and often competitive in terms of numerical expense,
to compute the forces from evaluating derivatives by explicit finite
differences.

\subsection{Comments on numerical solutions}

\subsubsection{Representation of wave functions and fields}

The valence electron cloud of alkaline clusters is rather smooth and
covers a limited range of length scales. That is a typical situation
which favors a direct representation of wave functions and local
fields on a grid in coordinate space, e.g. in three Cartesian
dimensions as
$
  f({\bf r})
  \;\longleftrightarrow\;
  f({\bf r}_{\bf n})
  \,,\,
  {\bf r}_{\bf n}\in\{(n_x\Delta x,n_y\Delta y,n_z\Delta z)\}
  \,.
$
A grid representation is also much preferable, if not compulsory, as
soon as electron emission becomes important in the course of the
dynamical evolution because that can be conveniently accounted for by
absorbing boundary conditions, see Sec.~\ref{sec:abs-bc}.
The operators of momentum and kinetic energy can be described in two
ways, finite differences or Fourier representation. 
The latter is
preferable on Cartesian grids because the swapping between coordinate
and momentum space can then be very efficiently realized by the
fast Fourier transform (FFT) \cite{Pre92}.

\esmod{It turns out that}
 many processes evolve close to axial symmetry.  Here
it is much advantageous to describe the electronic system on a grid in
cylindrical coordinates, in the spirit of the cylindrically averaged
pseudopotential scheme (CAPS) \cite{Mon94a,Mon95a}. Finite
differences are then the method of choice for the kinetic energy.
An extensive comparison of the gridding schemes and their performance
can be found in \cite{Blu92}. As a rule of thumb, Fourier techniques
are advantageous in three-dimensional Cartesian grids with a not too low
number of grid points (at and above 32 mesh points in each
direction) while finite differences perform better on small grids and
in restricted symmetries.

A major task is the evaluation of the Coulomb potential in the
Kohn-Sham Hamiltonian.  The Poisson equation is dominated by the
Laplacian, similar as the kinetic energy of the electrons. The solution
of the Poisson equation can be performed by FFT in the case of Fourier
representation or by iteration on a grid (similar to the stationary KS
solution, see Sec.~\ref{sec:stat-num}). The long-range part of the
Coulomb field requires special care. To that end, we separate the
long-range components up to the hexadecapole moment and treat it 
explicitly by analytically solvable multipole fields. The difference
to that long-range part is of sufficiently short range such that it
can be treated by inversion of the Laplacian on the grid, for details
of that method see \cite{Lau94}.

\subsubsection{Stationary state}
\label{sec:stat-num}

The stationary Kohn-Sham equations (\ref{eq:statKS}) constitute a
non-linear eigenvalue problem. This requires an iterative solution.
To that end, we employ the gradient step
\begin{subequations}
\begin{equation}
  \varphi_n^{(m+1)}
  =
  \hat{\cal O}\left\{
    \varphi_n^{(m)}
    -
    \hat{\cal D}\left(\hat{h}^{(m)}-
      (\varphi_n^{(m)}|\hat{h}^{(m)}|\varphi_n^{(m)})\right)
    \varphi_n^{(m)}
  \right\}
\label{eq:dampgrad}
\end{equation}
where $\hat{\cal O}$ stands for orthonormalization of the new set
$\{\varphi_n^{(m+1)}\}$, the upper index $(m)$ counts the
iteration, and $\hat{\cal D}$ is a convergence generating
operator. The simplest choice is just a sufficiently small number
$\hat{\cal D}=\eta$ which ensures convergence if it is smaller than
twice the maximum representable energy. Much faster convergence can be
achieved by appropriate tuning of $\hat{\cal D}$ (often called
pre-conditioning). For example, grid
representations as we use here have their largest conceivable
energies from the kinetic energy
operator. Here it is a good idea to use 
\begin{equation}
 \hat{\cal D}
  =
  \frac{\eta}{\hat{T}+E_0}
\end{equation}
\end{subequations}
where $\hat{T}$ is the operator of kinetic energy and $E_0$ is
typically the depth of the potential. This step can proceed very fast
with $\eta\approx 1$. For more details see \cite{Blu92}.

An obvious solution scheme for finding the ionic configurations would
be to follow the path along the steepest downhill gradient
$\vec{R}_I\longleftarrow\vec{R}_I-\eta_I\nabla_{\vec{R}_I}E$ where
$\eta_I$ is an appropriate step size. However, that simple downhill
method is very likely to get stuck in some local minimum possibly
still far away from the global minimum. One needs to couple that
stepping with stochastic methods which explore more thoroughly the
ionic energy landscape. The method of choice is here simulated
annealing, for details see \cite{Pre92,Rei03a}.

\subsubsection{Dynamical evolution}
\label{sec:dyn-num}

The choice of propagation scheme for the electronic wave functions
depends on the chosen numerical representation.  We are using a grid
in coordinate space and deal mostly with local mean fields.  For that
case, a very efficient stepping scheme is the time-splitting method
\cite{Fei82}
\begin{equation}
  \varphi_n({\bf r},t\!+\!\delta t)
  =
  \exp{\left(-\mathrm{i}\frac{\hbar\delta t}{2}U({\bf r},t\!+\!\delta t)\right)}
  \exp{\left(-\mathrm{i}\hbar\delta t \, \hat{T}\right)}
  \exp{\left(-\mathrm{i}\frac{\hbar\delta t}{2}U({\bf r},t)\right)}
  \varphi_n({\bf r},t)
  \quad.
\label{eq:timesplit}
\end{equation}
The action of the local operator 
$\exp{\left(-\mathrm{i}\frac{\hbar\delta t}{2}U({\bf r},t\!+\!\delta t)\right)}$
is trivial in coordinate space.  The workload consists in the kinetic
terms. They are evaluated most easily in Fourier representation, by
transforming $\varphi_n$ into momentum space, applying
$\exp{\left(-\mathrm{i}\hbar\delta t \,\hat{T}\right)}$ 
trivially there, and then transforming the result back into coordinate
space.  When one is using finite differences, the kinetic
propagator can be factorized into three (well manageable)
one-dimensional propagators along $x$-, $y$-, and $z$-directions.  A 
favorable feature of this time-splitting scheme is that the action of
potential propagation amounts to a phase factor which does not change
the local density. Thus the density $\rho({\bf r},t\!+\!\delta t)$ is
already known after the evaluation of the first kinetic
propagator. This allows one to compute $U({\bf r},t\!+\!\delta t)$
for the step without iteration.  Another advantage of the step
(\ref{eq:timesplit}) is that it is unitary and thus preserves
orthonormality of the set $\{\varphi_n\}$.
An often used alternative are the Crank-Nicholson or Peaceman-Rachford
steps. These rely on an approximate separation of the propagation into
a succession of three one-dimensional steps, for a general discussion,
see \cite{Pre92} and for applications in cluster dynamics, e.g.
\cite{Cal00}.

The classical equations of motion (\ref{eq:classEoM}) for ionic propagation
are usually solved by the velocity Verlet algorithm \cite{Ver67}. That method
exploits the symplectic structure of Hamiltonian equations and is thus 
particularly efficient in propagating them. More elaborate methods
\MDcomm{which would allow to use larger time steps} are rarely
needed because the electronic time scale sets the pace.
\MDcomm{This means that one can safely employ the velocity Verlet
algorithm for ionic propagation with conveniently shorter time steps
which are in any case enforced by electronic propagation.}

\subsubsection{Absorbing boundary conditions}
\label{sec:abs-bc}

Electrons escape the cluster for sufficiently strong excitations.  In
practice, they propagate across the grid until they hit the bounds of
the numerical box. Without further measures, they would be reflected,
travel back into the cluster, and so falsify the time evolution. One
needs to remove the electrons as soon as they approach the bounds. That
is achieved by imposing absorbing boundary conditions. These are
installed by using a mask function $\mathcal{M}(\vec{r})$ which is
$\mathcal{M}=1$ in the interior of the box and gently decreases to
$\mathcal{M}\longrightarrow 0$ towards the bounds \cite{Ull00b}. Each
time-step (\ref{eq:timesplit}) is augmented with one masking step
\begin{equation}
  \varphi_n({\bf r},t+\Delta t)
  \longleftarrow
  \mathcal{M}(\vec{r})
  \varphi_n({\bf r},t+\Delta t)
  \quad.
\label{eq:maskstep}
\end{equation}
A simple and robust choice for $\mathcal{M}$ is the (rectangular)
separable form
\begin{eqnarray}
  \mathcal{M}(x,y,z)
  &=&
  \mathcal{M}_x(x)
  \mathcal{M}_y(y)
  \mathcal{M}_z(z)
  \quad,
\\
  \mathcal{M}_i(r_i)
  &=&
  \left\{\begin{array}{ll}
     \cos^{1/4}\left(\frac{|r_{i,\mathrm{max}}-r_i|\pi}{b_\mathrm{abs}}\right)
     & \mbox{for} \quad |r_{i,\mathrm{max}}-r_i|\leq b_\mathrm{abs}\\
     1 & \mbox{else}
         \end{array}\right.
\nonumber
\end{eqnarray} 
where a symmetric grid is assumed for which
$r_{i,\mathrm{min}}=-r_{i,\mathrm{max}}$ in each direction.
A reliable analysis of angular distributions (see
Sec.~\ref{sec:obs-elec}) better uses absorbing bounds in spherical
shape. 
These are achieved with
\begin{equation}
  \mathcal{M}(\vec{r})
  =
  \left\{\begin{array}{lcl}
     1 &\mbox{for}& r<R_\mathrm{max}-b_\mathrm{abs}
   \\
     \cos^{1/4}\left(\frac{|R_\mathrm{max}-r|\pi}{b_\mathrm{abs}}\right)
     & \mbox{for} & R_\mathrm{max}-b_\mathrm{abs}\leq r<R_\mathrm{max}
   \\
     0 & \mbox{for} & R_\mathrm{max}\leq r
         \end{array}\right.
\end{equation}
where $r=|\vec{r}|$ and 
$R_\mathrm{max}=\mbox{min}(x_\mathrm{max},y_\mathrm{max},z_\mathrm{max})$
\,.
We typically use an absorbing margin of about $b_\mathrm{abs}\approx
4$--6 $a_0$. Larger margin improve the efficiency of absorption but
are growing quickly expensive, particularly in three dimensions.
The power of the mask function, here 1/4, has also some influence on
the quality. The optimum value, however, depends on the size of the
margin \cite{Rei06c}. In most applications, we use the 1/4 as
robust working prescription.

\PGcomm{Absorbing boundary conditions, although crucial for describing
dynamical scenarios with ionization, attenuate the wave functions in
the absorbing zone. There is the danger that the ground state becomes
unstable by probability loss through the absorbing bounds.  One needs thus
to use sufficiently large numerical boxes such that the ground state
wave functions have very small overlap with the absorbing bounds to
avoid that artefact.} \esmod{Large numerical boxes furthermore make 
absorbing boundary conditions even better physically. Indeed, although
absorbing  
bounds are closer to physical reality than reflecting ones, they are not fully 
realistic either,  as they arbitrarily suppress outgoing electrons at a certain distance from 
the system, thus \PGmod{reducing} some long range polarization effects. 
}

\subsection{Observables}

\subsubsection{Electronic observables}
\label{sec:obs-elec}

The KS calculations provide immediately the total energy $E$ and the
electron density $\rho(\vec{r},t)$. The energy is a key observable
allowing to distinguish, e.g., ground state from isomers. The separate
contributions to the energy, as disentangled in Sec.~\ref{sec:energ},
allow to gather insight into the energy balance in reaction
processes. The single electron energies $\varepsilon_n$ do not
belong to the ``guaranteed'' observables in the framework of density
functional theory \cite{Dre90}, but are useful and often looked at. In
fact, the SIC helps to put the $\varepsilon_n$ into the correct
relation to the continuum which enhances their value as an analyzing
instrument~\PGmod{\cite{Leg02}}.
The electron density $\rho(\vec{r},t)$ is very informative, but too
bulky for easy inspection. Useful reduced observables are the various
multipole moments from which the dipole momentum 
and the root-mean-square radius,\PGmod{
\begin{eqnarray}
\vec{D}(t)&=&e\int \textrm d \mathbf r\,\vec{r}\,\rho(\mathbf r,t)
\quad,\quad
r_{\rm rms}(t)=\sqrt{\int \textrm d\mathbf r\,r^2\,\rho(\mathbf r,t)/N_\mathrm{el}}  
\label{eq:diprms}
\end{eqnarray}
}
are the most prominent representatives. 

The time-dependent dipole moment does allow to deduce the dipole
spectra (optical absorption strength) by means of spectral analysis
\cite{Cal95a,Yab96,Cal97b}. To that end, one starts from a well
relaxed ground state and initializes dynamical evolution by an
instantaneous dipole boost
\begin{subequations}
\label{eq:spectr-anl}
\begin{equation}
  \varphi_n(\vec{r},t\!=\!0)
  =
  e^{\mathrm{i}\eta\hat{D}}
  \varphi_n^{(0)}(\vec{r})
\end{equation}
where $\varphi_n^{(0)}$ are the s.p. wave functions from the
ground state and $\eta$ is the boost momentum, chosen small enough to
run the analysis in the linear domain. Subsequent propagation with the
time-dependent equations (see Sec.~\ref{eq:dyn-EoM}) yields the set
$\varphi_n(\vec{r},t)$ and with it, the dynamical dipole momentum
$D(t)$. The dipole strength distribution is finally obtained from
Fourier transform in frequency domain as
\begin{equation}
  \mathcal{S}_D(\omega)
  =
  \sum_n\delta(\omega-\omega_n)|\langle\Phi_n|\hat{D}|\Phi_0\rangle|^2
  \propto
  \Im\{\tilde{D}(\omega)\}
  \quad,\quad
  \tilde{D}(\omega)
  =
  \int dt\,e^{\mathrm{i}\omega t}D(t)
  \quad.
\label{eq:strength}
\end{equation}
\end{subequations}
The simulation runs, of course, over a finite time span such that
$\Im\{\tilde{D}(\omega)\}$ is a sum of ``delta'' functions with finite
width. The spectral resolution increases with increasing simulation
time and should be carried as far as to distinguish the different
eigen-frequencies $\omega_n$ in the excitation spectrum.
The strength $\mathcal{S}_D(\omega)$ is, in fact, not the
photo-absorption cross-section, which is a similar expression, but with
weight $\omega_n$ in the summation \cite{Fai87}. It can be obtained in a
similar fashion by initializing with a dipole shift rather than a
boost \cite{Cal97b}.

Absorbing bounds allow to attain a further class of observables associated with
ionization. The absorption leads to a loss of norm of each
s.p. state. This, in turn, allows to compute the ionization
out of state $p$ as $n_{\mathrm{esc},p}$ and also the total
ionization $N_\mathrm{esc}$, i.e.\PGmod{
\begin{equation}
  n_{\mathrm{esc},p}(t)
  =
  1-(\varphi_p(t)|\varphi_p(t))
  \quad,\quad
  N_\mathrm{esc}(t)
  =
  \sum_{p=1}^N n_{\mathrm{esc},p}(t)
  \quad.
\end{equation}
}
These are averaged quantities as any observable computed from a mean
field state, e.g.\PGmod{,} the average number of emitted electrons
$N_\mathrm{esc}$ representing an ensemble averaged over many similar
measurements.
The $n_{\mathrm{esc},p}$ even allow to deduce the detailed
ionization probabilities with the help of some combinatorial analysis,
for details see \cite{Ull97a,Cal00}.

The absorbing bounds do also allow to keep track where electronic
density has been removed from, by accumulating
$\rho_{\mathrm{abs},p}(\vec{r})=
 \int \textrm dt\,|\varphi_p(\vec{r},t)|^2\left(1-\mathcal{M}(\vec{r})\right)$ 
over all masking steps. Integrating the loss
$\rho_{\mathrm{abs}(\vec{r}),p}$ in angular sectors from the origin
(practically over the absorbing margin) yields the angular
distribution of electrons $n_{\mathrm{esc},p}(\theta,\phi)$
emitted from state $p$, or of the total angular distribution
$N_\mathrm{esc}(\theta,\phi)$ 
when summed over all s.p. states. For details, see
\cite{Poh03a,Poh04b,Bae08a}.

Photo-electron spectra (PES), i.e. the kinetic energy spectrum of
emitted electrons, are a further crucial observable obtained from
electronic dynamics\PGmod{. They provide} information on the occupied s.p. states
before ionization when driven with one-photon processes
\cite{Fai87,Poh00} and \PGmod{allow} to study the subtle interplay between direct
and sequential emission in the multi-photon regime \cite{Poh01}.
Early studies explored the s.p. spectra of anions with laser light in
the visible range \cite{McH89}. The availability of high quality UV
sources allows meanwhile to analyze neutral clusters, see e.g.
\cite{Cam00,Hof01,Wri02}.
The PES are computed as follows.  We define a measuring point
$\vec{r}_\mathrm{bc}$ near the boundaries of the grid.  We record the
time evolution of the s.p. wave function
$\varphi_p(\vec{r}_\mathrm{bc},t)$ at that point.  Finally, we compute
the local frequency spectrum of the electronic wave functions by
Fourier transform
$\varphi_p(\vec{r}_\mathrm{bc},t)
 \longrightarrow\tilde{\varphi}_p(\vec{r}_\mathrm{bc},\omega)$\PGmod{.}
The absorbing boundary conditions guarantee that only outgoing waves
are passing by $\vec{r}_\mathrm{bc}$. The frequency spectrum
$|\tilde{\varphi}_p(\vec{r}_\mathrm{bc},\omega)|^2$ is thus
immediately the PES of electrons emitted from state
$p$. Altogether the total PES of all electrons together becomes
\begin{equation}
  n(E_{\rm kin})
  =
  \sum_{p=1}^{N_{\rm el}} |\tilde{\varphi}_p(\vec{r}_\mathrm{bc},E_{\rm
  kin})|^2 
  \quad.
\label{eq:nekin}
\end{equation}
It is found that this procedure maps the s.p. energies of the occupied
states directly into the PES with peaks at
$\varepsilon_p\longrightarrow\varepsilon_p+\nu\hbar\omega$
where $\nu$ is the number of photons involved in the process. It is
thus crucial for 
a correct description of experimental data that the position of the
$\varepsilon_p$ relative to the electron continuum is correct.
That feature is indeed established  when applying (AD)SIC, as we will do
in all following examples.

\subsubsection{Observables from ions and atoms}

The ionic/atomic configuration, which is fully classical, 
 is characterized by the set of
$\vec{R}_I(t)$. That again is a lot of information which can hardly be
viewed at once. \PGmod{We} will also look for reduced observables which
are obtained by averages over subsets (cluster ions or environment
atoms) or by cuts to watch trajectories in reduced dimensions.
The corresponding cuts and averages are self-explanatory and will be
introduced in connection with each application.
Note that vibrational spectra of cluster and/or environment can also be obtained
by spectral analysis similar to the case of electrons.  To that end,
one records various multipole moments during time evolution and
finally Fourier transforms them into the frequency domain
\cite{Rei02b,Feh06b}.

\PGcomm{Important further information is contained in the dipole
degrees of freedom of the environment.  The dipole } moments (which are
attained from the difference of localization between core and valence
clouds) are classical quantities. They can be characterized by an
amplitude and an orientation. \PGcomm{The squared amplitudes scale
the energy stored in the dipoles. In order to analyze the response of
the environment, we will look at the total dipole energy (summed squared
amplitudes) as well as at the spatial distribution of dipole energies,
see  for example Sec.~\ref{sec:sub-dip}.  }

\subsection{Limitations of the modeling}
\label{sec:limits}

\begin{SCfigure}[0.5][htbp]
{\includegraphics[width=0.6\linewidth]{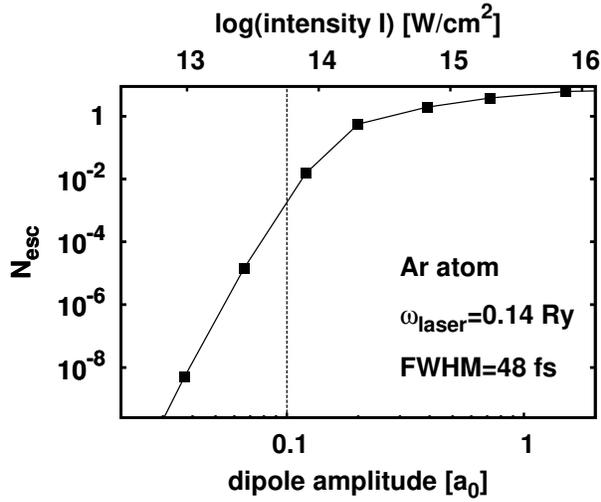}}
\hspace*{0.5em}
\caption{
Ionization rate for a free Ar atom under the influence of a
laser pulse with frequency of 1.9 eV, pulse length of 48 fs and
varying intensity (upper axis labeling). The lower $x$-axis shows the
corresponding maximum dipole amplitude of the motion induced by the
laser.
\label{fig:Ar-dipcrit}}
\end{SCfigure}
\PGcomm{The present QM/MM model includes the dynamical polarizability which
constitutes a big step forward to account for the dynamical response
of the environment. However, the description remains limited because
environment electrons are still encapsulated in the dipole moment and
will not be allowed to develop their own degrees of freedom. This may become
a problematic approximation if external fields grow sufficiently large 
to ionize the environment atoms. }
\EScomm{
At the level of our model, such a restriction directly shows up in
terms of a maximum amplitude \PGcomm{which the atomic dipole moment is
allowed to acquire,} corresponding to an excitation energy which would
lead to a sizable ionization probability. One thus needs to quantify
this effect in order to evaluate under which dynamical conditions our
approach is reliable. A simple way to quantify such an effect is to
perform a calculation on an Ar atom within treating its electrons
explicitely at LDA\PGmod{\&ADSIC} level
\PGmod{[see Eq.~(\ref{eq:ADSIC})]}. 
The Ar atom is then irradiated by a laser
and one records in TDLDA the net ionization.  }
\PGcomm{
Fig.~\ref{fig:Ar-dipcrit}
shows the result for an Ar atom irradiated by a short laser pulse
having a typical frequency in our applications
}
\EScomm{(the result 
depends a  bit on the laser frequency 
but \PGmod{but suffices for an order of magnitude estimate})}.
\PGcomm{Results are drawn versus laser intensity and, proportional to
that, the maximum dipole amplitude of the resulting atomic
oscillations. The vertical line indicates what we consider as a
critical intensity, or amplitude respectively. Below that line, the
ionization stays with maximally 0.001 electrons bearable. The critical
laser intensity of about 10$^{14}$W/cm$^2$ is above what we are going
to use in our explorations. But one should be careful. The actual
field strengths at the atom site can exceed the external field
strengths due to field amplification \cite{Rei98b}. The dipole
amplitude of the atom  (lower $x$-axis in Fig.~\ref{fig:Ar-dipcrit})
is the safer signal because it is determined by
the actual near-side field.  Thus a simple way to keep track of the
validity of the approach is to record the actual (time-dependent)
dipole moments at each atom.
We  have thus checked in our dynamical simulations that we were not
entering the dangerous domain, even also in the case of violent explosion
scenarios (see Sec.~\ref{sec:light}).  }
\EScomm{We are thus on a safe ground for exploring
a bunch of dynamical scenarios.  }


\section{Low energy properties}
\label{sec:struct}

In this section, we review structural properties and small amplitude
excitations (optical response) for metal atoms and clusters
in contact with an insulator environment. Numerous works exist for the
case of \MDcomm{an oxide surface~\cite{Cam97,Hen98b,Hen05}}, while rare
gas material was scarcely 
addressed. We perform the survey using our hierarchical
approach (as outlined in Sec.~\ref{sec:model}).  First, we discuss
ground state structures, stepping from atomic adsorbate to clusters.
Then we address the features of optical response.  At the side of
environment, we consider two polarizable and insulating materials, MgO
with large mechanical resistivity and substantial corrugation, and Ar
as a soft material with little corrugation.

\subsection{A single atom in an environment}
\label{sec:lowE_atom}

The simplest system in contact with an environment is \PGmod{a} single atom
deposited on a surface or embedded in a matrix. The topic has been
widely studied, mostly from two complementing points of view: one
adatom on a surface or a "mixed" cluster containing one atom in
contact with homonuclear clusters. Adatoms focus on surfaces and
infinite systems, while mixed clusters are usually attacked from the
chemical point of view and focus on small systems. Both approaches
bring complementing information, at various levels of detail.  As our
aim is to study clusters in contact with a (much) larger environment,
we will mostly discuss the case of atoms in/on extended systems but we
will also briefly consider the case of small mixed clusters (section
\ref{sec:smallchi}) for which direct comparisons between our
calculations and benchmark results are possible.

\subsubsection{Metallic adatoms in general}
\label{sec:lowenergy_adatoms}

\paragraph{Deposit on MgO(001) surface}
Adsorption of metal atoms on insulator surfaces has been extensively
\MDcomm{studied~\cite{Yud97,Mat99,Mus99b,Yan02b,Ney04,Bar05,Del05,Coq06,Bar07b,Xu08}.}
The case of MgO provides a standard example for a
simple, still realistic, insulating surface.  For instance
in~\cite{Yud97}, Cr, Mo, W, Ni, Pd, Pt, Cu, Ag, and Au adatoms on
MgO(001) are considered in DFT calculations with gradient corrections.
It is shown that the O site is always the preferred adsorption place.
This is attributed to the large polarizability of the O$^{2-}$ ion
which produces an attractive force on the metal atom electron cloud,
while Mg$^{2+}$ sites tend to repel the atom. More recently,
in~\cite{Bar07b}, DFT calculations on Au and Ag atoms adsorbed on
MgO(100) report similarly a site preference on top of an O site and an
atom diffusion by hopping from an O site to another O site via a
hollow site.

\paragraph{Deposit on a layer of MgO(001) supported on metal substrate}

It is interesting to note that the 
preferred site can be modified if the oxide surface is
itself supported on a metal substrate. Then, a charging of the adatom
can become possible, depending on the adsorbed metal. The first
experimental report of such a phenomenon was done for Au adatoms on
NaCl supported on Cu(111) and on Cu(100)~\cite{Rep04}. Different
charge states of Au were obtained after interaction with the voltage
of a STM \PGmod{(Scanning Tunneling Microscope)} tip. These results were in agreement with DFT calculations
(performed by the same authors) which found two different stable
states, a neutral one and a negatively charged one. This charging
effect was explained by inelastic electron tunneling. Later on, Pd and
Au adsorbed on thin MgO(100) films supported on Mo(100) were studied
by DFT calculations~\cite{Pac05}. For Pd, no significant change
between pure and supported MgO are observed, in accordance with a
small charge transfer. On the contrary, for Au, the Mg site is
preferred when deposited on MgO/Mo, because there is a large charge
transfer, explained by a reduction of work function of the mixed
substrate (compared with that of the bare Mo) and by  electronic
tunneling. That work was continued in~\cite{Gio05}, where adsorbed Pd,
Ag and Au are considered. Once again, the reported charge transfer is
very small for Pd, already significant for Ag, and largest for Au.
\MDcomm{This effect actually reflects increasing electronic affinities
when going from Pd ($-54.24$ kJ/mol) to Ag ($-125.86$ kJ/mol) and Au
($-222.75$ kJ/mol).}
These different
behaviors of Pd and Au atoms have been confirmed experimentally
in~\cite{Ste07a} when these species are adsorbed on thin MgO films
supported by Ag(001).  
The ranking on the preferred site for Au is modified when MgO is
supported on Mo~: hollow $>$ Mg$^{2+}$ $>$ bridge $>$ O$^{2-}$
(see Fig.~\ref{fig:PES_Na_MgO} for site terminology).
Finally, in~\cite{Hon07}, DFT calculations have been 
used to study Au adatom on thin MgO(100) films, with or without O
vacancies, supported on Mo or Ag. On Mo-supported MgO, Au prefers the
hollow site, with a negative charging of about 0.7--0.8, coming from a
fair share of electron loss between MgO and Mo. When MgO is supported
on Ag, the loss of electron seems to come mainly from MgO and much
less from Ag.

\subsubsection{Effects of deposition site}
\label{sec:Na_MgO_PES}

\begin{figure}[htbp]
\begin{center}
\begin{minipage}{40mm}
\fbox{
\epsfig{file=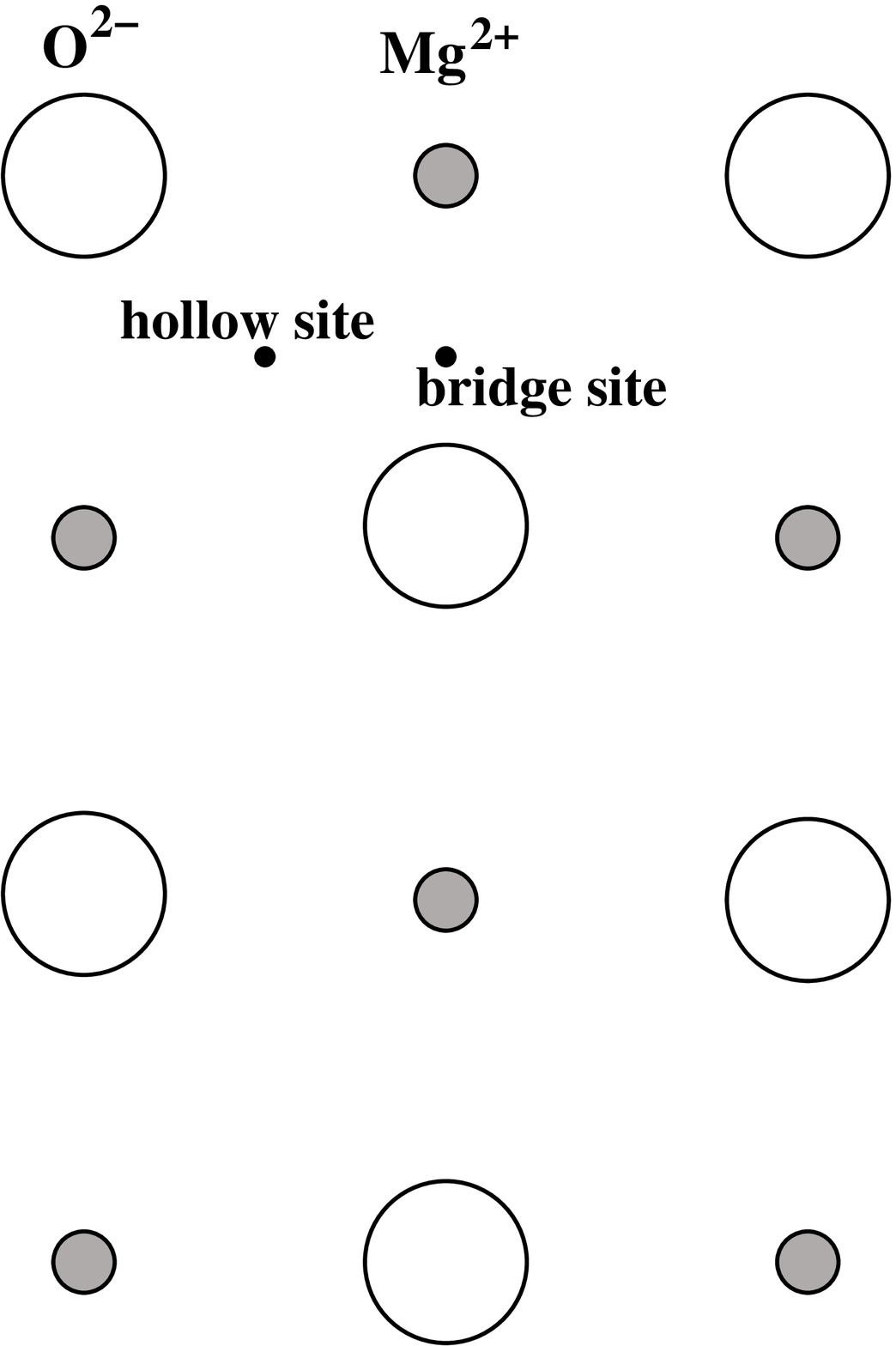,width=36mm}
}
\end{minipage}
\hspace*{9mm}
\begin{minipage}{76mm}
\noindent
\epsfig{file=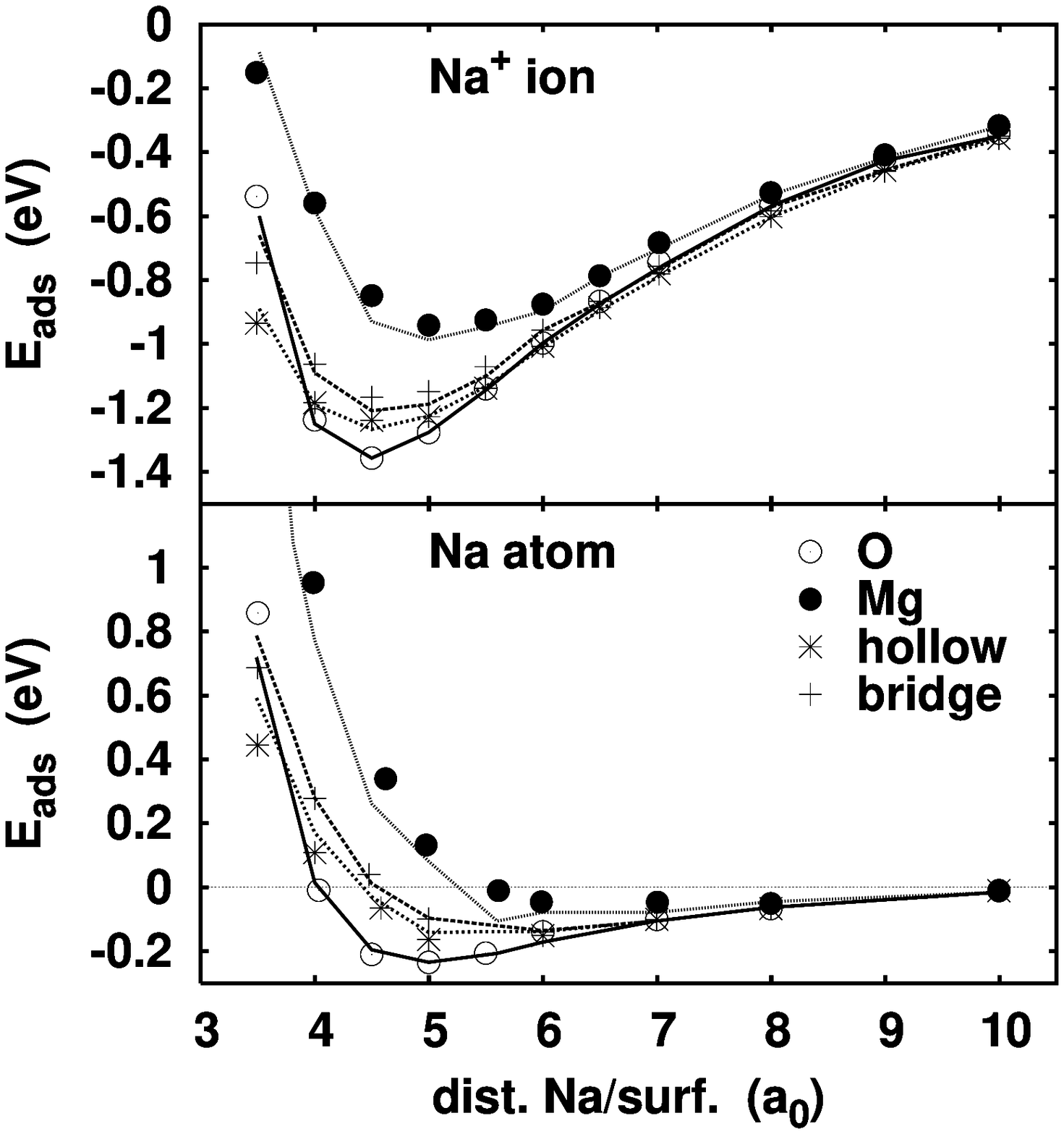,width=76mm}
\end{minipage}
\caption{
Left: Schematic view of the MgO(001) surface. O$^{2-}$ ions are
indicated by large open circles, Mg$^{2+}$ ions by small gray circles.
The four deposition sites are indicated.
Right:
Potential energy surface of a single Na atom (bottom) and a
  Na$^+$ ion (top) above O, Mg, hollow and bridge sites of a MgO(001)
  surface~\cite{Bae08a}. Lines come from quantum chemical
  calculations considered as benchmarks~\cite{Win06}, while points
  result from tuning of the MgO/Na pseudo-potentials in our
  hierarchical model.}
\label{fig:PES_Na_MgO}
\end{center}
\end{figure}

As a first test case of our hierarchical model, we consider the
influence of the surface sites 
on a Na atom adsorbed on a MgO(001) substrate.  That case of a
MgO(001) surface shows already a very rich surface structure.  We have
investigated four locations of the Na atom~: above an O$^{2-}$ site, a
Mg$^{2+}$ site, a hollow site (center of a square) and a bridge site,
see left panel of Fig.~\ref{fig:PES_Na_MgO}.
The right panel of Fig.~\ref{fig:PES_Na_MgO} 
shows the adsorption energy as a function of the Na atom distance
to the surface for the four sites.
The absorption energy is defined as
\begin{equation}
\label{eq:Eads}
E_{\rm ads} = E_{\rm Na/MgO} - E_{\rm Na} - E _{\rm MgO} \quad,
\end{equation}
and similarly for the ionic species Na$^+$.  The figure compares
benchmark quantum calculations~\cite{Win06} to results from our
hierarchical model.
First, one notes a very good agreement between our model and the ab
initio calculations.  Second, one observes a clear ranking in
the preferred adsorbed sites~: O$^{2-}$ $>$ hollow $>$ bridge $>$
Mg$^{2+}$, in accordance with the general discussion of
Sec.~\ref{sec:lowenergy_adatoms}. This affects both the value of the
energy 
binding the atom to the surface and the equilibrium distance at which
the adatom sits on the surface. The trend is monotonous, as
expected. Finally, an important aspect concerns the impact of
charge. In the top right panel of Fig.~\ref{fig:PES_Na_MgO}, we consider
the case of Na$^+$. One observes the same hierarchy as in the neutral
case (except for the point very close to the surface), but a large
quantitative difference in the values of the energies and equilibrium
distances to the surface. The Na$^+$ ion lies closer to the surface (1
$a_0$ or more) and is much better bound (between 0.8 and 1.2 eV
binding as compared to less than 0.2 in the neutral case). This
indicates that charge effects  play an important role, see also
Secs.~\ref{sec:depos_charge}, \ref{sec:sub-dip}, and
\ref{sec:light}.

The Ar(001) surface shows a much smaller site dependence. The
difference in binding between hollow and Ar site is less than 6 meV.
This material has much less corrugation because it consists of neutral
atoms instead of $\pm 2$ \PGmod{charged} ions and because Ar atoms are
more softly bound such they can more easily give way.

\subsubsection{Small composite systems}
\label{sec:smallchi}

At the other extreme, small clusters containing one metal atom
\MDcomm{or ion} 
constitute interesting test cases which have been investigated by
various quantum chemistry
methods, \MDcomm{for neutral NaAr$_n$
complexes~\cite{Boa94,Tso90,Tso92,Rho06a,Rho06b,Tut98,Rho04,Gro98},
M$^+$RG$_n$ (where ``M'' denotes a metal element and ``RG'' a rare
gas) for alcalines~\cite{Nag04,Fro00,Gij04,Rho06a}, Ni$^+$ and
Pt$^+$~\cite{Vel98b}, and Au$^+$~\cite{Zha09}. Indeed the study of
small rare gas metal-doped clusters provides better understanding of
micro-solvation effects and chromophore issues when metal
atoms/ions/clusters are in contact with ``inert'' solvents as rare gas
are.
Such systems also provide valuable}
benchmarks for the structure part of our modeling.  We thus will now
discuss mixed systems consisting in Ar clusters of various sizes with
\PGmod{one} Na atom deposited on, or embedded in it.  The questions of
interest concern the stability of the mixed system and the preferred
site of the Na atom, inside or outside the Ar cluster.  
We first study the simple case of a small  NaAr$_6$ system. The Ar
cluster is so small that the Na atom has no choice but to be on the
``surface'' of Ar$_6$.
\begin{SCfigure}[1.0][htbp]
\epsfig{file=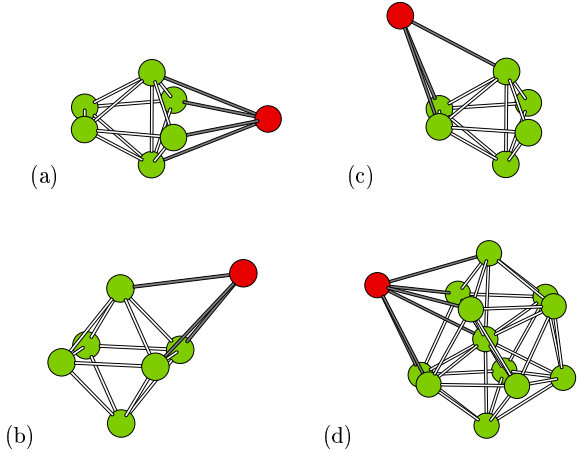,width=86mm}
\caption{Possible locations of Na (dark ball) on Ar$_6$ (gray balls)
  or Ar$_{12}$ (panel d).
  \label{fig:na_ar6}}
\end{SCfigure}
Fig.~\ref{fig:na_ar6} illustrates the three
possible configurations.
Table~\ref{tab:na_ar6_12} compares static results (distance between Na
and nearest neighbors, total energy) from different approaches.
\begin{SCtable}[0.5][htbp]
\begin{tabular}{|l|c|c|c|c|} \hline
NaAr$_6$ & References & 1NN \,$(a_0)$ & 2NN \,$(a_0)$ & E$_{\rm
  tot}$ \,(eV) \\ \hline
Config. a & this work     & 9.27  & 14.22 & $-$5.118 \\
          & \cite{Tso92}  &       && $-$5.289 \\
          & \cite{Boa94}  & 9.52  &&   \\ 
\hline
Config. b & this work     & 8.89  & 14.37 & $-$5.117 \\
          & \cite{Tso92}  &       && $-$5.287 \\ 
\hline
Config. c & this work     & 9.03  & 13.78 & $-$5.114 \\
          & \cite{Tso92}  &       && $-$5.283 \\ 
\hline\hline
NaAr$_{12}$ & this work   & 9.34  & 14.20 & $-$5.312 \\
            & \cite{Boa94}& 9.39  &&           \\
\hline
\end{tabular}
\caption{Comparison of our results for NaAr$_{6,12}$ complexes
  with other works, as indicated, for the various ionic configurations
  (a-d) presented in Fig.~\ref{fig:na_ar6}.
  The third and fourth columns report the mean distance between Na and
  nearest neighbours (1NN), and between Na and second nearest
  neighbours (2NN) respectively.
The last  row presents the case of Na on Ar$_{12}$.}
\label{tab:na_ar6_12}
\end{SCtable}
The comparison shows that our hierarchical model stays in good
agreement with results from more elaborate calculations.  This
indicates that our modeling allows to account for details of
structure, a feature which which will play a role even in the violent
dynamical scenarios to be discussed later on.

We now consider systems with an increasing number $N$ of Ar atoms.
This allows to hide the Na atom inside the Ar cluster.  To quantify
the question of the most stable location of Na (in or out), we define
the insertion/adsorption energy for Na in/on Ar in a way similar to the adsorption
energy for MgO, see equation  (\ref{eq:Eads}), as~:
\begin{equation}
  E_{\rm ins/ads}
  =
  E_{{\rm Na/Ar}_N} - E_{\rm Na} - E_{{\rm Ar}_N} 
\quad.
\label{eq:Eins}
\end{equation}
Note that this definition is biased towards an adsorption energy, see
Eq.~(\ref{eq:Eads}). A true insertion energy will be defined later on
in Eq.~(\ref{eq:Eins_Na8}). Here we want to have a unified definition
for better comparison of adsorption versus insertion. \PGmod{(}We have tested
also the more general definition (\ref{eq:Eins_Na8}) and it does not
alter the conclusions drawn here.\PGmod{)}
\begin{SCtable}[0.6][htbp]
\begin{tabular}{|l|r|r|r|r|}
\hline
$N$ & 6 & 12 & 20 & 24 \\
\hline
$E_{\rm ins}$ (eV) & & 1.09 & 0.32 & 0.21 \\
\hline
$E_{\rm ads}$ (meV) & $-$17.0 & $-$5.19 & $-$49.2 & $-$27.2 \\
\hline
\end{tabular}
\caption{\label{tab:Na_ArN_small}
Energy gain (\ref{eq:Eins}) through insertion (upper) or adsorption
(lower) of a Na atom in/on Ar$_N$ for various $N$.
}
\end{SCtable}
Table~\ref{tab:Na_ArN_small} shows results for sizes $N=6,12,20,24$.
Deposition at the surface is clearly preferred over embedding the Na
atom inside the Ar cluster. 
\MDcomm{This agrees with former findings of ground state structures of
NaAr$_{1-10}$ where the Na atom lies at the surface of the Ar
cluster~\cite{Rho04}. Moreover,}
embedding even gives positive insertion
energies which means that such configurations are asymptotically
unstable. We find them, however, locally stable. The embedded
configuration has to surmount some reaction barrier to release the Na
to a surface site.

\subsubsection{Towards larger environments}

We have just seen that for small
complexes, embedding is not favored. The question remains
to be checked for larger Ar clusters.
We thus consider larger systems and analyze the trend of
insertion energy
(in view of the numerous possibilities of
impinging sites for an adsorbed Na, we have not considered the case of
adsorption, which leads, anyhow to small binding on the few test cases 
we have explored, in agreement with the more detailed case of smaller
clusters, see table \ref{tab:Na_ArN_small} and discussion around it).
The systems are built by 
starting from bulk
fcc Ar structures of various sizes,
$N=$55 to 447 atoms, replacing the central Ar atom by Na, and
re-optimize the obtained structure.
\begin{SCfigure}[0.2][htbp]
\epsfig{file=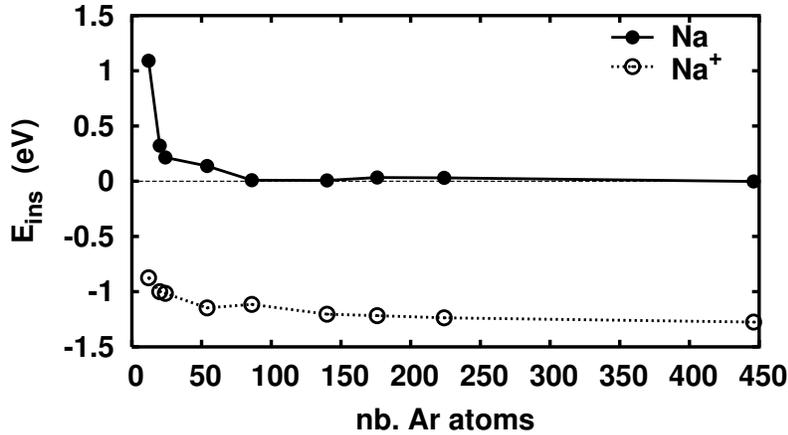,width=0.8\linewidth}
\caption{Insertion energy, Eq.~(\ref{eq:Eins}), of a single Na atom
  (black circles) or a Na$^+$ ion (white circles) in Ar matrices of
  different sizes.
\label{fig:na_nap_insert}}
\end{SCfigure}
Fig.~\ref{fig:na_nap_insert} shows insertion energies for
embedded Na atom and Na$^+$ ion. 
The insertion energy for the Na atom rapidly becomes independent of
matrix size and stays very close to zero, fluctuating about $\pm 20$
meV.  The various contributions to the total energy are: enhanced
electronic kinetic energy (positive), an enhanced Van der Waals
binding (negative), and the core repulsion (positive). They nearly
compensate each other with a tendency to remain slightly positive.
The situation is much different for the positively charged Na
ion. The polarization interaction with the Ar atoms is much stronger
and attractive (negative energy contributions).  Thus embedding an ion
always gives sizeable negative insertion energies, which quickly converges
towards a constant (bulk) value.

Let us finally step to adsorption on an infinite, planar surface.
\begin{SCfigure}[0.5]
\epsfig{file=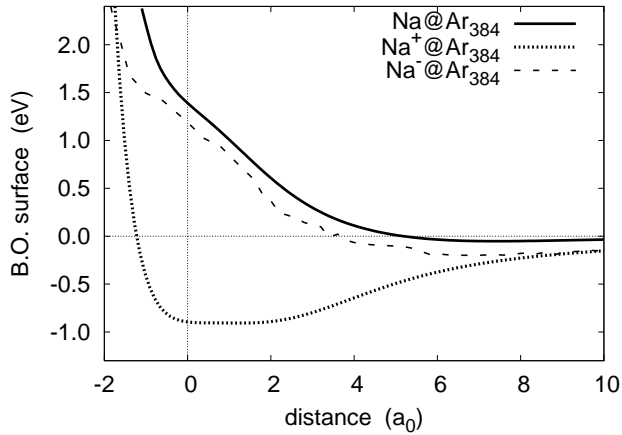,width=86mm}
\caption{
Potential energy curves of neutral Na, Na$^+$ and Na$^-$ on
Ar(001) as a function  of the distance of the monomer to the
first surface layer. The energies were calculated with fixed Ar atoms.
The surface is represented by  ${384}$ Ar atoms.
  \label{fig:bo}}
\end{SCfigure}
Fig.~\ref{fig:bo} shows the potential energy curves for
a Na atom, and Na$^+$ as well as Na$^-$ ions close to  an  Ar(001)
surface. The surface is modeled by 6 layers of 8$\times$8 Ar atoms,
all frozen at the optimized positions of a free Ar surface.
The neutral Na atom is faintly bound to the surface with a bond
distance of about 8 $a_0$. The Na$^+$ cation takes full advantage of
the strong, attractive polarization interaction and is practically
soaked into the surface. The Na$^-$ sees counteracting effects, the
attractive polarization interaction and the core repulsion on
the two Na valence electrons. This combines at the end to a significantly
better binding than for the neutral Na atom. But Na$^-$ is kept at a
safe distance from the surface, in contrast to Na$^+$. These potential
energy curves explain the different behaviors observed in deposition
of these three different monomers, see Sec.~\ref{sec:depos_charge}.

\subsubsection{Impact of environment on properties of embedded atoms}

In this section, we are going to analyze in more detail the impact of
the environment (especially its size) on atomic properties.  We
investigate the stationary \PGmod{state} in terms of a few global
observables: the r.m.s. radius $r_\mathrm{rms}$ of the electronic
cloud of Na, the distance of the Na atom to the first surrounding Ar
shell, and the dimensionless quadrupole deformation $\beta$ for the
shape of the Ar system. $\beta$ is defined from the quadrupole moment
normalized by particle number and r.m.s. radius
\cite{Rin80,Cal00,Rei03a}
\begin{equation} 
  \beta
  =
  \sqrt{\frac{\pi}{5}} \;\frac{1}{N r_\mathrm{rms,Ar}^2} \;
  \sum_{i=1}^{N} \; (3z_i^2-r_i^2)
  \quad,\quad
  r_\mathrm{rms,Ar}^2
  =
  \frac{1}{N} \sum_{i=1}^{N}  \; r_i^2 
  \quad.
\label{eq:rms_beta}
\end{equation}
Furthermore, we consider as spectroscopic
quantities, the 
${3s}\longrightarrow{3p}$ transition, which is especially relevant
for optical properties, and the energies of the $3s$- and $3p$-state
separately.
\begin{figure}[htbp]
\begin{center}
\epsfig{file=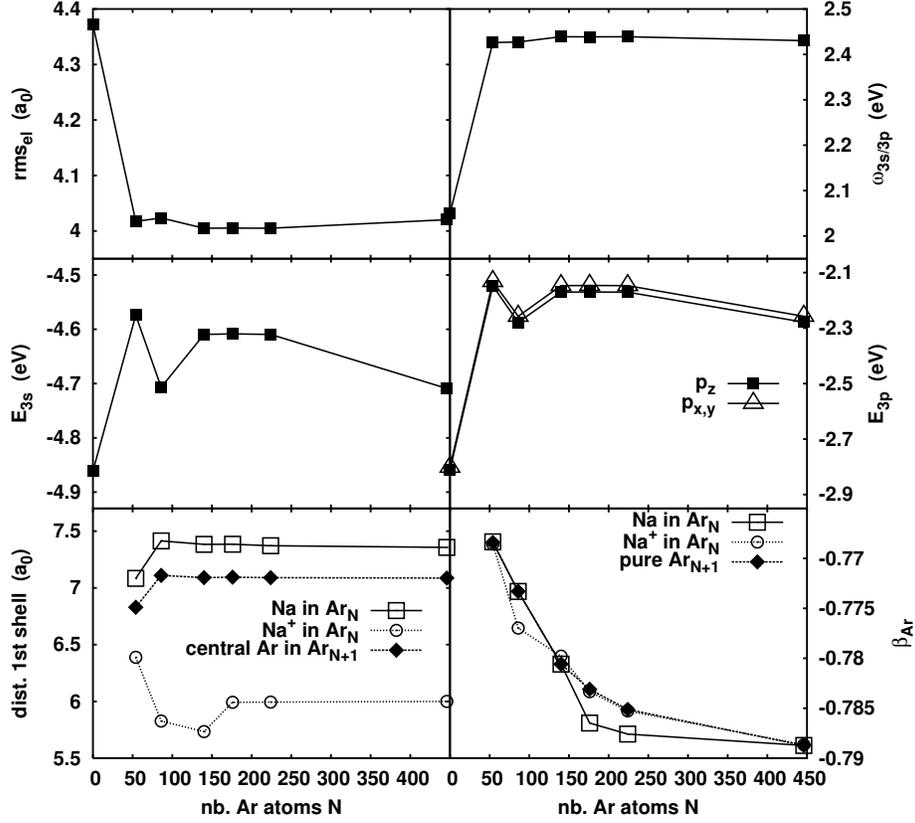,width=0.9\linewidth}
\caption{Global observables for a Na embedded in Ar$_N$, in a
  configuration where the Na replaces the center Ar atom.
  The four upper panels show electronic properties of the Na atom:
  lower left = distance from center (atom) to the first atomic shell,
  middle left = electronic r.m.s. radius, 
  lower right = energy is $3s$   state, 
  middle right = energy of $3p$ state distinguishing $x,y$- and
  $z$-mode, 
  upper right = $3s\rightarrow 3p$ transition energy.
  The matrix properties (lower left) are also shown for a Na$^+$ ion
  instead of the neutral Na atom, and for pure Ar clusters with the
  original Ar atom at the center (instead of the Na).
\label{fig:na_arN_struct}}
\end{center}
\end{figure} 
Fig.~\ref{fig:na_arN_struct} shows these observables as a function of
Ar system size. 
Most remarkable is that all electronic observables make a jump from
free to embedded Na atom. The variation amongst the different cluster
sizes is much smaller than the initial jump. Moreover, all observables
tend to stabilize for Ar sizes at and above $N_\mathrm{Ar}=176$.
The jump in the electronic $r_\mathrm{rms}$ (left upper panel) is
caused by the core repulsion from the surrounding first Ar shell which
squeezes the electron cloud to a smaller volume.  That repulsive
effect is correlated to an increase in electronic energies (middle
panels).  The more extended $3p$ state sees a larger repulsive effect
than the $3s$ state with the consequence that the $3s\longrightarrow
3p$ transition does also make a jump from free to embedded atom (right
upper panel). Note the fine level splitting of the $3p$ state.  The
different energies in $x,y$- and $z$-direction indicate that the
surrounding Ar is slightly deformed. The left
lower panel shows the radius of the first Ar shell.  It is increased
when replacing the central Ar atom by Na. This shows again the mutual
pressure between electron cloud and Ar shell and correlates nicely to
the squeezing of the electron radius (upper left). Inserting, however,
the charged and electron-free Na$^+$ cation induces a large
polarization attraction which leads to a much smaller shell radius.

The results altogether illustrate the subtle interplay between
long-range polarization attraction and short-range core repulsion.
In spite of the long range of the polarization effects, most
observables level off quickly with increasing system size such that a
reliable model for an Ar matrix is achieved starting from
$N_\mathrm{Ar}=176$.

\subsection{Metal clusters in contact with an insulator environment}

Now that the case of one atom adsorbed on or embedded in a matrix has
been presented, we are going to address the case of a whole metal
cluster and see how it  differs from a single atom. The new aspect is
that the cluster has an internal structure which may react sensitively
to the environment.  We will thus explore changes in structural
properties, either at the side of the cluster itself or of
the environment. As an illustration of these questions, we will present in this
section a systematic study of Na clusters deposited on MgO(001), and
of Na$_8$ embedded in Ar clusters. We shall discuss the influence of
system sizes and of the polarization interaction. We begin with an
overview of various studies on MgO surfaces.

\subsubsection{Wetting of metal clusters on MgO surface}

As in the case of metal atoms adsorbed on MgO, the number of
theoretical works on small metal clusters adsorbed to MgO surface is
large\MDcomm{(for a recent review, see~\cite{Fer09a})}, since they
represent model systems for heterogeneous catalysis \MDcomm{(e.g.,
see~\cite{Yoo05})} .
Energetics of small deposited metal clusters (from dimer to tetramer)
provides information on diffusion mechanisms during a film growth
process. Similar to the case of one atom (see
Sec.~\ref{sec:Na_MgO_PES}), ground states are always found adsorbed on
an O site. Various motions from an O site to another O site are possible,
e.g. hopping via a hollow site (for Cu~\cite{Mus98}, Au~\cite{Del05},
Au$_2$ and Ag$_2$~\cite{Bar07b}), rolling (for Pt$_2$~\cite{Gro03},
Cu$_2$ and Cu$_5$~\cite{Mus99b}, Ag$_2$~\cite{Bar07b},
Pd$_4$~\cite{Xu05,Bar05}), or walking (Ag$_3$ and
Au$_3$~\cite{Bar07b}, Cu$_3$~\cite{Mus99b}, Pd$_3$~\cite{Bar05}).

Large metal clusters on a surface bring wetting into play as a new
feature \cite{Sch92aB}. The question is whether larger amounts of
metal atoms prefer to cover the surface as a film (wetting) or
contract to a drop with minimal contact to the substrate. This
important question has been much investigated.
Na clusters on NaCl surface experience a strong interface
potential, favoring wetting.
They thus prefer planar ground state
shapes when deposited, although three dimensional configurations are
competitive isomers \cite{Koh97a,Koh98a}. This was shown in a model
using effective potentials between NaCl and the cluster constituents
\cite{Koh97a} tuned to ab initio calculations of small Na clusters on
NaCl surface \cite{Hak96}.
No wetting behavior is observed in the case of Cu$_{2-29}$ deposited
on MgO(001)~\cite{Mus99a}, but rather a three-dimensional growth. The
same result is reported in the case of Ag$_N$ grown on MgO(001)
simulated by Hartree-Fock calculations combined with a thermodynamic
treatment~\cite{Fuk02}. In this work, the observed high mobility of Ag
adatoms favors island growth. This has been confirmed experimentally
in~\cite{Men07}~:
Reflectance measurements of Ag film growth on
MgO(001) as a function of film thickness and temperature found a
percolation regime. In the case of Pd clusters deposited on MgO(001),
a theoretical study even shows that, in addition to a 3D growth, a
transition towards fcc truncated pyramids seems to take place around
11-13 Pd atoms~\cite{Bar07a}.
\MDcomm{In accordance to these calculations, a DFT investigation on
very small Ag clusters adsorbed on MgO(100) also reported that the
cluster is preferentially bound perpendicular to the
surface~\cite{Bon07}. For larger transition metal (Ag, Au, Pt, and Pd)
clusters (from 30 to a thousand atoms), three-dimensional deposited
geometries have also been reported in the framework of a tight-binding
method~\cite{Ros06,Fer09b,Gon09}.}

Closer to the kind of systems that we  study with our model, DFT
calculations on thin alkali layers on MgO(001) for different coverage
were reported in~\cite{Sny00}. As for the adatom case, the minimum in
energy corresponds to a coverage above O sites, while hollow sites
give saddle points and Mg sites maximum of energy. A low metal
coverage seems unstable or at best metastable, with respect to 3D
island or cluster coverage.
Interestingly, there exist theoretical investigations of layers of
Pd~\cite{Gon99}, Li, Na and K~\cite{Zav03}, and Cu and Ag~\cite{Zav05}
deposited on MgO(111), which presents alternating layers of O and Mg.
They show that adsorption is stronger than in the case of MgO(001), thus
giving probably a wetting behavior instead of the 3D growth on
MgO(001). While Pd, Cu and Ag films can be stabilized either on O- or
on Mg-terminated surfaces, alkali layers still prefer O-terminated
ones.

\begin{figure}
\begin{center}
\epsfig{file=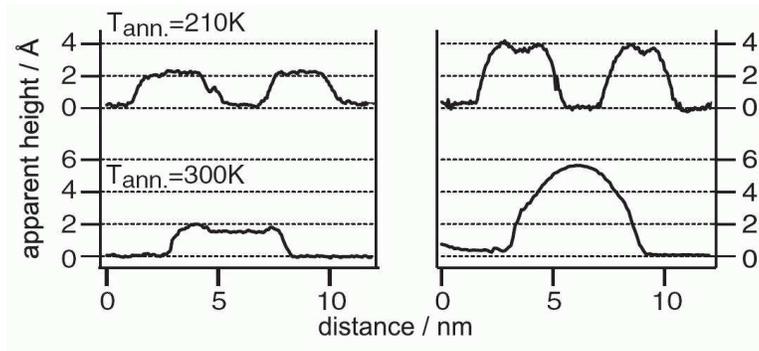,width=0.75\linewidth}
\caption{Height profiles of Au clusters deposited on 3 layer (left)
  and 8 layer (right) MgO(001) supported on Ag(001), for
  two different annealing temperatures as indicated, extracted from
  STM images. From~\cite{Ste07b}.
\label{fig:au_mgo_ag}
}
\end{center}
\end{figure}

A wetting behavior of metal clusters can be nevertheless observed when
MgO is itself a thin layer, supported by a metal substrate. Comparison
of deposited Au$_{8,16,20}$ on MgO(100) and MgO(100)/Mo(100) has been
performed in DFT calculations~\cite{Ric06}. Weak adhesion, small
exchanged charge and 3D structures for Au clusters deposited on pure
MgO are observed. On the contrary, when MgO is supported by Mo, a
sizable electronic charge at the interface between Au and MgO is
reported, producing a higher wettability of Au clusters with planar
geometries.
Deposition of Au atoms and film growth on MgO(001)/Ag(001) has been
recently explored experimentally by STM imaging for various
thicknesses of MgO and annealing temperatures~\cite{Ste07b}. 
Fig.~\ref{fig:au_mgo_ag} presents measured height profiles of Au
clusters deposited on 3 layer (left panel) and 8 layer (right panel)
MgO films, themselves supported on Ag(001), for two different
annealing temperatures. 
One sees preferably mono-layer islands for thin MgO films (left
panel), and thus a wetting behavior of Au clusters. In that case,
there is no strong preference between Mg and O sites. This reflects
the strong charging of the Au system. On thicker MgO films (right
panel), O sites are preferred and a 3D growth is observed, in
accordance with no charging of the adsorbate.
Hence,  playing with the thickness of metal-supported MgO(001)
allows to override the non-wetting trend of pure MgO(001) and to
switch experimentally a transition between film growth (wetting) and
3D growth of clusters. That has been predicted theoretically and can
be explained by a stronger charging of the deposited metal clusters
from electrons coming from the metal substrate via thin MgO layers.
Very recently, Au$_{1-6}$ and NO$_2$ on MgO(100) or Al$_2$O$_3$(0001)
supported on Mo, Ag, Pd, Au and Pt have been extensively studied
in~\cite{Fro08} by means of DFT simulations, with investigation on the
effect of oxide thickness, adsorbate coverage, choice of oxide, choice
of supporting metal, electron affinity of adsorbate. Contributions to
the stabilization of the metal clusters come from various effects,
such as the polarization of the substrates or the binding between the
metal adsorbate and the oxide. It seems however that no simple
correlation relates the stabilization energy of Au clusters to their
charging (of about 1 electron).

\subsubsection{Structure of sodium deposited clusters}
\label{sec:NaN_on_MgO}    

\begin{figure}
\begin{center}
\epsfig{file=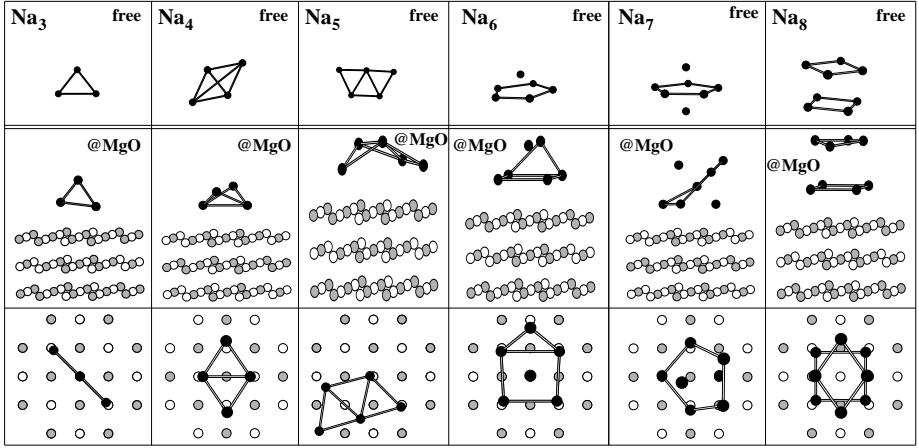,width=0.9\linewidth}
\caption{Na$_N$ structures for $N=3-8$ (black balls)~: Gas phase
  structures (top panels); geometries when deposited on MgO(001), with
  O$^{2-}$ appearing as white circles and Mg$^{2+}$ as gray ones, from
  a perspective (middle panels) or a top view (bottom
  panels). Adapted from \cite{Bae07a}. 
\label{fig:nan_mgo_struct}
}
\end{center}
\end{figure}

As a more detailed example, we discuss the structures of small Na
clusters deposited on a MgO(001) surface, described within our
hierarchical model~\cite{Bae07a}. Fig.~\ref{fig:nan_mgo_struct}
compares the free and deposited geometries obtained for these systems.
In contrast to NaCl, where the interface binding is
stronger~\cite{Hak96b,Koh97b}, the clusters show no clear preference
of planar structures, i.e. no trend to wet the surface.  The
clusters Na$_3$, Na$_4$, and Na$_5$ which are planar in free space are
bent into three dimensional structure to accommodate the strong
repulsion coming from the Mg$^{2+}$ sites.  \esmod{Note also that the 
MgO lattice constant does not match the Na one. In small Na clusters,
the effect is even stronger due to the different symmetries.}
A similar effect is seen
for the ring structures in Na$_6$ and Na$_7$.  \esmod{In turn,} the
very symmetric and 
compact Na$_8$ with its magic electron number remains almost
unchanged.
These findings agree with the above reported results that most metal
clusters on MgO(001) show no wetting.

\begin{figure}[htbp]
\begin{center}
\epsfig{figure=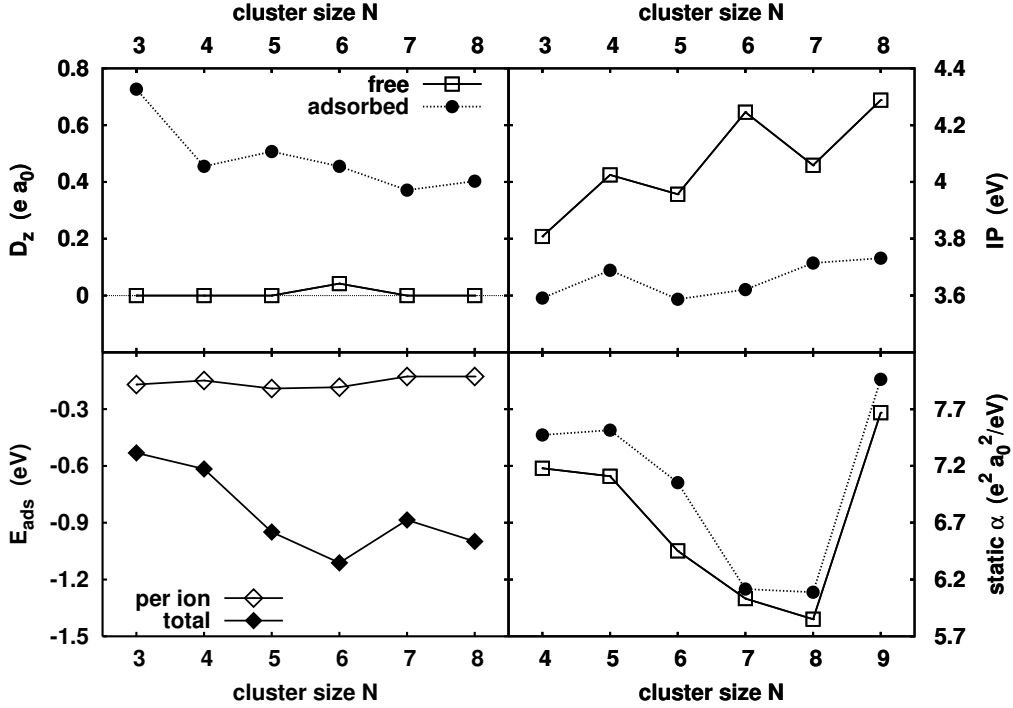,width=\linewidth}
\caption{\label{fig:nan_mgo_stat}
Electronic observables of free (open squares) and deposited on MgO
(full circles) Na$_N$ as a function of cluster size $N$.
Top left~: $z$-component (perpendicular to the optional surface) of the
electric dipole moment. Top right~: ionization potential.
Bottom right~: static polarizability. Bottom left~: adsorption energy,
Eq.~(\ref{eq:Eads}), of deposited Na$_N$. Adapted from~\cite{Bae07a}.
}
\end{center}
\end{figure}

In order to complement the structure view
(Fig.~\ref{fig:nan_mgo_struct}), we show in
Fig.~\ref{fig:nan_mgo_stat} some electronic observables of the free
Na$_N$ compared with that of the deposited clusters~\cite{Bae07a}.
The adsorption energy, as defined in Eq.~(\ref{eq:Eads}), is shown in
the bottom left panel.  It grows almost monotonously. There are
deviation\PGmod{s} from the monotonous trend in the step from Na$_6$ to Na$_7$.
The two larger clusters in the sample, Na$_7$ and Na$_8$, have a
relatively smaller contact area with the surface and so less
adsorption energy. 
Let us now turn towards electronic properties. The upper left panel of
Fig.~\ref{fig:nan_mgo_stat} shows the static dipole moment $D_z$
in the direction \esmod{normal} to the surface.  
While free clusters exhibit
vanishing\PGmod{ly small} static dipole moments, deposited clusters acquire
non-negligible values.  This happens even for the simplest case of
Na$_8$ which suffers least from ionic distortion when deposited. The
dipole moment then shows the direct influence of the MgO substrate on
the electronic charge distribution. The extension and deformation of
the electron cloud are well conserved, but the short-range repulsive
part of the interface potential pushes the electron cloud as a whole
away from the surface while the ions, staying farther away, feel much
less repulsion. This creates a (positive) static dipole moment.
The increase of $D_z$ occurs similarly for all clusters in this
survey and it is strongest for the weakest bound Na$_3$. 
The upper right panel of Fig.~\ref{fig:nan_mgo_stat} shows the
ionization potential for free and deposited clusters.  The substrate
induces a general reduction of the IP.  The effect is the same as had
been seen already in Fig.~\ref{fig:na_arN_struct} for the $3s$-state
of the Na atom embedded in Ar. It is again an upshift due to core
repulsion. Free clusters show the typical even-odd staggering of the
IP~\cite{Hee93} which is a spin effect \cite{Koh95}. This staggering
disappears for deposited clusters.  It seems that the strong
corrugation effects on cluster structure (see figure
\ref{fig:nan_mgo_struct}) overrule the spin effect.
The bottom right panel of Fig.~\ref{fig:nan_mgo_stat} shows static
polarizability $\alpha$ of Na$_N$, free and deposited.  Metal clusters
usually possess quite large polarizabilities, since the binding of the
valence electron cloud of a metal cluster extends over the whole
cluster and can respond with large amplitude. Sizes and trends are
very similar for both cases showing that the general extension of the
electron cloud is not much changed by the surface. The trend with $N$
reflects the sheöll structure. The polarizability shrinks towards the
shell closure at electron number $N=8$ and jumps up after closer
because the next electron has to go to the next shell which is much
more weakly bound.

\subsubsection{Embedded cluster -- structure and effects of matrix size}
\label{sec:na8_in_ar}

\begin{figure}
\begin{center}
\epsfig{file=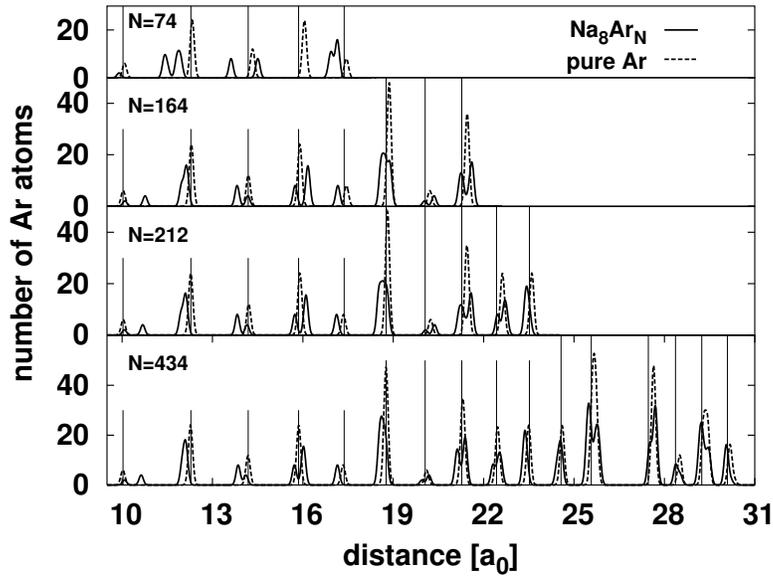,width=0.8\linewidth}
\caption{Radial distribution of Ar atoms in pure Ar$_{N+13}$
  (dashed lines) and Na$_8$Ar$_N$ systems (solid
  lines) after optimization of their structures. The vertical lines
  indicate the initial radial shells in the crystalline
  configuration. Adapted from~\cite{Feh06a}.
\label{fig:shell_na8_arN}
}
\end{center}
\end{figure}

In Sec.~\ref{sec:lowenergy_adatoms}, 
we have shown that embedding a Na atom in
Ar clusters has a significant influence on the static properties of
the mixed system. The observed modifications converge rapidly with the
number of surrounding Ar atoms such that only little changes are seen
above $N=164$. We address here the question of  Na$_N$ clusters
embedded in an Ar environment.
To that end, we consider large Ar clusters of various sizes as host
systems for an embedded Na$_8$~\cite{Feh06a}.  These clusters can be
arranged into radial atomic shells and we have chosen only cases with
closed radial shells to avoid artefacts from incomplete or uneven
surfaces. Similar as for the embedded atoms, the construction starts
from an infinite fcc crystals from which a sphere of wanted size is
cut. The structure of the pure Ar cluster thus defined is then
optimized by simulated annealing. The composite system is built by
excavating the 13 Ar atoms from the center and inserting the Na$_8$
cluster into the cavity.  Then the composite structures are again
optimized by simulated annealing.  The resulting system sizes still
exhibit closed ``atomic subshells'' (see dashes curves in
Fig.~\ref{fig:shell_na8_arN}). However they differ from the atomic
shell closures which are $N=13$, 55, 147, 309, obtained by freely
varying finite systems from scratch without reference to the bulk fcc
structure \cite{Hoa72a,Nor87a}.

Fig.~\ref{fig:shell_na8_arN} shows the structures in terms of radial
shells for various system sizes, full lines represent the radial Ar
distribution for the composite and dashed lines for the pure Ar
cluster. The Ar shells for bulk fcc are indicated by thin vertical bars.
The pure Ar clusters (dotted lines) come generally close to the bulk
fcc structure (vertical lines). The outer shells tend to be a bit more
extended than the corresponding bulk shell. This happens because the
pressure from the further shells in bulk is missing. That effect is
largest for the smallest system $N=74+13$ and becomes quickly
negligible for larger $N$. The inner shells are anyway always close to
bulk fcc shells.  Our mixed system can then be considered confidently
as a finite model of a metal cluster embedded in an ``infinite'' Ar
matrix.
The embedding of Na$_8$ induces significant changes which reach far
out, due to the long-range effects of the polarization potentials.
The changes are most visible for the smallest sample ($N=74$), but
persist for any size.  The well degenerated radial shells in pure Ar
matrices are often split into subshell structure with a double peak of
slightly different radii. This is due to the slight quadrupole
momentum of the Na$_8$ cluster which imprints its deformation onto the
surrouding Ar shells.
 
\begin{SCfigure}[0.5][htbp]
\epsfig{file=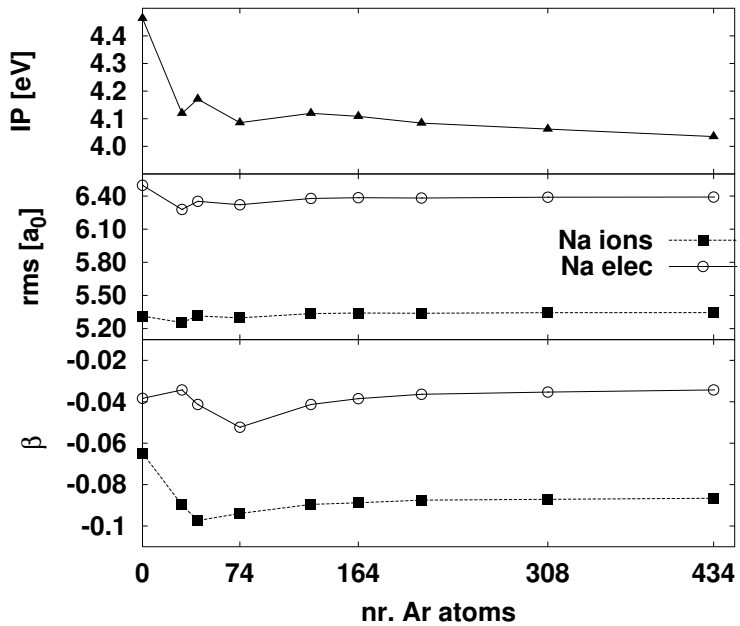,width=0.7\linewidth}
\hspace*{0.5em}
\caption{Basic observables of Na$_8$ embedded in Ar matrices of
  various sizes~: Ionization potential (top), ionic and
  electronic r.m.s. radii (middle), and ionic and electronic quadrupole
  deformation $\beta$ (bottom). Adapted from~\cite{Feh06a}.
\label{fig:na8_arN_stat}
}
\end{SCfigure}

We now turn to the effect of the matrix size on a few basic properties
of the embedded Na$_8$ clusters. Fig.~\ref{fig:na8_arN_stat} presents
the modifications in the IP, in the ionic and
electronic r.m.s. radii and in the electronic quadrupole deformation
$\beta$ (defined similarly as in Eq.~(\ref{eq:rms_beta}) for the ions).
As in the case of a single embedded Na atom, surrounding the Na$_8$ by
Ar atoms first reduces its electronic radius from the value of the
free cluster, due to the strong repulsive Ar core potentials. When the
matrix size increases further, the attractive polarization interaction
allows for some expansion of  the radii thus reducing the initial
shrinkage. The values level off again beyond $N=164$. The interplay
between core repulsion and polarization attraction leaves after all a
reduced electron radius and a slightly expanded ionic radius.  The
global oblate deformation of the Na$_8$ cluster undergoes very small
changes for the electrons and is somewhat enhanced for the ions, with
the variations \PGmod{equilibrating} again at $N=164$.
The IP behaves very similar to the $3s$ state of the embedded Na atom
(see Fig.~\ref{fig:na_arN_struct} \esmod{and mind the negative sign when comparing 
to Fig.~\ref{fig:na8_arN_stat}}). It drops substantially already
for the smallest environment and changes only very little for further
incre\PGmod{a}sing matrix size.  This trend is explained by the fact that the
IP is mostly affected by the short range, repulsive core where only
the first layer matters.

\subsubsection{Impact of rare gas polarizability}

The above results have shown that the polarizability of the matrix
atoms plays an important role, counterbalancing the repulsive short
range core interaction.  The impact of polarizability can be studied
by considering \esmod{various} rare gas species for the matrix ~\cite{Feh07a}. We will
consider as an example Na$_8$ embedded in Ne, Ar and Kr matrices. Going
to heavier rare gases, the polarizabitility increases while core
repulsion changes much less.  To produce the ground state
configurations, we proceed as in the previous section. Note that one
carves a cavity of 13 atoms in Ar and Kr matrices to insert the
Na$_8$, while in Ne matrices, one needs to remove 19 atoms. We should
also mention that the insertion of Na$_8$ in sufficiently large Kr
matrices does not change significantly their shell structures, as in
the case of Ar (see previous section). However the Ne cluster is
strongly modified in the presence of the embedded metal cluster. It
even loses its shell structure, because of the extremely weak binding
in bulk Ne. One thus expects different trends when comparing Ne to Ar
and Kr.

\begin{figure}[htbp]
\begin{center}
\epsfig{file=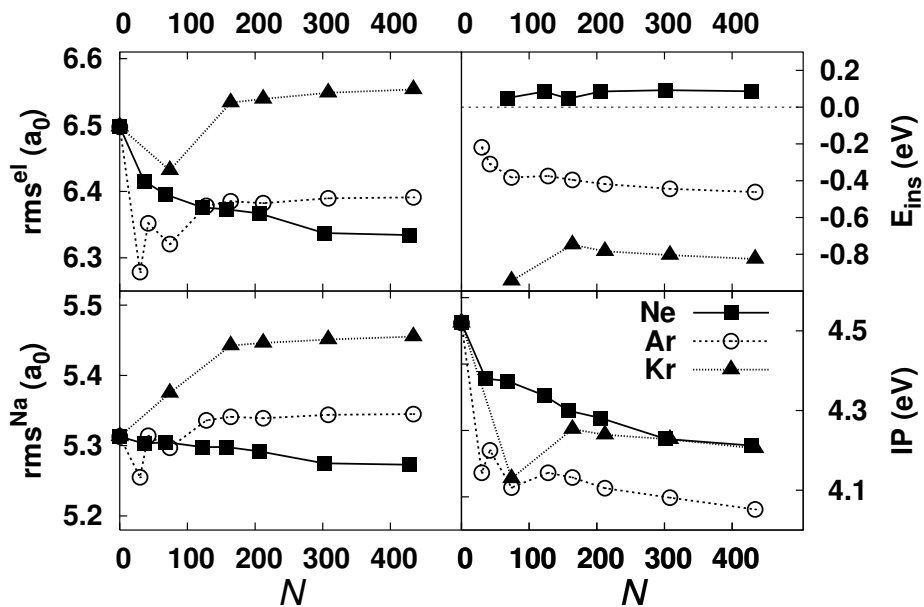,width=0.9\linewidth}
\caption{Ground state observables for Na$_8$ embedded in Ne (squares),
  Ar (circles), and Kr (triangles) matrices of various sizes $N$~:
  Electronic (top left) and ionic (bottom left) r.m.s. radii, insertion
  energy (top right), and ionization potential (bottom
  right). Adapted from~\cite{Feh07a}.
\label{fig:na8_nearkr}
}
\end{center}
\end{figure}

We will consider as observables the electronic and ionic radii, the IP
and the insertion energy of Na$_8$ in RG$_N$ defined as~:
\begin{equation}
  E_{\rm ins}
  =
  E_{{\rm Na}_8/{\rm RG}_N} + E_{{\rm RG}_N}- E_{\rm Na_8} - E_{{\rm
  RG}_{N+p}} \quad, 
\label{eq:Eins_Na8}
\end{equation}
with $p=19$ for Ne, and $p=13$ for Ar and Kr.
Fig.~\ref{fig:na8_nearkr} shows these four observables as a function
of matrix size.
The insertion energy (upper right panel) quantifies the binding of the
compound system.  The Ne environment has a too small polarizability
and does not capture successfully the Na$_8$, even for the largest
matrices. On the contrary, Ar, and even more so Kr, exhibit negative
insertion energies (i.e. good binding), even for the smallest matrix
($N=74$).
The IP (bottom right panel of
Fig.~\ref{fig:na8_nearkr}) shows in all cases the sudden drop from
free to embedded cluster which is due to the \esmod{effect of the} short-range core
repulsion on the cluster valence electrons. The value equilibrates
quickly for the two heavier rare gases because their shell structure
is rather robust. The Ne case shows a strong decrease with increasing
matrix size. This is due to the compression of the whole matrix when
further Ne shells are added which, in turn, brings the innermost shell
closer to the metal cluster and thus enhances the effect of the
repulsive core potentials.
This interaction also has an effect on the electronic extension (see
top left panel of Fig.~\ref{fig:na8_nearkr}).  Embedding Na$_8$
first compresses the valence cloud. Then for larger matrices the
electronic radius increases in the case of Ar and Kr. The asymptotic
value still shows some reduction for Ar but an enhanced radius for Kr.
That means that the larger polarizability of Kr overrides the core
repulsion from the first shell.  The case of Ne is again different in
that it shows monotonous decrease of the radius. This is again due to
the steady \PGmod{compression} of the Ne matrix with increasing system size.
The polarizability of Ne is too small to  counterweighting that trend.
The lower left panel  of Fig.~\ref{fig:na8_nearkr} shows
the trend of the ionic radii. \esmod{Ionic radii tend to increase with matrix size 
while electronic clouds experience rather have a tendancy to reduction. } 
\PGmod{Note that} \esmod{ ionic radii }are much smaller than the
electronic radii. Thus the ions see less of the core repulsion and
relatively more from the polarization  attraction. This explains
nicely the trends. There is little change for Ne, some increase for Ar
and a large increase for the highly polarizable Kr.
Altogether, the comparison of these four different observables and
three different rare gas species has clearly illustrated the subtle
interplay between core repulsion and polarization attraction. It shows
that careful modeling of these effects is crucial for an appropriate
description of these mixed metal/rare-gas systems.

\subsection{Optical analysis}
\label{sec:opt_resp}

Optical absorption spectra provide a key observable to analyze the
structure of metal clusters with their dominant Mie plasmon resonance
\cite{Kre93,Hee93,Bra93}. This holds also for embedded and deposited
clusters where one has as an additional agent the interface
interaction. We will investigate in that section the effects from
short-range core repulsion and the long-range polarization attraction
of the environment on the optical spectra, the consequences for the
Mie plasmon peak, its position  and fragmentation. 
\PGmod{The results presented in Secs.~\ref{sec:opt-emb} and
\ref{sec:opt-dep} were obtained with the technique of spectral
analysis as outlined in Eqs.~(\ref{eq:spectr-anl}),
Sec.~\ref{sec:obs-elec}.} 

\subsubsection{Basic mechanisms and competing effects}

There is an overwhelming amount of studies of optical response in
free Na clusters working out the various facets of underlying
cluster structure and spectral density.  The leading feature of the
Mie plasmon resonance is its dependence on cluster shape which allows
to conclude on cluster size \cite{Rei97c} and deformation
\cite{Bor93}. The width and sub-structure of the resonance peak
depends on several ingredients.  Spectral fragmentation (often called
Landau \esmod{fragmentation}) appears if the resonance peak lies in a
region of high 
density of $1ph$ excitations. This is interpreted as a ``wall friction''
from electrons colliding with the bounds of the mean field
\cite{Yan90}, an effect which sensitively depends  on the detailed
shape of the cluster \cite{Mon95d}. That spectral fragmentation
changes systematically with cluster size \cite{Rei96b,Rei99c} having a
maximum at about $N=1000$ and being extremely small for small
clusters and again for huge clusters \cite{Bab97}. Thermal agitation
of the clusters induces shape fluctuations and thus the resonance
width does increase significantly with increasing temperature
\cite{Mos01}. 
All these effects persist for clusters in contact with an environment.  The
advantage of an environment is to have better control over
temperature. On the other hand, the interface adds new effects which
need to be taken care of. We will consider in the following small
clusters to minimize spectral fragmentation and the theoretical
analysis will concentrate on ground state configurations at zero
temperature to eliminate complications from thermal broadening.

The simplest model of coupling between a metal cluster and a rare gas
environment is to consider the latter as a static dielectric medium
described only by its dielectric constant $\varepsilon$, while the
metal cluster is treated by TDDFT at the LDA level and jellium model,
as was done for metals as K~\cite{Rub93}, Na and Al~\cite{Kur96}, and
Ag~\cite{Ler98}. The limitations of such models is to account only for 
long-range polarization effects, thus producing a red-shift of the
cluster optical response.
 \PGmod{On the other side are
detailed calculations.} 
For example, the optical
response of Ag$_7$ embedded in Ar at 10 K~\cite{Con06} is in fair
agreement with ab-initio calculations performed on the free
cluster~\cite{Bon01}, provided a shift of 0.25 eV \esmod{added} between both
spectra.
\begin{SCfigure}
\epsfig{file=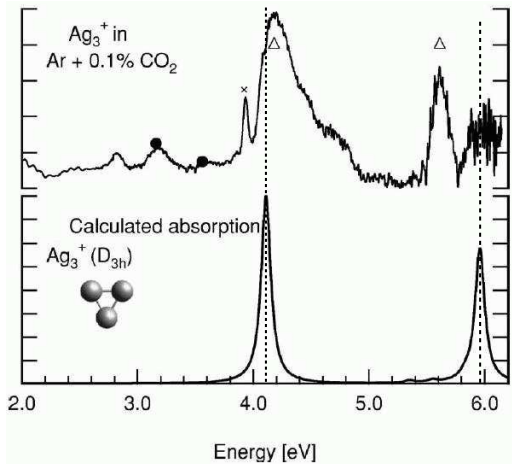,angle=90,width=0.5\linewidth}
\hspace*{0.5em}
\caption{Top~: \MDcomm{Fluorescence excitation} of Ag$_3^+$
  embedded in Ar matrix coembedded with CO$_2$ which is used as an
  electron scavenger~\cite{Lec07}.  
  Bottom panel~: Comparison with ab-initio calculations of free
  Ag$_3^+$ adsorption~\cite{Bon99}. The vertical dashed lines
  emphasize the position of the theoretical peaks.
\label{fig:Ag3+inAr}
}
\end{SCfigure}
However, the effect of an ``inert'' rare gas environment can lead to
non-trivial features, different from a mere red-shift. This is
exemplified in Fig.~\ref{fig:Ag3+inAr} which compares the fluorescence of
Ag$_3^+$ embedded in Ar matrix~\cite{Lec07} with ab-initio
calculations for a free Ag$_3^+$ cluster~\cite{Bon99}.  (The
experimental preparation of charged, embedded cluster uses
co-deposited CO$_2$ molecules as scavengers to prevent
re-neutralization.)  Now, when comparing the theoretical absorption of
the free Ag$_3^+$ with the experimental fluorescence of the embedded
cluster, one observes that the main peak at about 4 eV is only
slightly blue-shifted, while the secondary peak at about 6 eV is
strongly red-shifted.
\MDcomm{CI calculations of optical transition of Na$_2$ embedded in
Ar$_{54}$ also reported a blue or a red shift, depending on the
considered transition~\cite{Gro98}. Moreover, recent experimental
measurements on the optical absorption of Ag$_{4-12}$ embedded in Ar
show that (slight) blue or red shift of the spectrum, compared with
TDDFT calculations, also depends on the metal cluster
size~\cite{Har08}. On the opposite side, small metal clusters in
contact with a few (1 to 4) rare gas atoms exhibit slighly
blue-shifted photoabsorption spectra, demonstrating that, when adding
one by one rare gas atoms, the first predominant effect is core
repulsion, and not a dielectric effect~\cite{Col94,Ray99,Rho04,Sch03}.}
So the Ar matrix shows more than a mere
polarization effect which would simply red-shift the whole optical
response of the metal cluster.

Other TDDFT calculations of the optical response of Cu, Cu$_2$ and
Cu$_4$ deposited on MgO are reported in~\cite{Del04}. While the
spectrum of the dimer is almost unchanged with respect to the free
case (actually, its geometry also remains quasi identical), a strong
red-shift of 1.3 eV is observed for the atom and a slight red-shift of
0.3 eV for the tetramer. It was thus found that geometrical
effects are less determinant than the electronic polarization by the
substrate in the optical response of small Cu clusters.

\subsubsection{Na clusters embedded in rare gas matrices}
\label{sec:opt-emb}

We have seen in the previous sections about cluster structure that
electronic properties experience counteracting effects between
short-range core repulsion and long-range polarization attraction.
We expect that this continues for optical absorption. Our hierarchical
model takes properly into account both effects and is well suited for
the survey.
\begin{figure}
\begin{center}
\epsfig{file=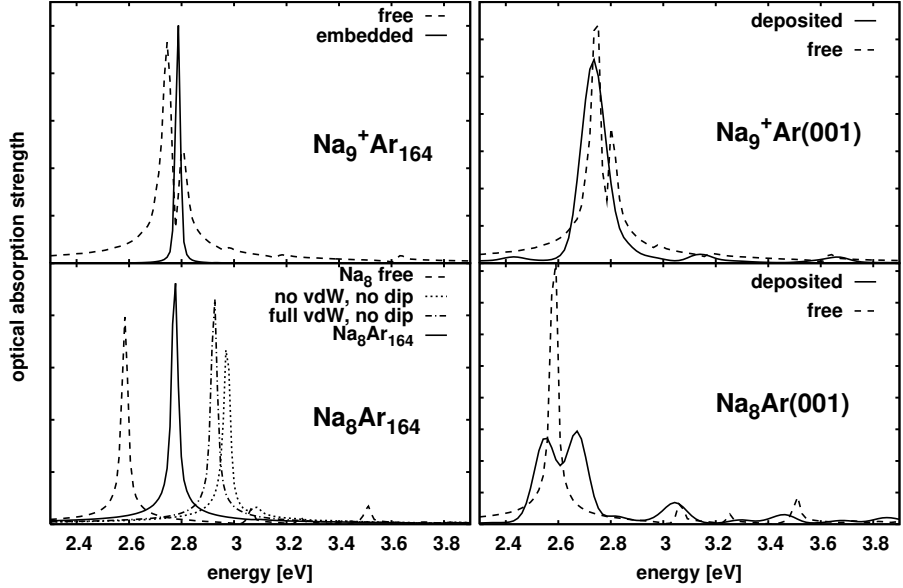,angle=-0,width=0.9\linewidth}
\end{center}
\caption{
Dipole strength distributions for Na$_8$ (lower panels) and
Na$_9^+$ (upper panels) comparing free with embedded clusters
(left panels) and with deposited ones (right panels).
The lower left panel shows the embedded case in three stages:
 polarization potentials and VdW force
neglected (dotted),  VdW included and  polarization potentials
neglected (dah-dotted), full interaction (solid line).
\label{fig:na8optical}
}
\end{figure}
We first discuss in detail the lower left panel of
Fig.~\ref{fig:na8optical} which illustrates the 
counteracting effects for the case of Na$_8$ embedded in
Ar$_{164}$. This magic cluster exhibits a clean plasmon peak, even
when embedded. The peak resides around 2.6 eV for the free
cluster (dashes). For the embedded cluster, we consider three
stages. First we take into account only the Ar core repulsion. This
yields a strong blue-shift of the plasmon peak by about 0.4 eV to 3
eV (see dotted curve).  Then we 
switch on the Van-der-Waals (VdW) interaction while \PGmod{still} keeping
polarization frozen.  This yields a small red-shift of less than 0.1
eV (dots-dashes). Finally, we step up to the full description by
activating the Ar 
dipole degrees of freedom in the calculation.  That adds a further
red-shift component which brings the final peak to 2.75 eV (full
curve), a much smaller blue-shift, \esmod{as compared to the free
  case,} than originally. 
The upper left panel of Fig.~\ref{fig:na8optical} shows the effect
of the matrix on optical response for the charged cluster Na$_9^+$.
In that case, the blue-shift from core repulsion is a bit smaller
because the electron cloud is more compact. Furthermore, the red-shift
from polarization is a bit larger because of the clusters charge.
The net effect amounts to  practically no net shift.
\PGcomm{ The right panels show, for comparison, results for the same
two clusters deposited on Ar(001) surface. The net shift of the mean
peak position is also very small in that cases. But symmetry breaking
in the direction perpendicular to the surface leads to change in
fragmentation pattern, generally pronouncing fragmentation.  }
Studies of other material combinations, Na$_{8,20}$ deposited on
NaCl~\cite{Koh98a}, Na$_8$ embedded in Ar matrices~\cite{Feh06a}, and
Na$_N$ for $N=3-8$ deposited on MgO~\cite{Bae07a}, all show that the
net interaction of the metal cluster with the environment only slightly
changes the mean plasmon frequency.
After all, we have a small net effect from initially two large effects.
The results of such a compensation is hardly foreseeable by simple
models. One better performs  detailed calculations for each case anew.

\begin{SCfigure}[0.7][htbp]
\epsfig{file=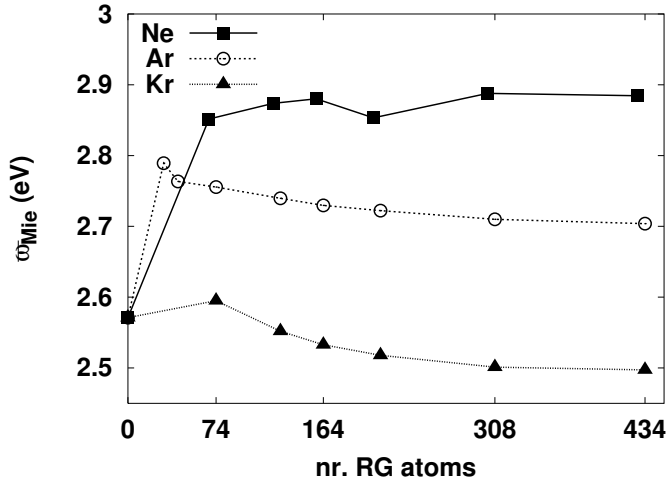,width=0.7\linewidth}
\hspace*{0.5em}
\caption{Plasmon peak frequencies of Na$_8$ embedded in matrices of
  different sizes and different rare gas material as indicated.
  Adapted from~\cite{Feh07a}.
\label{fig:na8_nearkr_opt}
}
\end{SCfigure}

The influence of polarizability can be worked out further by changing the rare
gas material.  Fig.~\ref{fig:na8_nearkr_opt} displays the plasmon
peak positions of Na$_8$ embedded in Ne, Ar and Kr matrices of various
sizes~\cite{Feh07a}.
We first note that the changes are generally small, somewhat at the
precision limits of our modeling (for which we estimate an uncertainty
of about 0.1 eV).  For all materials, the Mie peak first makes a jump
when going from free to embedded clusters. The further changes with
system size are small but systematic. The well polarizable Ar and Kr
show a small, but steadily increasin\PGmod{g}, red-shift with increasing system
size. This is due to the long range of the polarization interaction
which acknowledges every new dipole, even when added in a farther out
shell. The effect is, in principle, also present in a Ne environment. But
here it is outweighted by the steady compression of the Ne matrix
which yields increasing blue-shift.

Experimental data for the optical response of Na clusters embedded in
rare gas material unfortunately do not exist, but there exist some
results for small Ag clusters in rare gas material
\cite{Fel01,Fed93,Fed98,Con06,Die02}. These show that the effect of
the matrix on the optical response remains small whatever combination,
in good qualitative agreement with our findings. A more detailed
comparison between these experimental results and our calculations is
delicate for several reasons. First, there are significant differences
between Ag and Na concerning optical response because of the more
active $d$-shell core electrons in Ag.  Second, the two materials have
significantly different Wigner Seitz radii.  Third, one has to keep in
mind that experiments are often performed inside helium
droplets. Although helium admittedly interacts very little with the
cluster its presence slighty shifts the optical peak \cite{Nak02} as
compared to the true free case.

The Mie plasmon in free clusters is found to be an extremely robust,
collective excitation mode which persists for considerably large
amplitudes \cite{Cal95a,Cal98c}. This remains to be checked for
embedded clusters. To that end, we have studied the dipole spectrum by
spectral analysis after an instantaneous boost for a series of
different boost velocities, delivering different excitation energies.
\begin{SCfigure}[0.5]
\epsfig{figure=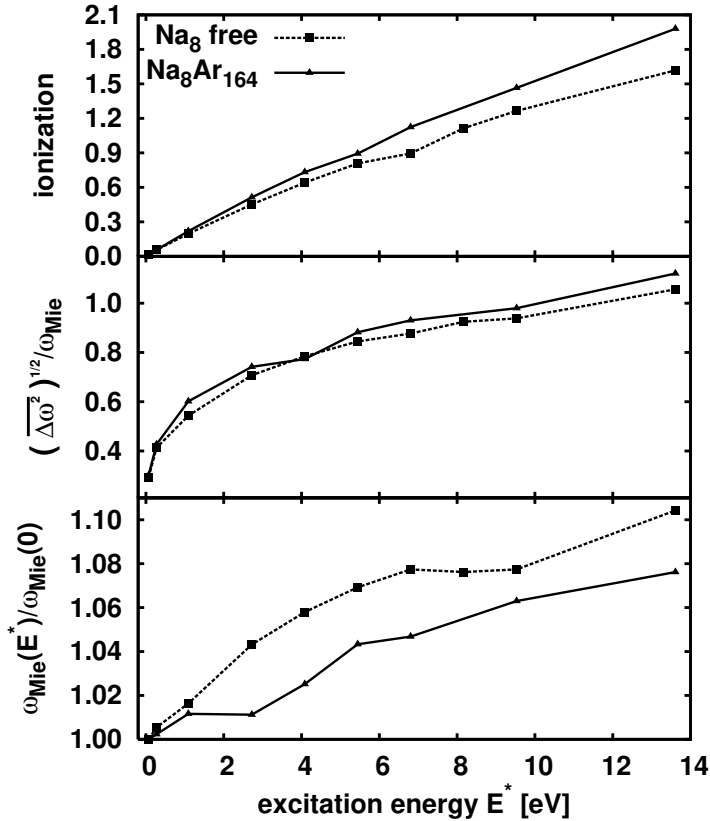,height=0.7\linewidth,angle=-90}
\caption{\label{fig:scan_boost}
\PGcomm{
Results for varying dipole boost excitation of Na$_8$, free and embedded,
drawn as function of average excitation energy.
Upper: ionization = number of emitted electrons.
Lower: average Mie resonance frequency relative to value
    in the linear regime ($E^*\rightarrow 0$).
Middle: average width of the optical absorption spectrum
  relative to peak energy. 
}}
\end{SCfigure}
%
Fig.~\ref{fig:scan_boost} shows trends with excitation energy $E^*$
for the key pattern of the dipole spectrum (peak position and width)
and for the net ionization induced by the excitation. Electron
emission (upper panel) shows in both cases a steady, almost linear
growth. That is the expected behavior for an instantaneous boost
\cite{Cal97b}. The yield is higher for the embedded cluster. That
complies with its lower IP, see Fig.~\ref{fig:na8_arN_stat}.
Furthermore, the slope of ionization decreases with increasing $E^*$
because the IP increases steadily with the degree of ionization.
The average peak positions ${\omega}_{\rm Mie}$ (lower panel)
drift to higher values with increasing $E^*$. That is due to the
increasing ionization of the system which provides a deeper and
more rigid potential well for the electron cloud. The trends are
similar for both cases \esmod{(free and embedded)}. 
The growth is smaller for the embedded cluster
because the Ar cage had already produced a rigid potential
initially. 
The width grows with increasing excitation, first rather quickly
and then levelling off to slower increase. The results from both
cases shows practically the same widths.

\subsubsection{Na clusters deposited on MgO(001) surface}
\label{sec:opt-dep}

In the previous section, we have discussed the effect of embedding on
optical response. 
Fig.~\ref{fig:na8optical} 
indicates that depositing has similarities and differences. 
The spectra of free Na clusters and that of clusters in
contact with a substrate can nevertheless differ dramatically in
detail. We are thus considering more deeply the case of small Na 
clusters deposited on MgO material. A 
systematic comparison of the spectra with the free case
has been done in~\cite{Bae07a}. The modifications
induced by the surface are quite involved. While in the case of
free clusters, the optical response of metal clusters is strongly 
correlated to their geometry, when deposited, the spectra exhibit
heavy fragmentations, especially in the direction perpendicular to the
surface (denoted in the following by $z$ direction, while the surface
is described by the directions $x$ and $y$). 
\begin{figure}[htbp]
\begin{center}
\epsfig{file=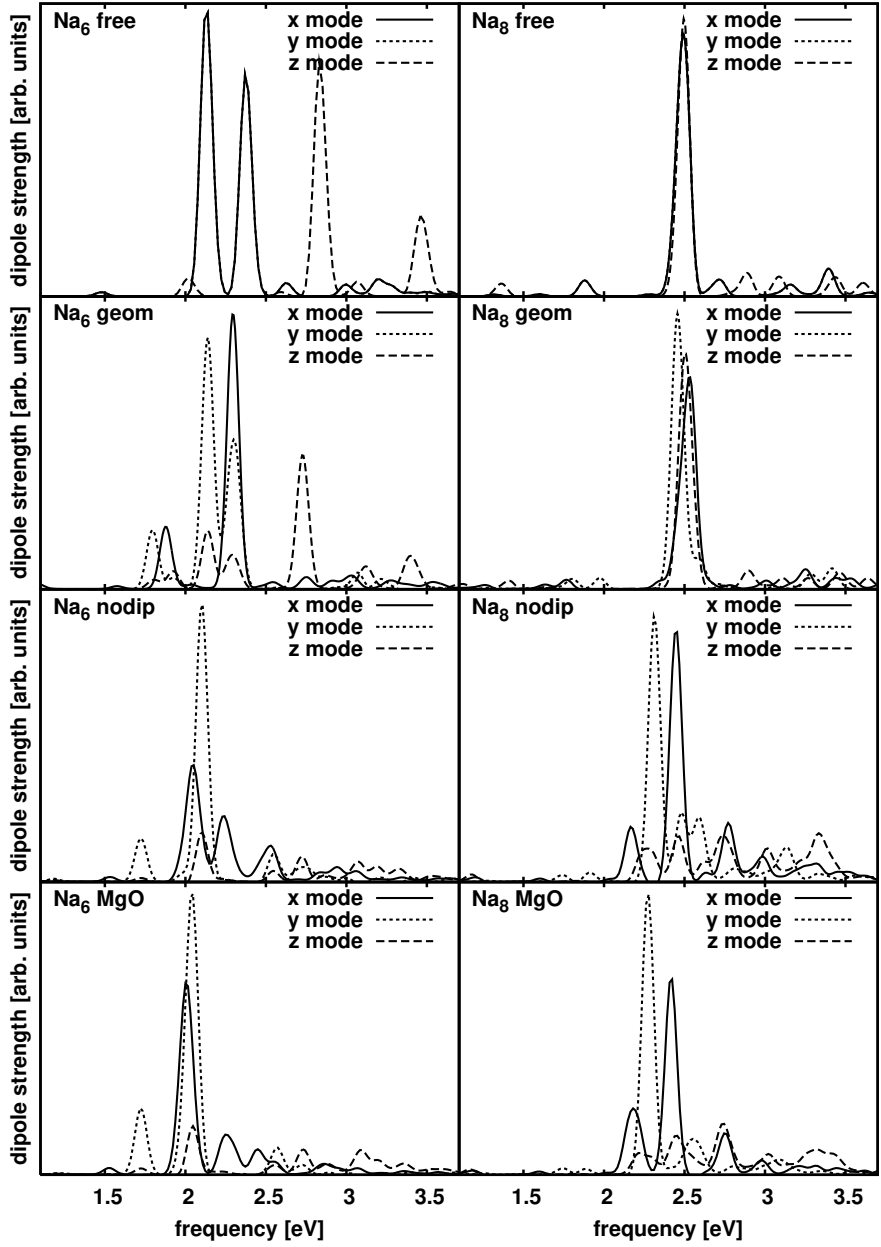,width=0.9\linewidth} 
\caption{Optical responses of Na$_6$ (left) and Na$_8$ (right). Free
  clusters (top); structure of deposited clusters on MgO but without
  surface interaction (second row from top); deposited clusters
  without dynamical polarization (third row from the top); deposited
  clusters with full MgO model. Adapted from~\cite{Bae07a}. 
\label{fig:nan_mgo_opt}
}
\end{center}
\end{figure}
To disentangle the
effect of modified geometry from the repulsive potentials due
Mg$^{2+}$ and O$^{2-}$ cores or from the dynamical polarizability of
the oxygen valence shells, we present in Fig.~\ref{fig:nan_mgo_opt}
optical absorption spectra
for Na$_6$@MgO(001) (left column) and  Na$_8$@MgO(001) (right column)
between free clusters (uppermost panels)
and the deposited ones, where each ingredient is
successfully switched on (full model calculations are presented in the
lowest panels).

Let us start the discussion with Na$_8$ (right panels) which has a
spherical symmetry and whose geometry only slightly changes under
adsorption at the surface (see Fig.~\ref{fig:nan_mgo_struct}). It
thus appears as an ideal tool to study the direct influence of MgO. As
already mentioned, the free cluster has a pronounced degenerated
plasmon resonance at about 2.5~eV (top right panel) due to its nearly
spherical shape. The spectrum of the deposited Na$_8$ (bottom right
panel) looks much different: the $x$-$y$-$z$ degeneracy is lost and a
strong fragmentation of the plasmon is observed, in particular for the
$z$ mode, which is almost completely dissolved.  The average peak
positions, however, remain almost unchanged.
In order to find out the reasons for these sizeable changes, we
proceed from the free to the deposited cluster in steps.  In the
second panel from the top, the optical spectrum is obtained from a
free Na$_8$ but using the ionic configuration of the deposited
cluster.  The spectrum almost maintains its initial structure, only a
small resonance splitting is observed. This is in accordance with the
small structural change of Na$_8$, see  Fig.~\ref{fig:nan_mgo_struct}.
In the next step (third panel from top), the O$^{2-}$ dipoles are still
kept frozen, but the static contributions of the interface potentials are
switched on. These consist in repulsive core potentials, but also in
the electrostatic potential of the crystal. 
This changes the spectra much towards the final  result.
The interface potentialö is obviously a key ingredient in the optical response of
the deposited cluster.
In particular, it causes the dramatic
fragmentation of the $z$
mode because it removes totally reflection symmetry.
This, in  turn, enhances the spectral density for $z$-modes
leading to strong spectral  fragmentation.
Symmetry violation in $x$-$y$-direction is much smaller and this \esmod{implies that} 
the density of $1ph$ states in the $x$ and $y$ directions does not
increase too much which explains why the $x$ and $y$ modes are less
affected.
\PGcomm{The effect on spectral fragmentation is particularly strong
for MgO(001) while it was much smaller for Ar(001), see figure
\ref{fig:na8optical}. The cluster is less bound for Ar and this
reduces the effect.
}
As a last step, the dynamical polarizability of the oxygen dipoles is
activated (bottom panel). The optical response of Na$_8$ does not
change significantly, except for a tiny red-shift and a slight
restoration of collectivity. Thus the polarization interaction seems
to play a minor role for the spectrum \esmod{in that case}.

But one should be cautious with this first conclusion. The tightly
bound Na$_8$ may be too robust to exemplify the direct and indirect
effect of the MgO polarizability.  Thus we perform the same study for
Na$_6$ (left panels of Fig.~\ref{fig:nan_mgo_opt}), whose structure
is significantly modified by deposition (see
Fig.~\ref{fig:nan_mgo_struct}) \esmod{due to lattice mismatch 
effects (Sec.~\ref{sec:NaN_on_MgO})}.  
The spectrum of free Na$_6$
displays two degenerate modes with a double peak structure between 2
and 2.5~eV. The twofold degeneracy is due to the axial symmetry of the
pentagonal pyramid. The $z$ mode, which corresponds to the axis of
symmetry here, is blue-shifted compared to the other modes owing to
the smaller extension in this direction. Changing the ionic structure
to that of the adsorbed cluster (leaving away the surface itself)
yields the spectrum in the left second panel from top. The
considerable change reflects the breaking of the cluster axial
symmetry. Moreover, since the Na$_6$ extension in the $z$ direction
has increased, the corresponding mode loses some strength and is
red-shifted by about 0.2~eV. Switching on the core repulsion (third
panel from top) has the same effect as for Na$_8$. The increase of the
density of accessible states causes spectral fragmentation,
particularly for the $z$ mode. Switching finally on the dynamical
polarization to the full description of the surface (lowest left
panel) hardly changes the spectrum, as in the previous case of
Na$_8$. Thus we see that the effect of polarization interaction is
here mediated indirectly through its strong modification of cluster
geometry. The direct influence is comparatively small.

The net result is to some extent similar as for embedded clusters.
The effect of the environment on the optical response pattern of
deposited clusters has many ingredients which act in different
directions. Simple estimates of trends are hardly possible. One needs
detailed modeling to cover all relevant effects.
\MDcomm{This agrees former TDDFT studies on the optical absorption
spectra of Ag$_{2,4,6,8}$ adsorbed on MgO(100) using the embedded
cluster model~\cite{Bon07}. While a global red-shift is
generally observed, some transitions can be not influenced by the
surface at all, and the details of the fragmentation pattern of the 
spectra is hardly predictable before complete and detailed
calculations.}


\section{Deposition processes}
\label{sec:depos}

We analyze \esmod{in this section} the dynamics of deposition of finite alkaline clusters on
rare gas and MgO surfaces.  Again, we work out the
key role played by substrate polarization, in relation with
experimental results which show internal excitations of
seemingly ``inert'' surfaces during the deposition
processes. We especially study the impact of cluster size and cluster
charge on the course of deposition. A cluster with
non-zero net charge induces large polarization of the substrate,
an effect unattainable with simple molecular dynamics approaches. We
also show how the deposition process can create "hot spots" in
the surface where sizeable amounts of energy are stored in internal
excitations of the substrate.

\subsection{Experimental context}
\label{sec:soft_landing}

Supported metal clusters have attracted much interest during the past fifteen
years for their potential applications to nano-structured materials. They have
thus motivated many experimental and theoretical investigations. Such studies
require well controlled conditions at the side of the deposited cluster (e.g.,
no fragmentation, conservation of shape and charge) and at the side of the
surface (e.g., only slight reorganization of the top layers).  The processes
we are interested in here are related to soft-landing in which the cluster
size is preserved as much as possible. Soft-landing actually requires
very low impact energies. It has been explored theoretically already more than
ten years for various metal clusters, e.g. Ag$_7$ on Pd(100)~\cite{Van96},
Ag$_{1,7,19}$ on Pd(100) and Pd(111)~\cite{Nac97}, Cu$_{13,55}$ and Au$_{55}$
on Cu(001)~\cite{Pal99}, Al$_{864}$ on Al(001)~\cite{Koh01}, Al$_{20-200}$ on
graphite~\cite{Xir01}. All these studies employ molecular dynamics (MD)
simulations with effective potentials.  
A DFT study of Na clusters on NaCl surface \esmod{can also be} found
in \cite{Ipa03}. Experimental studies exist, e.g., for
the deposition of Cr$_{1-10}$ on Ru(001)~\cite{Lau00a,Lau00b}, and Ag$_7$,
Au$_7$ and Si$_7$ on graphite~\cite{Pra03}.

Soft-landing can require to decrease the impact energy to hardly
manageable small values (less than a few eV per atom). For instance,
it has been shown by MD calculations that softly deposited
Al$_{50,200}$ on SiO$_2$ are always adsorbed on the surface, while Au
clusters of same size at the same impact energy are reflected unless
one uses extremely low impact energies \MDcomm{($<0.56$ eV/atom)} because of
the higher mass of Au compared with Al, Si and O~\cite{Tak01}.
\MDcomm{Experimentally, scanning tunneling microscopy (STM)
measurements on Ag$_{1,7,19}$ deposited on Pt(111) demonstrate that
decreasing substancially the impinging energy allows to deposit
clusters with larger average sizes and smaller standard
deviations~\cite{Bro96,Bro97}. For example, when going from 5 downto 1
eV/atom, the average size of Ag$_{19}$ increases by 23~\% and its
standard deviation is only 25~\% larger than that of single Ag
adatoms.}

An interesting \MDcomm{alternative or additional technique} is to use
a rare-gas layer in between the 
cluster and the substrate which serves mechanically as a soft stopper
while being chemically inert. The use of rare gas layers offers the
possibility to gently dissipate the kinetic energy of the deposited
cluster without causing damage to the cluster itself or to the
underlying metal surface.
This technique was suggested by MD simulations of the deposition of
(NaCl)$_{32}$ on Ne or Ar films adsorbed on a NaCl
surface~\cite{Che93}, and of Cu$_{147}$ on Ar and Xe films adsorbed on
Cu(111)~\cite{Che94}. It has been shown that the addition of rare gas
layers on the surface prevents damages that would occur on the bare
surface.
This method was first experimented in 1996 for the deposition of Ag
clusters on Pt(111) via Ar layers~\cite{Bro96,Bro97}, and relied on
former studies of deposition of metal clusters in rare gas
matrices~\cite{Har90,Hu91}. 
\MDcomm{After deposit, the rare gas material is eliminated by
increasing the sample temperature above the rare gas monolayer
desorption threshold while the deposited cluster remains tied to the
metal surface. 

As an example of the soft-landing technique with Ar buffer, we present in
Fig.~\ref{fig:FeNonAr} X-ray absorption measurements performed on
Fe$_{1-6}$ deposited on (2$\times$1)O/Ru(001) covered by various
thicknesses of Ar layers~\cite{Lau03}. 
\begin{SCfigure}[0.5][htbp]
\epsfig{file=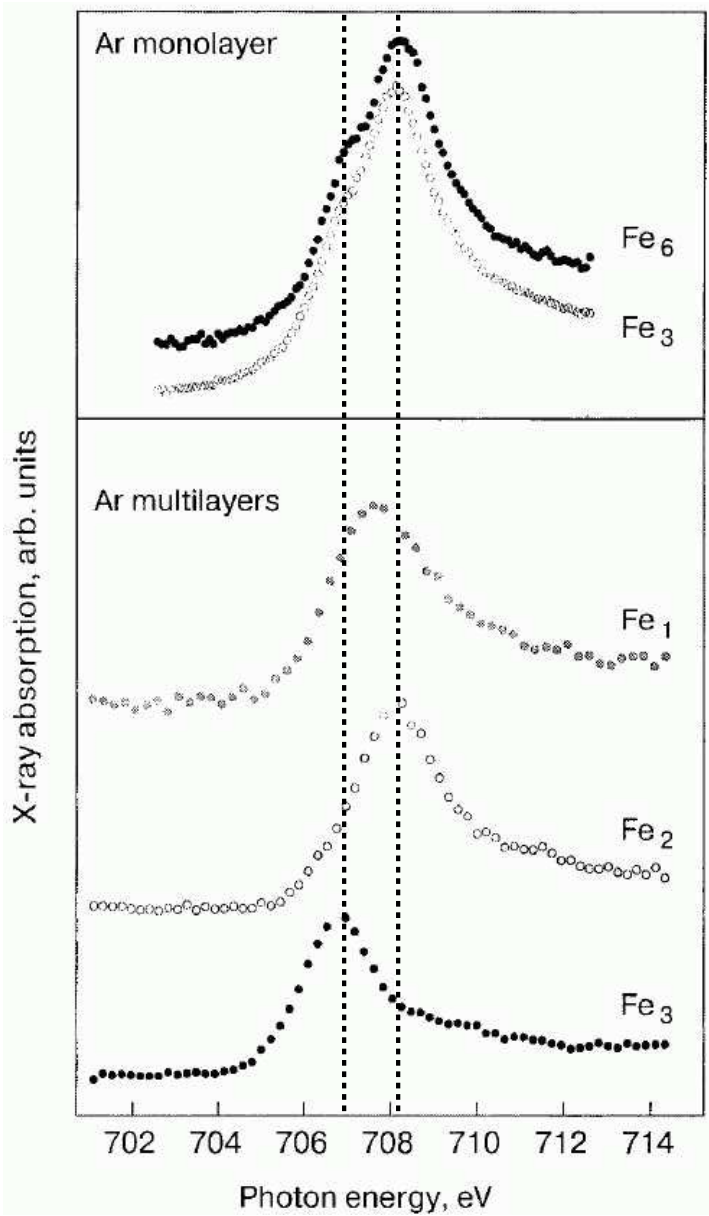,angle=90,width=0.7\linewidth}
\hspace*{0.3em}
\caption{X-ray absorption measurements of Fe$_N$ on
  Ar/(2$\times$1)O/Ru(001)~\cite{Lau03}. The vertical dashed lines
  represent the position of the peak of deposited Fe$_2$ and Fe$_3$
  from the bottom panel.
\label{fig:FeNonAr}
}
\end{SCfigure}
The bottom panel shows spectra of Fe$_{1-3}$ obtained after deposition
on approximately ten Ar layers. The observed shifts suggest that each
spectrum comes from a different cluster species, and that no
fragmentation had occurred. On the contrary, in the top panel, the
deposition of Fe$_3$ and Fe$_6$ on a Ar monolayer basically yields the
same absorption spectrum, with a main peak which coincides with that
of Fe$_2$ from the bottom panel, and a shoulder at lower energy
(approximately at the position of the Fe$_3$ peak in the bottom
panel). This indicates a fragmentation of Fe$_3$ and Fe$_6$ into at
least two pieces. Thus one single Ar layer is obviously not enough
to efficiently dissipate the translational kinetic energy of the iron
cluster.

More investigations have been performed on the fragmentation of
Ag$_2^+$ and Ag$_7^+$ into Ag atoms, when deposited on Ar, Kr, and Xe
surfaces, via fluorescence spectra~\cite{Fed98}. It was reported that
combining low impact energies (and not necessarily very small values,
that is $< 10$ eV/atom) with rare gas (Ar or Kr) buffer layers can
reduce fragmentation into adatoms below 10~\%.

At the side of the substrate, this soft-landing technique can also 
prevent formation of surface defects, as has been observed in STM
measurements~\cite{Bro96,Bro97}. This issue is actually relevant when
one wants to deposit clusters without implantation into the
surface. And indeed, soft-deposited Fe atoms on
Ar/(2$\times$1)O/Ru(001) with impact energy 
of $1-2$ eV/atom undergo an agglomeration process through annealing at
several hundreds of Kelvin~\cite{Lau03}, since the X-ray spectrum comes
closer to that of bulk iron as soon as the annealing temperature $T$ is
increased. This thus demonstrated a high mobility of the Fe adatoms
and thus, no implantation. Similarly, X-ray spectra for soft-deposited
Fe$_2$ and Fe$_3$ exhibit slight modifications with $T$ but remain
quantitatively different compared with bulk iron. This also shows that
no fragmentation into single atoms occured during the deposition
process, since no diffusion is observed (demonstrating at the same time
that Fe dimers and trimers are less mobile than monomers), so the
successful soft-landing of these species.

Note finally that, even if the cluster size seems preserved in a
soft-deposition on rare gas layers, the cluster geometry usually
differs from that of the free cluster because of its interaction,
although weak, with the rare gas substrate during the deposition or
the rare gas desorption process~\cite{Sch01,Har02}.}

\PGcomm{In the following, we will discuss in sections \ref{sec:depos-trends},
\ref{sec:internal_ar} and \ref{sec:large-ar} theoretical explorations
of soft deposition on inert substrates, soft rare gas surfaces in
comparison with more resistive MgO surfaces.  As started already in
the discussion of optical response (see Sec.~\ref{sec:opt_resp}),
we will check the impact of the substrate polarization, especially in
sections.~\ref{sec:depos_mecha} and \ref{sec:internal_ar}.  The issue of
cluster-geometry effects through Ar surface will be addressed in
Sec.~\ref{sec:wetting}. The remainder of the present subsection is
devoted to a few more typical experimental results.}
%

 
\subsection{Deposition on planar surfaces under varying conditions}
\label{sec:depos-trends}

In the following, we will present explorations of deposition of Na
clusters on MgO(001) or Ar(001) surfaces using our hierarchical model.
Both materials are insulators, though with a considerable polarization
interaction. Ar(001) is mechanically very soft with a rather smooth
surface while MgO(001) is more robust showing large surface
corrugation. It is instructive to discuss two such different systems
in parallel. In this subsection, we will investigate the dependence of
the dynamical evolution on initial conditions as, e.g., impact
energy or cluster orientation. The subsequent subsection is devoted to
a summary analysis of global observables.

\subsubsection{Influence of polarizability and core repulsion}
\label{sec:depos_mecha}

In this section, we analyze the influence of the modeling for the Ar
substrate on the deposition process.  Test case is the deposition of
neutral Na$_6$ on Ar(001). The Ar(001) surface is modeled by 6 layers of
8$\times$8 squares, containing altogether 384 Ar atoms. These squares
are copied periodically in both horizontal directions to simulate an
infinite surface. To stabilize the underlying (supposedly infinite)
crystal structure, the atomic positions in the lower two layers are
frozen at bulk positions.

The neutral cluster Na$_6$ consists in a pentagon ring as base and an extra
ion on the symmetry axis of the pentagon. It is initially positioned
with the top ion above a hollow site and facing away from the
surface. The center-of-mass of the cluster starts 15 $a_0$ above the
uppermost Ar layer. This initial configuration is illustrated in the upper
left part of Fig.~\ref{fig:na6_ar_top_down}.
The initial kinetic energy of the cluster is taken equal to $E_{\rm
kin}^0/N=$136 meV.  The corresponding initial velocity points down  along the
$z$ direction, perpendicular to the surface. The impact energy used
here lies in the regime of soft-landing if one extrapolates available
experimental data with proper scaling laws~\cite{Har90}.

\begin{figure}[htbp]
\begin{center}
\epsfig{file=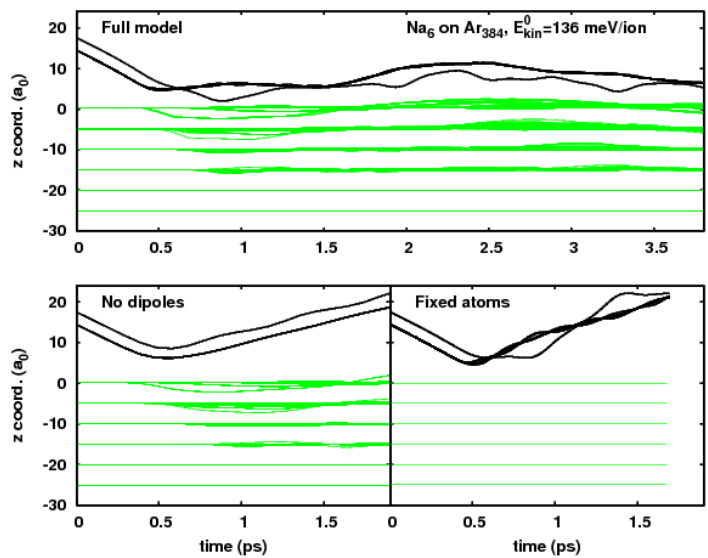,width=0.7\linewidth}
\caption{Dynamical deposition of neutral Na$_6$ on Ar$_{384}$. $z$ coordinates
  as a function of time, for three levels of substrate treatment~:
  fixed Ar cores (bottom right), no dynamical dipoles (bottom left),
  and full model (top)~\cite{Din07b}.
\label{fig:na6_depos_compar}
}
\end{center}
\end{figure}
Fig.~\ref{fig:na6_depos_compar} compares three different levels of
treatment of the Ar substrate~: mobile Ar cores but no Ar dipoles
\esmod{(i.e. Ar dipoles frozen at static, initial value, bottom left panel),}
fixed Ar cores but dynamical dipoles \esmod{(bottom right panel)}, 
and the full model \esmod{(upper panel)}.  Before impact (around 0.5
ps), the dynamics of the impinging neutral Na$_6$ looks similar in all three
cases. After the collision, the evolution differs significantly. 
In the case of fixed Ar cores (bottom right panel), the surface becomes
rather rigid and the cluster is reflected with strong internal
excitations where the top ion oscillates forth and back
through the pentagon.
In the case of frozen dipoles and mobile cores (bottom left panel), the
cluster is again reflected but dissipation in the substrate is
possible and much less internal excitation of internal motion of the
cluster is produced.
In the case where all model components are fully active (top panel), the
evolution behaves totally different from both previous cases, There is
more energy absorption from the Ar surface and there is dissipation of
energy into internal cluster degrees of freedom acting both together
to yield sticking of the (excited) cluster to the surface.  This
example demonstrates the importance of a full dynamical treatment of
the Ar surface, going beyond a mere MD of Ar positions in accounting
for their polarizabilities through dynamical dipoles.
Altogether, Fig.~\ref{fig:na6_depos_compar} demonstrates that the
elasticity of the surface plays a crucial role in the deposition
process. Ar is extremely soft and can serve as a true buffer material
for gentle deposition. The proper dynamical treatment of the surface
is thus essential.

In the top panel of Fig.~\ref{fig:na6_depos_compar}, one can notice
the propagation of the perturbation as a sound wave with approximately
the speed of sound in Ar bulk, about 20$-$30 $a_0$/ps. Because the 5th
Ar layer is kept fixed, this wave is reflected here and bounces
back. It reaches the surface again at about 1.6 ps and transfers some
momentum  back to the deposited Na$_6$. The effect is small
and does not modify the bound status of Na$_6$. And yet, this
reflection is an artefact of the limited description and would not
occur for an infinite number of layers.  In order to check the
validity of our description, we have redone the calculations (here and
in several \esmod{other cases} of the subsequent examples) with two more layers, namely
with Ar$_{512}$ composed of 8 layers (6 active, 2 frozen).  We have
still found good agreement with results from Ar$_{384}$
substrate. This shows that the 4+2 layers are sufficient in all
presented dynamical regimes.  One may even argue that this Ar
description with frozen bottom layers has a realistic touch, since it
could  simulate the set-up with Ar layers on metal surfaces used in
soft-landing techniques~\cite{Ira05} (in that case, however, we are
still missing the image potential from the metal underneath).

\subsubsection{The effect of initial cluster orientation}

\begin{figure}[htbp]
\centerline{\epsfig{figure=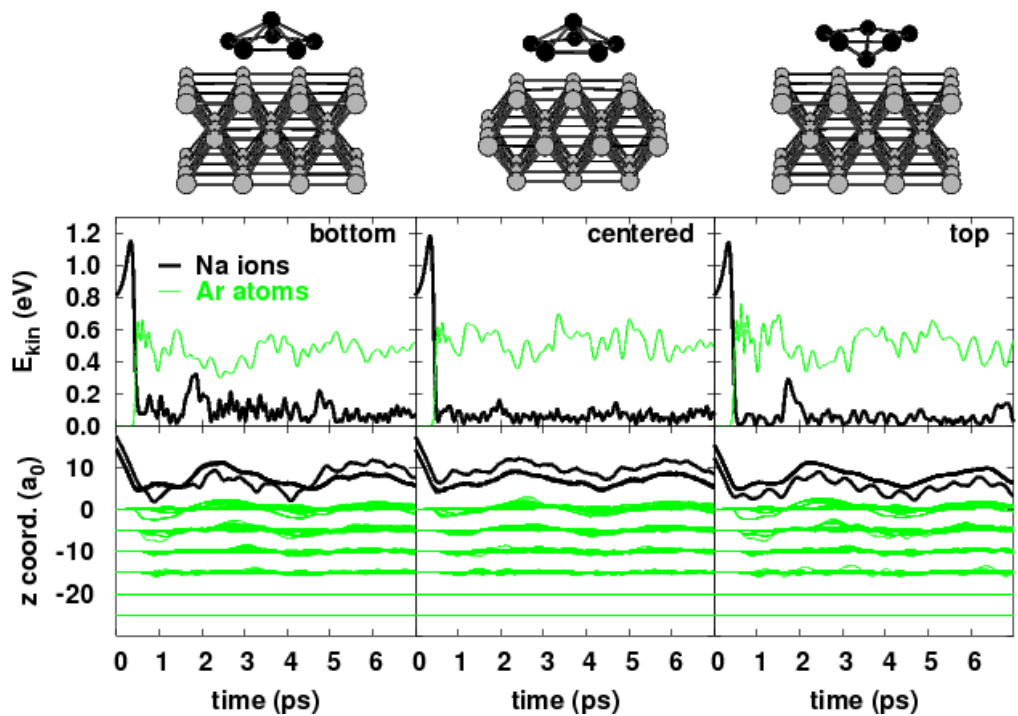,angle=90,
width=\linewidth}}
\caption{\label{fig:na6_ar_top_down}
Deposition of Na$_6$ on Ar$_{384}$ with initial kinetic energy
of 136 meV per Na atom. Time evolution of $z$ coordinates (lower
panels) and kinetic energies (middle panels) for three different
collision geometries as sketched in the upper most
panels, adapted from~\cite{Din07b}. 
} 
\end{figure}

The above example considered one fixed initial cluster orientation.
Experimental preparation will produce a mix of orientations.
We thus investigate here varied orientations and positioning relative
to the surface structure.
The results from Sec.~\ref{sec:depos_mecha} employ what will be
denoted the ``bottom'' geometry, that is the top ion faces away from
the surface and the cluster center axis is placed above an
interstitial position of the first Ar layer.  Two other cases are
presented together with this case in Fig.~\ref{fig:na6_ar_top_down},
that is the ``centered'' one (middle column), similar to the
``bottom'' configuration (left column) but with the axis exactly above
an Ar atom of the first layer, and the ``top'' one (right column)
obtained from the ``bottom'' configuration by reversing the top ion to
face towards the surface such that the top ion hits the Ar surface
first.
At first glance, we see that all three cases produce very similar
dynamical evolutions. Ar(001) is a smooth and soft, so to say
forgiving, substrate. Closer inspection reveals a few interesting
differences in details. The ``bottom'' configuration produces the
hardest collision, since the largest perturbation is seen in the Ar
surface and in the cluster itself (with the top ion oscillating fully
through the bottom ring).  We also note that the remaining average
kinetic energy of the cluster (black line in the middle panels) is in
the average largest for the ``bottom'' configuration. The smoothest
collision is seen for the ``centered'' configuration.  This difference
stems from the Ar core repulsion seen by the cluster. In ``bottom''
configuration, the collision axis lies on a hollow site and the Na
ions thus stay closer to the repulsive sites of the Ar atoms. In the
``centered'' case, the most repulsive Ar site coincides with the
(empty) center of the pentagon and the interaction with the Ar cores
is smaller. The top ion, directly above an Ar atom, is here also
hindered from diving through the pentagon. A similar situation is
observed for the ``top'' configuration~: the top ion meets now the
surface first and dives into the interstitial site with maximum
distance to the atoms. A minimization of core repulsion in this
configuration also explains why the top ion does not oscillate through
the pentagon in that case, and why the final distance is slightly
smaller than in the ``bottom'' or ``centered'' configuration.

\subsubsection{Dependence on the site of first contact}
\label{sec:Na_atom_on_MgO}

The example of Fig.~\ref{fig:na6_ar_top_down} has shown some
dependence on the surface sites which the cluster hits at
impact. Larger effects are to be expected for a material with larger
corrugation, as MgO(001). We start with the simplest case of a Na
monomer deposited on MgO~\cite{Bae08a}.
The geometry issue then reduces to the choice of the deposition site.
As has been discussed in Sec.~\ref{sec:Na_MgO_PES}, MgO(001) offers
a variety of sites (see figures \ref{fig:PES_Na_MgO}
and \ref{fig:nan_mgo_struct} for a view of
the MgO(001) surface). The most attractive is the O site, due to the
large polarizability of the O$^{2-}$ anion.  The Mg$^{2+}$ acts as a
repulsive site. We furthermore consider the hollow site at the center
of the square built by two O and two Mg sites.  The very different
binding properties let us expect a strong dependence of deposition
dynamics on the impinging site.

\begin{figure}[htbp]
\begin{center}
\epsfig{file=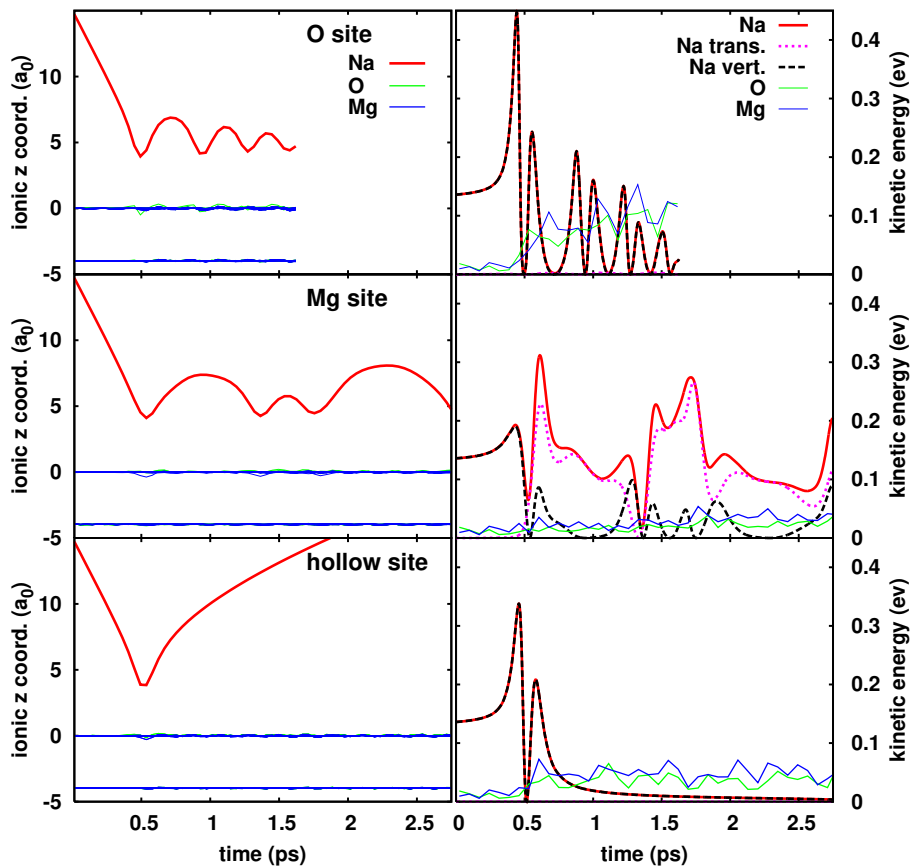,width=0.9\linewidth}
\caption{
Time evolution of the ionic coordinates and kinetic energies $E_{\rm
  kin}$ for the deposition of a Na monomer on MgO(001) with initial
kinetic energy $E_{kin}^0=0.136$ eV, impinging on various
sites~: O site (top), Mg site (center) and hollow site
(bottom). Left~: $z$ coordinates of Na (thick line), Mg (thin
curve), and O (gray line) cores. Right~: Total $E_{\rm kin}$ of Na
(thick gray or red curve), lateral $E_{\rm kin}$ of Na (dots), vertical
$E_{\rm kin}$ of Na (dashes), and total $E_{\rm kin}$ of Mg (thin
dark or blue curve) and O (thin light or green curve) cores~\cite{Bae08a}.
\label{fig:na_mgo_site}
}
\end{center}
\end{figure}

Fig.~\ref{fig:na_mgo_site} illustrates the time evolution for the
collision of a Na atom on a MgO surface, where the Na is started 15
$a_0$ above the surface with an initial kinetic energy of 136 meV.
The left panels show the $z$ coordinates and the right panels the
kinetic energies of the subsystems.
The simplest case is the impact on the O site (uppermost panels). The
atom is accelerated towards the surface up to a kinetic energy of 0.45
eV which is reached at the point of closest impact at a distance of
$4.5\,a_0$ and time of 500 fs.
At this time, it transfers very quickly
a large part of momentum to the substrate ions.  This
affects first the ions in the immediate vicinity of the atom
at closest impact. The perturbation quickly spreads over the
substrate, but not very deeply into it. After the first bounce, the Na atom
performs damped oscillations, transferring energy to the substrate with
each bounce and coming almost to rest within the first 2\,ps.  The
final distance approaches nicely the equilibrium distance of 5 $a_0$.
In order to check that the oscillations proceed only perpendicular to
the surface, the kinetic energy of the Na atom has been split into
contributions from perpendicular (or vertical) and parallel (or
transverse) motion. The transverse part of the energy is too small to
be visible in the right upper panel of Fig.~\ref{fig:na_mgo_site}.
The motion of Na on an O site proceeds strictly perpendicular to the
surface.
The kinetic energy transferred to the MgO can also be read off from
Fig.~\ref{fig:na_mgo_site} (see right upper panel).  The contributions
from oxygen and magnesium are given separately.  O ions are the
lighter species and therefore react first. Mg ions follow more slowly.
But about 100 fs later, the energy has already been distributed almost
equally over both ion sorts.

The dynamics behaves totally different if the atom impinges on the
repulsive Mg site, see middle panels of Fig.~\ref{fig:na_mgo_site}.
At first glance, the z-component of the Na trajectory looks quite
similar to the case before.  But one notes that the motion does not
become damped after the first reflection. The kinetic energies (middle
right panel) give a clue on the process. There is much less energy
transfer at first impact which is related to the fact that the
Mg$^{2+}$ ion is more inert. And there is a significant amount of
lateral kinetic energy for the Na atom creeping up after impact time
at 500 fs.  In fact, most of the kinetic energy is now in lateral
motion.  The atom is deflected by the Mg$^{2+}$ ion, bounces away in
sideward direction, hops over the surface several times changing
direction whenever it comes close to another surface ion. The motion
is almost undamped because little energy is transferred to the surface
after the first collision. The atom has thus still too much energy to
be caught by a certain site of the surface. But as energy loss is just
large enough that the atom cannot escape the surface as a whole, it
will continue to lose slowly energy and finally be attached to an
oxygen site, long after the simulation time of 3 ps.

The bottom panels of Fig.~\ref{fig:na_mgo_site} show the case of
impact at a hollow site. We see again the immediate reflection at
impact time associated with fast energy transfer. Less energy is
transferred than on the other sites (see upper and middle panels)
and thus the bounce-back has a much larger amplitude than in both
other cases. The Na motion remains strictly perpendicular to the
surface as practically no lateral kinetic energy can be seen. The
vertical kinetic energy is almost approaching zero because the
departing Na atom has to work against the polarization potential. The
case is at the limits of our box size and energy resolution such that
we cannot decide whether the atom will finally escape with extremely
small kinetic energy, or will bounce back and relax to an absorption
site on a very long time scale. Nevertheless, we find it noteworthy to
note that the hollow site seems sufficiently attractive to hinder
deflection towards the still more attractive oxygen site.

\subsubsection{Velocity dependence}

In this section, we study the dependence of the deposition dynamics on
the initial kinetic energy $E^0_{\rm kin}$ for two test cases: Na$_6$
on Ar(001) and Na$_8$ on MgO(001).
Varying $E^0_{\rm kin}$ amounts to change the
cluster's initial velocity along the $z$ direction.

\begin{SCfigure}[0.5][htbp]
\epsfig{file=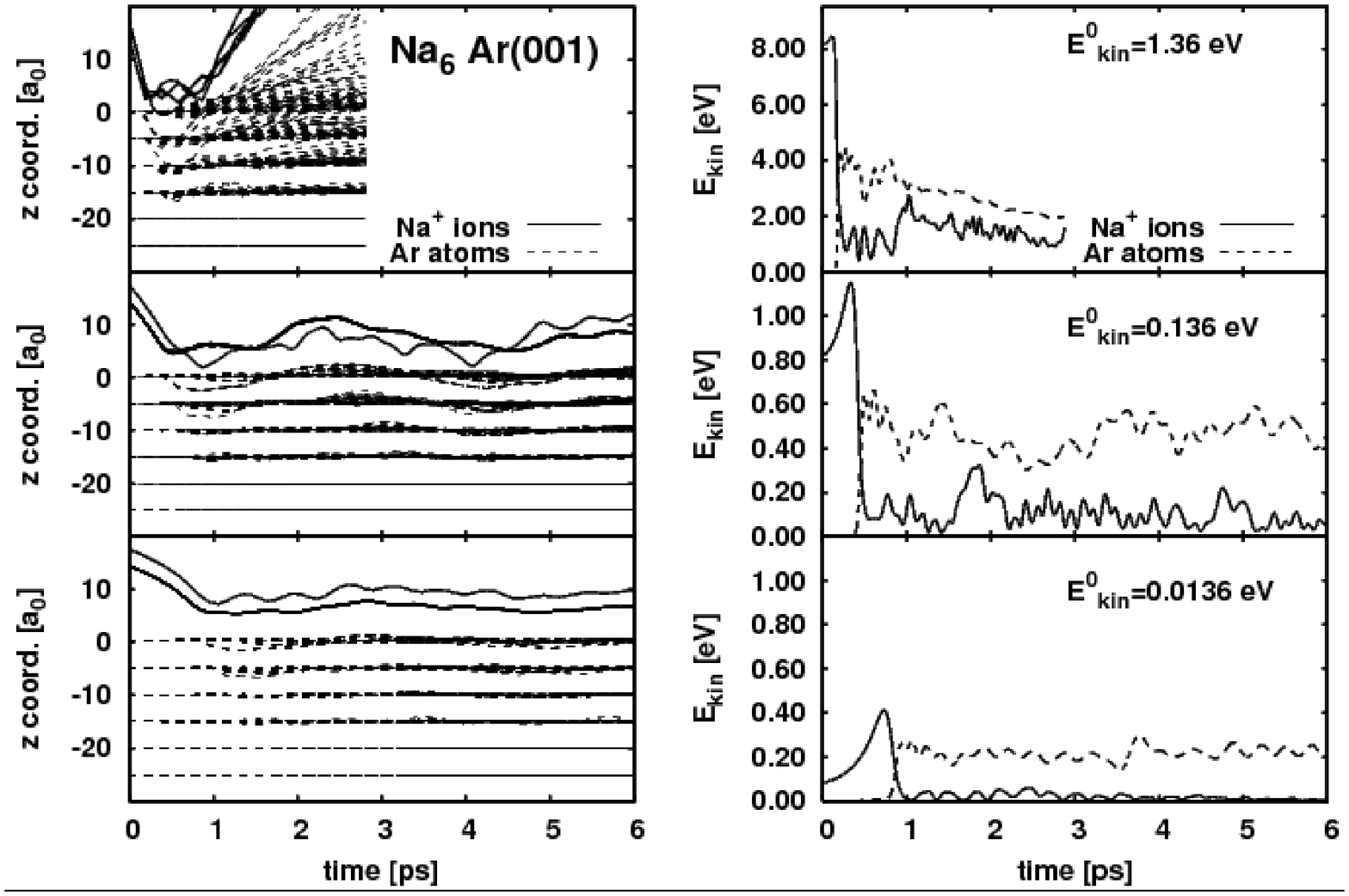,width=0.75\linewidth}
\hspace*{0.4em}
\caption{
Time evolution of deposition of Na$_6$ on Ar$_{384}$ for three
different initial kinetic energies $E^0_{\rm kin}$ as indicated.  
Left column: $z$ coordinates of Na ions (heavy solid lines) and of Ar
cores (faint dashed). Right column: Kinetic energies of the Na cluster
(solid lines) and of the Ar cores (dashes)~\cite{Din07b}.
\label{fig:na6_ar_zE}
}
\end{SCfigure}

Fig.~\ref{fig:na6_ar_zE} summarizes some results of Na$_6$ deposit
on Ar$_{384}$ for a broad range of initial kinetic energies
$E^0_\mathrm{kin}$ (a more extensive version is found in
\cite{Din07b}).
The evolution is similar for all cases, except the most violent one.
The cluster is accelerated first towards the surface through the
polarization attraction (Na$_6$ has a finite dipole moment),
abruptly stopped at first impact and tied to the surface keeping some
remaining oscillations. The substrate absorbs a huge fraction of the
cluster energy at first impact, on absolute scale the more the higher
the initial kinetic energy. Practically all substrate excitation is
collected at first impact. The initial perturbation propagates with
the velocity of sound throughout the whole substrate (note that the two
lowest layers are frozen, see the last paragraph in section
\ref{sec:depos_mecha}). The highest impact energy seems to show
cluster reflection. But that is achieved at the price of destroying
the surface, and to some extent the cluster, after all a very
inelastic collision.
These findings corroborate the fact that Ar is an extremely efficient
stopper material, as was observed experimentally in deposition of Ag
clusters in rare gas matrices~\cite{Har90}. Once deposition energies
are properly scaled (Ag heavier than Na, total deposition energy given
in experiments while $E^0_{\rm kin}$ given per Na atom here), our
results are in perfect agreement with the experimental findings. In
the latter, the destructive regime was however not attained. The
threshold in our simulations lies between 272 and 1360 meV per Na
atom.

The kinetic energies displayed in the right column of
Fig.~\ref{fig:na6_ar_zE} provide a few more quantitative results.
The additional acceleration of 
Na$_6$ depends on the initial $E^0_{\rm kin}$~: slower velocities yield more
relative energy gain since the cluster moves for a
longer time in the attractive polarization field.
The relative energy transfer at first impact is most efficient
for the lowest energy and decreases with increasing velocity
because the then faster cluster couples less efficiently to the
heavier Ar atoms. 
%

\begin{SCfigure}[0.5][htbp]
\epsfig{file=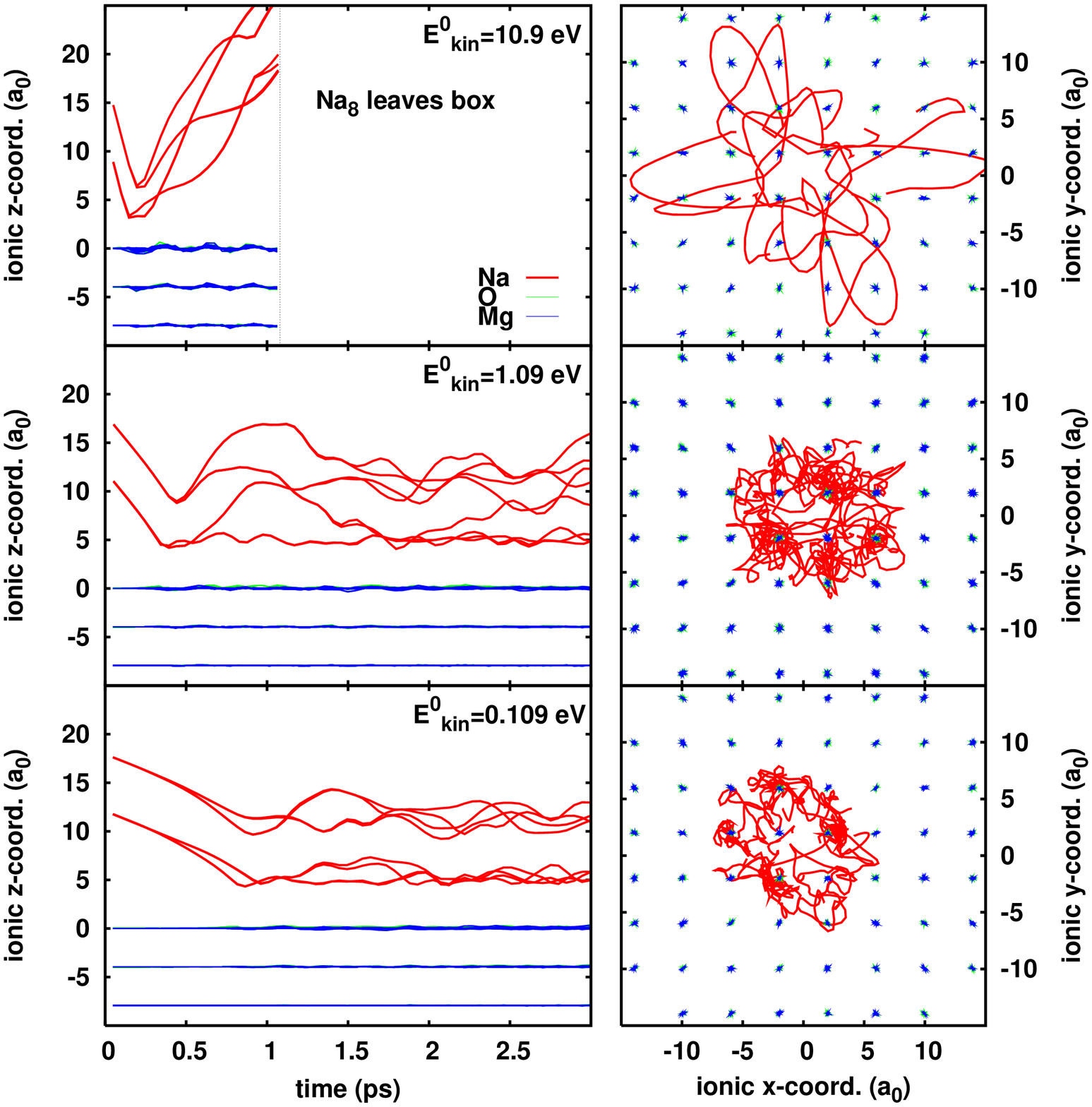,width=0.75\linewidth}
\hspace*{0.4em}
\caption{Dynamical deposition of Na$_8$ on MgO for three different
  initial kinetic energies $E_{\rm kin}^0$ as indicated. 
  Left panels~: $z$ coordinates as a function of time. Right panels~:
  top view~\cite{Bae08a}.
\label{fig:na8_mgo_ekin0}
}
\end{SCfigure}

We now turn to the case of the more rigid MgO(001) surface,
considering deposition of Na$_8$ on MgO~\cite{Bae08a}.
The Na$_8$ cluster consists of two rings of four ions each twisted
against each other by 45$^o$, see Fig.~\ref{fig:nan_mgo_struct}.
It has magic electron number $N=8$ and is thus a bit more densely
packed than Na$_6$.
Fig.~\ref{fig:na8_mgo_ekin0} displays the trajectories in the $z$
direction (left panels) and in the $x$-$y$ plane, parallel to the
surface (right panels) for three different impact energies.  The
cluster symmetry axis is initially above a hollow site with the lower
ring facing closer to bridge sites.
The regimes of attachment and reflection are similar to the previous
case. But elsewise there are remarkable differences to the soft Ar
material. Very little energy is transferred to the MgO surface as can
be seen from the rather small disturbances in the substrate. The
dissipation is achieved here through effective conversion of
translational kinetic energy into cluster internal motion as can be
seen from the strong internal oscillations within the cluster. The
case of reflection (largest impact energy, uppermost panels) does
little harm to the substrate (as opposed to the case of Ar) but
excites the cluster so much that later fragmentation is likely.
A few more detailed observations may be of interest. The
$z$-coordinates in the softest deposition (lowest left panel) start to
deviate already during the initial acceleration phase because the ions
in the lower ring approach different sites on the surface.  At the
same time, the cluster rotates in the $x$-$y$ plane to bring the four
ions of the lower ring closer to the attractive oxygen sites. One may
spot that from the top view of the trajectories in the lowest right
panel.
These trajectory projections in the right column show a further
interesting aspect. While the lowest impact energy leads, besides the
initial rotation, to localized oscillations in one well of the
substrate, the next higher energy (middle right panel) give the whole
cluster sufficient energy to allow for hopping to neighboring sites
which eventually ends up in a slow drift across the surface when
followed over longer times. The trajectories for the highest energy
(uppermost right panel) show the huge intrinsic perturbation of the
cluster, practically the precursor of a final fragmentation (which
cannot be assessed here due to limitations of box size).

\subsubsection{Charge effects}
\label{sec:depos_charge}

In oder to check the effects of different charges of the projectile,
we consider collision of simple monomer with Ar(001) surface.
\begin{SCfigure}
\epsfig{file=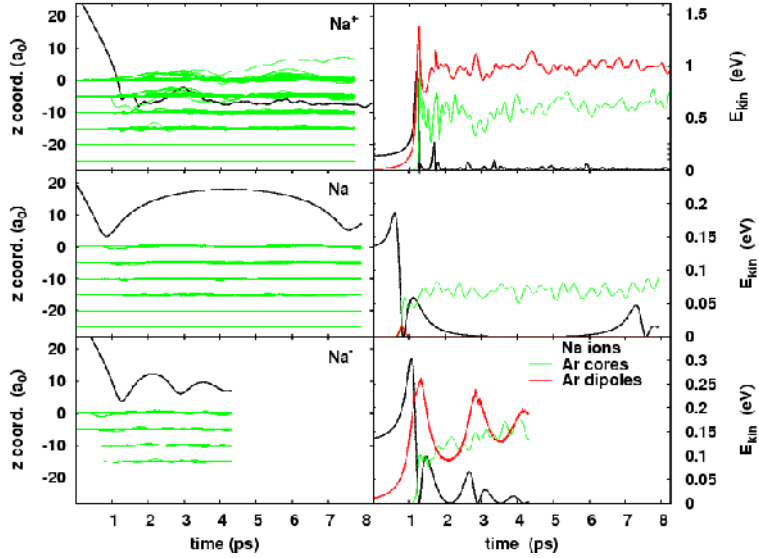,width=0.75\linewidth}
\hspace*{0.4em}
\caption{
Time evolution of $z$ coordinates (left panels) and typical energies
involved (right panels) in the dynamical 
deposition of a Na$^+$(top), neutral Na (middle) and Na$^-$ (bottom)
on Ar$_{384}$ as model for Ar(001) surface.
In all cases, the projectile has initial kinetic energies $E_0$ of 0.136 eV.
Adapted from~\cite{Din08a}.
\label{fig:napm_ar_zE}
}
\end{SCfigure}
Fig.~\ref{fig:napm_ar_zE}
shows the time evolution for $z$-coordinates and kinetic energies, for
collision of a Na$^+$ cation, neutral Na atom, and a Na$^-$ anion.
All three projectiles are captured by the substrate. But evolution and
final state differ dramatically.
The neutral atoms (middle panels) looses 2/3 of its kinetic energy at
first impact and uses the remaining energy to perform oscillations
perpendicular to the surface with large amplitude and cycle time.
The cation (upper panels) is immediately swallowed by the surface due
to the huge polarization interaction. Accordingly, a large amount of
energy goes into the Ar dipoles (upper right panel).
The anion (lower panels) sees also a strong polarization interaction
and strong excitation of the Ar dipoles (lower right panel). But the
two active electrons Na$^{-}$ do also explore the strong electron
repulsion from the Ar cores. Thus the Na$^-$ is not swallowed but
gently deposited in a safe distance from the surface.
Altogether, charge has a very strong influence on the interface
interaction of polarizable media. 
\PGmod{That is corroborated by potential energy surfaces shown in
  Fig.~\ref{fig:bo} and it}
will play a role again in
Sec.~\ref{sec:light} where we consider charging of a deposited (or
embedded) neutral cluster through a short laser pulse.

\subsection{Deposition on planar surfaces -- substrate excitations}
\label{sec:internal_ar}

In the previous section on deposit dynamics, we have \PGmod{discussed}
the general
scenarios under varying initial conditions and material combinations.
In this section, we want to proceed to a more quantitative analysis of
the excitation mechanisms \esmod{at the side of the environment}. 
This is done first in terms of global
observables in Sec.~\ref{sec:sub-global}. Our dynamical
hierarchical modeling allows to develop a detailed picture of the
polarization dynamics. That aspect is discussed in
Sec.~\ref{sec:sub-dip}.

\subsubsection{Substrate temperature and rearrangement}
\label{sec:sub-global}

In the previous section, we have seen for cluster deposition on
Ar(001) that most of the initial cluster kinetic energy is transferred
almost instantaneously at first impact to the substrate.  At the same
time, the cluster, bound or reflected, is internally excited.
It is interesting to see how the excitation energy is distributed over
the various components. 
\begin{figure}[htbp] 
\begin{center}
\epsfig{figure=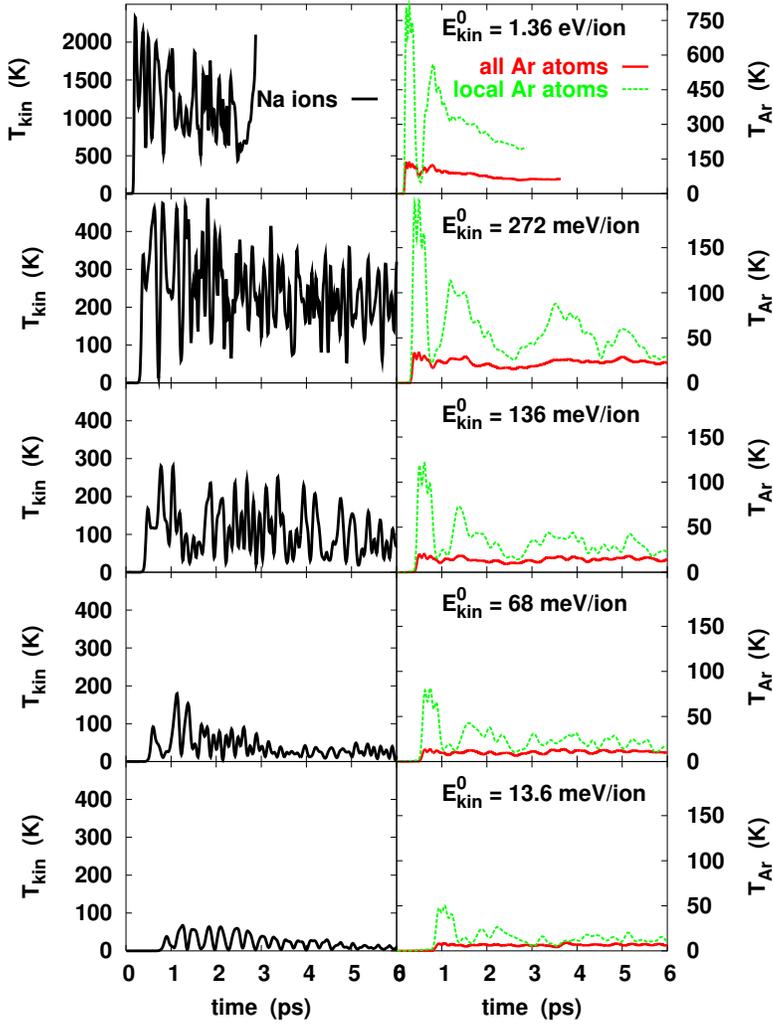,width=0.8\linewidth}
\caption{\label{fig:na6ekint}
Time evolution of kinetic energies for the
deposition of Na$_6$ on Ar\PGmod{(001)} for three
different initial kinetic energies per atom $E^0_{\rm kin}$, as
indicated.
Left column:  kinetic temperature of Na ions.
Right column: kinetic temperature of the Ar substrate
denoted ``all'' (full lines) and
restricted to 
the impact hemisphere (defined in the text),
denoted ``local'' (dashes).
}
\end{center}
\end{figure}
Fig.~\ref{fig:na6ekint} analyzes the distribution of kinetic
energies for deposit of Na$_6$ on Ar\PGmod{(001)} and for three different
initial kinetic energies. Results are presented in terms of kinetic
temperatures. These are evaluated from the total kinetic
energies. First one deduces an intrinsic kinetic energy
$E^\mathrm{(int)}_\mathrm{kin}$ by subtracting the center-of-mass
motion (which is particularly relevant for the small Na cluster).
Then one produces a comparable scale by dividing through the particle
number, yielding the kinetic temperature as $T_{\rm kin} = 2 E^{\rm
int}_{\rm kin}/3N$, where $N=6$ for the cluster and 384 for the
substrate.
The left column of Fig.~\ref{fig:na6ekint} displays the cluster
temperature.  The initial few 100 fs are purely center-of-mass motion
with no intrinsic excitation. The temperature jumps at impact due to
the large perturbation of all constituents. The jump produces almost
the final temperature while there remains some slow and moderate
relaxation to thermal equilibrium.  
The right column of Fig.~\ref{fig:na6ekint} analyzes the temperature
in the substrate, taking the substrate as a whole or restricting the
analysis to an ``impact hemisphere'', that is the Ar atoms in the
vicinity of the impact point (4$\times$4 in first layer, 3$\times$3 in
second, and 2$\times$2 in third). The differences between the two
systems are huge. Most of the initial energy transfer at impact time
goes to the impact hemisphere. The corresponding temperature shows
recurrent bumps related to slow oscillations within the substrate (see
Fig.~\ref{fig:na6_ar_zE}) and associated energy exchanges between
potential and kinetic energy. The temperature then slowly relaxes
towards that of the total system and the details of the relaxation
process strongly depend on the initial energy. The relaxation times
range from about 5 ps for the weakly excited cases to even longer
times (outside the shown time span) for the heftier processes. It is
interesting to note that the average temperature for the highest
energy lies above the melting point of Ar (of about 84
K~\cite{Wah12,Pol64}), which indicates a strong perturbation of the
substrate, in accordance with the graphical impression of figure
\ref{fig:na6_ar_zE}.
Finally, it should be reminded that the electronic excitation during
the collision only amounts to small dipole oscillations (amplitude of
about 0.05 $a_0$) and associated small energy content of about a few
meV, corresponding to a few 10 K temperature. The energy relaxation
between ions/atoms and electrons is extremely slow, see also figure
\ref{fig:na8_ar_temp}.

We have seen from Fig.~\ref{fig:na6ekint} that the dynamics tends
towards asymptotic states with well defined energy share between the
various constituents.  This motivates an analysis of these
``asymptotic states''. To that end, the values of the ionic kinetic,
the atomic potential and the atomic kinetic energies after 6 ps are
recorded and are normalized to the maximum kinetic energy, ${E_{\rm
kin}}^{\rm max}$, reached just before impact. 
\begin{SCfigure}[0.6][htbp]
\epsfig{file=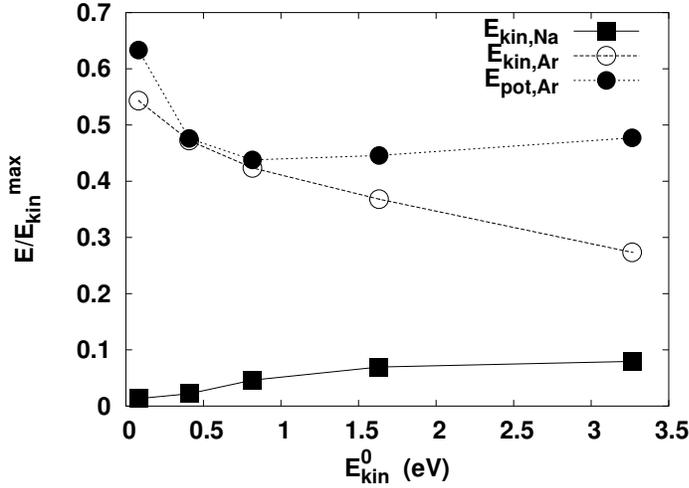,width=0.7\linewidth}
\hspace*{0.5em}
\caption{Partitioning of final energies over cluster kinetic energy,
substrate kinetic, and substrate potential energy \esmod{(see equation. (\ref{eq:Esubstr}))},
normalized to the maximum kinetic energy of the Na$_6$ before impact on
the Ar(001) surface, as a function of the total initial kinetic
energy~\cite{Din07b}.
\label{fig:ar_ekin0}
}
\end{SCfigure}
The resulting ratios are shown in Fig.~\ref{fig:ar_ekin0} as a
function of the initial kinetic energy $E^0_{\rm kin}$.
The relative energy share for the Na cluster increases with increasing
deposition energy, but the energy loss at the side of the Na cluster
is always dramatic, even for the most violent case. In the Ar
substrate, one observes an equal share between potential and kinetic
energies, except for the highest initial energy. The gain in potential
energy reflects the spatial rearrangements experienced in the Ar
substrate after Na impact. At moderate impact energy, the
perturbation generated by the Na does not affect the structure of the
substrate itself. It mostly provokes vibrations of Ar atoms around
their original position.  However, with increasing impact energy, the
collision produces enhanced rearrangement of the substrate structure
and correlatively a larger contribution from the potential energy.
The detailed internal excitations of the Ar atoms as such will be
discussed in Sec.~\ref{sec:sub-dip}.

\begin{SCfigure}[0.5][htbp]
\epsfig{file=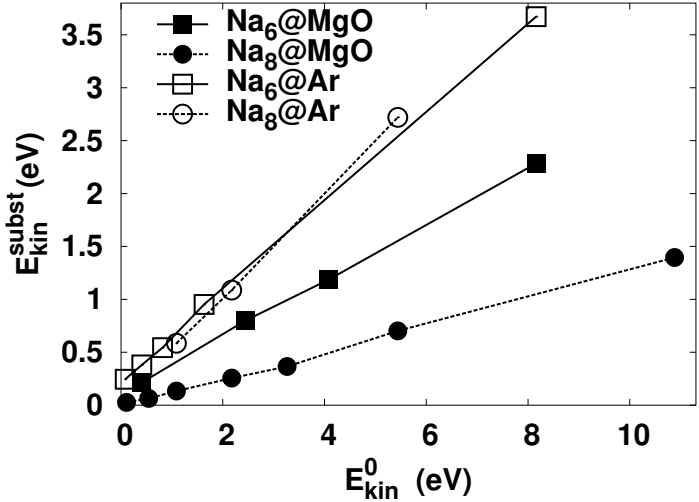,width=0.7\linewidth}
\hspace*{0.4em}
\caption{Kinetic energy transfered within the first 2 ps
 from initial kinetic energy $E_{\rm
    kin}^0$ of the projectile (Na$_6$ or Na$_8$) to the MgO or Ar
  substrate~\cite{Bae08a}. 
\label{fig:mgo_ar_ekin0}
}
\end{SCfigure}
Fig.~\ref{fig:mgo_ar_ekin0} tries as a complement analysis of
kinetic energies in the early stages of the deposition process,
comparing both Ar and MgO substrates, and Na$_6$ and Na$_8$.  It shows
the kinetic energy of the substrate soon after the collision, namely
averaged over the first 2 ps after impact, again plotted as a function
of the initial kinetic energy of the cluster $E_{\rm kin}^0$.
In all cases, we find that the energy absorbed by the substrate is
proportional to $E_{\rm kin}^0$. The slope depends very much on the
cluster/surface combination.  The softer Ar substrate absorbs much
more energy than MgO, typically \esmod{about}  50\% of the
initial kinetic energy.  Furthermore, the softness and the rather
small surface corrugation of Ar make the absorption insensitive to
cluster structure. 
The situation is different for MgO(001) in which the cluster structure
makes a difference. Remind that Na$_6$ does not match very well to the
MgO surface while Na$_8$ does (see Sec.~\ref{sec:NaN_on_MgO}).  In
any case, there is much less energy absorption because of the more
rigid nature of MgO as compared to Ar.

All in all we see that, both at short and long times, the deposition
process leads to sizeable energy transfers (kinetic and potential
energies) to substrate. Actual quantitative details do depend, of
course, on the cluster-substrate combinations.  The energies analyzed
up to now represent global quantities.  In the following subsection,
we will take a closer look on the internal atom excitations and their
spatial distributions.

\subsubsection{Excitations of internal degrees of freedom}

\label{sec:sub-dip}

In our dynamical hierarchical approach, the internal degrees of freedom
of the environment are described by one single quantity, namely the
dipole moment of each substrate atom. A simple measure of the degree
of excitation of the substrate is thus provided by the amplitudes
$\mathbf d$ of the atom dipoles, or the associated energy which is
defined as
\begin{equation}
E_{\rm dip} = \frac{1}{2} \, e^2 \, \frac{{q_{\rm Ar}}^2}{\alpha_{\rm
    Ar}} \, \mathbf d^2,
\label{eq:Edip}
\end{equation}
where $\alpha_{\rm Ar}$ is the static polarizability of bulk Ar, and
$q_{\rm Ar}$ the effective charge of the Ar cores~\cite{Feh05}). This
dipole energy can be 
analyzed globally (summed over all substrate atoms) or locally, by
considering its values at each Ar site. As a first step, we consider
it globally and look at how it evolves in time in typical deposition
scenarios.

A first indication was given in Fig.~\ref{fig:napm_ar_zE} proving
that sizable dipole energies show up as soon as one considers
deposition of charged species~\cite{Din08a}.
\begin{SCfigure}[0.5]
\epsfig{file=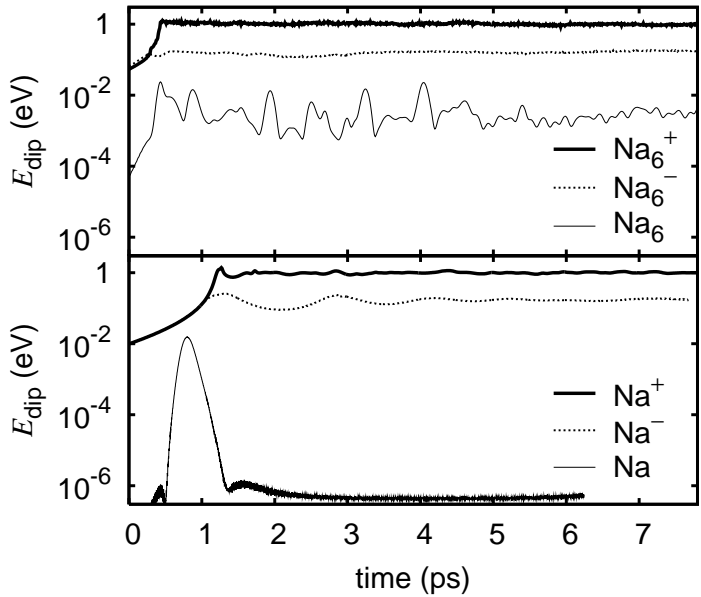,width=0.7\linewidth}
\hspace*{0.4em}
\caption{Time evolution of total dipole excitation energy in the substrate
  Ar$_{384}$   following deposition of various clusters (upper panel)
  or monomers (lower panel),
  all with initial kinetic energy $E_0=0.136$ eV/ion.
 Adapted from \cite{Din08c}.} 
\label{fig:depos_edip}
\end{SCfigure}
This is confirmed by Fig.~\ref{fig:depos_edip} showing the time
evolution of the total dipole energy (summed over all the substrate
atoms) for deposition of various projectiles, neutral and charged as
well as Na$_6$ clusters and Na monomers~\cite{Din08c}.  In all cases,
we observe a similar behavior with a short time transient regime
leading after typically 1-2 ps to an asymptotic value. The largest
dipole energy is attained in the case of Na$_6^+$.  The anions have
still large dipole energies due their charge. But the larger \esmod{equilibrium  distance
of the deposited cluster} to the substrate reduces the effect.  Neutral systems produce orders
of magnitude smaller dipoles, still visible only for Na$_6$ with its
finite dipole moment, and negligible values for the Na atom.
Concerning the sharing between the Ar core kinetic energy and the
internal excitation energy stored in the Ar dipoles, we found the
following.  The ratio between dipole energy and Ar core energy is large
and comparable for all charged species.  Both cations reach a ratio of
1.7 while the anions have about half of that value.  Neutral systems,
on the other hand, have generally a very small fraction of dipole
energy, orders of magnitude smaller than the core energies.


The marked role of charge in the dipole energy discussed above
suggests to explore in more detail the spatial distribution of the
dipole excitation.
\begin{SCfigure}[0.5]
\epsfig{file=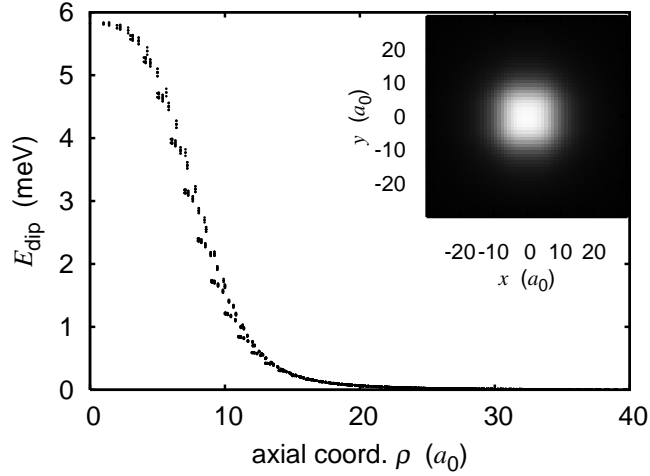,width=0.65\linewidth}
\hspace*{0.4em}
\caption{Axial distribution of the atomic dipole energies in the first layer
of the Ar(001) substrate at impact time, for the deposition of
Na$^+$ with initial kinetic energy $E_0=0.136$ eV. The insert shows
a the full distribution in the first layer ($x-y$
plane) as gray scale plot. Adapted from \cite{Din08c}.}
\label{fig:nap_hotspot}
\end{SCfigure}
Fig.~\ref{fig:nap_hotspot} shows a snapshot of the distribution of
the dipole energies at impact time for the Ar atoms in the uppermost
layer as a function of the axial coordinate $\varrho=\sqrt{x^2+y^2}$.
The impact point corresponds to $\varrho=0$. The insert shows the full
distribution in the surface plane.  The results were produced for
deposition of Na$^+$ with initial kinetic energy of 0.136 eV. An
analysis of the deeper lying layers yields a qualitatively similar
picture, although the effects are quantitatively much suppressed. The
figure shows a high excitation of the dipoles, strongly located around
the impact point.  The pattern remains stable in time, as could be
expected from the time behavior of the total dipole energy, see figure
\ref{fig:depos_edip}.

\begin{SCfigure}[0.5]
\epsfig{file=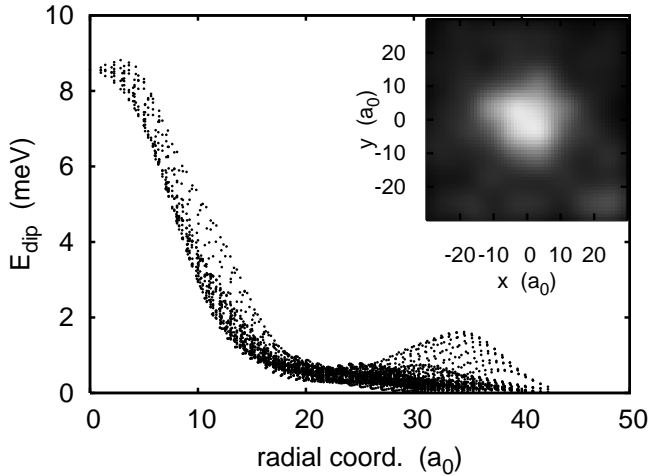,width=0.65\linewidth}
\hspace*{0.4em}
\caption{Same as Fig.~\ref{fig:nap_hotspot} but for Na$_6^+$.}
\label{fig:na6p_hotspot}
\end{SCfigure}
Fig.~\ref{fig:na6p_hotspot} shows the dipole distributions for
collision with Na$_6^+$ in similar fashion
as Fig.~\ref{fig:nap_hotspot}.
The overall picture is similar. But there appear some noticeable
differences. 
The spatial extension of the polarization spot is much
larger for Na$_6^+$, because it exhibits a larger
cross-section with respect to the surface and because it
is surrounded by a large electron cloud.
The detailed $x$-$y$-distribution even reflects the fivefold 
structure of the lower ring of Na$_6^+$.
More surprising might be the revival at larger distance. This is
caused by the extension of Na$_6^+$ and particularly by the large
range of its electron cloud.  This leads to a screened charge for the
Ar atoms close to the center and a revival at larger distances.  As
for the cation, the dipole energy distribution changes very little
with time.

To summarize, we have seen that the substrate exhibits a sizeable
response both in terms of atomic motion and atomic internal
excitations (dipoles). We have analyzed these effects as a function of
time and found that asymptotic values are quickly attained after
impact. The internal excitation is mostly dependent on charge of the
deposited species indicating that most of the effect is a static
induced polarization of the substrate. Such a static polarization is
confirmed by detailed analysis.


\subsection{Collisions with large Ar clusters}
\label{sec:large-ar}

The present treatment of planar surfaces uses layers of small
plaquettes of 8$\times$8 sites in the case of Ar. Although large, this
limits shape relaxation of the substrate in lateral direction. In
order to investigate the shape response of Ar substrate in a more
freely variable situation, we go to the other extreme of finite Ar
clusters as models for the substrate. We will discuss in this section
the collision of small Na cluster with larger Ar clusters with an
emphasis on the shape evolution.

\subsubsection{Trends with system size}
\label{sec:trends-ar}

First, we investigate the effect of target size, considering collision
of Na$_6$ with Ar$_{7}$, Ar$_{43}$, and Ar$_{87}$~\cite{Din07c}. The
initial kinetic energy of Na$_6$ is $E_{\rm kin}^0=68$ meV per Na
atom, and the separation between the centers of mass of the two
clusters is initially 30~$a_0$. \esmod{When considering such collision 
processes between two finite systems, it is interesting to remind some orders of 
magnitude of typical binding energies: 800 meV for
Na$_2$ dimer to be compared to 50 meV for Ar bulk and 5 meV to Na-Ar dimer. } 
%
\begin{SCfigure}[0.5][htbp]
\epsfig{file=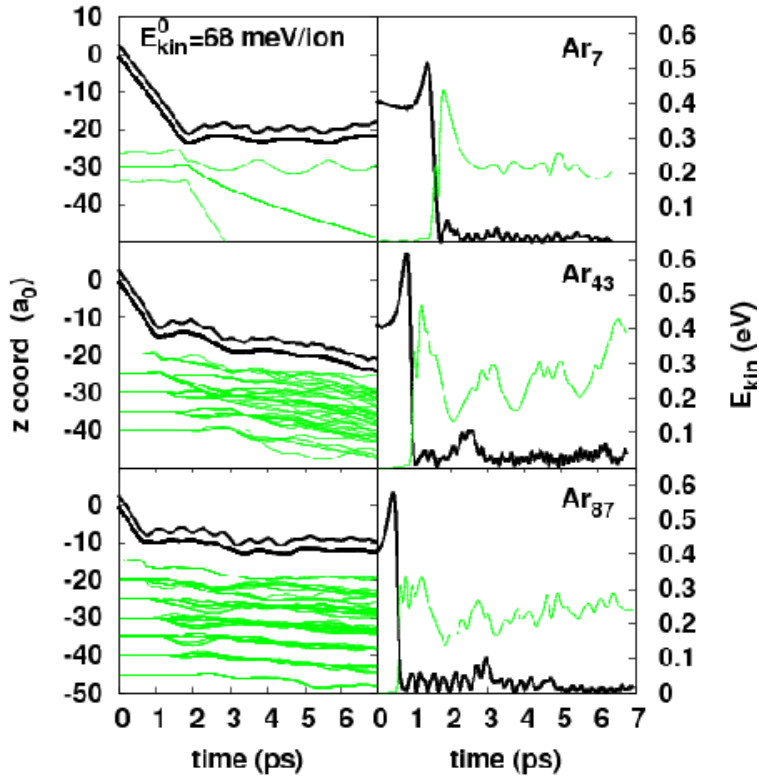,width=0.75\linewidth}
\hspace*{0.5em}
\caption{Collision of Na$_6$ (thick lines) with an initial kinetic
  energy of 68 meV/ion, on Ar$_N$ (thin curves) for $N=7$ (top), $N=43$
  (middle) and $N=87$ (bottom). $z$-coordinates (left) and kinetic
  energies of Na ions and Ar cores, as a function of
  time~\cite{Din07c}.
\label{fig:na6_arNcl}
}
\end{SCfigure}
Fig.~\ref{fig:na6_arNcl} shows the time evolution of the $z$
coordinates (along the symmetry axis of the system Na$_6$/Ar$_N$,
which is also the direction of the collision) and kinetic energies of
the Ar cores and the Na ions.  Ar$_7$ is not massive enough to survive
the collision.  The larger Ar$_{43}$ remains intact but is strongly
perturbed probably above the melting point, while Ar$_{87}$ has enough
capacity to absorb the collisional energy without changing its atomic
shell structure.
%
%
After the collision, almost the same total kinetic energy of about 0.3
eV is transferred in all three cases, independent from the Ar$_N$ size
and the energy absorbed by the Ar system is also similar.  These
findings coincide with the energy transfer for the planar surface, see
Fig.~\ref{fig:na6_ar_zE}.  If we convert the kinetic energy in the
Ar system into a typical temperature for the Ar system, we obtain 200
K for Ar$_7$, 30 K for Ar$_{43}$, and 10 K for Ar$_{87}$, in
accordance with the observation of break-up, melting and
stability. Deposition on true Ar bulk would yield even smaller
temperatures because of the  larger heat capacity. The difference to
10 K, may be even 30 K, is not so dramatic such that Ar$_{87}$ can be
considered as a reliable representative of a bulk surface, and to some
extent also Ar$_{43}$.

The times scales which can be read off from Fig.~\ref{fig:na6_arNcl}
also agree nicely with those found with the planar surface model:  The
energy transfer from the projectile to the Ar cluster after typically
0.5 ps is very fast. The energy distribution into the Ar cluster
proceeds as a sound wave with speed of 20$-$30 $a_0$/ps, as the
perturbation propagates like a straight line through the Ar layers
(see left panels below 2 ps and also top panel of
Fig.~\ref{fig:na6_depos_compar}). The relaxation operates at a
longer time scale, typically of order of 10 ps.


\subsubsection{Shape dynamics of the Ar system}
\label{sec:wetting}

\begin{SCfigure}[0.5][htbp]
\epsfig{file=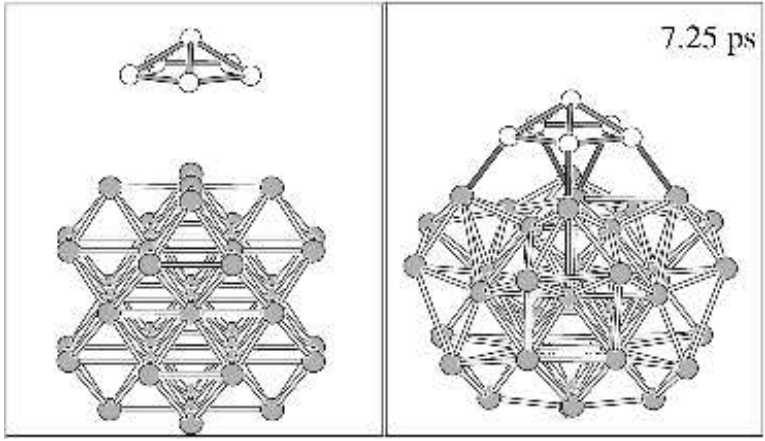,angle=90,width=0.75\linewidth}
\hspace*{0.5em}
\caption{Configuration of the Na$_6$Ar$_{43}$ system
  at initial time (left panel) and after successful landing
  and relaxation (right panel). Na ions are represented by open
  circles and Ar atoms by gray circles. Adapted from~\cite{Din07c}.
\label{fig:Na6-Ar43}
}
\end{SCfigure}
Fig.~\ref{fig:Na6-Ar43} illustrates the shape response of the Ar
system during a soft landing process of Na$_6$ on Ar$_{43}$. The
smaller Ar cluster has  been chosen for better graphical oversight.
The initial configuration was prepared such that the Ar cluster
present to the Na$_6$ a nice flat side, almost like a flat surface.
The substantial reshaping of the target through the impinging cluster
is obvious. The cluster modifies the surface to optimize its vicinity
which, in turn, leads to a global deformation of the whole system.

The result indicates that there may be substantial rearrangements at
the side of the Ar substrate, possibly also some for the Na cluster.
In order to quantify these effects, we perform a shape analysis in
terms of the first three multipole moments both for Na$_6$ and
Ar$_N$. These moments are given by\PGmod{
\begin{eqnarray*}
\sqrt{\langle r^2 \rangle} &=& \sqrt{\langle x^2 + y^2 + z^2 \rangle} 
 =  \left( \frac{1}{p} \sum_{i=1}^p ({x_i}^2
     + {y_i}^2 + {z_i}^2) \right)^{1/2},\cr 
\beta_2 &=& \sqrt{\frac{\pi}{5}} \frac{1}{\langle r^2 \rangle}
\langle 2 z^2 - x^2 - y^2\rangle, \cr
\beta_3 &=& \left(\frac{2}{5 \langle r^2 \rangle} \right)^{3/2}
\langle z\left(z^2 - \frac{3}{2}(x^2+y^2) \right)\rangle,
\end{eqnarray*}
}
where $p$ is either the number of Na atoms $N_{\rm Na}$ or the number
of Ar atoms $N_{\rm Ar}$, and $x$, $y$ and $z$ are the coordinates of
the Na (Ar) atom with respect to the center of mass of the Na (Ar)
cluster.  The r.m.s. radius $r$ stands for the overall extension
(monopole moment) and the deformations are parameterized as
dimensionless quantities which have immediate geometrical meaning
independent of system size. For example, a value of $|\beta_2|\approx
0.8$ is a large quadrupole deformation with axis ratio of about 2:1.
\begin{SCfigure}[0.8][htbp]
\epsfig{file=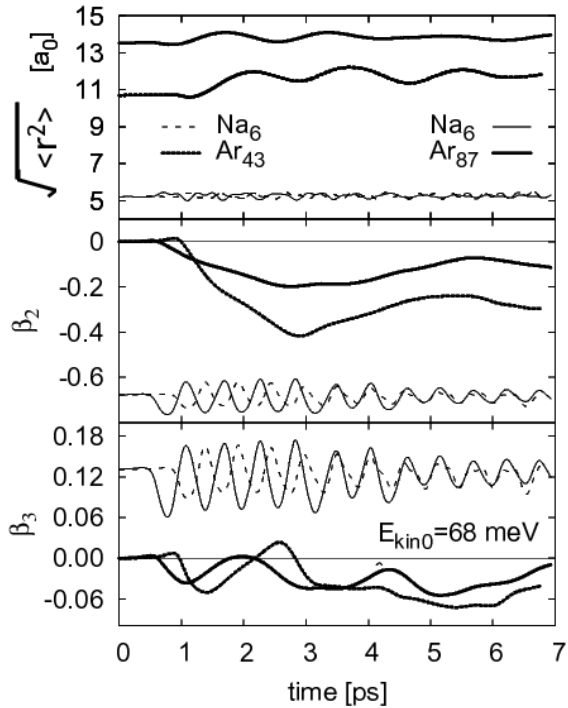,angle=90,width=0.65\linewidth}
\caption{Three first multipole moments $\sqrt{\langle r^2 \rangle}$,
  $\beta_2$ and $\beta_3$, as a function of time, for Na$_6$ (thin
  dashes and lines) deposited on Ar$_{43}$ (thick dots) and Ar$_{87}$
  (thick full lines) for $E_{\rm kin}^0=68$ meV/ion~\cite{Din07c}.
\label{fig:na6_arNcl_mom}
}
\end{SCfigure}
Fig.~\ref{fig:na6_arNcl_mom} shows the three moments for the Na and
Ar subsystem in the cases Na$_6$@Ar$_{43}$ and Na$_6$@Ar$_{87}$ for
the moderate initial kinetic energy $E_{\rm kin}^0=68$ meV per Na ion.
The shape of Na$_6$ is rather rigid in any case.  There are some
deformation oscillations short after impact which relax within about 3
ps. These oscillations are predominantly caused by the outer ion. The
ring is tightly bound and stays more robust.  The relaxation of these
cluster internal oscillations \PGmod{are} much faster than for the overall bouncing
oscillations of the cluster relative to the Ar part. This is
explained by the fact that the NaAr binding is softer than the Na$_6$
internal binding.
The Ar clusters, after the impact with Na$_6$, increase slightly in
size due to their heating.  The growth is relatively larger for the
smaller Ar$_{43}$ which acquires a higher temperature as discussed
above (see Sec.~\ref{sec:trends-ar}).
The Ar clusters undergo a strong persistent change in deformation
towards a sizeable oblate (negative $\beta_2$) and somewhat pear-like
(non-zero $\beta_3$) shape. They obviously accommodate their
configuration as to establish a most compact combined system.
The global deformations shrink with Ar system size. They will converge
to zero for an infinite substrate. What remains to be learned from the
study with finite Ar cluster is that there may emerge substantial
reconfiguration of the Ar surface near the deposited cluster. A
thorough analysis requires larger systems from both side: larger
finite clusters as well as larger plaquettes in the modeling of a
planar surface.


\subsubsection{Influence of temperature}

Another parameter which may influence the deposition process is the
temperature of the substrate. The binding energy of Ar substrate as
such and between the metal cluster and the Ar is low which requires
low temperatures.  Low temperatures are also needed to avoid cluster
diffusion and coalescence. Indeed deposition experiments of metal
clusters on Ar coated metal surfaces were performed at temperatures
between 20 and 30~K \cite{Bro96,Fed98,Lau00a,Lau00b,Lau03}.

In this section, we explore the influence of Ar \PGmod{temperature} on the collision, 
starting from four different 
Ar temperatures, namely 0, 20, 50 and 100~K. The results are presented in
\begin{figure}[htbp]
\begin{center}
\epsfig{file=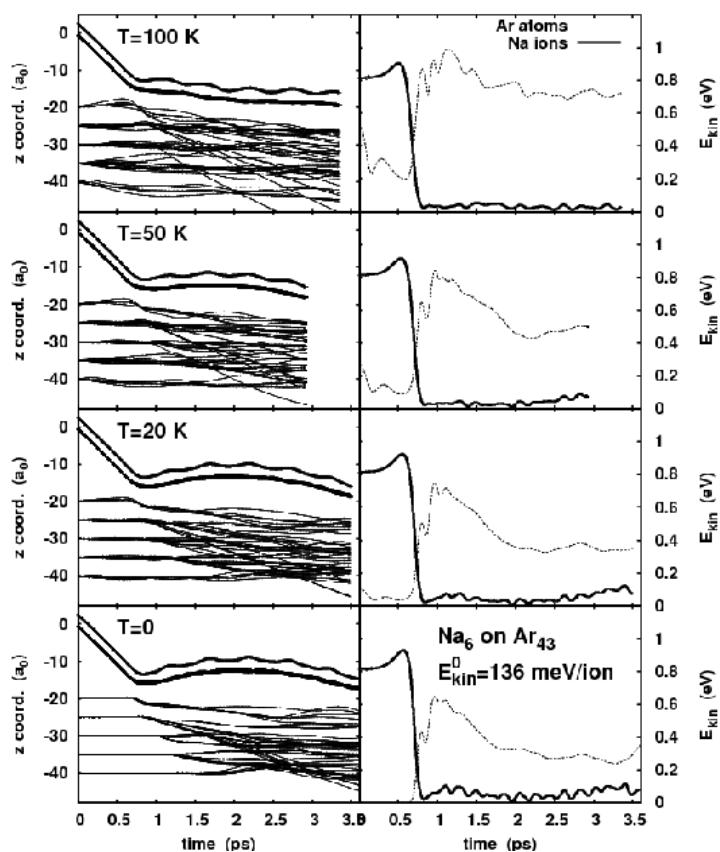,width=0.7\linewidth}
\caption{Collision between Na$_6$ and Ar$_{43}$ with cluster initial
  kinetic energy of 0.136 eV/ion, for three different temperatures of
  the Ar cluster, as indicated. Left column~: $z$ coordinates; right
  column~: ionic and atomic kinetic energies. The quantities are
  plotted as a function of time.
\label{fig:na6ar43_therm}
}
\end{center}
\end{figure}
Fig.~\ref{fig:na6ar43_therm} which shows the results of a collision
between Na$_6$ and the finite Ar$_{43}$ with initial kinetic energy of
\PGmod{$E_\mathrm{kin}^0=136$} meV per Na atom and for different
initial temperatures of the Ar system.  The temperature does not seem
to affect very much the dynamics, at the side of either the Na cluster
or the Ar one. The melting of the Ar$_{43}$ is already observed when
one starts from 0~K; it becomes all the more important for higher
starting temperatures. From the energetic point of view, the larger
the Ar temperature, the higher its kinetic energy after the
impact. However, the energy transfer from the metal cluster to the Ar
system is independent of the initial Ar temperature. Indeed, once one
has subtracted the kinetic energy of the Ar cluster due to its initial
temperature, one observes in all cases the same quick transfer of two
thirds (about 0.6 eV) of the kinetic energy of Na$_6$ before
impact. The asymptotic value of the Ar kinetic energy also seems to
simply be the addition of its initial kinetic energy to the amount
which is gained after the deposition.

\section{Coupling to light}
\label{sec:light}

Metal clusters with their pronounced plasmon resonance are very
responsive to electromagnetic excitations. Thus the phenomena emerging
for clusters under the influence of strong electromagnetic fields have
been much studied in several respects.  One way to exert substantial
perturbations is the collision of a cluster with highly charged ions
which leads to strong electronic excitations, ionization, and
subsequently often fragmentation, see e.g. \cite{Cha95,Ced00,Nta02}.
The majority of studies deal with laser irradiation, which is widely
tunable in frequency, pulse shape and strength which, in connection
with the selective Mie plasmon resonance, provides a world of
scenarios for laser induced non-linear dynamics, \PGcomm{for reviews in
different regimes of laser-cluster interaction see e.g.
\cite{Cal00,Pos01B,Kra02aR,Rei03a,Rei04c,Bel04a,Kou05a,Kou05aR,Saa06aR,Rei06aR,Fen08a}.}
The plasmon serves also as a versatile handle for various scenarios in
pump and probe experiments, e.g. \cite{Bes99,And02,Doe05b}, and it
provides the key mechanism in driving the hefty Coulomb explosion of
large clusters \cite{Dit96,Buz96}.

An even richer variety of phenomena emerges when considering metal
clusters in contact with other materials (embedded in a matrix or
deposited at the surface of a substrate), as e.g. second harmonic
generation \cite{Goe95,Bal00,Koh00} or dedicated shaping of clusters
with intense laser pulses \cite{Sei00,Oua05a}.  
Moreover, the
environment simplifies the handling such that many interesting ongoing
experiments can only be done with clusters in contact with a carrier
material, see e.g. \cite{Nil00,Leh00,Gau01,Die02}. Last but not least,
metal clusters in contact with insulators are a versatile model system
for chromophores which can be used, e.g, for studies of radiation
damage in materials \cite{Bar02b,Niv00} or as indicators in
biological tissues \cite{May01,Dub02}.

\PGcomm{ An example for dynamics of embedded clusters induced by
strong laser pulses was already given in the left lower panel of
Fig.~\ref{fig:examples}. It shows the dramatic change of optical
absorption strength through laser irradiation of Ag clusters embedded
in glass \cite{Sei00}. Large Ag clusters in glass are produced by
inserting the glass into molten Ag and waiting until a sufficient
amount of Ag atoms has diffused into the glass, replaced its Ag ions,
and coagulated to clusters (for a theoretical Molecular Dynamics simulation
of that process, see \cite{Tim97}). These clusters are preferably
spherical with diameter between 5 nm to 50 nm. Accordingly, they show
one clean Mie plasmon peak as seen from the solid line in the left
lower panel of Fig.~\ref{fig:examples}. The sample is then
irradiated with a strong laser (wavelength = 400 nm $\equiv$ frequency
= 3.1 eV, fluence 10 mJ/cm$^2$). After a sufficiently long pause for
full relaxation of the shaken system (about 1 s), the optical spectra
are recorded again using two different laser polarizations. The
results (dashed and dotted lines) show a substantial red-shift and
broadening, independent of the laser polarization. The general
red-shift indicates substantial growth of cluster radius.
The peak from the both polarization are shifted somewhat differently
which indicates that a moderate quadrupole deformation has developed
together with the global expansion.
The strong broadening of both peaks suggests that a somewhat
diffuse environment has been produced by the violent laser
excitation. The most probable scenario is that the laser heats the
cluster up and ejects a lot of electrons initially. The electrons are
stuck in the vicinity of the cluster due to the poor conduction of
glass. The heat stored in the cluster leads to evaporation of
monomers, Ag ions, and perhaps larger fragments which diffuse away
through the glass. The rigidity of the glass confines the diffusion
processes.  At the end, we have large Ag clusters surrounded by a halo
of small Ag clusters. This setup still shows one strong (but broad)
resonance peak with some red-shift due to the effectively larger
radius.

The example above uses a late second pulse to probe the final changes
caused by the first pulse. True pump-and-probe analysis (PPA) aims at a
time-resolved protocol of the process.  It has been extensively
exploited to track molecular motion, a rich field of research, often
called femto-chemistry \cite{Zew94B,Gar95,Zew00a}. PPA is also an
advanced tool in the context of laser-cluster interaction, see
e.g. \cite{Kle97,Vor97a,Wen99a,Per00a,Sei00,Ver04,Doe05b}. The
prominent Mie surface plasmon in metal clusters serves as a
particularly useful doorway for the analysis. The unique relation
between cluster extension and peak frequency allows to map the
deformation dynamics of a cluster, of the evolution of radial shape
\cite{And02}, of quadrupole deformations \cite{And04}, and of
elongation in a fission process~\cite{Din05,And06}. It has been
employed for the reasoning in the above discussion of the left
lower panel of Fig.~\ref{fig:examples}. An instructive example is
the time-resolved analysis of explosion dynamics for Ag clusters in
\cite{Doe05a,Doe05b}. 
}
\begin{figure}[htbp]
\begin{center}
\epsfig{file=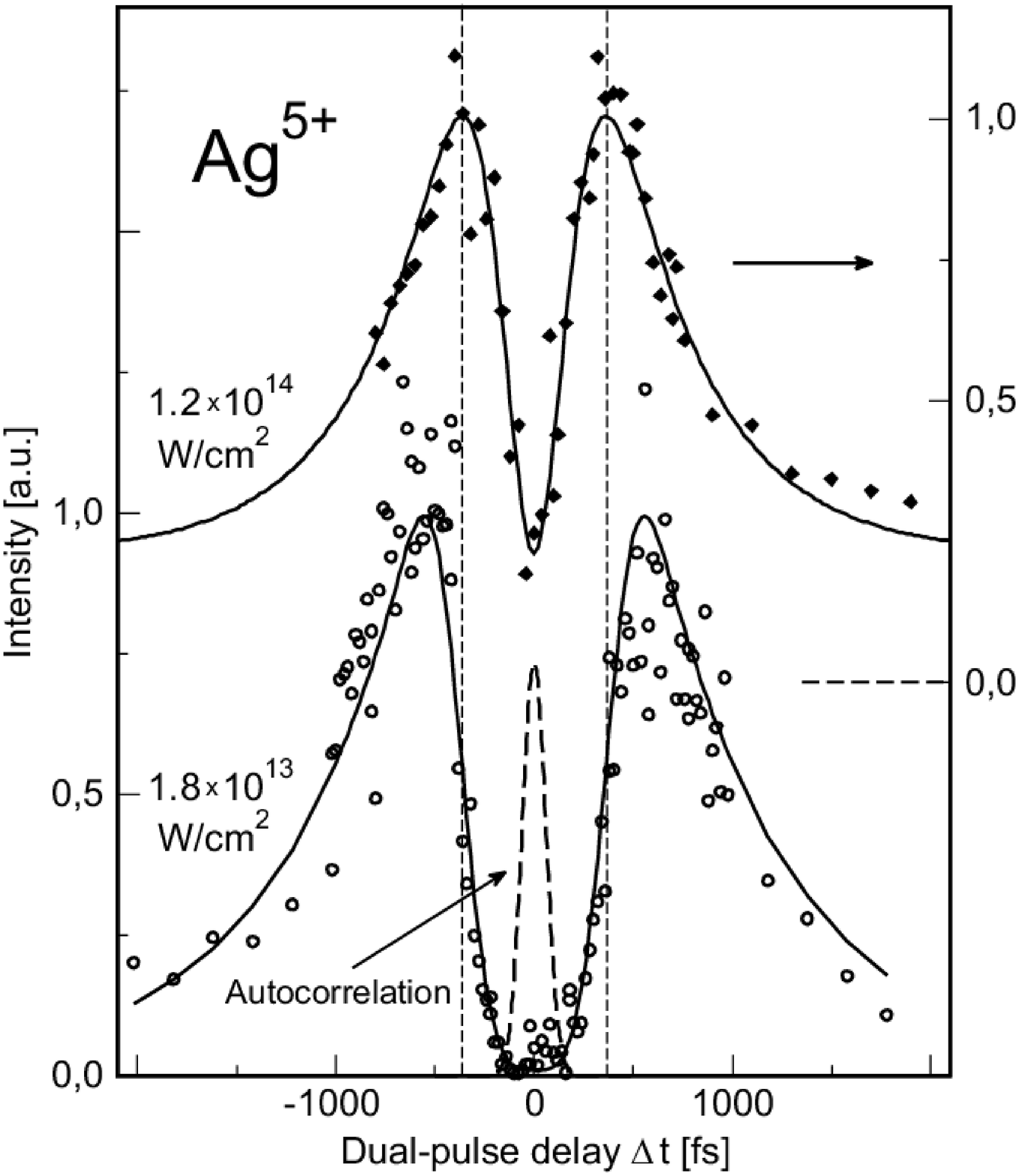,width=0.7\linewidth}
\end{center}
\caption{\label{fig:Ag5_PP} 
Yield of Ag$^{5+}$ ions as a function of
pulse delay in a dual pulse measurement of Coulomb explosion of Ag
clusters embedded in a He droplet. Typical cluster sizes were
$N\approx 2\times 10^4$. Excitation and probing was done by laser
pulses of 1.5 eV energy and 100 fs length. Two different intensities were
used as indicated. The left $y$-axis applies to the lower intensity
and the right axis to the higher. The dashed line shows the
auto-correlation signal from the laser to indicate the time
resolution.  Adapted from \cite{Doe05b}.
}
\end{figure}
{
Fig.~\ref{fig:Ag5_PP} shows the net yield of Ag$^{5+}$ ions as a
measure for the violence of the reaction and thus for the net light
absorption. Both results show a pronounced maximum at a definite delay
time. What happens is that the first pulse produces a strong
ionization and so triggers a Coulomb explosion of the cluster. The
cluster radius grows and the Mie plasmon resonance frequency shrinks
accordingly. The second pulse couples maximally if it comes \PGmod{at} a time
where it is just in resonance with the actual Mie frequency. Now one can see
in the figure that the maximum appears at an earlier delay time for
the more intense laser. The then larger initial ionization causes a
faster expansion and the cluster meets the resonance conditions
earlier.}  
{These measurements were performed for clusters embedded  in 
a helium droplet but \PGmod{the} effect of the He environment does not play a decisive
role at the present qualitative level of the discussion, all the more that the 
perturbation of the system is quite sizable.}
{The above two examples belong to the more violent events caused by
laser irradiation. Lower laser intensities, of course, probe different
features and observables. The weakest pulses are related to the linear
response regime of optical absorption studies as discussed in section
\ref{sec:opt_resp}. Just above that regime comes the more detailed
analysis of electronic properties by means of Photo-Electron
Spectroscopy (PES) and Photo-Electron Angular Distributions (PAD).
Further up, the coupling to ionic motion comes into play which can be
explored, e.g., by PPA. And finally, we reach the regime of violent
excitations with Coulomb explosion and associated fragment analysis.
}
In the present section, we will discuss briefly the broad range of scenarios
emerging when going from "gentle" to "strong" irradiation
processes. We especially analyze chromophore effects in the moderate
energy domain and hindered, or delayed, Coulomb explosion in the high
energy regime, both examples being related to ongoing experiments.

\subsection{Basic mechanisms and chromophore effects}

\begin{figure}[htbp]
\begin{center}
\epsfig{file=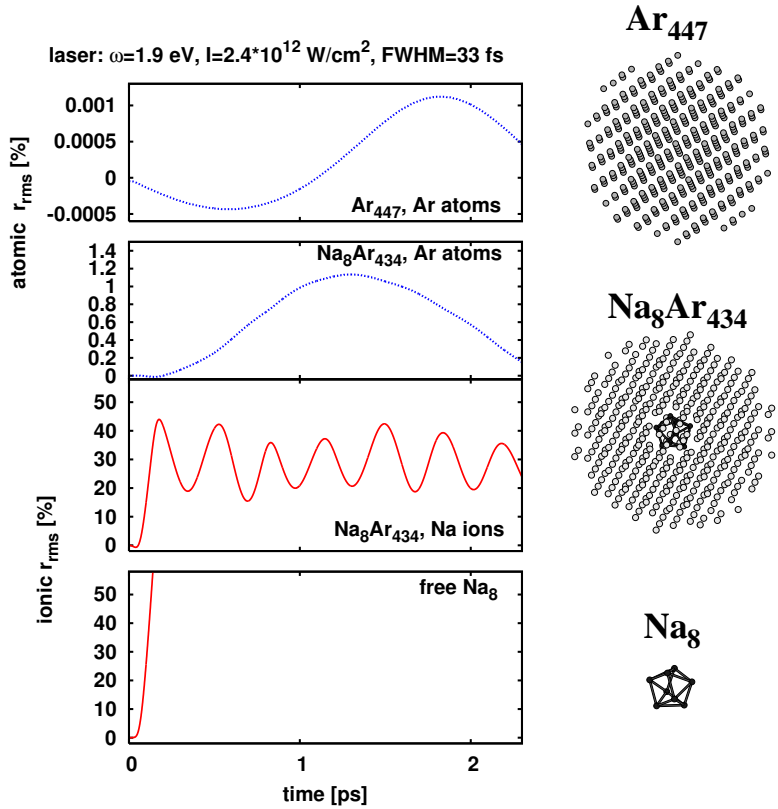,width=0.8\linewidth}
\caption{\label{fig:na8_compar}
Time evolution of the root mean square (r.m.s.) radius of free Na$_8$
(bottom), Na$_8$ embedded in Ar$_{434}$ (middle panels), and pure
Ar$_{447}$ (top), as a function of time, after irradiation by a laser
of intensity $2.4\times 10^{12}$ W/cm$^2$, frequency of 1.9 eV, FWHM
of 33 fs. 
\PGcomm{
The corresponding ionic/atomic structures are shown to the right of
the corresponding figures. Na ions are drawn as black circles and
Ar atoms in gray. Note that the composite cluster Na$_8$Ar$_{434}$
has Na$_8$ embedded in a sufficiently large cavity inside the
Ar$_{434}$ environment. Collected} from~\cite{Feh08a}.}
\end{center}
\end{figure}

In a first step, it is enlightening to compare the response of the various
combinations, free cluster, embedded cluster and pure substrate, to a
strong laser pulse. To that end, we show in Fig.~\ref{fig:na8_compar}
the result of irradiation of the various species by the same laser.
Test cases are: free Na$_8$ as an example for a small metal
cluster, Na$_8$ embedded in Ar$_{434}$ (for structure and construction,
see Sec.~\ref{sec:na8_in_ar})
as a metal cluster embedded in an inert material, and a pure Ar$_{447}$
cluster to countercheck the material's response.  The laser parameters
have been tuned to lead to a charge state 3+ of the Na$_8$ cluster. In
the free case, this leads to an immediate explosion of the cluster as
can be seen from the quickly diverging ionic radii (lowest panel). The
situation comes out quite different when Na$_8$ is embedded in an
Ar$_{434}$ matrix. The metal cluster, which acts then as a chromophore
inside the "matrix", is again highly excited but its explosion is
hindered by the Ar atoms (second panel from below) whose strong
Na$^+$-Ar repulsion together with their large inertia keeps the Na$^+$
ions in its Ar cage and allows the Ar shells to absorb the excitation
energy of the system. The Ar matrix as a whole is perturbed and
exhibits predominantly monopole oscillations, but of much smaller
amplitude than Na$_8$ (second panel from above).  The uppermost panel
of Fig.~\ref{fig:na8_compar} finally shows the case of the pure Ar
cluster. The size is now 447 corresponding to the original cluster
before the hole was drilled to make space for Na$_8$ (the 13 central
atoms of Ar$_{447}$ were extracted to accommodate Na$_8$, 
see Sec.~\ref{sec:na8_in_ar}).  Under the
same laser conditions, one can see that the Ar$_{447}$ remains
essentially unperturbed. No electron is emitted (the intensity
threshold for electron emission is about two orders of magnitude
larger than the intensity of the laser used in this case, 
see Sec.~\ref{sec:limits}) and the Ar
cluster experiences very low amplitude monopole oscillations (mind the
vertical scale), which
reflect its negligible coupling to the laser field in the frequency
range of the cluster's plasmon resonance.  For these purposes, the
rare gases are useful representatives of transparent and inert
substrates (the relations may change, of course, for UV light at
about 25 eV, the Ar \MDcomm{atom electronic} resonance frequency).

\subsection{Photoelectron spectra and photoelectron angular
  distributions}

Optical absorption measurements allowed to collect rich information on
structure and dynamics of clusters, see Sec.~\ref{sec:struct}.  More
information can be gathered when additionally measuring reaction
products \PGcomm{which are more and more produced with increasing
laser intensity. The reaction channel which shows up first is electron
emission. The simplest observable in that channel is the net
ionization induced by a laser pulse. More information is retrieved
with} photoelectron spectroscopy (PES) where the distribution of
kinetic energies of the emitted electrons is recorded.  That method
has been applied for free clusters since long, see
e.g. \cite{McH89,Lic91}. \PGcomm{It allows to deduce the 
energy of the occupied single-electron states $\varepsilon_\alpha$
in the cluster ground state. A known number $\nu$ of photons  is
used to lift the bound electron into the continuum which thus leaves
the cluster with a kinetic energy 
$\varepsilon_\mathrm{kin}=\varepsilon_\alpha+\nu\hbar\omega_\mathrm{phot}$
\,.
Each emitting state $\alpha$ and its energy $\varepsilon_\alpha$ can
thus easily be identified in the PES, see e.g. \cite{Poh00}.}
Stepping further in refinement, one can also determine the
Photo-electron Angular Distribution (PAD) of the outgoing electrons
for which meanwhile several measurements exist on free clusters and
which are mostly done simultaneously together with PES
\cite{Pin99,Bag01,Ver04,Kos07a,Skr08}.
\begin{SCfigure}[0.65]
\begin{picture}(70,80)
\put(-5,-45){\epsfig{file=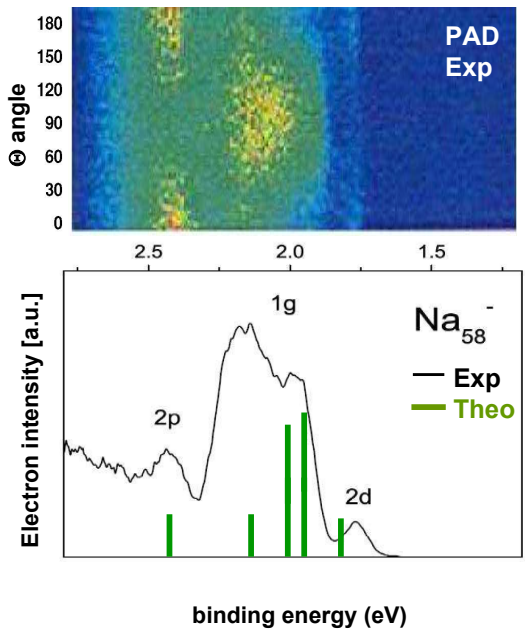,width=0.95\linewidth}}
\end{picture}
\hspace*{5mm}
\caption{\label{fig:na58m_pes_pad}
Upper part: 
Double differential photoelectron angular distribution
for Na$_{58}^-$ produced with a laser of frequency 2.48 eV.
The horizontal axis represents the kinetic energy of
the outgoing electrons. The labeling of the axis is shifted
by the laser frequency to represent the single-particle
binding energies.
The vertical axis is the angle relative to laser
polarization. \esmod{Light (or yellow)} spots 
stand for the highest yields and \esmod{dark (or blue)} for
the lowest.
Lower part: Photo-electron spectra for  Na$_{58}^-$ produced with a 
laser of frequency 4.02 eV. The upper axes label the
corresponding single-electron binding energies.
The peaks can be associated to quantum numbers of the spherical
harmonic oscillator as indicated. The bars indicate theoretical
results from a jellium model. 
Adapted from \cite{Kos07a}.
}
\end{SCfigure}
{Fig.~\ref{fig:na58m_pes_pad} shows an example of a recent measurement
of combined PES \& PAD in Na cluster anions \cite{Kos07a}, each one
from a one-photon process ($\nu=1$). The lower panel shows a PES with
an energy scale where the photon energy has already been subtracted
such that the binding energy
$\varepsilon_\alpha=\varepsilon_\mathrm{kin}-\nu\hbar\omega_\mathrm{phot}$
can immediately be read off. One sees distinct peaks which can be
nicely correlated to the single-particle states in a spherical
harmonic oscillator, the Nilsson-Clemenger model \cite{Cle85,Hee93}.
Theoretical results from a DFT calculations with soft jellium 
background \cite{Mon94b} are indicated by bars, pretty well in
agreement with the experimental data.
A broad scan for Na clusters has shown that even the simple
Nilsson-Clemenger model provides a fairly good description of PES
\cite{Wri02}, although careful final state analysis is required for
quantitative success with  deeply lying states \cite{Mun06b}.
The upper panel of Fig.~\ref{fig:na58m_pes_pad} shows the combined
PES \& PAD for the same system using for the $x$-axis the same energy
scale as in the lower panel. One recognizes the peaks from PES.
Each peak is associated with a different angular distribution 
nicely in accordance with the angular momentum assignment shown in
the upper panel ($2p$ has angular momentum $l=1$, $2d$ has $l=2$, and
$1g$ has $l=4$). 

The above example dealt with a free cluster, as all presently
available measurements of PAD do. The case is \PGmod{a priori}
more involved  for systems in
contact with a substrate. In case of embedded clusters, the
emitted electrons will be much perturbed by the environment
such that PAD are probably too much blurred. PES and measurement of net
ionization can still deliver useful information if the substrate 
thickness stays below the mean free path of the electrons. 
All observables are accessible and useful in case of deposited
clusters. The substrate has the advantage to give the cluster a well
defined orientation} \EScomm{(while the analysis of free clusters 
requires orientation
averaging, which also somewhat blurs the signal itself).} 
\PGcomm{On the other hand, it overlays the PAD by electrons which
are re-scattered from the surface. }
\EScomm{This in turns blurs the PAD signal which becomes more 
complex to analyze. All in all, the case of embedded/deposited clusters, 
thus requires some more detailed analysis. }
We shall thus briefly discuss theoretical explorations of
PAD for laser excited metal
clusters deposited on insulating surfaces, MgO(001) as well as
Ar(001).  Test case is Na$_8$ in the various combinations and also as
a free cluster for comparison~\cite{Bae08a}. 

To compute the angular distribution of emitted electrons,
the density which is eliminated at the
absorbing bounds is accumulated for each  
(absorbing) grid point as
\begin{subequations}
\begin{eqnarray}
  \Gamma(\vec r)
  &=&
  \sum_{i=1}^{N_\mathrm{el}}
  \int_0^\infty \textrm dt\,
  \left|
  \left(1-{\cal M}(\vec r)\right)\,\hat U^{\rm TV}\varphi_i(t)
  \right|^2
  \;.
\end{eqnarray}
where $\hat U^{\rm TV}\varphi_i(t)$ stands for the unitary step
(\ref{eq:timesplit}) before the mask step (\ref{eq:maskstep}) is
applied. By definition of $\cal M$, the field $\Gamma(\vec r)$ is
non-vanishing only in a spherical shell. The angular distribution of
emitted electrons is finally gathered by dividing the spherical shell
into bins $A_i$, each bin being associated to spherical angles
$\theta,\phi$, and integrating $\Gamma(\vec r)$ along a bin.  The
angular distribution then becomes
\begin{equation}
  \frac{\textrm dN_{\rm esc}(\theta,\phi)}{\textrm d\Omega}
  \sim \frac{1}{||A_i(\theta,\phi)||}
  \int_{A_i} \textrm d \mathbf r\, \Gamma(\vec r)
  \;,
\label{eq:crossect}
\end{equation}
\end{subequations}
where $||A_i(\theta,\phi)||$ denotes the area of the 
\EScomm{elementary surface element} $A_i$ on
the surface of a unit sphere.

\begin{figure}[htbp]
\begin{center}
\epsfig{file=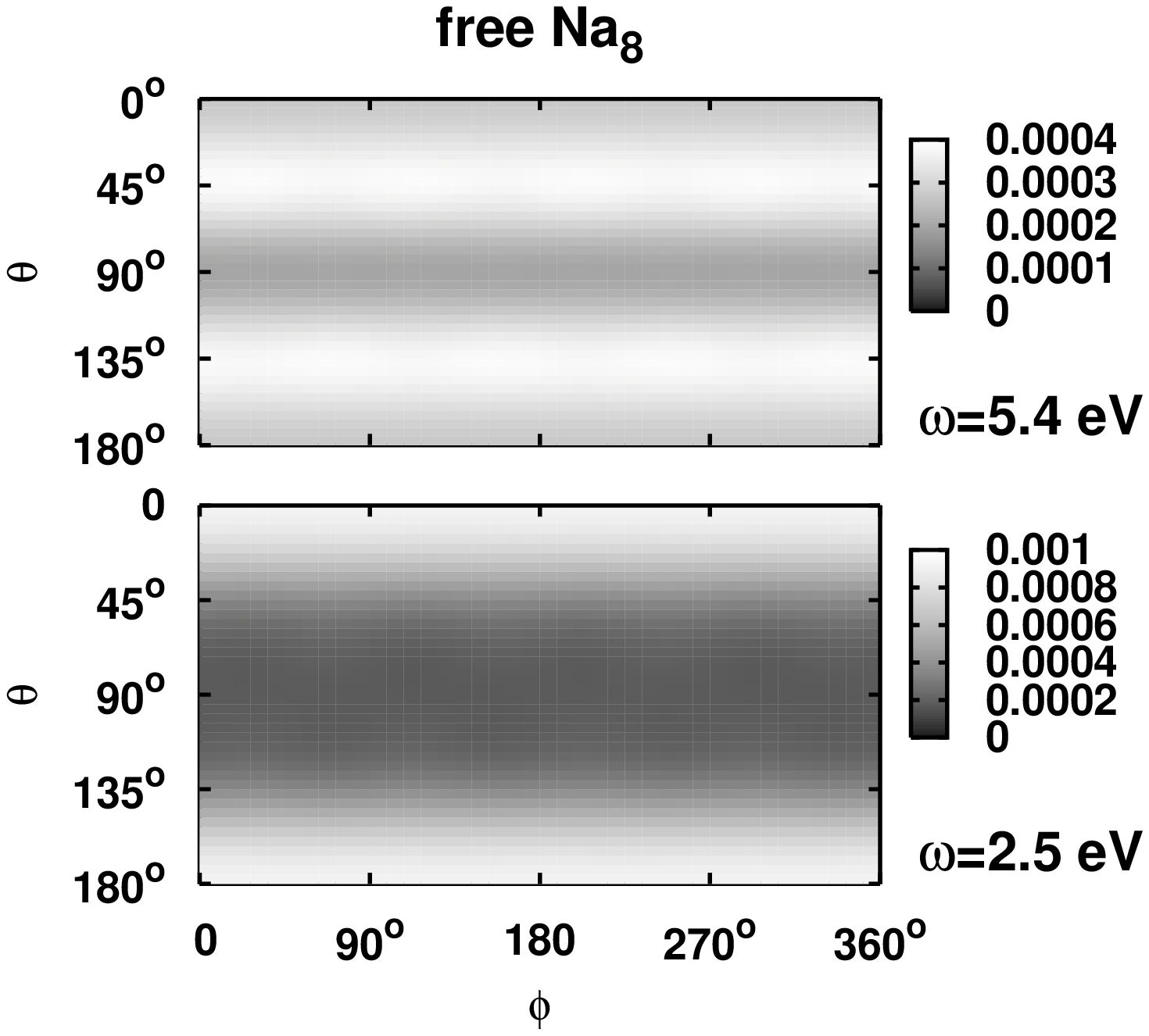,width=0.49\linewidth}
\epsfig{file=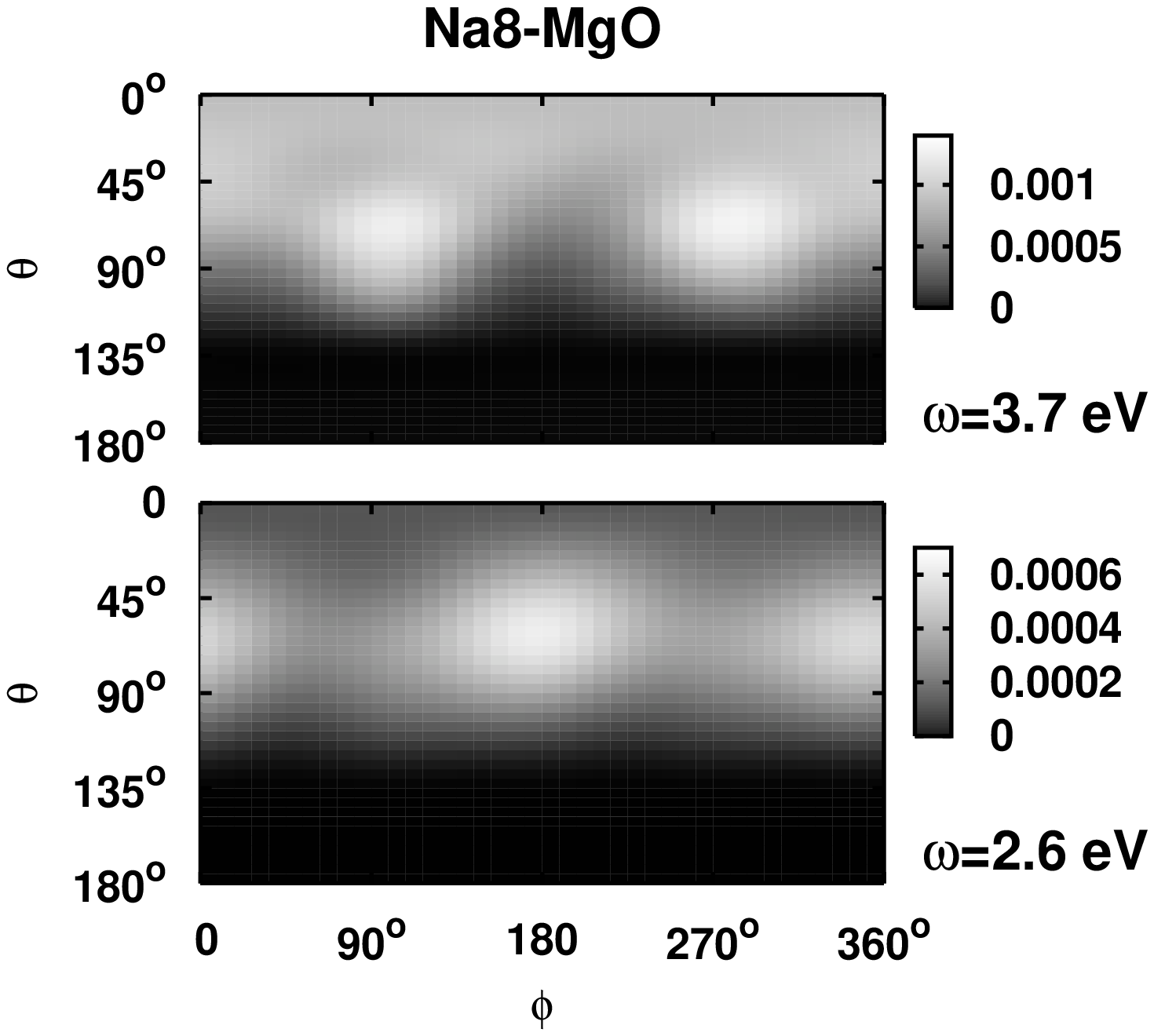,width=0.49\linewidth}
\\
\epsfig{file=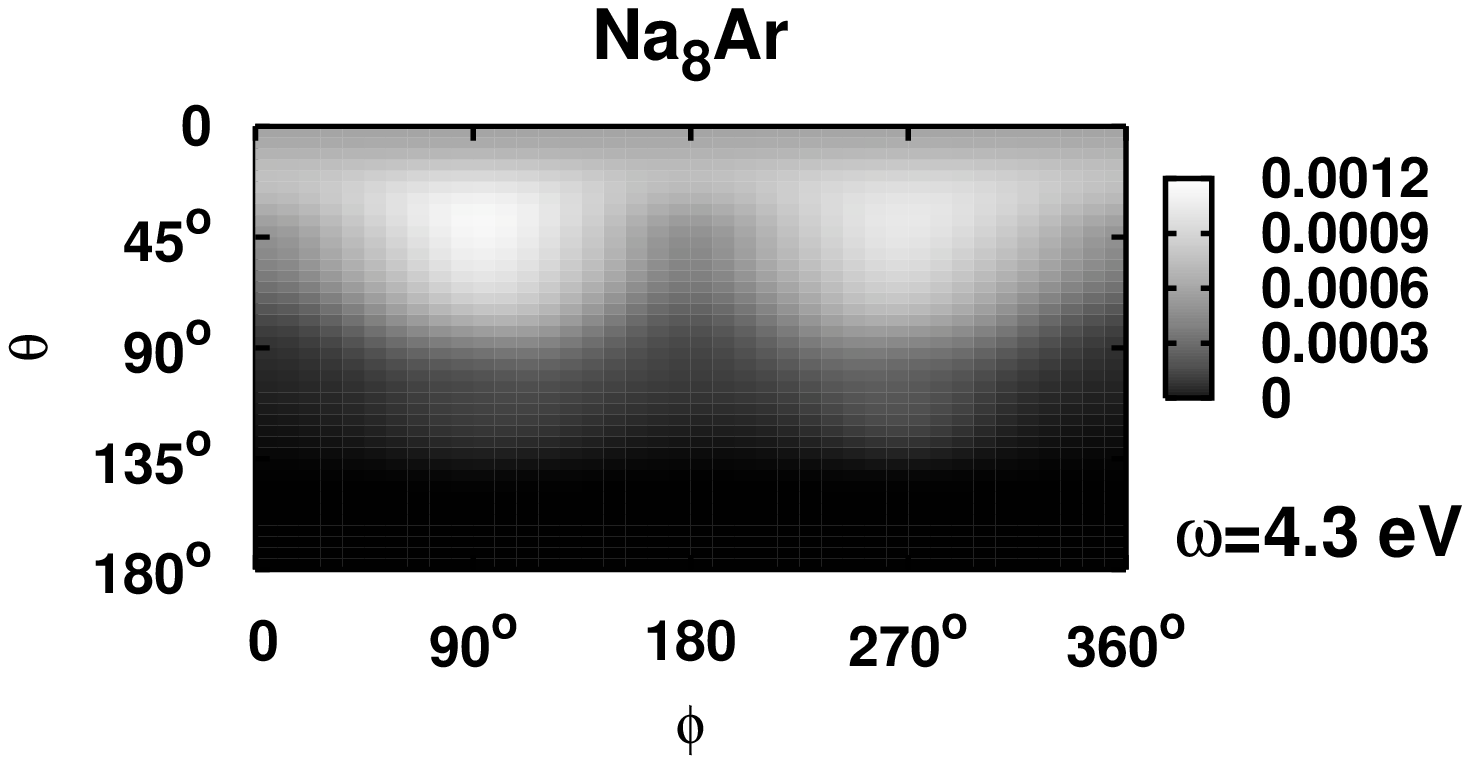,width=0.49\linewidth}
\begin{picture}(65,45)
\put(0,-7){\epsfig{file=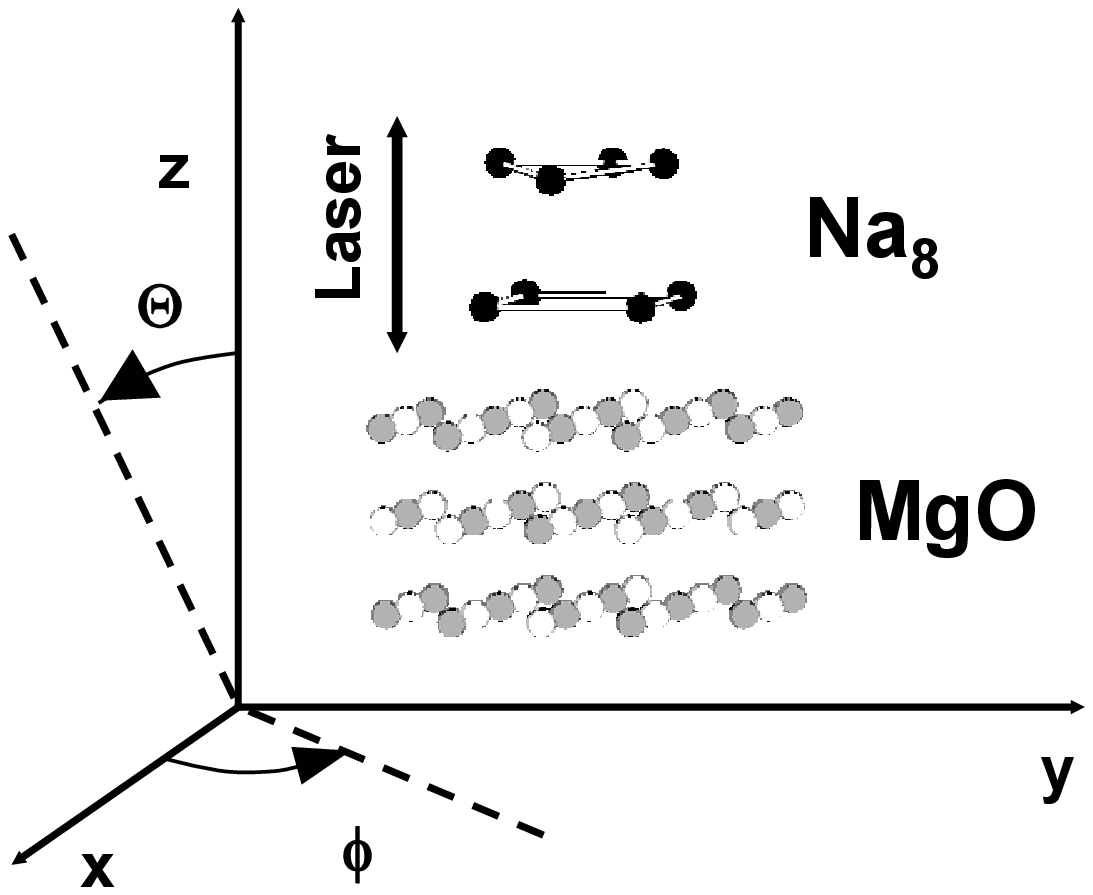,width=0.78\linewidth,clip}}
\end{picture}
\end{center}
\caption{\label{fig:na8_depos_pad}
Double differential photoelectron angular distributions
for various systems and frequencies as indicated. 
All cases had been irradiated by a laser with intensity of $10^9$
W/cm$^2$, 
a pulse with $\sin^2$ envelope having a FWHM of 60 fs, and
polarization along $z$-axis (the axis perpendicular to the surface).
The right lower insert sketches the geometry of the Na$_8$ in relation
to the surface, indicates the laser polarization direction and
the definition of the angles.
Adapted from~\cite{Bae08a}.}
\end{figure}

Surfaces keep the deposited cluster in a well defined orientation.
The combined system of cluster and surface has usually no specific
symmetry and thus the angular distribution of escaped electrons 
varies all around the emission angles $\vartheta$ and $\varphi$, where
 $\vartheta$ is the polar angle with respect to the $z$-axis 
(perpendicular to the surface) and $\varphi$ the corresponding
azimuthal angle. 
Fig.~\ref{fig:na8_depos_pad} shows the double differential
PAD $\textrm dN_\mathrm{esc}/ \textrm d\vartheta\, \textrm d\varphi$
for Na$_8$ deposited on MgO or on Ar, in comparison to the free
cluster. The latter is kept 
at the same orientation relative to the laser polarization as the
deposited ones, where the $z$-axis is aligned with the symmetry axis
of the Na$_8$ cluster. Note that a fixed orientation is hard to
realize in gas phase experiments of free clusters. We use that
configuration here for reasons of comparison.
All angular distributions have been calculated for a broad range of
frequencies, part of which are shown in the panels of
Fig.~\ref{fig:na8_depos_pad}. It is obvious that the PAD sensitively depend
on laser frequency. The reason is that the PAD depend very much on the
single-electron states from which they are emitted and that the
dominantly emitting single-electron states change with the relations
between emission threshold and frequency. Both features together
produce the frequency sensitivity observed in all three cases.  
The free Na$_8$, which is reflection and nearly axially symmetric,
produces always distributions which have
$\vartheta\rightarrow\pi-\vartheta$ symmetry and depend only very
faintly on $\varphi$. The peak strength in $\vartheta$ depends
strongly on frequency with maxima changing from forward--backward
dominance ($\omega=2.5$ eV for Na$_8$) over diagonal emission
($\omega=5.4$ eV for Na$_8$) to sideward emission under
$\vartheta=90^\circ$ (not shown).
The results for Na$_8$@MgO(001) demonstrate the very strong influence
of the substrate.  Backward emission in the segment
$90^\circ<\vartheta<180^\circ$ is, of course, totally suppressed by the
presence of the insulating substrate.  The now emerging strong
azimuthal dependence is due to symmetry breaking by the substrate
which removes the degeneracy of the $1p_x$ and $1p_y$ states. As soon
as these states are close to the emission threshold, the now small
energy difference has a large effect on the relative emission
strengths. One of the two states dominates emission and its pattern
shine through in the total distribution.
For the polar distributions, we see that the emission cones are neither
perfectly along $\theta=90^\circ$ nor along $\theta=45^\circ$ as was
the case for free Na$_8$.  Electron repulsion near the surface and
long-range polarization attraction modify the outwards cones.
The lower panel of Fig.~\ref{fig:na8_depos_pad} shows one angular
distribution for Na$_8$@Ar(001). The effects (backward suppression,
pronounced azimuthal structures) are very similar to the case of MgO
surface.  There is, however, one difference to the case of MgO. The
diagonal emission angle of $\theta=45^\circ$ remains close to the free
case. That is probably due to the smaller interface interaction of the
Ar substrate.

\begin{figure}[htbp]
\begin{center}
\epsfig{file=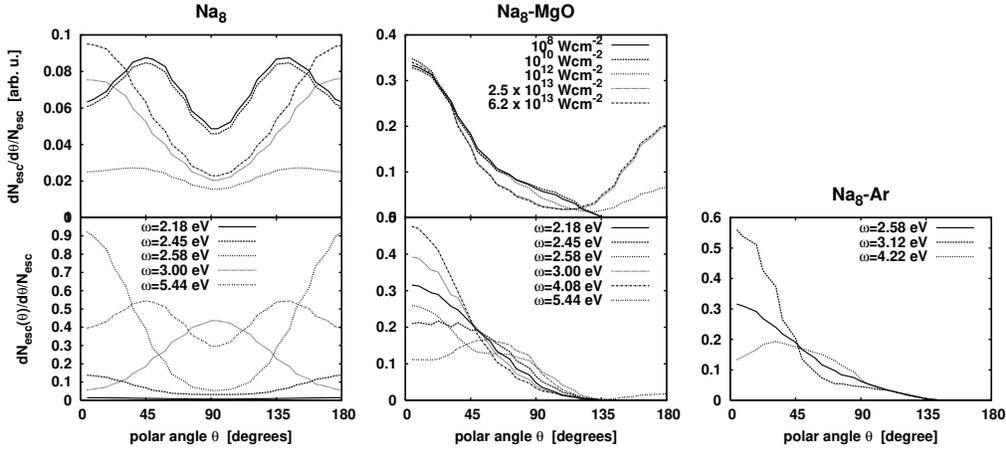,width=0.99\linewidth}
\end{center}
\caption{\label{fig:na8_depos_theta}
Photoelectron angular distributions integrated over angle $\varphi$
for free Na$_8$ (left), Na$_8$@MgO(001) (middle)
and Na$_8$@Ar(001) (right). The lower panels show a variation 
of frequency for fixed intensity $I=10^9$ W/cm$^2$.
The upper panels show variation of intensity for fixed frequency,
$\omega=5.44$ eV in the case of free Na$_8$ (upper left)
and $\omega=4.76$ eV in the case of Na$_8$@MgO(001) (upper middle).
Adapted from~\cite{Bae08a}.}
\end{figure}

Fig.~\ref{fig:na8_depos_theta} shows PAD for a broader variation of
laser conditions, averaged over azimuthal angle to allow better
comparison. \EScomm{We consider again free (left panels) and deposited 
species (on MgO \PGmod{upper two right} panels, on Ar in \PGmod{lower} right panel).}
Frequency dependence is demonstrated in the lower panels.  We see for
free Na$_8$ the three possible outcomes, forward-backward emission,
sideward emission (peak at 90$^\circ$) and diagonal emission.  Both
deposited cases (lower middle, lower right) show the strong
suppression of backward emission through the substrate. In the forward
segment, we see again the variation of pattern with frequency
qualitatively similar to the free case.  The pattern are, however,
more involved than simply copying the reflected electrons as
$\theta\longrightarrow 180^\circ-\theta$. The surface interaction modifies
diagonal and sidewards emission.  A first glimpse of transmission into
the substrate can be seen for $\omega_\mathrm{las}=5.44$ eV in case of
Na$_8$@Mg(001). That frequency is very close to the transmission
threshold into the material at about 5.58~eV.

The upper panels of Fig.~\ref{fig:na8_depos_theta} show variation
of laser intensity. The case of free Na$_8$ (upper left) demonstrates
nicely the transition from the frequency-dominated regime of moderate
intensities where varying pattern can be seen (here diagonal emission)
to  the field-dominated (highly non-linear) regime
\cite{Rei99a,Zwi99,Cal00,Fen08a} with simple forward-backward
dominance.  When the field becomes very strong, the pulling force
along the $z$-axis overrules any subtle quantum-mechanical shell
effects and produces emission simply 
along the laser polarization axis.  That is also the
regime where semi-classical approaches perform very well
\cite{Gig02,Gig03,Fen04}.
A similar trend is seen for deposited Na$_8$@MgO(001), now with
suppression of backward emission by the substrate. But high field
strength starts to overrule the backward suppression. Mind however that
we run into a regime where the present hierarchical model needs
revision because electron emission from the substrate is not
ignorable anymore (see Sec.~\ref{sec:limits}).

\subsection{Coulomb explosion}

{We now turn to large perturbations of the system by applying rather intense 
perturbations leading to the emission of several electrons.}
We will discuss as an example the case of an embedded metal
cluster and the effect the surrounding, inert, \PGmod{environment} has 
on the cluster dynamics.
Test case is Na$_8$ embedded in an Ar substrate for which we take a
finite Ar$_{434}$ system. The set-up simulates a typical scenario for a
chromophore in an inert environment. Mind that Ar is, in principle, a very soft
material having a weak interface energy. It exerts only a very small
perturbation on the cluster which, in turn, maintains basically its
structure and optical properties, see Sec.~\ref{sec:struct}. The
weak-perturbative situation will change if we now consider violent
dynamics where cluster electrons and ions are driven to heftier
encounters with the surrounding substrate atoms.
We excite the embedded cluster by intense and short laser pulses to a
high charge state and follow the subsequent dynamical evolution
over several ps~\cite{Feh07b,Feh08a}.

\subsubsection{General trends}

\begin{figure}[htbp]
\begin{center}
\epsfig{file=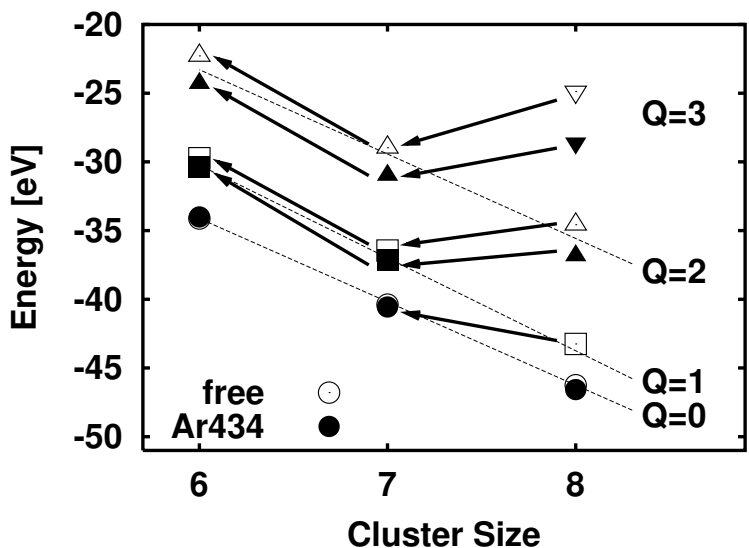,width=0.8\linewidth}
\caption{\label{fig:na8_ar434_basic} 
Binding energies for \PGmod{Na$_{N}^{Q+}$} clusters, free (open symbols) or
embedded in Ar$_{434}$ (filled symbols). 
The different symbols characterize the charge state~:
$Q=0\leftrightarrow$ circles,
$Q=1\leftrightarrow$ squares,
$Q=2\leftrightarrow$ up-triangles,
$Q=3\leftrightarrow$ down-triangles.
The faint dotted lines
connect clusters along same charge states $Q$. 
The arrows show the energetic path for emission of one Na$^+$ ion.
All clusters have been fully relaxed into their optimal
configuration. Adapted from~\cite{Feh07b}.
}
\end{center}
\end{figure}

Before proceeding to the dynamical simulation, we briefly look at the
asymptotic stability.  Fig.~\ref{fig:na8_ar434_basic} shows the
ground-state binding energies of various Na clusters (free and
embedded) for several charge states.  
The steady down-slope of the
dotted lines shows that the binding energy for fixed charge state $Q$
increases in almost constant amounts with the \PGmod{cluster size $N$.}
The energy difference in vertical direction represents the
(adiabatic) ionization potentials which naturally increase with
increasing $Q$.  Stability is checked by following the
decay paths  going along emission of a
Na$^+$ ion (arrows).  Down slope means energy gain and thus asymptotic
instability, which still may allow local stability combined
with long life time. Free Na$_8^{++}$ is indeed
found to be asymptotically unstable. The energy difference shrinks to
a negligible amount for embedded Na$_8^{++}$. This system will then be
\EScomm{(meta)}stable. We see generally that embedding has a stabilizing effect for
charged clusters. For example, the case of Na$_8^{3+}$ which was
clearly explosive for the free cluster is ``downgraded'' to a
situation which is comparable to free Na$_8^{++}$.
Energy differences alone however are not fully conclusive for the
stability times. These also depend on the phase space of decay
channels and on the reaction barriers which are surely larger for
embedded clusters due to the huge Ar cage around. \EScomm{
A deeper analysis thus requires an actual \PGmod{tracking} of the whole
excitation/deexcitation process in the course for example of 
a laser irradiation.}

\begin{figure}[htbp]
\begin{center}
\epsfig{file=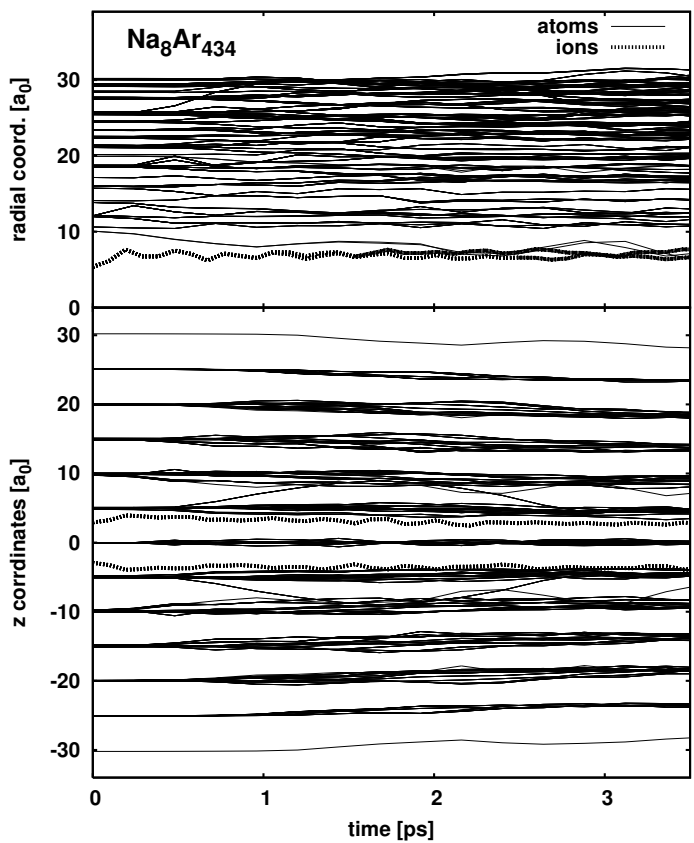,width=0.7\linewidth}
\caption{\label{fig:na8_ar434_coord}
Time evolution of the atomic (full lines) and ionic (dotted lines)
$z$-coordinates (lower panel) and radial distances
$r=\sqrt{x^2+y^2+z^2}$ (upper panel) for Na$_8$@Ar$_{434}$ excited
with a laser of frequency $\omega=1.9$ eV, intensity $2 \times 10^{12}$
W/cm$^2$, and a cos$^2$ pulse profile with FWHM = 50 fs. 
The laser was polarized along the $z$ axis which is also
the symmetry axis of the system. Adapted from \cite{Feh08a}.}
\end{center}
\end{figure}

Fig.~\ref{fig:na8_ar434_coord} shows an example of time evolution after a
hefty laser excitation. 
The initial reaction 
of the system is dominated by electronic response (not shown) leading
here to a direct emission of 3 electrons which escape before ionic
motion plays a role \cite{Cal98c,Rei99a,Cal00}.
The thus produced large Coulomb pressure leads first, up to about 200
fs, to an attempted Coulomb explosion which, however, is abruptly
stopped when the ions hit the repulsive cores of the first shell of Ar
atoms. The ionic motion turns to damped oscillations around a (r.m.s.)
radius of about 7 $a_0$.
The Ar core dynamics shows also two stages, although on longer time
scales. The first phase is a spreading of the momentum acquired from
stopping of Na ions into the various Ar shells. This is especially
visible along the $z$ axis (laser polarization axis)
which allows to read off the propagation
speed of this perturbation as 20-30 $a_0$/ps, close to the sound
velocity in the Ar system \cite{Feh08a}, which suggests an
interpretation as a sound wave sent by the initial bounce of the Na
ions.
Once transferred to a given Ar shell, the perturbation
generates oscillations combined with some diffusion which,
after about 1.5 ps, has spread over all shells.
Even the outermost shell acquires such oscillations. 
This is an effect of the finite size of the ``matrix''.  Larger
matrices distribute the given energy more widely and would thus lead to  smaller
oscillations while smaller matrices show more intense response at the
Ar side, possibly even with some Ar emission \cite{Feh08a}.
The relaxation of Ar oscillations is much slower than that of the
Na ions and beyond the time scale computed here.  These long time
scales for full relaxation and evaporation of Ar atoms are well known
from experiments of dimer molecules embedded in Ar clusters, see
e.g. \cite{Vor96a}.

\begin{figure}[htbp]
\begin{center}
\epsfig{file=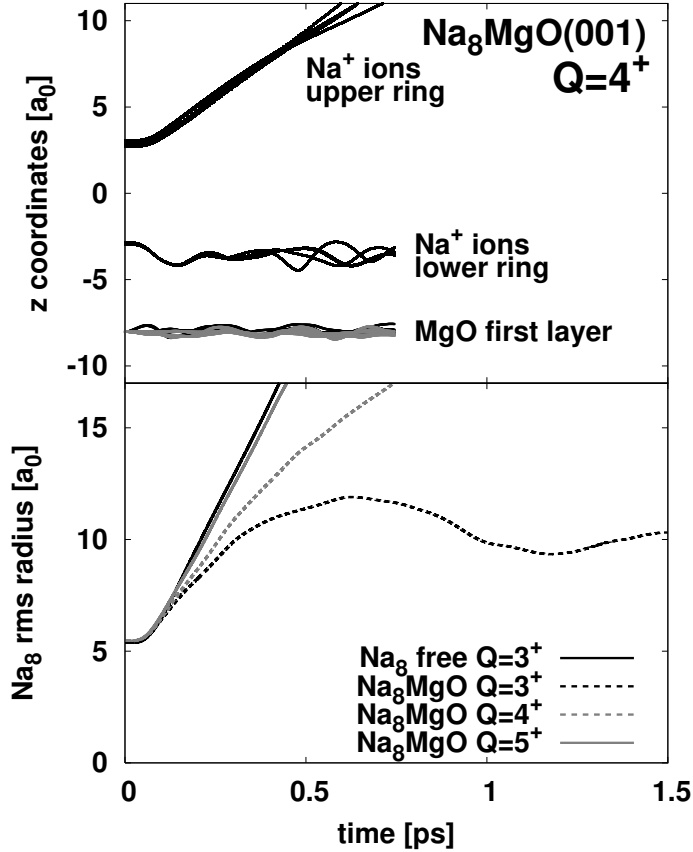,width=0.7\linewidth}
\caption{\label{fig:Na8-MgO-explo-rad-zQ4}
\PGcomm{
Upper panel:
Time evolution of $z$ coordinates of Na$_8$ deposited on MgO(001) and
irradiated by a laser of frequency $\omega=1.9$ eV, pulse duration
60 fs and intensity $I=1.21 \times 10^{12}$ W/cm$^2$ such that a
charge state $Q=4^+$ is reached. 
Lower panel:
time evolution of the cluster r.m.s. radius for  Na$_8$ deposited on MgO(001)
and irradiated by a laser of frequency of $\omega=1.9$ eV, pulse duration
of 60 fs and different intensities leading to different charge states
as indicated.
}}
\end{center}
\end{figure}

Stabilization of high charge states also takes  place for deposited
clusters \cite{Bae08a}. In particular, the MgO(001) surface with its
strong polarization potential becomes increasingly attractive with
increasing charge state of the
cluster. Fig.~\ref{fig:Na8-MgO-explo-rad-zQ4} illustrates this point 
for a variety of (average) charge states around the critical 
point for cluster explosion in the combination Na$_8$@MgO(001).
\EScomm{Here again, we consider irradiation by lasers which produce 
the high ionization states under study. In all cases, we start from the
equilibrium,  deposited, \PGmod{Na$_8$} cluster on MgO.}
\PGcomm{A global view in terms of the cluster radius is shown in the
lower panel of Fig.~\ref{fig:Na8-MgO-explo-rad-zQ4}.  The state
$Q=3^+$ leads to straightforward Coulomb explosion for the free
cluster \PGmod{(solid black line)}.  Note that already $Q=2^+$ is
Coulomb unstable for the free cluster.  The presence of the substrate
stabilizes the states up to $Q=3^+$ \PGmod{(dashed black line)}. It
requires $Q=4^+$ to drive instability  \PGmod{(dashed gray line)}
and only after one step more
\PGmod{at $Q=5^+$}, we see a heftiness comparable to free $Q=3^+$
\PGmod{(compare black and gray solid lines)}. Loosely speaking, the
polarization potential from the surface shifts the appearance size by
two units for that material combination.  But that is a rather global
statement. Even when explosion takes place, the reaction takes a much
different path due to the presence of the substrate.  That is
demonstrated in the upper panel of 
Fig.~\ref{fig:Na8-MgO-explo-rad-zQ4} showing detailed coordinates
for the case of $Q=4^+$.  The upper ring of four ions is blown away by
Coulomb pressure, while the lower ring (carrying the majority of
remaining electrons) remains tied close to the substrate.  }
\EScomm{One thus observes in that case an interesting scenario in
which the double layer \PGmod{Na$_8$} cluster is literally sliced into
two layers, one sticking to the surface, the other one exploding. }

\subsubsection{Energy balance and relaxation times}
\label{sec:energybalance}

Let us come back to the case of hindered Coulomb explosion of
Na$_8$@Ar$_{434}$ for the charge state $Q=3$ and have a look
at the energetic relations and associated relaxation
times~\cite{Feh07b}. Total energies are not the best measure for
internal excitation. Energy per particle is better suited and we
express it in terms of the kinetic temperature
$T_{\rm kin}=2E_{\rm kin}/(3N)$.
The definition is a bit more involved for the electrons.  We have to subtract
the offset from the Pauli principle as well as possible collective
flow contribution to define an intrinsic electron temperature,
following the procedure described in \cite{Cal00}.
\begin{SCfigure}[0.6]
\epsfig{file=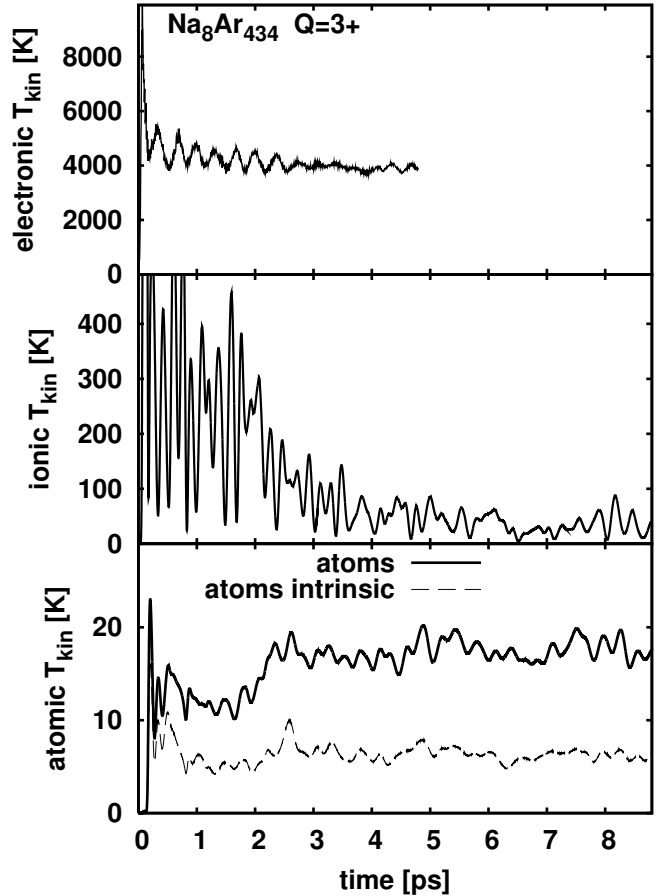,width=0.65\linewidth}
\caption{\label{fig:na8_ar_temp}
Time evolution of electronic, ionic and atomic kinetic
temperatures for Na$_8$@Ar$_{434}$ at charge state $Q=3^+$.
For the atoms, we show the temperature from total kinetic energy and
from ``intrinsic'' kinetic energy for comparison
(the kinetic energy from collective radial motion is then subtracted).
From~\cite{Feh07b}.}
\end{SCfigure}
Fig.~\ref{fig:na8_ar_temp} shows the various kinetic temperatures in
the case of an irradiation to charge state $Q=3^+$ for
Na$_8$@Ar$_{434}$.  The electronic temperature is much higher than
any other one and there is no sign of relaxation towards the other parts,
on the time span explored here. This shows that from a thermal point
of view electrons and ions/atoms are, to a large extent, decoupled
with mutual relaxation times far beyond our analysis. The large
electronic temperature calls, in fact, for a description beyond pure
mean field. Electron-electron collisions should play a role in that
regime. These could be included by switching to a semi-classical
Vlasov-Uehling-Uhlenbeck description of the electron cloud
\cite{Gig01a}. A first attempt to extend the dynamical QM/MM modeling
to a semi-classical description for the case Na in contact with Ar is
found in \cite{Feh05}. But the strongly repulsive Ar cores make the
semi-classical sampling of the phase-space distribution function much
more delicate than for previous implementation in pure simple metals.

The ionic temperature is one order of magnitude lower than the
electronic one and  relaxes to thermal equilibrium with the Ar
system within a few ps.  Even lower temperature scales appear for the
Ar matrix. A large and immediate energy transfer is seen at about 200
fs when the Ar cage stops the Na explosion. In the further evolution,
the atomic kinetic energy grows in accordance with decreasing Na
temperature.  For Ar, we also show an ``intrinsic'' temperature which
is obtained by subtracting the contribution from the collective
breathing oscillations of the matrix.  That is twice lower.  Thus there
remains a substantial amount of regular motion in the matrix, again
with a thermal relaxation time far beyond our simulation time.

\subsubsection{Dipole polarization}

\begin{figure}[htbp]
\begin{center}
\epsfig{file=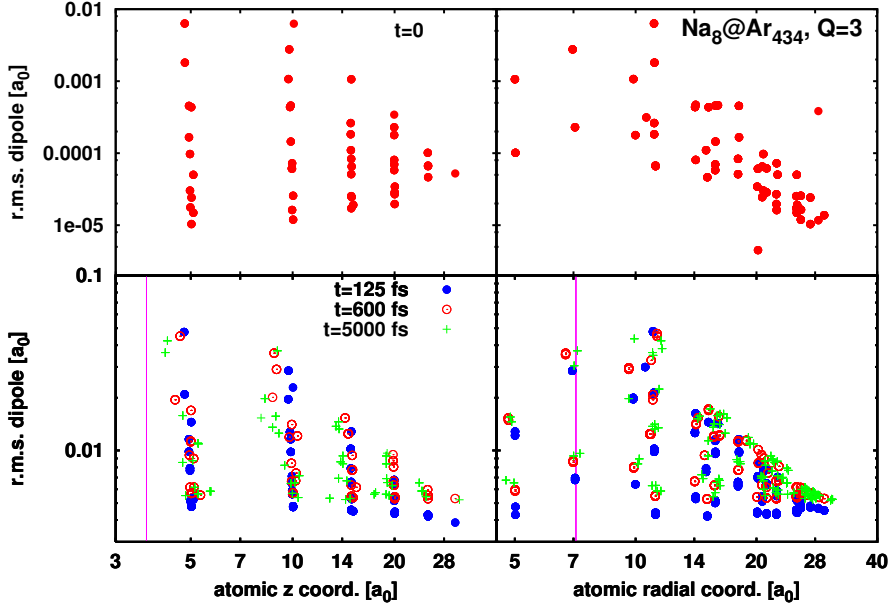,width=0.9\linewidth}
\caption{\label{fig:na8_ar_dip}
Root mean square Ar dipoles for the hindered Coulomb explosion
  of Na$_8$ embedded in Ar$_{434}$, exposed to a laser of intensity
  $2\times 10^{12}$ W/cm$^2$, frequency $\omega=1.9$ eV, and FWHM=33
  fs. Left panels~: distribution as a function of the Ar $z$
  distance of the Na$_8$ center-of-mass; right panels~: that as a
  function of the Ar radial distance $\rho=\sqrt{x^2+y^2}$ to the
  Na$_8$ c.o.m. Top panels~: at initial time;
  bottom panels~: for three subsequent times as indicated. The maximum
  of excitation energy observed in the bottom right panel at 11 $a_0$
  is due to the oblate deformation of the created Na$_8^{3+}$. The
  vertical lines in the bottom panels indicate the corresponding
  coordinates of the Na$_8$ outermost ions.}
\end{center}
\end{figure}

\EScomm{The analysis of Sec.~\ref{sec:energybalance} took into 
account the degrees of freedom of the environment but only in terms of 
kinetic energies\PGmod{, not} mentioning\PGmod{, e.g.,} potential energy transfers 
connected to matrix rearrangements\PGmod{. One} should also take into account the
matrix response in terms of its internal degrees of freedom, namely 
Ar dipoles. We thus follow here the same path as explored in
Sec.~\ref{sec:sub-dip} for the case of deposition scenarios, but now
considering  
the impact of much higher charges as attained in the course of 
irradiation processes.} We thus 
consider here the spatial distributions of dipoles at various
instants again in the case of a charging to a $Q=3^+$ state.  
Fig.~\ref{fig:na8_ar_dip} shows the r.m.s. dipoles along
($z$ direction, left panels) and perpendicular 
(axial direction, right panels) to the laser
polarization.  
The cluster is initially neutral. This yields a rather
"democratic" distribution of small dipoles all over the matrix (upper
panels).  The lower panels show snapshots at later times where the
cluster is highly charged. This leads to much larger dipoles and a
clear dependence on the distance from the center in both directions.
The distribution is, however, not monotonously decreasing due to the
finite (large) extension of the Na cluster after irradiation. A
sizable fraction of Ar sites overlaps with the Na cluster electron
cloud and thus see a screened charge, whence the reduced dipole
polarization. In order to exemplify the point, we have also indicated
by vertical lines the actual position of the outermost Na ions. The
effect is seen only along the radial coordinate (lower right panel)
and is due to the strongly oblate shape of the charged cluster
\cite{Feh07b,Feh08a}.  Apart from that detail, the pattern are much
similar to the case of deposited, charged clusters~\cite{Din08c}.

\subsubsection{Pump and probe analysis}

\begin{figure}[htbp]
\begin{center}
\epsfig{file=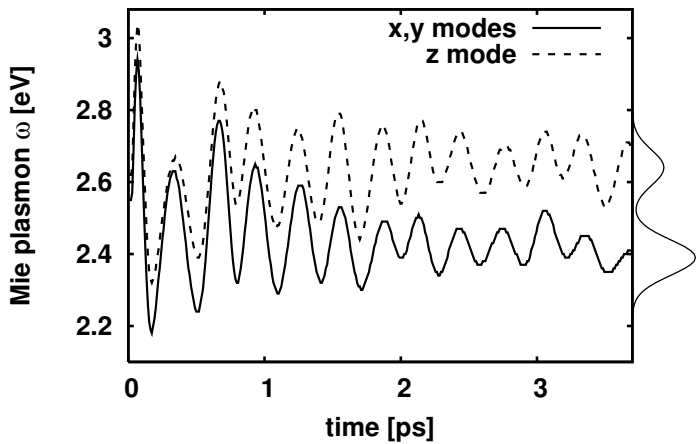,width=0.7\linewidth}
\caption{\label{fig:na8_ar_pp}
Time evolution of the estimated Mie plasmon frequencies in $x$,
$y$ and $z$ directions for the embedded Na$_8$ cluster. The cluster
remains almost axially symmetric all time such that the $x$ and $y$
modes are nearly  degenerate. 
\PGcomm{
The full spectral distribution is shown for the last time slot
to the right of the figure. 
}
Adapted from~\cite{Feh08a}.
}
\end{center}
\end{figure}

The most interesting effect in the dynamics of the Na$_8$@Ar$_{434}$
system was the hindered explosion of the imprisoned Na cluster and its
subsequent shape oscillations in the Ar cavity.  Such shape
oscillations and eventual relaxation to deformed shape have been
produced and observed experimentally for Ag clusters embedded in glass
\cite{Sei00} or deposited on a substrate \cite{Wen99a}.  
\PGcomm{As outlined in the introduction to this section, pump and
probe analysis allows to map a time-dependent picture of global
shape oscillations of metal clusters.}
The (time-dependent) cluster
deformation is analyzed by probe pulses measuring the actual optical
response of the metal cluster, exploiting the fact that the Mie
plasmon  strongly couples to light at very specific frequencies and
these plasmon frequencies are uniquely related to the cluster shape.
Key ingredient is the time evolution of the Mie frequencies due to the
slowly changing ionic configuration.
Fig.~\ref{fig:na8_ar_pp} shows that for the case of Na$_8$
embedded in Ar$_{434}$ in the case of the irradiation to the \PGmod{$Q=3^+$} state.  
Note that there are, in principle, three
modes, one for each principal direction. But Na$_8$ maintains nearly
axial symmetry such that the $x$- and $y$-modes are degenerate.  The
evolution starts out with clean radial oscillations (all modes close
to each other).  A then evolving oblate deformation leads to a
splitting of the resonance peak where the shorter extension along $z$
is associated to a relative blue-shift of the mode and the larger
extension in orthogonal direction to a relative red-shift.  These
pattern can be tracked in pump-and-probe measurements.  Note
here that the experimental results obtained for Ag clusters in glass
\cite{Sei00} are in qualitative agreement with our findings (see
Fig.~\ref{fig:examples}, bottom right panel).



\section{Conclusion}
\label{sec:concl}

We have discussed in this paper the dynamics of metal clusters in
contact with inert (insulating) \esmod{environment}s.  The studies considered the
complementing aspects of the cluster itself as well as of its
environment.
A dynamical treatment of internal degrees of freedom of the
environment has proven to be an essential ingredient for an
appropriate description of the combined system (cluster +
environment), especially for truly dynamical scenarios.
To that end, we have presented in this paper a hierarchical modeling
in the spirit of a quantum-mechanical/molecular-mechanical (QM/MM)
method. It describes the valence electrons of the metal cluster fully
quantum-mechanically with time-dependent density-functional theory and
the cluster ions as well as the degrees of freedom of the \esmod{environment} by
classical molecular dynamics.  We
emphasize that our implementation of QM/MM includes properly the
dynamical polarization potentials of the environment.  This represents
a crucial improvement over usual treatments in which the internal
degrees of freedom of the \esmod{environment} atoms are hidden in (static)
effective potentials or at best in stationary polarization effects as,
e.g., in conventional QM/MM methods. The many examples reported
here have demonstrated the importance of these explicitly dynamical
polarization potentials. 
We have explained our dynamical QM/MM modeling in detail, on the
basis of the modification of existing static interaction
potentials. The model parameters were calibrated with respect to
existing reference calculations and/or experimental data so that the
final modeling leads to a realistic reproduction of existing data.
We have also worked out the ranges of applicability. Excitations
should not lead to ionization of the atoms in the \esmod{environment}.  This
relates to a maximally allowed local electrical field in the \esmod{environment}
corresponding, e.g., for Ar to a laser intensity of about
10$^{13}$-10$^{14}$ W/cm$^2$. 
The applications were developed in three steps: first, we checked the
performance for static properties and optical response; second, we
discussed the dynamics of cluster deposition; and third, we considered
dynamics of embedded or deposited clusters following excitation by
short laser pulse of varying intensity. Test cases for the detailed
examples were Na atoms or Na clusters in contact with Ar \esmod{environment}s or
MgO(001) surfaces. Each one of these three parts is headed by a more
general summary covering different system combinations and trying to
work out the basic effects to be studied in the actual test cases.


We have first applied our model to check static properties of
deposited and embedded Na clusters in contact with Ar or MgO substrate
and find widely varying results: clusters which are strongly deformed
by the interface (e.g. Na$_6$ on MgO(001)) and those which are almost
unaffected (e.g. Na$_8$ embedded in Ar matrix), substrates which stay
robust (generally MgO(001)) and those which are much modified
(e.g. Na$_6$ attached to small Ar systems as Ar$_{43}$).
There are, in fact, counteracting effects leading to the final shape~:
Electronic shell in the cluster, 
symmetry breaking due to the \esmod{environment} (particularly for surfaces),
short-range repulsion of cluster electrons from the \esmod{environment},
long-range polarization attraction from the \esmod{environment},
geometrical matching of cluster and substrate bond lengths
(MgO differing from Na binding while Ar complies better),
surface corrugation (MgO having strong and Ar very weak corrugation). 
The final compromise is thus hard to predict by simple rules.
Detailed calculations are required which properly take into account
all the crucial ingredients.

Optical response was found to provide a sensitive test to environment
properties. The dominant Mie plasmon resonance in metal clusters is a
collective dipole mode and thus very sensitive to dynamical (dipole)
polarization effects. Again, we find counteracting effects~: the short
range repulsion confines the electronic vibrations delivering a strong
blue-shift ($\approx$ 0.5 eV) while the dynamical polarization induces
a comparatively strong red-shift. The net effect is in most cases an
extremely small net shift of the Mie plasmon resonance.
Larger effects are seen for the detailed spectral fragmentation of the
resonance (due to single-electron excitations in the vicinity). They
depend very much on the details of the surface and the surface
interaction. One effect, however, is generic and always observed in
case of a deposited cluster~: the mode in the direction \esmod{normal} to
the surface is strongly fragmented.  This is caused by the strong symmetry breaking
through the surface which enhances dramatically the density of dipole
excitations \esmod{normal}  to the surface.

Cluster deposition has been studied at the threshold of soft landing
below which the impinging cluster is captured by the surface without
cluster or surface destruction. That threshold is found to be similar
for Ar(001) and MgO(001) surfaces. But the dissipation mechanisms,
responsible for successful landing, are much different. Ar substrate
is a very efficient stopper which absorbs almost all energy from the
cluster and quickly installs a rather calm, deposited cluster. MgO as a
mechanically robust material acts mainly as a converter of energy within
the cluster, from initial translational energy to intrinsic ionic
motion. This keeps deposited cluster at first highly excited and
leaves a long time for cooling. 
The behavior above landing threshold is also much different. The
cluster is immediately reflected from MgO in a highly excited state
prone to final fragmentation while the substrate stays almost
unaffected, almost as a spectator. The reflection from Ar is achieved
at the price of sever\PGmod{e} damage of the surface around the contact point.
A technical study varying the ingredients of the model showed clearly
the importance of polarization effects in the description of the
deposition dynamics.

Finally, we have discussed the dynamics of embedded and deposited
clusters following irradiation by a short laser pulse.
At the side of moderate laser intensities, we have considered
photo-electron spectroscopy, either angle-averaged or angle-resolved,
which constitutes a detailed and sensitive observable.  Deposition on
a surface provides a unique means to work with well defined cluster
orientation. We see strong perturbations of electron flow through
immediate reflection from the repulsive surface and a weak deflection
of the final electron path towards the surface due to polarization
attraction. Taking into account these propagation effects, the
pattern can be understood from an interplay of single-electron binding
energy, ionization potential, and laser frequency, similar as is done
for free clusters.
Strong laser pulses lead to a fast (following the pulse within 2-10
fs) initial ionization of the cluster which, in turn, creates an
enormous Coulomb pressure. But the \esmod{environment}s very efficiently hinder
a Coulomb explosion and stabilize charge states which would be highly
unstable in a comparable free cluster. For embedded clusters, the
stabilization is established by two effects~: first, the surrounding
environment acts as an inert cage which hinders the cluster ions from
direct escape, and second, the polarization interaction adds a
substantial amount of binding energy thus shifting the asymptotic
stability (appearance size) to higher charge states. The second effect
remains active in case of deposited clusters and thus we also see an
impressive stabilization of high charge states also here. It is to be
noted that this stabilization is most probably a transient effect
because slow ion diffusion through the medium may still be possible
and because the infinite charge supply of infinite media may lead to a
slow re-neutralization of the cluster.


Throughout all the above collected detailed results, we emphasize
that the 
analysis of the environment response constitutes an important 
ingredient in our treatment. This concerns 
the explicit account for  internal degrees of freedom of 
the environment by means of dynamical atomic polarizabilities.
This  has 
allowed us to understand many subtle aspects, otherwise
overlooked. We summarize theme here under this special viewpoint. 
We have seen that, both in deposition and irradiation 
scenarios, the \esmod{environment} is significantly modified. A good indicator 
is the amount of transferred energy which is in turn stored as both 
kinetic and potential energy. At low energy, the \esmod{environment} accommodates
the perturbation by standard vibrations, little affecting the overall
environment structure. Most effects are then seen at the side of atom
kinetic energy. In the case of larger perturbations, especially when
charges are present (deposit of charged species, irradiation of
embedded clusters), the environment also experiences spatial
rearrangement leading to substantial potential energy variations. The
impact of charge, generally speaking, has proven sizeable on many
environment properties. This is especially true at the side of
internal degrees of freedom (polarizability) of the environment
atoms. The effect is particularly visible in the case of deposition
processes where it was shown that the charge creates a strong
localized polarization of the substrate around the deposition
point. The same holds true in the case of irradiation of embedded
clusters in which the net charge generated by the irradiation leads to
a sizeable (again rather localized) polarization of the
medium. Generally speaking, many observables presented in this paper
have proven extremely (sometimes even unexpectedly) sensitive to the
dipole polarizability. This is true whatever dynamical regime, and both
quantitatively and qualitatively. This aspect is extremely interesting because
that sort of effects had not yet, to the best of our knowledge, been
analyzed before, due to the usual neglect of the dynamics of internal
degrees of freedom of the environment. The many examples shown all
along this paper have on the contrary demonstrated the key importance
of this effect. The impact is particularly important in dynamics,
especially when charges (as is often the case in practice) are
involved.

Altogether, the collection of results shows that the dynamical QM/MM
approach constitutes a pertinent and efficient description of cluster
and \esmod{environment} dynamics including electronic response and transport.
The present test cases of metal clusters in/on insulating \esmod{environment}
will still provide much space for further interesting research, e.g.,
by going further in system sizes and establishing trends thereof, by
extending the test cases to multi-layered material, or by transferring
the methods to describe dedicated devices as quantum dots.  The
success also motivates to proceed in the still more demanding domain
of bio-chemical compounds as, e.g., reactive molecules in water
environment. This will bring back the QM/MM approach to where it
stems from, but now with an emphasis on non-linear dynamics with
electron and ion transport.

\bigskip

\noindent
Acknowledgment: 
We thank our colleagues M. B\"ar, G. Bousquet, B. Faber,
F. Fehrer, T. Fennel, C. F\'elix, B. Gervais, E. Giglio, A. Ipatov, B. von
Issendorf, S. K\"ummel, K.-H. Meiwes-Broer, J. Messud, M. Mundt,
L. V. Moskaleva, U. F. Ndongmouo, N. R\"osch, J. Tiggesb\"aumker,
S. Vidal, Z. P. Wang, and P. Wopperer for helpful contributions and
discussions.
This work was supported by the DFG, project nr. RE
322/10-1, the French-German exchange program PROCOPE nr. 07523TE, the
CNRS Programme ``Mat\'eriaux'' (CPR-ISMIR), Institut Universitaire de
France, the Humboldt foundation, a Gay-Lussac price, the French
computational facilities CalMip (Calcul en Midi-Pyr\'en\'ees), IDRIS and
CINES, and the Regional Computing Center of the University Erlangen.

\section{Appendix}
\label{sec:app}

\subsection{Details for the MgO(001) substrate}

MgO is an insulating material with a rather large band gap of 6.9 eV
\cite{Tje90a}. It is built as an ionic crystal in fcc configuration
(similar to NaCl) with lattice parameter $a=7.94~a_0$. To a very
good approximation, one can consider it as a composition of O$^{2-}$
anions and Mg$^{2+}$ cations sorted in interlacing simple cubic
lattices \cite{Ash76}. We consider the MgO(001) surface which exposes 
interlaced regular squares of O$^{2-}$ and Mg$^{2+}$.
\begin{table}[htbp]
\centerline{
\begin{tabular}{|l|c|c||c|c|c|}
\hline
\multicolumn{3}{|c||}{\bf Constituents of MgO(001) substrate} 
& charge [$e$] & mass [$m_n$] & spring const. [Ry ${a_0}^{-2}$]
\\
\hline
O$^{2-}$ ions & cores $\leftrightarrow(\mathrm{O}\,c)$
 & $N_{(\mathrm{O}\,c)}=N_{(\mathrm{O}\,v)}$
 & \hspace{0.15cm} 0.21 & 16.0 &
\\
 & valence clouds $\leftrightarrow(\mathrm{O}\,v)$
 & $N_{(\mathrm{O}\,c)}=N_{(\mathrm{O}\,v)}$
 & $-2.21$ & $3.42 \times 10^{-3}$ 
 &
\begin{picture}(5,5)(10,-4)
$\Big\} \kappa_{\mathrm{O}c,\mathrm{O}v}=0.56192$
\end{picture}
\\
\cline{6-6}
Mg$^{2+}$ ions & cores $\leftrightarrow(\mathrm{Mg}\,c)$
 & $N_{(\mathrm{Mg}\,c)}=N_{(\mathrm{O}\,c)}$
 & 2.0 & 24.3 
\\
 & no valence cloud &   $N_{(\mathrm{Mg}\,v)}=0$
 & $-$ & $-$ 
\\
\cline{1-5}
\end{tabular}
}
\vspace*{0.8em}
\centerline{
\begin{tabular}{|c|r|r|r|}
\hline
\multicolumn{4}{|c|}{
$
  \displaystyle
  f_{\alpha \beta }^\mathrm{(short)}(r) 
  =
  a_{\alpha\beta} e^{-r/\lambda_{\alpha\beta}} 
  -\frac{b_{\alpha\beta}}{r^{6}}  
$
}
\\[6pt]
\hline \hline
$(\alpha,\beta)$  &  $a_{\alpha\beta}\;[\rm Ry]$ &
$\lambda_{\alpha\beta}\;[a_0]$ & $b_{\alpha\beta}\;[{\rm
    Ry}\,{a_0}^6]$\\ 
\hline
$((\mathrm{O}\,v),(\mathrm{O}\,v))$   &      1673.8456     &  0.28157
&     93.353658 \\
\hline
$((\mathrm{Mg}\,c),(\mathrm{O}\,v))$   &        90.4147     &  0.57123
&  \multicolumn{1}{|c|}{0}       \\
\mbox{all other}
& \multicolumn{1}{|c|}{0} & \multicolumn{1}{|c|}{--} & \multicolumn{1}{|c|}{0}
\\
\hline
\end{tabular}
\qquad
\begin{tabular}{|l|r|r|r|}
\hline
\multicolumn{4}{|c|}{
$
  \displaystyle
  f_{\alpha\,\mathrm{el}}^\mathrm{(short)}(r) 
  =
  \frac{A_{\alpha}}
       {1+\exp \left[(|\vec{r}-\vec{R}_{{{i}^{(\alpha)}} }|
        -\omega_{\alpha})/C_{\alpha} \right]}
$
}
\\[6pt]
\hline\hline
$\alpha\!\equiv(s\tau)$ &  $A_\alpha\; \rm[Ry]$   &   $w_\alpha\;
    [a_0]$  &   $C_\alpha\;[a_0]$ 
\\
\hline
$(\mathrm{O}\,c)$      & 0.0591  & 1.515 & 0.901 
\\
$(\mathrm{O}\,v)$     & \multicolumn{1}{|c|}{0} &
\multicolumn{1}{|c|}{0} & \multicolumn{1}{|c|}{--}
\\
$(\mathrm{Mg}\,c)$      & 0.0643  & 1.731 & 0.961 
\\
\hline
\end{tabular}
}
\vspace*{0.8em}
\centerline{
\begin{tabular}{|l|r|r|r|r|r|}
\hline
\multicolumn{6}{|c|}{
$
  f_{\alpha\,\mathrm{Na}}^\mathrm{(short)}(r) 
  =
\displaystyle
  \frac{A_{\alpha}^{\prime}}
       {1+e^{(r-w_{\alpha}^{\prime})/C_{\alpha}^{\prime}}} 
  -
  D_{\alpha}^{\prime}
  \left(\frac{\mbox{erf}(r/\Lambda_{\alpha}^{\prime})}{r}\right)^{8}
$
}
\\[6pt]
\hline\hline
$\alpha\!\equiv(s\tau)$ &  $A'_\alpha\; \rm[Ry]$   &   $w'_\alpha\;
    [a_0]$  &   $C'_\alpha\;[a_0]$  & $D'_\alpha\;[{\rm Ry}\,{a_0}^8]$
    & $\Lambda'_\alpha\;[a_0]$   
\\
\hline
$(\mathrm{O}\,c)$      &  10.58806      &   1.18         &   0.65905
& 35959.129    &   4.99             
\\
$(\mathrm{O}\,v)$      &  \multicolumn{1}{|c|}{0}  &
\multicolumn{1}{|c|}{--} & \multicolumn{1}{|c|}{--} & 
\multicolumn{1}{|c|}{0} & \multicolumn{1}{|c|}{--}
\\
$(\mathrm{Mg}\,c)$      &   8.12304      &   1.10         &   0.56032      &
15359.304    &   4.83             
\\
\hline
\end{tabular}
}
\caption{\label{tab:MgOparams}
Summary of the model for MgO(001) substrate. The uppermost block
explains the constituents and gives the parameters entering
Eq.~(\ref{eq:Esubstr}). Masses are given in terms of nuclear mass
units $m_n$. 
The three other block lists the functional form of the short-range
potentials, and the corresponding parameters, in Eq.~(\ref{eq:Vab}),
which couple the constituents of the MgO(001) substrate among
themselves [$f_{\alpha\beta}^\mathrm{(short)}$], these constituents
to the Na valence electrons [$f_{\alpha\,{\rm el}}^\mathrm{(short)}$],
and to the Na ions [$f_{\alpha\,\mathrm{Na}}^\mathrm{(short)}$].
}
\end{table}
The constituents, the short-range potentials, and the parameters for
the QM/MM model for Na clusters on MgO(001) are summarized in Table
\ref{tab:MgOparams}. The uppermost block shows the constituents.  The 
O$^{2-}$ anions have, of course, a large and soft valence electron
cloud while the Mg$^{2+}$ cations are very rigid, exclusively of
``core'' type, such that no valence cloud is associated with them.
The functional forms of the short-range potentials are taken over from
previous modeling of MgO(001) and Na@MgO(001) \cite{Nas01a,Win06}.
The same holds for the pure Mg\&O parameters in the third and fourth
block of table \ref{tab:MgOparams}. 
However the parameters for the coupling between MgO and the Na cluster
are calibrated from scratch to accommodate the modeling where only the
Na cluster is in the QM regime, while all substrate belongs to the MM
regime. The tuning was performed using Born-Oppenheimer surfaces (the
corresponding curves are presented in Sec.~\ref{sec:Na_MgO_PES},
Fig.~\ref{fig:PES_Na_MgO}) 
for Na@MgO(001) from \cite{Win06}. The latter were computed in the
shell model of \cite{Nas01a} where the MgO was treated
quantum-mechanically in a large vicinity of the Na contact point. Care
was also taken to reproduce basic dynamical properties as IP, band
gap, and optical response of the Na atom, for details see
\cite{Bae07a,Bae08a}.

\subsection{Details for rare gas substrates}
\label{sec:RG}

The constituents, the short-range potentials, and the parameters for
the QM/MM model for Na clusters in contact with rare gas (Rg)
substrates are summarized in Table~\ref{tab:Rg-params}. 
\begin{table}[htbp]
\centerline{
\begin{tabular}{|c|c||l|l|l|l|l|l|l|}
\hline
\multicolumn{2}{|c||}{\bf Constituents of rare gas}
& Rg & charge & 
\multicolumn{2}{c|}{mass} & Gaussian & spring const.
\\
\cline{5-6}
\multicolumn{2}{|c||}{\bf  (Rg) substrate} 
& & [$e$] & $M_{(\mathrm{Rg}c)}$ [$m_n$] & $M_{(\mathrm{Rg}v)}$ [$m_{\rm el}$] 
& width [$a_0$] & [Ry ${a_0}^{-2}$] 
\\
\hline
cores $\leftrightarrow(\mathrm{Rg}\,c)$
 & $N_{(\mathrm{Rg}\,c)}=N_{(\mathrm{Rg}\,v)}$
& Ne & 9.555   & 20.2 & 9.555  & 0.8834   & 35.223
\\
valence clouds $\leftrightarrow(\mathrm{Rg}\,v)$
 & $N_{(\mathrm{Rg}\,c)}=N_{(\mathrm{Rg}\,v)}$
& Ar & 6.119  & 40.0 & 4.38   & 1.43     & 6.758          
\\
&& Kr & 6.935  & 83.8  & 4.266  & 1.648    & 5.729
\\
\hline
\end{tabular}
}
\vspace*{0.8em}
\centerline{
\begin{tabular}{|c|c|c|}
\hline
\multicolumn{3}{|c|}{
$
\displaystyle
  f_{\alpha \beta }^\mathrm{(short)}(r) 
  =
  \delta_{\tau_\alpha\,\mathrm{c}}\,\delta_{\tau_\beta\,\mathrm{c}}\,
  \delta_{s_\alpha s_\beta}
  e^2\epsilon_\mathrm{Rg}\left[ \left(\frac{A}{R}\right)^{12} 
                              -\left(\frac{A}{R}\right)^6 \right]
$
}
\\[6pt]
\hline
\hline
$(\alpha,\beta) \equiv$ (Rg $c$,Rg $c$) & $\epsilon_\mathrm{Rg}$  &
$A$ [$a_0$] \\ \hline 
Ne & $3.4184 \times 10^{-4}$      &  5.426         \\
Ar & $1.3670  \times 10^{-3}$      &  6.501         \\
Kr & $1.8802 \times 10^{-3}$      &  6.917         \\ \hline
\end{tabular}
\qquad
\begin{tabular}{|c|c|c|c|}
\hline
\multicolumn{4}{|c|}{
$
  \displaystyle
  f_{\alpha\,\mathrm{el}}^\mathrm{(short)}(r) 
  =
  \delta_{\tau_\alpha\,\mathrm{c}}
  \frac{e^2A_\mathrm{el}}{1+e^{\beta_\mathrm{el}(r-r_\mathrm{el})}}
$
}
\\[6pt]
\hline
\hline
$\alpha\equiv$ (Rg $c$) & $A_\mathrm{el}$   & $\beta_\mathrm{el}$
    [${a_0}^{-1}$]  & $r_\mathrm{el}$ [$a_0$]
\\ \hline
Ne & 0.55  & 2.60368  & 1.8  
\\
Ar & 0.47  & 1.6941  & 2.2  
\\   
Kr & 0.555 & 1.56068  & 2.2  
\\
\hline
\end{tabular}
}
\vspace*{0.8em}
\centerline{
\begin{tabular}{|l|l|l|l|l|l|l|}
\hline
\multicolumn{7}{|c|}{$
\begin{array}{rcl}
  f_{\alpha\,\mathrm{Na}}^\mathrm{(short)}(r) 
  &=&
\displaystyle
  \delta_{\tau_\alpha\,\mathrm{c}}
  e^2\left[
  \frac{B {\rm e}^{-b_1R}}{R} -f_c(R) \left(\frac{C_6}{R^6}+\frac{C_8}{R^8} \right)
  \right]
\\[12pt]
\displaystyle
  f_c(R)
  &=&
  \left\{ \begin{array}{l@{\qquad}l}
   \frac{2}{1+{\rm e}^{b_2/R}}     & \hbox{for Rg=Ne,Ar}\\
   \frac{1}{1+{\rm e}^{(b_3-R)/b_2}}& \hbox{for Rg=Kr}
  \nonumber \end{array} \right.
\end{array}
$}
\\[6pt]
\hline
\hline
$\alpha \equiv$ (Rg $c$) & B  [$a_0$] & b$_1$ [${a_0}^{-1}$] & C$_6$
    [${a_0}^6$] & C$_8$ [${a_0}^8$] & b$_2$ [$a_0$] & b$_3$ [$a_0$] 
\\ \hline
Ne    & 171.8  & 2.1391  & 6.419  & 1358.7  & 10.4161
& \multicolumn{1}{c|}{$-$}     
\\
Ar    & 334.85 & 1.7624  & 52.5   & 1383    & 1.815  
& \multicolumn{1}{c|}{$-$}     
\\   
Kr    & 157.88 & 1.537   & 97.0   & 2691    & 0.7022 & 5.0764  
\\
\hline
\end{tabular}
}
\caption{\label{tab:Rg-params}
Summary of the model for rare gas substrate. The uppermost block
explains the constituents and gives the parameters entering
Eq.~(\ref{eq:Esubstr}). Core and valence cloud masses are given in
terms of nuclear mass unit $m_n$ and electron mass unit
$m_\mathrm{el}$ respectively.
The three other block lists the functional form of the short-range
potentials, and the corresponding parameters, in Eq.~(\ref{eq:Vab}),
which couple the cores of the rare gas substrate among
themselves [$f_{\alpha\beta}^\mathrm{(short)}$], the Rg cores
to the Na valence electrons [$f_{\alpha\,{\rm el}}^\mathrm{(short)}$],
and to the Na ions [$f_{\alpha\,\mathrm{Na}}^\mathrm{(short)}$].
}
\end{table}
We are
considering pure substrates from one rare gas species only, either Ne,
Ar or Kr.  Each rare gas atom is neutral and will be described by a
core and a valence cloud as indicated in the upper most block of table
\ref{tab:Rg-params}. The functional forms of the various short-range
potentials is taken over from previous treatments.
The Rg-Rg interaction is of standard Lennard-Jones type and taken from
\cite{Ash76}. This interaction provides a complete description of pure
rare gas compounds. The dipole polarizability is inactive in case of
pure rare gas systems without external electrons or Coulomb fields
such that the distinction between core and valence cloud becomes
obsolete and the (soft) Coulomb interaction between Rg atoms
disappears.
Rare gas polarization comes into play as soon as other materials are
around, in our case Na ions and valence electrons. The description in
terms of polarization potentials was initiated in \cite{Mue84} (for
alkaline cores) and applied to rare gas atoms in \cite{Ker95}.  The
present modeling takes up the more recent implementations from
\cite{Tso90,Dur97,Gro98}. It was worked out in detail for Ar in
\cite{Ger04b} and extended to Ne and Kr in \cite{Feh07b}, for a
detailed protocol see also \cite{Feh05,Feh06a}.

The parameters of the Rg valence cloud (right part of the top block in
Table~\ref{tab:Rg-params}) are adjusted to the (dynamical) polarization
properties of the Rg atoms. We use the static dipole polarizabilties
$\alpha_D({\rm Ne})=2.67\,a_0$ \cite{Ric91},
$\alpha_D({\rm Ar})=11.08\,a_0$ \cite{Dal61}, 
$\alpha_D({\rm Kr})= 16.79\,a_0$ \cite{Kum85}, and add
information from optical response \cite{Cha91,Cha92}, namely sum rule
and energy centroids $\bar{\omega}$. For the latter, we adopt
$\bar{\omega}=1.92, 1.76,$ and 1.64 Ry.
The Na$^+$-Rg potential is calibrated by a fit to scattering data
\cite{Ahm95,Vie03}.  The combined Na$^+$-Rg and electron-Rg potentials
are finally tuned to ground-state and excitation properties of NaRg
dimers taken from experimental as well as theoretical work, for NaAr
from \cite{Sax77,Las81,Sch00,Sch03}, for NaNe \cite{Lap80,Hli85},
and for NaKr \cite{Bru91}.
\begin{table}[htbp]
\begin{center}
\begin{tabular}{|c|c|c||c|c|}
\hline
\multicolumn{5}{|c|}{\bf Parameters for an effective Van-der-Waals
  model of Ar substrate} 
\\
\hline
\multicolumn{5}{|c|}{
$\displaystyle
  f_{{\rm Ar} c,\mathrm{el}}^\mathrm{(short)}(r) 
  =
   \frac{e^2A_\mathrm{el}}{1+e^{\beta_\mathrm{el}(r-r_\mathrm{el})}}
   -
   e^2V_\mathrm{VdW}^{\mbox{}}\,
   r^2V_\mathrm{soft}^8(r,\sigma_\mathrm{VdW}^{\mbox{}})
  \;,\;
  E_\mathrm{VdW}^{\mbox{}}
  =
  0
$
}
\\[6pt]
\hline
\hline
\quad $A_\mathrm{el}\quad$   & $\quad\beta_\mathrm{el}$\;
    [${a_0}^{-1}\mbox{]}\quad$ 
 & $\quad r_\mathrm{el}$\; [$a_0\mbox{]}\quad$
 & $\quad\sigma_\mathrm{VdW}$\;  [$a_0\mbox{]}\quad$
 & $\quad V_\mathrm{VdW}\quad$
\\ 
\hline
0.14  & 1.515502  & 2.2  &  6.0 & 201.25 
\\
\hline
\end{tabular}
\end{center}
\caption{\label{tab:effVdW}
Modified Ar-electron interaction which allows to incorporate the
Van-der-Waals energy effectively.
}
\end{table}
We have argued in section \ref{sec:calib} that one can approximate the
expensive treatment of the full Van der Waals (VdW) energy
(\ref{eq:EvdW}) by setting this explicit contribution to zero and to
build the effect implicitely into re-tuned model parameters.
Table~\ref{tab:effVdW} shows these effective parameters for a model
which omits the VdW term as such.

\bibliographystyle{review}
\bibliography{master}

\newcommand{\etalchar}[1]{$^{#1}$}
\begin{thebibliography}{DDTMB01}

\bibitem[]{}


\bibitem[AAR95]{Ahm95}
G.~R. Ahmadi, J.~Alml\"of and I.~R{\o}eggen.
\newblock Chem. Phys. {\bf 199} 33 (1995)

\bibitem[ADRS06]{And06}
K.~Andrae, P.~M. Dinh, P.-G. Reinhard and E.~Suraud.
\newblock Comp. Mat. Sci. {\bf 35} 169 (2006)

\bibitem[AKK08]{Arm08a}
R.~Armiento, S.~K\"ummel and T.~Korzd\"orfer.
\newblock Phys. Rev. B {\bf 77} 165106 (2008)

\bibitem[Alo06]{Alo06aB}
J.~A. Alonso.
\newblock {\em {Structure and properties of atomic clusters}\/} (Imperial
  College Press, London, 2006)

\bibitem[AM76]{Ash76}
N.~W. Ashcroft and N.~D. Mermin.
\newblock {\em {Solid State Physics}\/} (Saunders College, Philadelphia, 1976)

\bibitem[ARS02]{And02}
K.~Andrae, P.-G. Reinhard and E.~Suraud.
\newblock J. Phys. B {\bf 35} 1 (2002)

\bibitem[ARS04]{And04}
K.~Andrae, P.-G. Reinhard and E.~Suraud.
\newblock Phys. Rev. Lett. {\bf 92} 173402 (2004)

\bibitem[ASBG07]{Alt07aR}
P.~Altoè, M.~Stenta, A.~Bottoni and M.~Garavelli.
\newblock Theor. Chem. Acc. {\bf 118} 219 (2007)

\bibitem[AT87]{All87}
M.~P. Allen and D.~J. Tildesley.
\newblock {\em Computer Simulation of Liquids\/} (Oxford University Press, New
  York, 1987)

\bibitem[Bae08]{Bae08a}
M.~Baer.
\newblock {\em Non-linear dynamics of metal clusters on insulating
  substrates\/}.
\newblock Ph.D. thesis, Universit\"at Erlangen-N\"urnberg (2008)

\bibitem[BB99]{Bjo99}
S.~Bjornholm and J.~Borggreen.
\newblock Phil. Mag. {\bf 79} 1321 (1999)

\bibitem[BBF{\etalchar{+}}97]{Bro97}
K.~Bromann, H.~Brune, C.~F\'elix, W.~Harbich, R.~Monot, J.~Buttet and K.~Kern.
\newblock Surf. Sci. {\bf 377-379} 1051 (1997)

\bibitem[BBRS04]{Bel04a}
M.~Belkacem, M.~Bouchenne, P.-G. Reinhard and E.~Suraud.
\newblock Encycl. Nanosc. Nanotechn. {\bf 8} 575 (2004)

\bibitem[BCK{\etalchar{+}}93]{Bor93}
J.~Borggreen, P.~Chowdhury, N.~Kebaili, L.~Lundsberg-Nielsen,
  K.~Luetzenkirchen, M.~B. Nielsen, J.~Pedersen and H.~D. Rasmussen.
\newblock Phys. Rev. B {\bf 48} (1993)

\bibitem[Bec84]{Bec84}
D.~E. Beck.
\newblock Sol. St. Comm. {\bf 49} 381 (1984)

\bibitem[BF94]{Boa94}
J.~A. Boatz and M.~E. Fajardo.
\newblock J. Chem. Phys. {\bf 101} 3472 (1994)

\bibitem[BF07]{Bar07b}
G.~Barcaro and A.~Fortunelli.
\newblock New J. Physics {\bf 9}(2) 22 (2007)

\bibitem[BFB{\etalchar{+}}96]{Bro96}
K.~Bromann, C.~F\'elix, H.~Brune, W.~Harbich, R.~Monot, J.~Buttet and K.~Kern.
\newblock Science {\bf 274} 956 (1996)

\bibitem[BFNF05]{Bar05}
G.~Barcaro, A.~Fortunelli, F.~Nita and R.~Ferrando.
\newblock Phys. Rev. Lett. {\bf 95}(24) 246103 (2005)

\bibitem[BFR{\etalchar{+}}07]{Bar07a}
G.~Barcaro, A.~Fortunelli, G.~Rossi, F.~Nita and R.~Ferrando.
\newblock Phys. Rev. Lett. {\bf 98}(15) 156101 (2007)

\bibitem[BGS02]{Bar02b}
M.~Bargheer, M.~Guhr and N.~Schwentner.
\newblock J. Chem. Phys. {\bf 117} 5 (2002)

\bibitem[BHS82]{Bac82}
G.~B. Bachelet, D.~R. Hamann and M.~Schl\"uter.
\newblock Phys. Rev. B {\bf 26} 4199 (1982)

\bibitem[Bin01]{Bin01}
C.~Binns.
\newblock Surf. Sci. Rep. {\bf 44} 1 (2001)

\bibitem[BJR00]{Bal00}
F.~Balzer, S.~D. Jett and H.-G. Rubahn.
\newblock Solid Films {\bf 372} 78 (2000)

\bibitem[BKBK{\etalchar{+}}07]{Bon07}
V.~Bonacic-Koutecky, C.~B\"urgel, L.~Kronik, A.~E. Kuznetsov and R.~Mitric.
\newblock Euro. Phys. J. D {\bf 45} 471 (2007)

\bibitem[BKFK89]{Bon89}
V.~Bona\^{c}i\`{c}-Kouteck\`{y}, P.~Fantucci and J.~Kouteck\`{y}.
\newblock J. Chem. Phys. {\bf 91} 3794 (1989)

\bibitem[BKM05]{Kou05aR}
V.~Bona\^{c}i\`{c}-Kouteck\`{y} and R.~Mitri\`{c}.
\newblock Chem. Rev. {\bf 105} 11 (2005)

\bibitem[BKPBF99]{Bon99}
V.~Bona\v{c}i\'{c}-Kouteck\'{y}, J.~Pittner, M.~Boiron and P.~Fantucci.
\newblock J. Chem. Phys. {\bf 110}(8) 3876 (1999)

\bibitem[BKVM01]{Bon01}
V.~Bonacic-Koutecky, V.~Veyret and R.~Mitric.
\newblock J. Chem. Phys. {\bf 115} 10450 (2001)

\bibitem[BKZ91]{Bru91}
R.~Br\"uhl, J.~Kapetanakis and D.~Zimmermann.
\newblock J. Chem. Phys. {\bf 94} 5865 (1991)

\bibitem[BLMR92]{Blu92}
V.~Blum, G.~Lauritsch, J.~A. Maruhn and P.-G. Reinhard.
\newblock J. Comp. Phys {\bf 100} 364 (1992)

\bibitem[BLW{\etalchar{+}}99]{Bes99}
B.~Bescos, B.~Lang, J.~Weiner, V.~Weiss, E.~Wiedemann and G.~Gerber.
\newblock Euro. Phys. J. D {\bf 9} 399 (1999)

\bibitem[BMBK05]{Bue05a}
C.~B\"urgel, R.~Mitri\^{c} and V.~Bona\^{c}i\`{c}-Kouteck\`{y}.
\newblock Appl. Phys. A {\bf 82} (2005)

\bibitem[BMW{\etalchar{+}}07]{Bae07a}
M.~B\"{a}r, L.~V. Moskaleva, M.~Winkler, P.-G. Reinhard, N.~R\"{o}sch and
  E.~Suraud.
\newblock Eur. Phys. J. D {\bf 45} 507 (2007)

\bibitem[BPBB01]{Bag01}
B.~Baguenard, J.~C. Pinar, C.~Bordas and M.~Broyer.
\newblock Phys. Rev. A {\bf 63} 023204 (2001)

\bibitem[BR97]{Bab97}
J.~Babst and P.-G. Reinhard.
\newblock Z. f. Physik~D {\bf 42} 209 (1997)

\bibitem[Bra93]{Bra93}
M.~Brack.
\newblock Rev. Mod. Phys. {\bf 65} 677 (1993)

\bibitem[BSB98]{Ben98a}
L.~X. Benedict, E.~L. Shirley and R.~Bohn.
\newblock Phys. Rev. Lett. {\bf 80} 4514 (1998)

\bibitem[BSC{\etalchar{+}}96]{Buz96}
S.~A. Buzza, E.~M. Snyder, D.~A. Card, D.~E. Folmer and A.~W.~C. Jr.
\newblock J. Chem. Phys. {\bf 105} 7425 (1996)

\bibitem[CAL{\etalchar{+}}94]{Col94}
B.~A. Collins, K.~Athanassenas, D.~Lacombe, D.~M. Rayner and P.~A. Hackett.
\newblock J. Chem. Phys. {\bf 101} 3506 (1994)

\bibitem[Cam97]{Cam97}
C.~T. Campbell.
\newblock Surface Science Reports {\bf 27} 1 (1997)

\bibitem[CCG{\etalchar{+}}92]{Cha92}
W.~Chan, G.~Cooper, X.~Guo, G.~Burton and C.~Brion.
\newblock Phys. Rev. A {\bf 46} 149 (1992)

\bibitem[CCGB91]{Cha91}
W.~Chan, G.~Cooper, X.~Guo and C.~Brion.
\newblock Phys. Rev. A {\bf 45} 1420 (1991)

\bibitem[CCR07]{Cas07a}
M.~Cascella, M.~Cuendet and I.~T.~U. Rothlisberger.
\newblock J. Phys. Chem. {\bf 111} 10239 (2007)

\bibitem[CDR{\etalchar{+}}98]{Cal98c}
F.~Calvayrac, A.~Domps, P.-G. Reinhard, E.~Suraud and C.-A. Ullrich.
\newblock Euro. Phys. J. D {\bf 4} 207 (1998)

\bibitem[CFH{\etalchar{+}}00]{Ced00}
H.~Cederquist, A.~Fardi, K.~Haghighat, A.~Langereis, H.~T. Schmidt, S.~H.
  Schwartz, J.~C. Levin, I.~A. Sellin, H.~Lebius, B.~Huber, M.~O. Larsson and
  P.~Hvelplund.
\newblock Phys. Rev. A {\bf 61} 022712 (2000)

\bibitem[CGH{\etalchar{+}}95]{Cha95}
F.~Chandezon, C.~Guet, B.~A. Huber, D.~Jalabert, M.~Maurel, E.~Monnand,
  C.~Ristori and J.~C. Rocco.
\newblock Phys. Rev. Lett. {\bf 74} 3784 (1995)

\bibitem[CHH{\etalchar{+}}00]{Cam00}
E.~E.~B. Campbell, K.~Hansen, K.~Hoffmann, G.~Korn, M.~Tchaplyguine,
  M.~Wittmann and I.~V. Hertel.
\newblock Phys. Rev. Lett. {\bf 84} 2128 (2000)

\bibitem[CHTW06]{Coq06}
R.~Coquet, G.~J. Hutchings, S.~H. Taylor and D.~J. Willock.
\newblock J. Mater. Chem. {\bf 16} 1978 (2006)

\bibitem[CL93]{Che93}
H.-P. Cheng and U.~Landmann.
\newblock Science {\bf 326} 1304 (1993)

\bibitem[CL94]{Che94}
H.-P. Cheng and U.~Landman.
\newblock J. Phys. Chem. {\bf 98} 3527 (1994)

\bibitem[Cle85]{Cle85}
K.~Clemenger.
\newblock Phys. Rev. B {\bf 32} 1359 (1985)

\bibitem[CM82]{Cat82}
C.~R.~A. Catlow and W.~C. Mackrodt.
\newblock {\em {Computer Simulation of Solids}\/} (Springer, Berlin, 1982)

\bibitem[CRL{\etalchar{+}}06]{Con06}
F.~Conus, V.~Rodrigues, S.~Lecoultre, A.~Rydlo and C.~F\'elix.
\newblock J. Chem. Phys. {\bf 125} 024511 (2006)

\bibitem[CRS95]{Cal95a}
F.~Calvayrac, P.-G. Reinhard and E.~Suraud.
\newblock Phys. Rev. B {\bf 52} R17056 (1995)

\bibitem[CRS97]{Cal97b}
F.~Calvayrac, P.-G. Reinhard and E.~Suraud.
\newblock Ann. Phys. (NY) {\bf 255} 125 (1997)

\bibitem[CRSU00]{Cal00}
F.~Calvayrac, P.-G. Reinhard, E.~Suraud and C.~A. Ullrich.
\newblock Phys. Rep. {\bf 337} 493 (2000)

\bibitem[DDR{\etalchar{+}}96]{Dit96}
T.~Ditmire, T.~Donnelly, A.~M. Rubenchik, R.~W. Falcone and M.~D. Perry.
\newblock Phys. Rev. A {\bf 53} 3379 (1996)

\bibitem[DDS97]{Dur97}
G.~Durand, P.~Duplaa and F.~Spiegelmann.
\newblock Z. f. Physik~D {\bf 40} 177 (1997)

\bibitem[DDTMB01]{Die01}
T.~Diederich, T.~D\"oppner, J.~Tiggesb\"aumker and K.-H. Meiwes-Broer.
\newblock Phys. Rev. Lett. {\bf 86} 4807 (2001)

\bibitem[DFB{\etalchar{+}}07]{Din07c}
P.~M. Dinh, F.~Fehrer, G.~Bousquet, P.-G. Reinhard and E.~Suraud.
\newblock Phys. Rev. A {\bf 76} 043201 (2007)

\bibitem[DFD{\etalchar{+}}05]{Doe05b}
T.~D\"oppner, T.~Fennel, T.~Diederich, J.~Tiggesb\"aumker and K.~Meiwes-Broer.
\newblock Phys. Rev. Lett. {\bf 94} 013401 (2005)

\bibitem[DFR{\etalchar{+}}05]{Doe05a}
T.~D\"oppner, T.~Fennel, P.~Radcliffe, J.~Tiggesb\"aumker and K.-H.
  Meiwes-Broer.
\newblock Euro. Phys. J. D {\bf 36} 165 (2005)

\bibitem[DFRS07]{Din07b}
P.~M. Dinh, F.~Fehrer, P.-G. Reinhard and E.~Suraud.
\newblock Euro. Phys. J. D {\bf 45} 415 (2007)

\bibitem[DFRS08]{Din08a}
P.~M. Dinh, F.~Fehrer, P.-G. Reinhard and E.~Suraud.
\newblock Surf. Sci. {\bf 602} 2699 (2008)

\bibitem[DG90]{Dre90}
R.~M. Dreizler and E.~K.~U. Gross.
\newblock {\em {Density Functional Theory: An Approach to the Quantum Many-Body
  Problem}\/} (Springer-Verlag, Berlin, 1990)

\bibitem[DGRS00]{Dom00c}
A.~Domps, E.~Giglio, P.-G. Reinhard and E.~Suraud.
\newblock J. Phys. B {\bf 33} L333 (2000)

\bibitem[dH93]{Hee93}
W.~A. de~Heer.
\newblock Rev. Mod. Phys. {\bf 65} 611 (1993)

\bibitem[DK61]{Dal61}
A.~Dalgarno and A.~Kingston.
\newblock Proc. Phys. Soc. A {\bf 251} 424 (1961)

\bibitem[DLRS97]{Dom97b}
A.~Domps, P.~L'Eplattenier, P.-G. Reinhard and E.~Suraud.
\newblock Ann. Phys. (Leipzig) {\bf 6} 455 (1997)

\bibitem[DO58]{Dic58}
B.~G. Dick and A.~W. Overhauser.
\newblock Phys. Rev. {\bf 112} 90 (1958)

\bibitem[DRS98]{Dom98b}
A.~Domps, P.-G. Reinhard and E.~Suraud.
\newblock Phys. Rev. Lett. {\bf 81} 5524 (1998)

\bibitem[DRS05]{Din05}
P.~M. Dinh, P.-G. Reinhard and E.~Suraud.
\newblock J. Phys. B {\bf 38} 1637 (2005)

\bibitem[DRS09]{Din08c}
P.~M. Dinh, P.-G. Reinhard and E.~Suraud.
\newblock Surf. Science {\bf 603} 400 (2009)

\bibitem[DSN{\etalchar{+}}02]{Dub02}
B.~Dubertret, P.~Skourides, D.~J. Norris, V.~Noireaux, A.~H. Brivanlou and
  A.~Libchaber.
\newblock Science {\bf 298} 1759 (2002)

\bibitem[DTMB02]{Die02}
T.~Diederich, J.~Tiggesb\"aumker and K.~H. Meiwes-Broer.
\newblock J. Chem. Phys. {\bf 116} 3263 (2002)

\bibitem[DVP05]{Del05}
A.~{D}el {V}itto and G.~Pacchioni.
\newblock J. Phys. Chem. B {\bf 109} 8040 (2005)

\bibitem[DVSIP04]{Del04}
A.~{D}el {V}itto, C.~Sousa, F.~Illas and G.~Pacchioni.
\newblock J. Chem. Phys. {\bf 121}(15) 7457 (2004)

\bibitem[Eka84]{Eka84}
W.~Ekardt.
\newblock Phys. Rev. Lett. {\bf 52} 1925 (1984)

\bibitem[Eka99]{Eka99}
W.~Ekardt, ed.
\newblock {\em {Metal Clusters}\/} (Wiley, New York, 1999)

\bibitem[FA34]{Fer34}
E.~Fermi and E.~Amaldi.
\newblock Accad. Ital. Rome {\bf 6} 117 (1934)

\bibitem[Fai87]{Fai87}
F.~H.~M. Faisal.
\newblock {\em {Theory of Multiphoton Processes}\/} (Plenum Press, New York,
  1987)

\bibitem[FAL{\etalchar{+}}02]{Fel02}
N.~Felidj, J.~Aubard, G.~Levi, J.~R. Krenn, M.~Salerno, G.~Schider,
  B.~Lamprecht, A.~Leitner and F.~R. Aussenegg.
\newblock Phys. Rev. B {\bf 65} 075419 (2002)

\bibitem[FBMB04]{Fen04}
T.~Fennel, G.~F. Bertsch and K.-H. Meiwes-Broer.
\newblock Eur. Phys. J. D {\bf 29} 367 (2004)

\bibitem[FDB{\etalchar{+}}07]{Feh07b}
F.~Fehrer, P.~M. Dinh, M.~Baer, P-G, Reinhard and E.~Suraud.
\newblock Euro. Phys. J. D {\bf 45} 447 (2007)

\bibitem[FDPG{\etalchar{+}}07]{Feh07a}
F.~Fehrer, P.~M. Dinh, P-G, Reinhard and E.~Suraud.
\newblock Phys. Rev. B {\bf 75} 235418 (2007)

\bibitem[FDPG{\etalchar{+}}08]{Feh08a}
F.~Fehrer, P.~M. Dinh, P-G, Reinhard and E.~Suraud.
\newblock Comp. Mat. Sci. {\bf 42} 203 (2008)

\bibitem[FDZ{\etalchar{+}}02]{Fuk02}
D.~Fuks, S.~Dorfman, Y.~F. Zhukovskii, E.~A. Kotomin and A.~M. Stoneham.
\newblock Surf. Sci. {\bf 499} 24 (2002)

\bibitem[Feh06]{Feh06c}
F.~Fehrer.
\newblock {\em Metallcluster in Kontakt mit Edelgaassubstraten\/}.
\newblock Ph.D. thesis, Universit\"at Erlangen/N\"urnberg (2006)

\bibitem[FF09]{Fer09a}
R.~Ferrando and A.~Fortunelli.
\newblock Journal of Physics: Condensed Matter {\bf 21} 264001 (2009)

\bibitem[FFS82]{Fei82}
M.~D. Feit, J.~A. Fleck and A.~Steiger.
\newblock J. Comp. Phys. {\bf 47} 412 (1982)

\bibitem[FFV00]{Fro00}
G.~E. Froudakis, S.~C. Farantos and M.~Velegrakis.
\newblock Chem. Phys. {\bf 258}(1) 13 (2000)

\bibitem[FHB93]{Fed93}
S.~Fedrigo, W.~Harbich and J.~Buttet.
\newblock Phys. Rev. B {\bf 47} 10706 (1993)

\bibitem[FHB98]{Fed98}
S.~Fedrigo, W.~Harbich and J.~Buttet.
\newblock Phys. Rev. B {\bf 58}(11) 7428 (1998)

\bibitem[FHH{\etalchar{+}}08]{Fro08}
P.~Frondelius, A.~Hellman, K.~Honkala, H.~H\"{a}kkinen and H.~Gr\"{o}nbeck.
\newblock Phys. Rev. B {\bf 78}(8) 085426 (2008)

\bibitem[FJJ08]{Fer08}
R.~Ferrando, J.~Jellinek and R.~L. Johnston.
\newblock Chem. Rev.  (2008)

\bibitem[FMBT{\etalchar{+}}08]{Fen08a}
T.~Fennel, K.-H. Meiwes-Broer, J.~Tiggesb\"aumker, P.~M. Dinh, P.-G. Reinhard
  and E.~Suraud.
\newblock preprint, subm. Rev.Mod.Phys.  (2008)

\bibitem[FMRS05]{Feh05}
F.~Fehrer, M.~Mundt, P.-G. Reinhard and E.~Suraud.
\newblock Ann. Phys. (Leipzig) {\bf 14} 411 (2005)

\bibitem[FRL{\etalchar{+}}09]{Fer09b}
R.~Ferrando, G.~Rossi, A.~C. Levi, Z.~Kuntov\'{a}, F.~Nita, A.~Jelea,
  C.~Mottet, G.~Barcaro, A.~Fortunelli and J.~Goniakowski.
\newblock J. Chem. Phys. {\bf 130} 174702 (2009)

\bibitem[FRS06a]{Feh06b}
F.~Fehrer, P.-G. Reinhard and E.~Suraud.
\newblock Appl. Phys. A {\bf 82} 145 (2006)

\bibitem[FRS{\etalchar{+}}06b]{Feh06a}
F.~Fehrer, P.-G. Reinhard, E.~Suraud, E.~Giglio, B.~Gervais and A.~Ipatov.
\newblock Appl. Phys. A {\bf 82} 151 (2006)

\bibitem[FSH{\etalchar{+}}01]{Fel01}
C.~F\'elix, C.~Sieber, W.~Harbich, J.~Buttet, I.~Rabin, W.~Schulze and G.~Ertl.
\newblock Phys. Rev. Lett. {\bf 86}(14) 2992 (2001)

\bibitem[FSSB96]{Fel96}
V.~I. Feldman, F.~F. Sukhov, N.~A. Slovokhotova and V.~P. Bazov.
\newblock Radiat. Phys. hem. {\bf 48} 261 (1996)

\bibitem[G\"06]{Goe06b}
A.~G\"orling.
\newblock Lecture Notes in Physics {\bf 706} 137 (2006)

\bibitem[GB03]{Gro03}
H.~Gr\"{o}nbeck and P.~Broqvist.
\newblock J. Chem. Phys. {\bf 119}(7) 3896 (2003)

\bibitem[GBD{\etalchar{+}}95]{Goe95}
T.~Goetz, M.~Buck, C.~Dressler, F.~Eisert and F.~Traeger.
\newblock Appl. Phys. A {\bf 60} 607 (1995)

\bibitem[GBP05]{Gio05}
L.~Giordano, M.~Baistrocchi and G.~Pacchioni.
\newblock Phys. Rev. B {\bf 72} 115403 (2005)

\bibitem[GDP96]{Gro96}
E.~K.~U. Gross, J.~F. Dobson and M.~Petersilka.
\newblock Top. Curr. Chem. {\bf 181} 81 (1996)

\bibitem[GGJ{\etalchar{+}}04]{Ger04b}
B.~Gervais, E.~Giglio, E.~Jaquet, A.~Ipatov, P.-G. Reinhard and E.~Suraud.
\newblock J. Chem. Phys. {\bf 121} 8466 (2004)

\bibitem[GGJ{\etalchar{+}}05]{Ger05}
B.~Gervais, E.~Giglio, E.~Jacquet, A.~Ipatov, P.-G. Reinhard, F.~Fehrer and
  E.~Suraud.
\newblock Phys. Rev. A {\bf 71} 015201 (2005)

\bibitem[GJM{\etalchar{+}}09]{Gon09}
J.~Goniakowski, A.~Jelea, C.~Mottet, G.~Barcaro, A.~Fortunelli, Z.~Kuntov\'{a},
  F.~Nita, A.~C. Levi, G.~Rossi and R.~Ferrando.
\newblock J. Chem. Phys. {\bf 130} 174703 (2009)

\bibitem[GLC{\etalchar{+}}01]{Gau01}
M.~Gaudry, J.~Lerm\'e, E.~Cottancin, M.~Pellarin, J.-L. Vialle, M.~Broyer,
  B.~Pr\'evel, M.~Treilleux and P.~M\'elinon.
\newblock Phys. Rev. B {\bf 64} 085407 (2001)

\bibitem[GN99]{Gon99}
J.~Goniakowski and C.~Noguera.
\newblock Phys. Rev. B {\bf 60}(23) 16120 (1999)

\bibitem[GPGP99]{Gre99a}
N.~Gresh, O.~Parisel and C.~Giessner-Prettre.
\newblock THEOCHEM {\bf 458} 27 (1999)

\bibitem[GRGL04]{Gij04}
K.~T. Giju, S.~Roszak, R.~W. Gora and J.~Leszczynski.
\newblock Chem. Phys. Lett. {\bf 391} 112 (2004)

\bibitem[GRS01]{Gig01a}
E.~Giglio, P.-G. Reinhard and E.~Suraud.
\newblock J. Phys. B {\bf 34} 1253 (2001)

\bibitem[GRS02]{Gig02}
E.~Giglio, P.-G. Reinhard and E.~Suraud.
\newblock Ann. Phys. (Leipzig) {\bf 11} 291 (2002)

\bibitem[GRS03]{Gig03}
E.~Giglio, P.-G. Reinhard and E.~Suraud.
\newblock Phys. Rev. A {\bf 67} 43202 (2003)

\bibitem[GS95]{Gar95}
B.~M. Garraway and K.-A. Suominen.
\newblock Rep. Prog. Phys. {\bf 58} 365 (1995)

\bibitem[GS98]{Gro98}
M.~Gross and F.~Spiegelmann.
\newblock J. Chem. Phys. {\bf 108} 4148 (1998)

\bibitem[GT02]{Gao02aR}
J.~Gao and D.~Truhlar.
\newblock Ann. Rev. Phys. Chem. {\bf 53} 467 (2002)

\bibitem[GTH96]{Goe96}
S.~Goedecker, M.~Teter and J.~Hutter.
\newblock Phys. Rev. B {\bf 54} 1703 (1996)

\bibitem[GVB{\etalchar{+}}07]{Gle07a}
T.~Gleitsmann, M.~Vaida, T.~Bernhardt, V.~Bona\v{c}i\'{c}-Kouteck\'{y},
  C.~B\"urgel, A.~Kuznetsov and R.~Mitric.
\newblock Eur. Phys. J. D {\bf 45} 477 (2007)

\bibitem[Hab94a]{Hab94a}
H.~Haberland, ed.
\newblock {\em {Clusters of Atoms and Molecules 1- Theory, Experiment, and
  Clusters of Atoms}\/}, vol.~52 (Springer Series in Chemical Physics, Berlin,
  1994)

\bibitem[Hab94b]{Hab94b}
H.~Haberland, ed.
\newblock {\em {Clusters of Atoms and Molecules 2- Solvation and Chemistry of
  Free Clusters, and Embedded, Supported and Compressed Clusters}\/}, vol.~56
  (Springer Series in Chemical Physics, Berlin, 1994)

\bibitem[HBM85]{Hli85}
M.~Hliwa, J.~C. Barthelat and J.~P. Malrieu.
\newblock J. Phys. B {\bf 18} 2433 (1985)

\bibitem[Hen98]{Hen98b}
C.~R. Henry.
\newblock Surface Science Reports {\bf 31} 231 (1998)

\bibitem[Hen05]{Hen05}
C.~R. Henry.
\newblock Progress in Surface Science {\bf 80} 92 (2005)

\bibitem[HF02]{Har02}
W.~Harbich and C.~F\'elix.
\newblock C. R. Physique {\bf 3} 289 (2002)

\bibitem[HFM{\etalchar{+}}90]{Har90}
W.~Harbich, S.~Fedrigo, F.~Meyer, D.~Lindsay, J.~Lignires, J.~C. Rivoal and
  D.~Kreisle.
\newblock J. Chem. Phys. {\bf 93} 8535 (1990)

\bibitem[HH07]{Hon07}
K.~Honkala and H.~H\"akkinen.
\newblock J. Phys. Chem. C {\bf 111} 4319 (2007)

\bibitem[HIM93]{Hab93}
H.~Haberland, Z.~Insepov and M.~Moseler.
\newblock Z. f. Physik~D {\bf 26} 229 (1993)

\bibitem[HM96a]{Hak96}
H.~H\"akinnen and M.~Manninen.
\newblock Europhys. Lett. {\bf 34} (1996)

\bibitem[HM96b]{Hak96b}
H.~H\"akkinen and M.~Manninen.
\newblock J. Chem. Phys. {\bf 105} 10565 (1996)

\bibitem[HP72]{Hoa72a}
M.~Hoare and P.~Pal.
\newblock Nature {\bf 236} 35 (1972)

\bibitem[HRS{\etalchar{+}}08]{Har08}
M.~Harb, F.~Rabilloud, D.~Simon, A.~Rydlo, S.~Lecoultre, F.~conus, V.~Rodrigues
  and C.~F\'elix.
\newblock J. Chem. Phys. {\bf 129} 194108 (2008)

\bibitem[HSZ{\etalchar{+}}91]{Hu91}
Z.~Hu, B.~Shen, Q.~Zhou, S.~Deosaran, J.~R. Lombardi, D.~M. Lindsay and
  W.~Harbich.
\newblock The Journal of Chemical Physics {\bf 95}(3) 2206 (1991)

\bibitem[HWvI{\etalchar{+}}01]{Hof01}
M.~A. Hoffmann, G.~Wrigge, B.~v~Issendorff, J.~Muller, G.~Gantefor and
  H.~Haberland.
\newblock Euro. Phys. J. D {\bf 16} 9 (2001)

\bibitem[IBG{\etalchar{+}}05]{Ira05}
T.~Irawan, D.~Boecker, F.~Ghaleh, B.~v~Issendorf and H.~H\"ovel.
\newblock Appl.~Phys.~A {\bf 82} 81 (2005)

\bibitem[IRS03]{Ipa03}
A.~Ipatov, P.-G. Reinhard and E.~Suraud.
\newblock Int. J. Mol. Sci. {\bf 3} 301 (2003)

\bibitem[ISS77]{Iss1}
{\em Proceedings of 1th International Symposium on Small Particles and
  Inorganic Clusters\/}, vol.~38 (1977).
\newblock J. Phys.

\bibitem[ISS81]{Iss2}
{\em Proceedings of 2th International Symposium on Small Particles and
  Inorganic Clusters\/}, vol. 106 (1981).
\newblock Surf. Sci.

\bibitem[ISS85]{Iss3}
{\em Proceedings of 3th International Symposium on Small Particles and
  Inorganic Clusters\/}, vol. 156 (1985).
\newblock Surf. Sci.

\bibitem[ISS89]{Iss4}
{\em Proceedings of 4th International Symposium on Small Particles and
  Inorganic Clusters\/}, vol.~12 (1989).
\newblock Z. Phys. D

\bibitem[ISS91]{Iss5}
{\em Proceedings of 5th International Symposium on Small Particles and
  Inorganic Clusters\/}, vol.~19 (1991).
\newblock Z. Phys. D

\bibitem[ISS93]{Iss6}
{\em Proceedings of 6th International Symposium on Small Particles and
  Inorganic Clusters\/}, vol.~26 (1993).
\newblock Z. Phys. D

\bibitem[ISS96]{Iss7}
{\em Proceedings of 7th International Symposium on Small Particles and
  Inorganic Clusters\/}, vol.~3 (1996).
\newblock Surf. Rev. Lett.

\bibitem[ISS97]{Iss8}
{\em Proceedings of 8th International Symposium on Small Particles and
  Inorganic Clusters\/}, vol.~40 (1997).
\newblock Z. f. Phys. D

\bibitem[ISS99]{Iss9}
{\em Proceedings of 9th International Symposium on Small Particles and
  Inorganic Clusters\/}, vol.~9 (1999).
\newblock Euro. Phys. J. D

\bibitem[ISS01]{Iss10}
{\em Proceedings of 10th International Symposium on Small Particles and
  Inorganic Clusters\/}, vol.~16 (2001).
\newblock Euro. Phys. J. D

\bibitem[ISS03]{Iss11}
{\em Proceedings of 11th International Symposium on Small Particles and
  Inorganic Clusters\/}, vol.~24 (2003).
\newblock Euro. Phys. J. D

\bibitem[ISS05]{Iss12}
{\em Proceedings of 12th International Symposium on Small Particles and
  Inorganic Clusters\/}, vol.~34 (2005).
\newblock Euro. Phys. J. D

\bibitem[ISS07]{Iss13}
{\em Proceedings of 13th International Symposium on Small Particles and
  Inorganic Clusters\/}, vol.~43 (2007).
\newblock Euro. Phys. J. D

\bibitem[JBD{\etalchar{+}}00]{Jan00}
D.~B. Janes, M.~Batistuta, S.~Datta, M.~R. Melloch, R.~P. Andres, J.~Liu, N.-P.
  Chen, T.~Lee, R.~Reifenberger, E.~H. Chen and J.~M. Woodall.
\newblock Superlattices and Microstructures {\bf 27} 555 (2000)

\bibitem[Jel99]{Jel99}
J.~Jellinek.
\newblock {\em {Theory of Atomic and Molecular Clusters: with a Glimpse at
  Experiments}\/} (Springer, Berlin, 1999)

\bibitem[JFA{\etalchar{+}}01]{Jac01}
T.~Jacob, B.~Fricke, J.~Anton, S.~Varga, T.~Bastug, S.~Fritzsche and W.~Sepp.
\newblock Euro. Phys. J. D {\bf 16} 257 (2001)

\bibitem[JMG{\etalchar{+}}98]{Bru98}
M.~B. Jr, M.~Moronne, P.~Gin, S.~Weiss and A.~P. Alivisato.
\newblock Science {\bf 281} 2013 (1998)

\bibitem[KB62]{Kad62}
L.~P. Kadanoff and G.~Baym.
\newblock {\em Quantum Statistical Mechanics\/} (Benjamin, New York, 1962)

\bibitem[KBR99]{Kue99}
S.~K\"ummel, M.~Brack and P.-G. Reinhard.
\newblock Euro. Phys. J. D {\bf 9} 149 (1999)

\bibitem[KBS{\etalchar{+}}07]{Kos07a}
O.~Kostko, C.~Bartels, J.~Schwobel, C.~Hock and B.~v~Issendorff\.
\newblock J. Phys.~: Conf. Ser. {\bf 88} 012034 (2007)

\bibitem[KCRS98]{Koh98a}
C.~Kohl, F.~Calvayrac, P.-G. Reinhard and E.~Suraud.
\newblock Surf. Sci. {\bf 405} 74 (1998)

\bibitem[Ker95]{Ker95}
C.~Kerner.
\newblock {\em Definition von Rumpf-Polarisations-Potentialen f{\"u}r Edelgase
  {\"u}ber die Berechnung von Elektronstreuphasen und Anwendung auf
  Alkali-Edelgas-Wechselwirkungspotentiale\/}.
\newblock Ph.D. thesis, Universit\"at Kaiserslautern D368 (1995)

\bibitem[KF96]{Kur96}
L.~I. Kurkina and O.~V. Farberovich.
\newblock Phys. Rev. B {\bf 54}(20) 14791 (1996)

\bibitem[KKP04]{Kue04a}
S.~K\"ummel, L.~Kronik and J.~Perdew.
\newblock Phys. Rev. Lett. {\bf 93} 213002 (2004)

\bibitem[KKS05]{Kra05a}
P.~Krause, T.~Klamroth and P.~Saalfrank.
\newblock J. Chem. Phys. {\bf 123} (2005)

\bibitem[KM85]{Kum85}
A.~Kumar and W.~Meath.
\newblock Can. J. Chem. {\bf 63} 1616 (1985)

\bibitem[KM07]{Kue07}
J.~K\"upper and J.~M. Merritt.
\newblock Int. Rev. Phys. Chem. {\bf 26} 249 (2007)

\bibitem[KMB{\etalchar{+}}05]{Kou05a}
V.~B. Kouteck\'y, R.~Mitri\'c, T.~Bernhardt, L.~Wöste and J.~Jortner.
\newblock Adv. Phys. Chem. {\bf 132} 179 (2005)

\bibitem[KMR95]{Koh95}
C.~Kohl, B.~Montag and P.-G. Reinhard.
\newblock Z. f. Physik~D {\bf 35} 57 (1995)

\bibitem[Koh97]{Koh97b}
C.~Kohl.
\newblock Ph.D. thesis, F.-A. Universit\"at Erlangen-N\"urnberg (1997)

\bibitem[KPS{\etalchar{+}}01]{Koh01}
K.~Kholmurodov, I.~Puzynin, W.~Smith, K.~Yasuoka and T.~Ebisuzaki.
\newblock Comput. Phys. Commun. {\bf 141} 1 (2001)

\bibitem[KR97]{Koh97a}
C.~Kohl and P.-G. Reinhard.
\newblock Z. f. Physik~D {\bf 39} 225 (1997)

\bibitem[KRS00]{Koh00}
C.~Kohl, P.~G. Reinhard and E.~Suraud.
\newblock Euro. Phys. J. D {\bf 11} 115 (2000)

\bibitem[KS37]{Kra02aR}
V.~P. Krainov and M.~B. Smirnov.
\newblock Phys. Rep. {\bf 370} 2002 (237)

\bibitem[KSK06]{Kri06}
S.~Krischok, P.~Stracke and V.~Kempter.
\newblock Appl. Phys. A {\bf 82} 167 (2006)

\bibitem[KV93]{Kre93}
U.~Kreibig and M.~Vollmer.
\newblock {\em Optical Properties of Metal Clusters\/}, vol.~25 (Springer
  Series in Materials Science, 1993)

\bibitem[KWSR97]{Kle97}
J.-H. Klein-Wiele, P.~Simon and H.-G. Rubahn.
\newblock Phys. Rev. Lett. {\bf 80} 45 (1997)

\bibitem[KZM{\etalchar{+}}06]{Khl06a}
B.~Khlebtsov, V.~Zharov, A.~Melnikov, V.~Tuchin and N.~Khlebtsov.
\newblock Nanotechnology {\bf 17} 5167 (2006)

\bibitem[LABM{\etalchar{+}}80]{Lap80}
W.~P. Lapatovich, R.~Ahmad-Bitar, P.~E. Moskowitz, I.~Renhorn, R.~A. Gottscho
  and D.~E. Pritchard.
\newblock J. Chem. Phys. {\bf 73} 5419 (1980)

\bibitem[LAW00a]{Lau00b}
J.~T. Lau, A.~Achleitner and W.~Wurth.
\newblock Surf. Sci. {\bf 467} L834 (2000)

\bibitem[LAW00b]{Lau00a}
J.~T. Lau, A.~Achleitner and W.~Wurth.
\newblock Chem. Phys. Lett. {\bf 317} 269 (2000)

\bibitem[LC85]{Lew85}
G.~V. Lewis and C.~R.~A. Catlow.
\newblock J. Phys. C {\bf 18} 1149 (1985)

\bibitem[LK05]{Lei05a}
M.~Lein and S.~K\"ummel.
\newblock Phys. Rev. Lett. {\bf 94} 143003 (2005)

\bibitem[LLS82]{Las81}
B.~C. Laskowski, S.~R. Langhoff and J.~Stallcop.
\newblock J. Chem. Phys. {\bf 75} 815 (1982)

\bibitem[LMP{\etalchar{+}}00]{Leh00}
J.~Lehmann, M.~Merschdorf, W.~Pfeiffer, A.~Thon, S.~Voll and G.~Gerber.
\newblock Phys. Rev. Lett. {\bf 85} 2921 (2000)

\bibitem[LNH{\etalchar{+}}06]{Liu06}
B.~Liu, S.~B. Nielsen, P.~Hvelplund, H.~Zettergren, H.~Cederquist, B.~Manil and
  B.~A. Huber.
\newblock Phys. Rev. Lett. {\bf 97} 133401 (2006)

\bibitem[LNR{\etalchar{+}}91]{Lic91}
D.~L. Lichtenberger, K.~W. Nebesny, C.~D. Ray, D.~R. Huffman and L.~D. Lamb.
\newblock Chem. Phys. Lett. {\bf 176} 203 (1991)

\bibitem[LPP{\etalchar{+}}98]{Ler98}
J.~Lerm\'e, B.~Palpant, B.~Pr\'evel, M.~Pellarin, M.~Treilleux, J.~L. Vialle,
  A.~Perez and M.~Broyer.
\newblock Phys. Rev. Lett. {\bf 80}(23) 5105 (1998)

\bibitem[LR94]{Lau94}
G.~Lauritsch and P.-G. Reinhard.
\newblock Int. J. Mod. Phys. C {\bf 5} 65 (1994)

\bibitem[LRF07]{Lec07}
S.~Lecoultre, A.~Rydlo and C.~F\'{e}lix.
\newblock J. Chem. Phys. {\bf 126}(20) 204507 (2007)

\bibitem[LSR02]{Leg02}
C.~Legrand, E.~Suraud and P.-G. Reinhard.
\newblock J. Phys. B {\bf 35} 1115 (2002)

\bibitem[LWEA03]{Lau03}
J.~T. Lau, W.~Wurth, H.-U. Ehrke and A.~Achleitner.
\newblock Low Temperature Physics {\bf 29}(3) 223 (2003)

\bibitem[MB00a]{Mei00aB}
K.~H. Meiwes-Broer, ed.
\newblock {\em {Metal clusters at surfaces}\/} (Springer, Berlin, 2000)

\bibitem[MB00b]{Mei00}
K.~H. Meiwes-Broer.
\newblock {\em Metal Clusters at Surfaces: Structure, Quantum Properties,
  Physical Chemistry\/} (Springer, New York, 2000)

\bibitem[MB06]{Mei06aB}
K.-H. Meiwes-Broer, ed.
\newblock {\em Clusters at Surfaces: Electronic Properties and Magnetism\/},
  vol.~82 (2006).
\newblock {Applied Phys. A, special issue}

\bibitem[MBB07]{Mei07aB}
K.-H. Meiwes-Broer and R.~Berndt, eds.
\newblock {\em {Atomic Clustes at Surfaces and in Thin Films}\/}, vol.~45
  (2007).
\newblock {Eur. Phys. J. D, topical issue}

\bibitem[MDCS99]{Mus99a}
V.~Musolino, A.~Dal~Corso and A.~Selloni.
\newblock Phys. Rev. Lett. {\bf 83}(14) 2761 (1999)

\bibitem[MDRS08a]{Mes08b}
J.~Messud, P.~M. Dinh, P.-G. Reinhard and E.~Suraud.
\newblock Chem. Phys. Lett. {\bf 461} 316 (2008)

\bibitem[MDRS08b]{Mes08a}
J.~Messud, P.~M. Dinh, P.-G. Reinhard and E.~Suraud.
\newblock Phys. Rev. Lett. {\bf 101} 096404 (2008)

\bibitem[MEL{\etalchar{+}}89]{McH89}
K.~M. McHugh, J.~G. Eaton, G.~H. Lee, H.~W. Sarkas, L.~H. Kidder, J.~T.
  Snodgrass, M.~R. Manaa and K.~H. Bowen.
\newblock J. Chem. Phys. {\bf 91} 3792 (1989)

\bibitem[MFM84]{Mue84}
W.~M\"uller, J.~Flesch and W.~Meyer.
\newblock J. Chem. Phys. {\bf 80} 3297 (1984)

\bibitem[MI99]{Mil99}
P.~Milani and S.~Iannotta.
\newblock {\em {Cluster Beam Synthesis of Nanostructured Materials}\/}
  (Springer, Berlin, 1999)

\bibitem[Mil01]{Mil01aB}
R.~Miller, ed.
\newblock {\em {Helium Nanodroplets: A Novel Medium for Chemistry and
  Physic}\/}, vol. 115 (2001).
\newblock J. Chem. Phys., special issue

\bibitem[MK05]{Mun05b}
M.~Mundt and S.~K\"ummel.
\newblock Phys. Rev. Lett. {\bf 95} 1 (2005)

\bibitem[MKHM06]{Mun06b}
M.~Mundt, S.~K\"{u}mmel, B.~Huber and M.~Moseler.
\newblock Phys. Rev. B {\bf 73}(20) 205407 (2006)

\bibitem[MLY01]{Mos01}
M.~Moseler, U.~Landman and C.~Yannouleas.
\newblock Phys. Rev. Lett. {\bf 87} 053401 (2001)

\bibitem[MNPR99]{Mat99}
A.~V. Matveev, K.~M. Neyman, G.~Pacchioni and N.~R\"osch.
\newblock Chem. Phys. Lett. {\bf 299} 603 (1999)

\bibitem[MP07]{Men07}
F.~Meng and A.~Pucci.
\newblock Phys. Stat. Sol. (b) {\bf 244} 3739 (2007)

\bibitem[MPBS01]{May01}
C.~Mayer, R.~Palkovits, G.~Bauer and T.~Schalkhammer.
\newblock J. Nanoparticle Res. {\bf 3} 361 (2001)

\bibitem[MR94]{Mon94a}
B.~Montag and P.-G. Reinhard.
\newblock Phys. Lett. A {\bf 193} 380 (1994)

\bibitem[MR95a]{Mon95a}
B.~Montag and P.-G. Reinhard.
\newblock Z. f. Physik~D {\bf 33} 265 (1995)

\bibitem[MR95b]{Mon95d}
B.~Montag and P.-G. Reinhard.
\newblock Phys. Rev. B {\bf 51} 14686 (1995)

\bibitem[MRM94]{Mon94b}
B.~Montag, P.-G. Reinhard and J.~Meyer.
\newblock Z. f. Physik~D {\bf 32} 125 (1994)

\bibitem[MSC98]{Mus98}
V.~Musolino, A.~Selloni and R.~Car.
\newblock J. Chem. Phys. {\bf 108}(12) 5044 (1998)

\bibitem[MSC99]{Mus99b}
V.~Musolino, A.~Selloni and R.~Car.
\newblock Phys. Rev. Lett. {\bf 83}(16) 3242 (1999)

\bibitem[MTG{\etalchar{+}}05]{Mor05}
M.-E. Moret, E.~Tapavicza, L.~Guidoni, U.~R\"ohrig, M.~Sulpizi, I.~Tavernelli
  and U.~Rothlisberger.
\newblock Chimia {\bf 59} 493 (2005)

\bibitem[NAI04]{Nag04}
T.~Nagata, M.~Aoyagi and S.~Iwata.
\newblock J. Phys. Chem. A {\bf 108} 683 (2004)

\bibitem[NBG00]{Niv00}
M.~Y. Niv, M.~Bargheer and R.~B. Gerber.
\newblock J. Chem. Phys. {\bf 113} 6660 (2000)

\bibitem[NEF00]{Nil00}
N.~Nilius, N.~Ernst and H.-J. Freund.
\newblock Phys. Rev. Lett. {\bf 84} 3994 (2000)

\bibitem[NHC{\etalchar{+}}02]{Nta02}
G.~E. Ntamack, B.~A. Huber, F.~Chandezon, M.~G.~K. Njock and C.~Guet.
\newblock J. Phys. B {\bf 35} 2729 (2002)

\bibitem[NIN{\etalchar{+}}04]{Ney04}
K.~Neyman, C.~Inntam, V.~Nasluzov, R.~Kosarev and N.~R\"osch.
\newblock Appl. Phys. A {\bf 78} 823 (2004)

\bibitem[NMF97]{Nac97}
B.~Nacer, C.~Massobrio and C.~F\'elix.
\newblock Phys. Rev. B {\bf 56}(16) 10590 (1997)

\bibitem[Nor87]{Nor87a}
J.~Northby.
\newblock J. Chem. Phys. {\bf 87} 6166 (1987)

\bibitem[NRG{\etalchar{+}}01a]{Nas01a}
A.~Nasluzov, V.~V. Rivanenkov, A.~B. Gordienko, K.~Neyman, U.~Birkenheuer and
  N.~R\"osch.
\newblock J. Chem. Phys. {\bf 115} 8157 (2001)

\bibitem[NRG{\etalchar{+}}01b]{Nas01}
V.~A. Nasluzov, V.~V. Rivanenkov, A.~B. Gordienko, K.~M. Neyman, U.~Birkenheuer
  and N.~R\"{o}sch.
\newblock J. Chem. Phys. {\bf 115}(17) 8157 (2001)

\bibitem[NYB02]{Nak02}
T.~Nakatsukasa, K.~Yabana and G.~F. Bertsch.
\newblock Phys. Rev. A {\bf 65} 032512 (2002)

\bibitem[OHHT05]{Oua05a}
H.~Ouacha, C.~Hendrich, F.~Hubenthal and F.~Tr\"ager.
\newblock Appl. Phys. B {\bf 81} 663 (2005)

\bibitem[Par75]{Par75a}
D.~E. Parry.
\newblock Surf. Sci. {\bf 49} 433 (1975)

\bibitem[Par76]{Par76a}
D.~E. Parry.
\newblock Surf. Sci. {\bf 54} 195 (1976)

\bibitem[PBBB99]{Pin99}
J.~Pinar\'e, B.~Baguenard, C.~Bordas and M.~Broyer.
\newblock Eur. Phys. J. D {\bf 9} 21 (1999)

\bibitem[PDMT03]{Par03a}
J.~S. Parker, B.~J.~S. Doherty, K.~J. Meharg and K.~T. Taylor.
\newblock J. Phys.B {\bf 36} (2003)

\bibitem[PGB05]{Pac05}
G.~Pacchioni, L.~Giordano and M.~Baistrocchi.
\newblock Phys. Rev. Lett. {\bf 94} 226104 (2005)

\bibitem[PGM{\etalchar{+}}00]{Per00a}
M.~Perner, S.~Gresillon, J.~M\"arz, G.~von PLessen, J.~Feldmann,
  J.~Porstendorfer, K.-J. Berg and G.~Berg.
\newblock Phys. Rev. Lett. {\bf 85} 792 (2000)

\bibitem[PH05]{Pav05a}
Y.~Pavlyukh and W.~H\"ubner.
\newblock Appl. Phys. A {\bf 82} (2005)

\bibitem[PInLA99]{Pal99}
F.~J. Palacios, M.~P. I\~niguez, M.~J. L\'opez and J.~A. Alonso.
\newblock Phys. Rev. B {\bf 60}(4) 2908 (1999)

\bibitem[PN66]{Pin66}
D.~Pines and P.~Nozi\`eres.
\newblock {\em {The Theory of Quantum Liquids}\/} (W A Benjamin, New York,
  1966)

\bibitem[Poh03]{Poh03a}
A.~Pohl.
\newblock {\em Der doppelt-differentielle Wirkungsquerschnitt f\"ur
  Photoionisation von Metallclustern\/}.
\newblock Ph.D. thesis, Friedrich-Alexander-Universit\"at, Erlangen/N\"urnberg
  (2003)

\bibitem[Pol64]{Pol64}
G.~L. Pollack.
\newblock Rev. Mod. Phys. {\bf 36} 748 (1964)

\bibitem[Pos01]{Pos01B}
J.~Posthumus, ed.
\newblock {\em {Molecules and Clusters in Intense Laser Fields}\/} (Cambrige
  University Press, Cambridge, 2001)

\bibitem[PPX{\etalchar{+}}03]{Pra03}
S.~Pratontep, P.~Preece, C.~Xirouchaki, R.~E. Palmer, C.~F. Sanz-Navarro, S.~D.
  Kenny and R.~Smith.
\newblock Phys. Rev. Lett. {\bf 90}(5) 055503 (2003)

\bibitem[PRS00]{Poh00}
A.~Pohl, P.-G. Reinhard and E.~Suraud.
\newblock Phys. Rev. Lett. {\bf 84} 5090 (2000)

\bibitem[PRS01]{Poh01}
A.~Pohl, P.-G. Reinhard and E.~Suraud.
\newblock J. Phys. B {\bf 34} 4969 (2001)

\bibitem[PRS04]{Poh04b}
A.~Pohl, P.-G. Reinhard and E.~Suraud.
\newblock Phys. Rev. A {\bf 70} 023202 (2004)

\bibitem[PSD{\etalchar{+}}01]{Por01a}
H.~Portales, L.~Saviot, E.~Duva, M.~Fujii, S.~Hayashil, N.~D. Fatti and
  F.~Vall\'ee.
\newblock J. Chem. Phys. {\bf 115} 3444 (2001)

\bibitem[PTVF92]{Pre92}
W.~H. Press, S.~A. Teukolsky, W.~T. Vetterling and B.~P. Flannery.
\newblock {\em {Numerical Recipes}\/} (Cambridge University Press, Cambridge,
  1992)

\bibitem[PW92]{Per92}
J.~P. Perdew and Y.~Wang.
\newblock Phys. Rev. B {\bf 45} 13244 (1992)

\bibitem[PZ81]{Per81}
J.~P. Perdew and A.~Zunger.
\newblock Phys. Rev. B {\bf 23} 5048 (1981)

\bibitem[RAC{\etalchar{+}}99]{Ray99}
D.~Rayner, K.~Athanassenas, B.~A. Collings, S.~Mitchell and P.~A. Hackett.
\newblock p. 371 (Springer, Berlin, 1999)

\bibitem[RBLS04]{Rho04}
M.~B. E.~H. Rhouma, H.~Berriche, Z.~B. Lakhdar and F.~Spiegelman.
\newblock Intern. J. Quant. Chem. {\bf 99} 495 (2004)

\bibitem[RBPL06]{Ric06}
D.~Ricci, A.~Bongiorno, G.~Pacchioni and U.~Landman.
\newblock Phys. Rev. Lett. {\bf 97}(3) 036106 (2006)

\bibitem[RCK{\etalchar{+}}99]{Rei99a}
P.-G. Reinhard, F.~Calvayrac, C.~Kohl, S.~K\"ummel, E.~Suraud, C.~A. Ullrich
  and M.~Brack.
\newblock Euro. Phys. J. D {\bf 9} 111 (1999)

\bibitem[RCS06]{Rho06a}
M.~B. E.~H. Rhouma, F.~Calvo and F.~Spiegelman.
\newblock J. Phys. Chem. A {\bf 110} 5010 (2006)

\bibitem[RESH97]{Rei97c}
T.~Reiners, C.~Ellert, M.~Schmidt and H.~Haberland.
\newblock Phys. Rev. Lett. {\bf 74} 1558 (1997)

\bibitem[RG06]{Roh06a}
S.~Rohra and A.~G\"orling.
\newblock Phys. Rev. Lett. {\bf 97} 013005 (2006)

\bibitem[RGB96]{Rei96b}
P.-G. Reinhard, O.~Genzken and M.~Brack.
\newblock Ann. Phys. (Leipzig) {\bf 5} 1 (1996)

\bibitem[RHG{\etalchar{+}}06]{Ric06a}
H.~H. Richardson, Z.~N. Hickman, A.~O. Govorov, A.~C. Thomas, W.~Zhang and
  M.~E. Kordesch.
\newblock Nano Lett. {\bf 6} (2006)

\bibitem[RLBS06]{Rho06b}
M.~B. E.~H. Rhouma, Z.~B. Lakhdar, H.~Berriche and F.~Spiegelman.
\newblock The Journal of Chemical Physics {\bf 125}(8) 084315 (2006)

\bibitem[RMNF06]{Ros06}
G.~Rossi, C.~Mottet, F.~Nita and R.~Ferrando.
\newblock J. Phys. Chem. B {\bf 110} 7436 (2006)

\bibitem[RMOP04]{Rep04}
J.~Repp, G.~Meyer, F.~E. Olsson and M.~Persson.
\newblock Science {\bf 305} 493 (2004)

\bibitem[RNN{\etalchar{+}}04]{Roe04aR}
N.~R\"osch, V.~A. Nasluzov, K.~M. Neyman, G.~Pacchioni and G.~N. Vayssilov.
\newblock In {\em Computational Material Science\/} (J.~Leszczynski, ed.),
  Theoretical and Computational Chemistry Series, Vol. 15, p. 367 (Elsevier,
  Amsterdam, 2004)

\bibitem[RS80]{Rin80}
P.~Ring and P.~Schuck.
\newblock {\em {The Nuclear Many-Body Problem}\/} (Springer, Berlin, 1980)

\bibitem[RS93]{Rub93}
A.~Rubio and L.~Serra.
\newblock Phys. Rev. B {\bf 48} 18222 (1993)

\bibitem[RS98]{Rei98b}
P.-G. Reinhard and E.~Suraud.
\newblock Euro. Phys. J. D {\bf 3} 175 (1998)

\bibitem[RS99]{Rei99c}
P.-G. Reinhard and E.~Suraud.
\newblock {\em Resonance dynamics in metal clusters and nuclei\/}, p. 211
  (Wiley, New York, 1999)

\bibitem[RS02]{Rei02b}
P.-G. Reinhard and E.~Suraud.
\newblock Euro. Phys. J. D {\bf 21} 315 (2002)

\bibitem[RS03]{Rei03a}
P.-G. Reinhard and E.~Suraud.
\newblock {\em Introduction to Cluster Dynamics\/} (Wiley, New York, 2003)

\bibitem[RS04]{Rei04c}
P.-G. Reinhard and E.~Suraud.
\newblock Encycl. Nanosc. Nanotechn. {\bf 2} 717 (2004)

\bibitem[RS06]{Rei06aR}
P.-G. Reinhard and E.~Suraud.
\newblock In {\em Time-dependent density functional theory\/} (M.~A.~L.
  Marques, C.~A. Ullrich and F.~Nogueira, eds.), vol. 706 of {\em Lecture Notes
  in Physics\/}, p. 391 (Springer, Berlin, 2006)

\bibitem[RSA{\etalchar{+}}06]{Rei06c}
P.-G. Reinhard, P.~D. Stevenson, D.~Almehed, J.~A. Maruhn and M.~R. Strayer.
\newblock Phys. Rev. E {\bf 73} 036709 (2006)

\bibitem[RTLA91]{Ric91}
J.~E. Rice, P.~R. Tylor, T.~J. Lee and J.~Alml\"of.
\newblock J. Chem. Phys. {\bf 94} 4972 (1991)

\bibitem[SAH{\etalchar{+}}99]{San99}
A.~Sanchez, S.~Abbet, U.~Heiz, W.-D. Schneider, H.~H\"akkinen, R.~N. Barnett
  and U.~Landman.
\newblock J. Phys. Chem. {\bf 103} 9573 (1999)

\bibitem[SBF{\etalchar{+}}04]{Sok04a}
A.~A. Sokol, S.~T. Bromley, S.~A. French, C.~R.~A. Catlow and P.~Sherwood.
\newblock Int. J. Quant. Chem. {\bf 99} 695 (2004)

\bibitem[Sch92]{Sch92aB}
M.~Schrader.
\newblock {\em {Modern Approaches to Wattability}\/} (Plenum Press, New York,
  1992)

\bibitem[Sch00]{Sch00}
D.~Schwarzhans.
\newblock {\em Hochaufl\"osende Laserspektroskopie der elektronischen
  Zust\"ande $X^2\Sigma, A^2\Pi$ und $B^2\Sigma$ von Natrium-Argon\/}.
\newblock Ph.D. thesis, TU Berlin (2000)

\bibitem[SDP{\etalchar{+}}07]{Sim07a}
D.~Simberga, T.~Duzaa, J.~H. Park, M.~Esslera, J.~Pilcha, L.~Zhanga, A.~M.
  Derfus, M.~Yang, R.~M. Hoffman, S.~Bhatiai, M.~J. Sailor and E.~Ruoslahti.
\newblock Proc. Nat. Acad. Sci. USA {\bf 104} 932 (2007)

\bibitem[SFY05]{Sch05a}
D.~M. Schaadt, B.~Feng and E.~T. Yu.
\newblock Appl. Phys. Lett. {\bf 86} 063106 (2005)

\bibitem[SJG{\etalchar{+}}00]{Sny00}
J.~A. Snyder, J.~E. Jaffe, M.~Gutowski, Z.~Lin and A.~C. Hess.
\newblock J. Chem. Phys. {\bf 112}(6) 3014 (2000)

\bibitem[SKBG00]{Sei00}
G.~Seifert, M.~Kaempfe, K.-J. Berg and H.~Graener.
\newblock Appl. Phys. B {\bf 71} 795 (2000)

\bibitem[SKvIH01]{Sch01}
R.~Schlipper, R.~Kusche, B.~von Issendorff and H.~Haberland.
\newblock Appl. Phys. A {\bf 72} 255 (2001)

\bibitem[SL06]{Sti06}
F.~Stienkemeier and K.~K. Lehmann.
\newblock J. Phys. B {\bf 39}(8) R127 (2006)

\bibitem[SL08]{Kue08}
K.~S. and K.~L.
\newblock Rev. Mod. Phys. {\bf 80} 3 (2008)

\bibitem[SOL77]{Sax77}
R.~P. Saxon, R.~E. Olson and B.~Liu.
\newblock J. Chem. Phys. {\bf 67} 2692 (1977)

\bibitem[SPP{\etalchar{+}}08]{Skr08}
S.~Skruszewicz, J.~Passig, A.~Przystawik, J.~Tiggesb\"aumker and K.-H.
  Meiwes-Broer {\bf to be published} (2008)

\bibitem[SRH{\etalchar{+}}07]{Ste07b}
M.~Sterrer, T.~Risse, M.~Heyde, H.-P. Rust and H.-J. Freund.
\newblock Phys. Rev. Lett. {\bf 98} 206103 (2007)

\bibitem[SRP{\etalchar{+}}07]{Ste07a}
M.~Sterrer, T.~Risse, U.~M. Pozzoni, L.~Giordano, M.~Heyde, H.-P. Rust,
  G.~Pacchioni and H.-J. Freund.
\newblock Phys. Rev. Lett. {\bf 98} 096107 (2007)

\bibitem[SSL07]{Sch07a}
H.~Schlegel, S.~Smith and X.~Li.
\newblock J. Chem. Phys. {\bf 126} (2007)

\bibitem[SSR06]{Saa06aR}
U.~Saalmann, C.~Siedschlag and J.~M. Rost.
\newblock J. Phys. B {\bf 39} R39 (2006)

\bibitem[Sug98]{Sug98}
S.~Sugano.
\newblock {\em {Microcluster Physics}\/} (Springer, Berlin, 1998)

\bibitem[SV01]{Sti01}
F.~Stienkemeier and A.~F. Vilesov.
\newblock J. Chem. Phys. {\bf 115} 10119 (2001)

\bibitem[SZ03]{Sch03}
D.~Schwarzhans and D.~Zimmermann.
\newblock Eur. Phys. J. D {\bf 22} 193 (2003)

\bibitem[Sza85]{Sza85}
L.~Szasz.
\newblock {\em {Pseudopotential Theory of Atoms and Molecules}\/} (Wiley, New
  York, 1985)

\bibitem[TES90]{Tso90}
C.~Tsoo, D.~A. Estrin and S.~J. Singer.
\newblock J. Chem. Phys. {\bf 93} 7187 (1990)

\bibitem[TES92]{Tso92}
C.~Tsoo, D.~A. Estrin and S.~J. Singer.
\newblock J. Chem. Phys. {\bf 96} 7977 (1992)

\bibitem[TM98]{Tut98}
A.~B. Tutein and H.~R. Mayne.
\newblock J. Chem. Phys. {\bf 108} 308 (1998)

\bibitem[TSG97]{Tim97}
D.~Timpel, K.~Scheerschmidt and S.~H. Garofalini.
\newblock J. Non-Cryst. Solids {\bf 221} 187 (1997)

\bibitem[TSKM01]{Tak01}
S.~Takami, K.~Suzuki, M.~Kubo and A.~Miyamoto.
\newblock J. Nanoparticle Res. {\bf 3} 213 (2001)

\bibitem[TVS90]{Tje90a}
L.~H. Tjeng, A.~R. Vos and G.~A. Sawatzky.
\newblock Surf. Sci. {\bf 235} 269 (1990)

\bibitem[UG97]{Ull97a}
C.~A. Ullrich and E.~K.~U. Gross.
\newblock Comm. At. Mol. Phys. {\bf 33} 211 (1997)

\bibitem[Ull00]{Ull00b}
C.~A. Ullrich.
\newblock J. Mol. Struct. (THEOCHEM) {\bf 501-502} 315 (2000)

\bibitem[VBK{\etalchar{+}}04]{Ver04}
J.~R.~R. Verlet, A.~E. Bragg, A.~Kammrath, O.~C. y and D.~M. Neumark.
\newblock J. Chem. Phys. {\bf 121} 10015 (2004)

\bibitem[VCN{\etalchar{+}}96]{Vor96a}
V.~Vorsa, P.~Campagnola, S.~Nandi, M.~Larsson and W.~Lineberger.
\newblock J. Chem. Phys. {\bf 105} 2298 (1996)

\bibitem[Ver67]{Ver67}
L.~Verlet.
\newblock Phys. Rev. {\bf 159} 98 (1967)

\bibitem[VFF98]{Vel98b}
M.~Velegrakis, G.~E. Froudakis and S.~C. Farantos.
\newblock J. Chem. Phys. {\bf 109} 4687 (1998)

\bibitem[VFM96]{Van96}
G.~Vandoni, C.~F\'elix and C.~Massobrio.
\newblock Phys. Rev. B {\bf 54}(3) 1553 (1996)

\bibitem[vGKS{\etalchar{+}}98]{Gis98}
A.~J.~A. van Gisbergen, F.~Kootstra, P.~R.~T. Schipper, O.~V. Gritsenko, J.~G.
  Snijders and E.~J. Baerends.
\newblock Phys. Rev. A {\bf 57} 2556 (1998)

\bibitem[VLS{\etalchar{+}}03]{Vie03}
L.~A. Viehland, J.~Lozeille, P.~Soldan, E.~P.~F. Lee and T.~G. Wright.
\newblock J. Chem. Phys. {\bf 119} 3729 (2003)

\bibitem[VNC{\etalchar{+}}97]{Vor97a}
V.~Vorsa, S.~Nandi, P.~Campagnola, M.~Larsson and W.~Lineberger.
\newblock J. Chem. Phys. {\bf 106} 1402 (1997)

\bibitem[Wah12]{Wah12}
W.~Wahl.
\newblock Proc. Roy. Soc. (London) ~A {\bf 87} 371 (1912)

\bibitem[WBGT99]{Wen99a}
T.~Wenzel, J.~Bosbach, A.~Goldmann and F.~Tr\"ager.
\newblock Appl. Phys. B {\bf 69} 513 (1999)

\bibitem[Wei37]{Wei37}
V.~Weisskopf.
\newblock Phys. Rev. {\bf 52} 295 (1937)

\bibitem[WHvI02]{Wri02}
G.~Wrigge, M.~A. Hoffmann and B.~von Issendorff.
\newblock Phys. Rev. A {\bf 65} 063201 (2002)

\bibitem[Win06]{Win06}
M.~Winkler.
\newblock {\em Density functional studies of small metal species supported on
  magnesium oxide\/}.
\newblock Diploma Thesis, Technische Universit\"at M\"unchen (2006)

\bibitem[WL76]{War76a}
A.~Warshel and M.~Levitt.
\newblock J. Mol. Biol. {\bf 103} 227 (1976)

\bibitem[XH08]{Xu08}
L.~Xu and G.~Henkelman.
\newblock Phys. Rev. B {\bf 77} 205404 (2008)

\bibitem[XHCJ05]{Xu05}
L.~Xu, G.~Henkelman, C.~T. Campbell and H.~J\'{o}nsson.
\newblock Phys. Rev. Lett. {\bf 95}(14) 146103 (2005)

\bibitem[XP01]{Xir01}
C.~Xirouchaki and R.~E. Palmer.
\newblock Vaccuum {\bf 66} 167 (2001)

\bibitem[XP02]{Xir02}
C.~Xirouchaki and R.~E. Palmer.
\newblock Vacuum {\bf 66} 167 (2002)

\bibitem[YB96]{Yab96}
K.~Yabana and G.~F. Bertsch.
\newblock Phys. Rev. B {\bf 54} 4484 (1996)

\bibitem[YHL{\etalchar{+}}05]{Yoo05}
B.~Yoon, H.~H\"akkinen, U.~Landman, A.~S. W\"orz, J.-M. Antonietti, S.~Abbet,
  K.~Judai and U.~Heiz.
\newblock Science {\bf 307} 403 (2005)

\bibitem[YPB90]{Yan90}
C.~Yannouleas, J.~M. Pacheco and R.~A. Broglia.
\newblock Phys. Rev. B {\bf 41} 41 (1990)

\bibitem[YPNR97]{Yud97}
I.~Yudanov, G.~Pacchioni, K.~Neyman and N.~Rosch.
\newblock J. Phys. Chem. B {\bf 101}(15) 2786 (1997)

\bibitem[YWZG02]{Yan02b}
Z.~Yang, R.~Wu, Q.~Zhang and D.~W. Goodman.
\newblock Phys. Rev. B {\bf 65} 155407 (2002)

\bibitem[Zew94]{Zew94B}
A.~H. Zewail.
\newblock {\em {Femtochemistry:Ultrafast Dynamics Of The Chemical Bond, Vol.
  1}\/} (World Scientific, Singapore, 1994)

\bibitem[Zew00]{Zew00a}
A.~H. Zewail.
\newblock J. Phys. Chem. A {\bf 104} 5660 (2000)

\bibitem[ZK03]{Zav03}
V.~G. Zavodinsky and A.~Kiejna.
\newblock Surf. Sci. {\bf 538} 240 (2003)

\bibitem[ZKK05]{Zav05}
V.~Zavodinsky, M.~Kuz'menko and A.~Kiejna.
\newblock Surf. Sci. {\bf 589} 114 (2005)

\bibitem[ZTR99]{Zwi99}
G.~Zwicknagel, C.~Toepffer and P.-G. Reinhard.
\newblock Phys. Rep. {\bf 309} 117 (1999)

\bibitem[ZZH{\etalchar{+}}09]{Zha09}
P.-X. Zhang, Y.-F. Zhao, F.-Y. Hao, X.-D. Song, G.-H. Zhang and Y.~Wang.
\newblock J. Mol. Struct. (THEOCHEM) {\bf 899} 111 (2009)

\end{thebibliography}

\end{document}